\documentclass[fleqn,usenatbib]{mnras}
\usepackage[T1]{fontenc}
\usepackage{newtxtext,newtxmath}
\DeclareRobustCommand{\VAN}[3]{#2}
\let\VANthebibliography\thebibliography
\def\thebibliography{\DeclareRobustCommand{\VAN}[3]{##3}\VANthebibliography}
\usepackage{subcaption}
\usepackage{graphicx}	% Including figure files
\usepackage{amsmath}	% Advanced maths commands
\usepackage{breqn}
\usepackage[section]{placeins}
\usepackage{float}
\restylefloat{table}
\makeatletter
\renewcommand{\fnum@table}{\textbf{\tablename~\thetable}}
\makeatother
\usepackage{longtable}
\usepackage{rotating,tabularx}
\usepackage{multirow}
\usepackage[dvipsnames]{xcolor}

\newcommand{\aqlx}{{Aql\,X--1}}
\newcommand{\Msun}{{M$_{\odot}$}}
\newcommand{\Rsun}{{R$_{\odot}$}}
\newcommand{\kmps}{{km\,s$^{-1}$}}

\newcommand{\vsys}{\upsilon_{\rm sys}}
\newcommand{\vpec}{\upsilon_{\rm pec}}

\newcommand{\vorb}{\upsilon_{\rm rel}}

\newcommand{\vnk}{\upsilon_{\rm NK}}

\newcommand{\vmlk}{\upsilon_{\rm MLK}}
\newcommand{\thetank}{\theta_{\rm NK}}
\newcommand{\phink}{\phi_{\rm NK}}

\newcommand{\vnkconf}{\upsilon_{\rm NK}^{68\%}}

\newcommand{\cphi}{\phi_{\mathrm{c}}}
\newcommand{\stheta}{\theta_{\mathrm{s}}}

\newcommand{\Mc}{M_{\mathrm{CO}}}

\newcommand{\mtwo}{M_{\rm 2}}
\newcommand{\porb}{P_{\rm orb}}
\newcommand{\monei}{$M_{\rm 1, i}$}
\newcommand{\mtwoi}{$M_{\rm 2, i}$}
\newcommand{\porbi}{$P_{\rm orb, i}$}
\newcommand{\sepi}{$r_{\rm orb, i}$}
\newcommand{\ei}{$e_{\rm i}$}
\newcommand{\monepre}{$M_{\rm 1, SN}$}
\newcommand{\mtwopre}{$M_{\rm 2, SN}$}
\newcommand{\porbpre}{$P_{\rm orb, SN}$}
\newcommand{\seppre}{$r_{\rm orb, SN}$}
\newcommand{\epre}{$e_{\rm SN}$}
\newcommand{\monepost}{$M_{\rm 1, post-SN}$}
\newcommand{\mtwopost}{$M_{\rm 2, post-SN}$}
\newcommand{\porbpost}{$P_{\rm orb, post-SN}$}
\newcommand{\seppost}{$r_{\rm orb, post-SN}$}
\newcommand{\epost}{$e_{\rm post-SN}$}
\newcommand{\moneobs}{$M_{\rm 1, obs}$}
\newcommand{\mtwoobs}{$M_{\rm 2, obs}$}
\newcommand{\porbobs}{$P_{\rm orb, obs}$}
\newcommand{\eobs}{$e_{\rm obs}$}
\newcommand{\mdot}{$\dot{M}$}

\newcommand{\fouru}{{4U\,1700--37}}

\newcommand{\gaia}{{\em Gaia}}
\newcommand{\cosmic}{{\tt COSMIC}}

\defcitealias{Zhao2023}{Z23}
\defcitealias{dashwoodbrown2026i}{DB26}

\title[Natal Kicks of Galactic Compact-Object Binaries]{Natal Kicks of Compact Binaries:\\ II. Neutron-star and black-hole systems share similar kick distributions}

\author[C. Dashwood Brown et al.]{Cordelia Dashwood Brown,$^{1}$\thanks{E-mail: c.dashwood.brown@soton.ac.uk},
Poshak Gandhi,$^{1}$
Christian Knigge,$^{1}$
Yue Zhao,$^{1}$
Phil Charles,$^{1,2,3}$
\newauthor
Katelyn Breivik,$^4$
Pornisara Nuchvanichakul$^{1}$
and Siddharth Malu$^6$
\\
$^{1}$School of Physics \& Astronomy, University of Southampton, Highfield, Southampton SO17 1BJ, UK\\
$^{2}$Astrophysics, Department of Physics, University of Oxford, Keble Road, Oxford OX1 3RH, UK\\
$^{3}$Department of Physics, University of the Free State, PO Box 339, Bloemfontein 9300, South Africa\\
$^{4}$McWilliams Center for Cosmology \& Astrophysics, Department of Physics Carnegie Mellon University, Pittsburgh, PA 15213, USA\\
$^{5}$Department of Physics, University of Oxford, Keble Road, Oxford OX1 3RH, UK\\
$^{6}$AI Center of Excellence, Shiv Nadar Institution of Eminence, NCR Delhi, India}

\date{Received 2026 Nov 05; in original form 2026 Aug 20.}
\pubyear{2026}

\begin{document}
\label{firstpage}
\pagerange{\pageref{firstpage}--\pageref{lastpage}}
\maketitle

% Abstract of the paper
\begin{abstract}
Natal kicks --- kinematic impulses to neutron stars and black holes at birth --- play a decisive role in the end-stage evolution of binaries hosting massive stars, in determining the global phase-space properties of compact objects, and in setting expected merger-rates detectable through gravitational-wave observations. Inferring the kick ($\vnk$) for individual systems requires well constrained binary system properties and centre-of-mass kinematics, which must then be modelled to constrain likely system characteristics at supernova. As a result, population-wide studies of kicks remain few and far between. Here, we model plausible evolutionary histories for 68 accreting binaries, encompassing almost all black-hole and neutron-star X-ray binaries with kinematically useful 3D measurements, constraining allowed $\vnk$ distributions for each source using a simulation framework presented in \citeauthor{dashwoodbrown2026i} (Paper I). The resultant $\vnk$ distribution for the joint sample can be simply represented by a broad Gamma distribution with mean $\vnk$\,=\,$128^{+17}_{-15}$\,\kmps\ and scatter $\sigma$\,=\,117\,\kmps. Intriguingly, the distributions for black-hole vs. neutron-star systems are found to be formally indistinguishable. Despite still being limited by small-number statistics, we find that black holes can receive large natal kicks, and that kicks for all remnant types span similar ranges ($\sim$\,0--700\,km\,s$^{-1}$). There is a hint, but no statistically significant evidence at present, for $\vnk$ scaling inversely with black-hole mass. Our work suggests either that similar physics drives kicks and shows evidence for effects beyond simplistic fallback-moderated scaling for black holes, or that selection effects conspire to erase differences between remnant types in current accreting-binary samples. Larger samples are needed to provide robust discriminatory tests between these possibilities. 
\end{abstract}

% Select between one and six entries from the list of approved keywords.
% Don't make up new ones.

%%%%%%%%%%%%%%%%%%%%%%%%%%%%%%%%%%%%%%%%%%%%%%%%%%

\section{Introduction}
\label{Section:Intro}
Determining natal nicks --- recoil impulses to black holes and neutron stars at birth --- is a key outstanding problem in compact-object formation and evolution. Both the magnitude ($\vnk$) and the direction of the kick influence the compact remnant's subsequent evolution, particularly in the case of compact objects formed by large stars in binary systems. Natal kicks may eject the compact object \& stellar companion from their stellar clusters \citep{Pfahl2002a, Strader2012, Giesler2018}, or otherwise thrust the compact object into atypical Galaxy orbits \citep{Repetto2015,Gandhi2020, Jonker2021}, which has implications for the observability of such systems (given that extinction is a function of Galactic location, and black holes are best studied using observations of a stellar companion). A strong or sufficiently fine-tuned kick can unbind the binary, producing isolated neutron stars and black holes, together with a population of fast {\em runaway} ($\vpec \geq 30$\,\kmps) or slower {\em walk-away} ($\vpec \leq 10$\,\kmps) companions e.g. \citep{Nelemans1999,Renzo2019}, where $\vpec$ is a system's peculiar velocity (its motion in excess of that induced by the Galactic gravitational potential). Isolated black holes can essentially only be detected through microlensing, and constraining their demographics and mass function is an emerging frontier \citep{Wyrzykowski2016,Mroz2021,Sahu2022,Lam2022,Koshimoto2024}. A kick can also do the opposite: tighten an orbit and drive the components towards merger, altering compact-binary merger rates \citep{Belczynski1999,Giacobbo2018,Belczynski2020}. In surviving binaries, the kick changes the orbital separation and eccentricity as well as the systemic velocity \citep{Brandt1995a,Kalogera1996,Nelemans1999,tauris98,Gandhi2020}, and can introduce spin--orbit misalignments \citep{Callister2021}. This determines whether and when mass transfer occurs, and how likely the system is to be observed.

Instantaneous mass loss during a supernova changes the orbital angular momentum of the entire system. This alone can substantially alter the post-supernova period and eccentricity and give the surviving binary's centre-of-mass a recoil  \citep{Blaauw1961, Nelemans1999}. This {\em Blaauw} or mass-loss kick was first proposed to explain massive runaway stars, and it operates by carrying away the instantaneous orbital momentum of the ejected matter, even when material is ejected  perfectly spherically from the compact-object progenitor. Any additional impulse requires asymmetry, though the exact origin and nature of this remains the subject of much discussion. Candidate mechanisms include: hydrodynamic momentum transfer at core bounce \citep{Burrows1996,Lai2001};
anisotropic neutrino emission from the cooling proto-neutron star \citep{Tamborra2014,Janka2024};
asymmetric ejecta seeded by neutrino-driven convection and the standing accretion-shock instability \citep{Dorofeev1985, Blondin2003,Wongwathanarat2013}; and large-scale magnetic fields threading rapidly rotating collapsing cores \citep{li2026}. Inhomogeneities may be present in the progenitor's density structure, in the neutrino-heated region behind the shock, and in the shock itself. Rayleigh–Taylor instabilities, for example, are a well-established feature of supernova hydrodynamics. These perturbations grow with time, so a slower onset of explosion tends to produce larger asymmetries and therefore larger kicks. Natal kicks thus offer a direct probe of how the supernova engine operates throughout the volume of the exploding star.

There is no agreed first-principles theory of natal kicks, so observations provide the key constraints. Neutron-star (NS) kicks are the best studied so far. Decades of radio astrometry and pulsar timing have found mean peculiar velocities of several hundred \kmps\ for isolated pulsars \citep{Lai2001, Hobbs2005, Disberg2025}. Black holes (BH) are much harder to study, because they can usually only be detected when bound to a luminous companion whose motion traces that of the system \citep{Charles2006}. Very-long-baseline interferometry has constrained the kinematics of a handful of accreting black holes \citep{Mirabel01,Dhawan2007,Miller-Jones2009,Reid2011,Atri2020}. Steady gains in optical astrometry over the years, culminating in the \gaia\ mission \citep{Gaia2016}, now provide proper motions and parallaxes, and hence peculiar velocities, for a growing sample of X-ray binaries (XRBs) \citep{chevalier1998,mirabel02, Mirabel2017, Gandhi2019, atri2019, Arnason2021, Fortin2022, Zhao2023, Nuchvanichakul25, disberg2026}. A new population of non-accreting compact-object binaries is also emerging, identified from the astrometric wobble of the companion about the centre-of-mass by leveraging {\em Gaia}'s astrometric precision. \citep{ElBadry23a, chakrabarti2023, ElBadry23b, elbadry24, Gaia2024, gaians2}. The study of such systems will leap forward with improved binary-star processing and data-quality thresholding in upcoming {\em Gaia} data releases \citep{Halbwachs2023,Gaia2024}. 

Studies show that at least some compact objects have $\vpec$ of several hundred \kmps. However, $\vpec$ is not a direct measure of the natal kick. The companion's inertia and changes in orbital angular momentum at the supernova both moderate any impulse, so a given kick can produce a widely scattered range of systemic velocities. Recovering the kick requires modelling of the source characteristics as they may have been manifest at the time of compact-object formation. This is non-trivial, but studies of current-day binaries present a major advantage in this regard -- the physical state of the binary is imprinted with the effects of the kick, which it carries for the rest of its life and which can potentially be disentangled with careful modelling. Based on such studies investigating detailed plausible evolutionary pathways for XRBs, there is now increasing evidence that at least some black holes likely receive strong kicks at birth, over and above the baseline impulse associated with mass-loss in a binary system \citep[e.g.][]{Willems2005, Fragos2009, Kimball2023, DashwoodBrown2023}. This builds on proxy inferences based on current kinematic constraints \citep[][and references therein]{Mirabel2017, atri2019, Gandhi2019, Zhao2023, Nagarajan2025}, as well as similar indirect results based on many statistical studies of the spatial and parameter-space scatter of Galactic XRBs \citep{Jonker2004, Repetto2015, Gandhi2020, Wagg2025}. The assumed natal-origin scenario here is of an isolated-binary that was born somewhere in, or near, the Galactic thin disc. An origin in stellar streams \citep{Gaia2024} and in globular clusters \citep{Strader2012} is known for some, though it remains less clear whether a significant proportion of the current population of known compact-object binaries originated outside the disc. 

Whilst the notion of strong natal kicks being imparted to neutron stars is uncontroversial, it is widely thought that black-hole natal kicks ought to scale down with fallback factor $f_{\rm fb}$. This denotes the fraction of progenitor material that is initially ejected away during SN explosion, but which ultimately fails to escape and falls back on to the proto-neutron star core, boosting the compact-object mass while also moderating its asymptotic kick. In the most widely adopted fallback-moderated scalings, substantial kicks are not expected for progenitor core mass $M_{\rm CO}$\,$\gtrsim$\,7\,\Msun\ \citep{Fryer2012}. Should it transpire that such systems have received significant natal kicks, the description of fallback and compact-object formation would require revision, with consequences spanning binary evolution, gravitational-wave source predictions, and supernova physics. Investigating this issue is the primary motivation of the present work. 

In a series of studies, our team is endeavouring to better understand the natal kicks impacting X-ray binary systems. While our previous works have focused on smaller samples of objects (e.g., first \gaia\ inferences on black-hole X-ray binaries [\citealt{Gandhi2019}]; testing donor-classification-based segregation of kicks in high-mass X-ray binaries [\citealt{Nuchvanichakul25}]; investigating kicks in individually [\citealt{Rao2020, DashwoodBrown2023}], and also statistically [\citealt{Gandhi2020}]), the present work presents comprehensive modelling of potential natal kicks impacting the full sample of X-ray binaries with kinematically useful constraints currently known. Our efforts are underpinned by progressive improvements in astrometric measurements (cited above and culminating in our published sample of \gaia\ DR3 [\citealt{Gaia2023}] kinematic constraints on X-ray binaries [\citealt{Zhao2023}; hereafter Z23]), coupled with improved codes for rapid binary population synthesis (BPS) (e.g., \cosmic; \citealt{cosmic}) to investigate the influence of natal kicks on binary evolution pathways. 

In \citet[][formally referred to as `Paper~I' hereafter]{dashwoodbrown2026i}, we introduced the core binary synthesis methodology, its benefits, limitations, and applications on a few testbed systems chosen to showcase the framework and explore evolutionary pathways in depth. In this work (Paper~II), we expand our modelling to the full \citetalias{Zhao2023} sample. Encompassing 66 kinematically constrained systems, this study is the largest of its kind and the first to apply the same population synthesis methodology to neutron stars as well as black holes. We present the overall $\vnk$\ distribution of compact objects in accreting binaries, and conduct statistical comparisons of the distributions of natal kicks depending on compact-object type, before discussing implications for explosion models. Appendix \ref{Section:App_Terminology} summarises the notations and terminology used in this study.

\section{Methodology}
\label{Section:Methods}

Our work follows the methods detailed in Paper I; salient points are outlined here to aid the reader. 

\citetalias{Zhao2023} computed present-day peculiar velocities ($\vpec$) following the formulation in \citet{Reid2009}, and these are used as conservative estimates of the resultant perturbation to the binary motion, following supernova. We then simulate a wide range of possible progenitor binaries within the \cosmic\ code \citep{cosmic}, which is based on BPS algorithms. The main advantage of this workflow is the capability to simulate very large ensembles of potential progenitors and evolve them through all key stages of stellar and binary lifecycles in a tractable computational time. The major disadvantage is the loss of fidelity, with physical processes and evolutionary schemes being based on pre-determined scaling laws, either empirical or based on theory. These issues are discussed at length in Paper I. 

Individual binaries are subject to natal kicks with magnitudes 0-1000\,\kmps, in steps of 1\,\kmps. Natal kicks are assumed to be isotropic, and are sampled uniformly in solid angle, with a natal kick of a given magnitude assigned 50 distinct directions. Therefore, every simulated binary is subjected to 50,000 distinct kicks. These priors are independent of any physical mechanisms driving the kicks, and are unrelated to system parameters such as remnant mass. Whilst the choice of priors will always have some impact on the resultant posterior distribution, opting for the same natal kicks for all systems allows for a fairer comparison of black hole and neutron star populations.

When matching to the observed peculiar velocity, we subtract $\sigma_{\rm dis}$\,=\,40\,\kmps in quadrature, accounting for any intrinsic Galactic dispersion velocity and ensuring random scatter in stellar velocities not associated with natal kicks is accounted for (for example, the velocity of systems that formed within stellar clusters may be partially attributed to the peculiar velocity associated with those clusters, as is seemingly the case for Cyg X–1 \citep{Rao2020}). Thus, the natal kicks presented here are, in general, lower limits - for example, the minimum natal kick Cyg X–1 could have received is calculated as 36\,\kmps when considering its nominal peculiar velocity is $\sim20-30$\,\kmps, but the system could have formed without a natal kick if its systemic velocity is due to an intrinsic velocity dispersion of objects within the Galactic disc. In practice, the stellar dispersion velocity is a function of Galactic location. Where the velocity dispersion of the Galactic disc is $\sim$40\,\kmps\ \citep{Carlberg1985}, the dispersion in the Galactic halo is larger: 50–120\,\kmps\ \citep{Battaglia2005,Brown2010}. When examining evidence for natal kicks in a sample of black holes, \citet{Nagarajan2025} use the local stellar dispersion velocity for each system, quoting values from 40-125\,\kmps. Here, the Galactic dispersion velocity is treated as the same for each system. This choice is justified given that the primary focus here is on the comparison between the kicks applied to black holes vs neutron stars - applying the same corrections to each system makes for a more meaningful direct comparison, however, it must be acknowledged that the natal kick estimates for individual systems are subject to error based on this assumption. Within this sample, there is no obvious difference in Galactic location or scale height between the neutron star and black hole populations, therefore, consistency between each system, whilst technically erroneous, is arguably more appropriate for the nature of this study. Additionally, whilst 40\,\kmps describes the velocity distribution of late-type stars, this may be an over-estimate for the velocity dispersion for compact objects, which may have been slowed due to interactions with other celestial bodies and the interstellar medium.

We simulated the evolutionary histories of all objects in \citetalias{Zhao2023}, spanning black-hole as well as neutron-star systems, with high- or low-mass donor stars. The neutron-star systems also include pulsars, which need not necessarily be undergoing efficient disc-mediated accretion at present --- many are wind-accretors --- or could be subject to mode-switching behaviour as seen in sources like PSR\,J1023+0038. We also note that there remains some ambiguiuty in classification of a handful of sources, e.g. SS\,433; and in one case, for \aqlx, there is fresh doubt whether \gaia\ detects the correct counterpart. We checked that such uncertainties did not impact our core inferences below, and details are discussed in the appendices.

The sample is not {\em complete} in any inherent sense of physical parameter space coverage, but it is representative of the current kinematically-constrained census of accreting binaries --- i.e. those with optical or radio astrometrically measured proper motions, centre-of-mass spectroscopic radial velocities, and a published estimate of the source distance. Biases inherent to the sample selection are outlined in Paper I. For the purposes of this paper, the most important question relates to any biases impacting one class of compact object relative to another; further discussion on this can be found in Section\,\ref{Section:Biases} below.

\section{Results}\label{Section:Results}

We are able to successfully model the evolution of 66 binaries hosting compact objects so that their simulated characteristics match present-day source properties, allowing the distributions of their natal kicks to be inferred. The final sample comprises 18 black holes in X-ray binaries (15 BH-LMXB \& 3 BH-HMXB), 33 neutron stars in X-ray binaries (15 NS-LMXB \& 18 NS-HMXB), and 15 neutron star pulsars (14 NS-LM-PSR \& 1 NS-HM-PSR); detailed source properties and classifications can be found in \citetalias{Zhao2023} - we consider systems with stellar companions greater than 2\,\Msun\  to be `high-mass' (though we acknowledge the split between `high'- and `low'-mass X-ray binaries is somewhat subjective, and one may argue for a third classification of `intermediate' mass companions \cite{Pfahl2003}). 

\begin{figure*}
    \centering
    \includegraphics[width=0.9\linewidth]{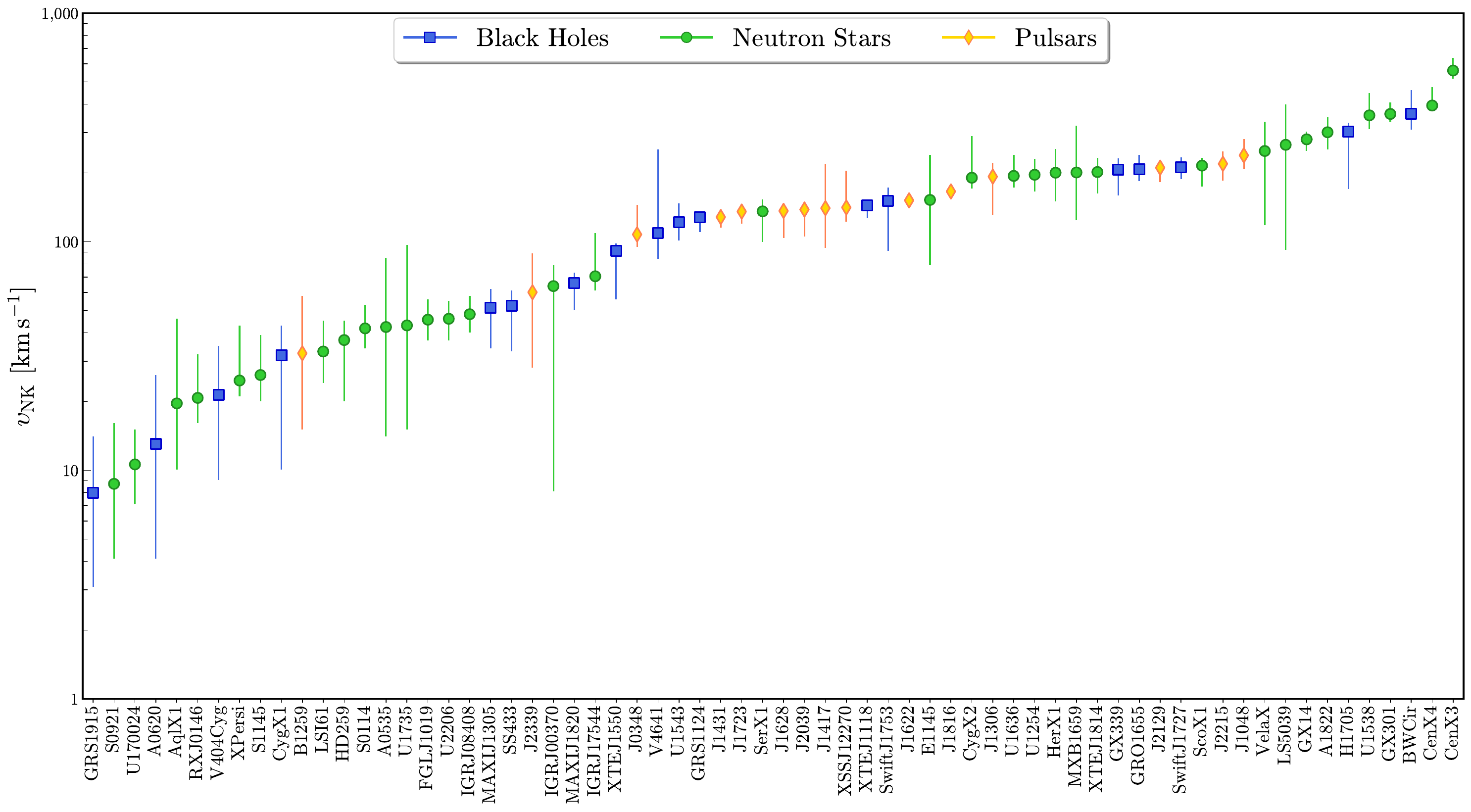}
    \caption{Natal kick magnitudes for all systems in this study. Points indicate the MAP value; errorbars indicate the 68\% HDI confidence intervals; thin lines indicate minimum and maximum allowed values. System types are distinguished by colour \& marker type: blue squares for black holes; green circles for neutron stars; and orange diamonds for pulsars.}
    \label{Figure:NKRanges}
\end{figure*}

\begin{figure*}
    \centering
    \includegraphics[width=0.9\textwidth]{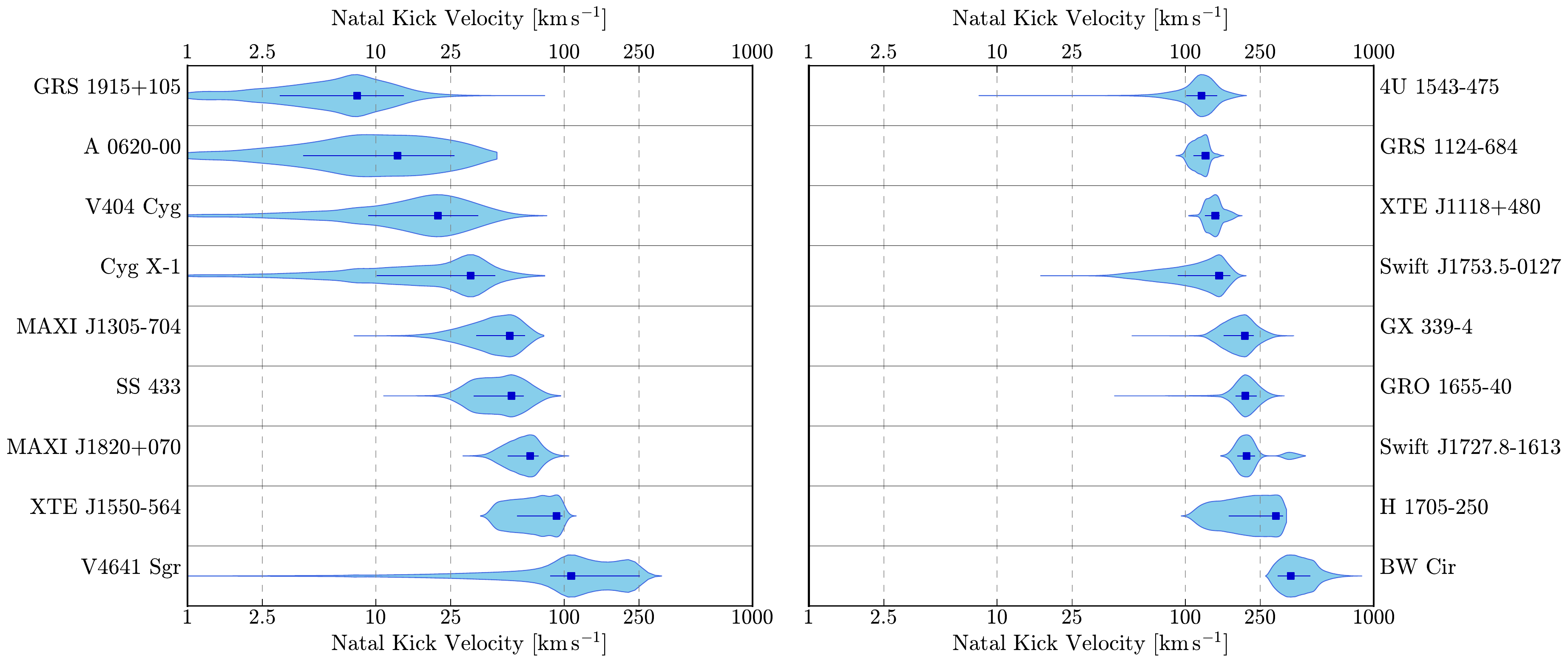}
    \caption{Violin plots showing distribution of all viable natal kick $\vnk$ magnitudes for each black hole system in this study. Maximum aposteriori probability (MAP) values are indicated by small square markers; internal horizontal lines indicate the 68\% highest-density probability intervals (HDI) of the distribution.}
    \label{Figure:NKViolinsA}
\end{figure*}

\begin{figure*}
    \centering
    \includegraphics[width=0.9\textwidth]{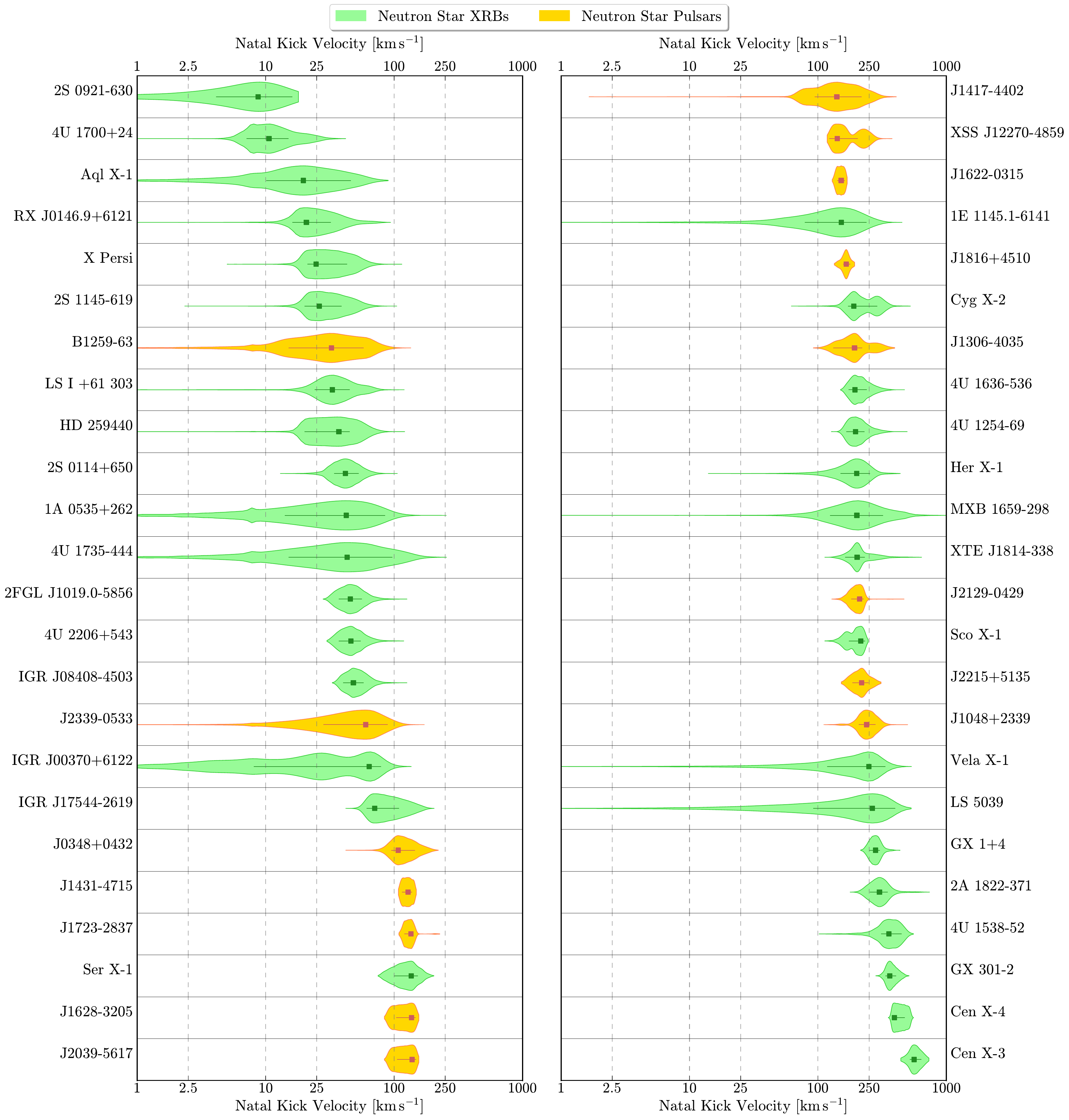}
    \caption{Violin plots showing distribution of all viable natal kick magnitudes for each neutron star system in this study. Maximum aposteriori probability (MAP) values are indicated by small square markers; internal horizontal lines indicate the 68\% highest-density probability intervals (HDI) of the distribution. Systems types are distinguished by their colour: green for neutron stars in X-ray binaries and orange for neutron stars in pulsars binaries.}
    \label{Figure:NKViolinsB}
\end{figure*}

Figure\,\ref{Figure:NKRanges} shows the viable natal kicks for each system, arranged in increasing magnitude, illustrating how the object classes are inter-mingled with no obvious preference of $\vnk$ values with specific source class. Figures\,\ref{Figure:NKViolinsA} \& \ref{Figure:NKViolinsB} shows the $\vnk$ distributions for all sources as violin plots. Both figures denote the maximum aposteriori probability (MAP) values for the natal-kick point estimators, together with the 68\% highest-density probability intervals (HDI). All kick estimators are also summarised in Appendix Table\,\ref{Table:AllSystems}.

Appendix\,\ref{Section:App_AllResults} shows the distribution of natal kicks ($\vnk$), mass-loss kicks ($\vmlk$), and initial (ZAMS) binary characteristics for each of the 66 sources. In some cases, one can identify various evolutionary channels through which a given system may form from the multimodal distribution of the above parameters. 

Within our sample, 28 systems have published natal kick estimates from previous studies.\footnote{though in some cases, these are published estimates are $\vpec$ proxies, and not natal kicks.} For the remaining 37, our results provide the first constraints. In general, our estimates broadly agree with previous works based on complete population modelling. Where discrepancies exist, these can either be attributed to improved astrometric measurements from {\em Gaia} (as presented by \citetalias{Zhao2023}) and/or incomplete evolutionary modelling in previous works. Detailed comparison with previous estimates and the origin of any differences is discussed for individual sources in Appendix\,\ref{Section:App_PrevEstimates}.

\begin{figure*}
    \centering
    \includegraphics[width=0.95\textwidth]{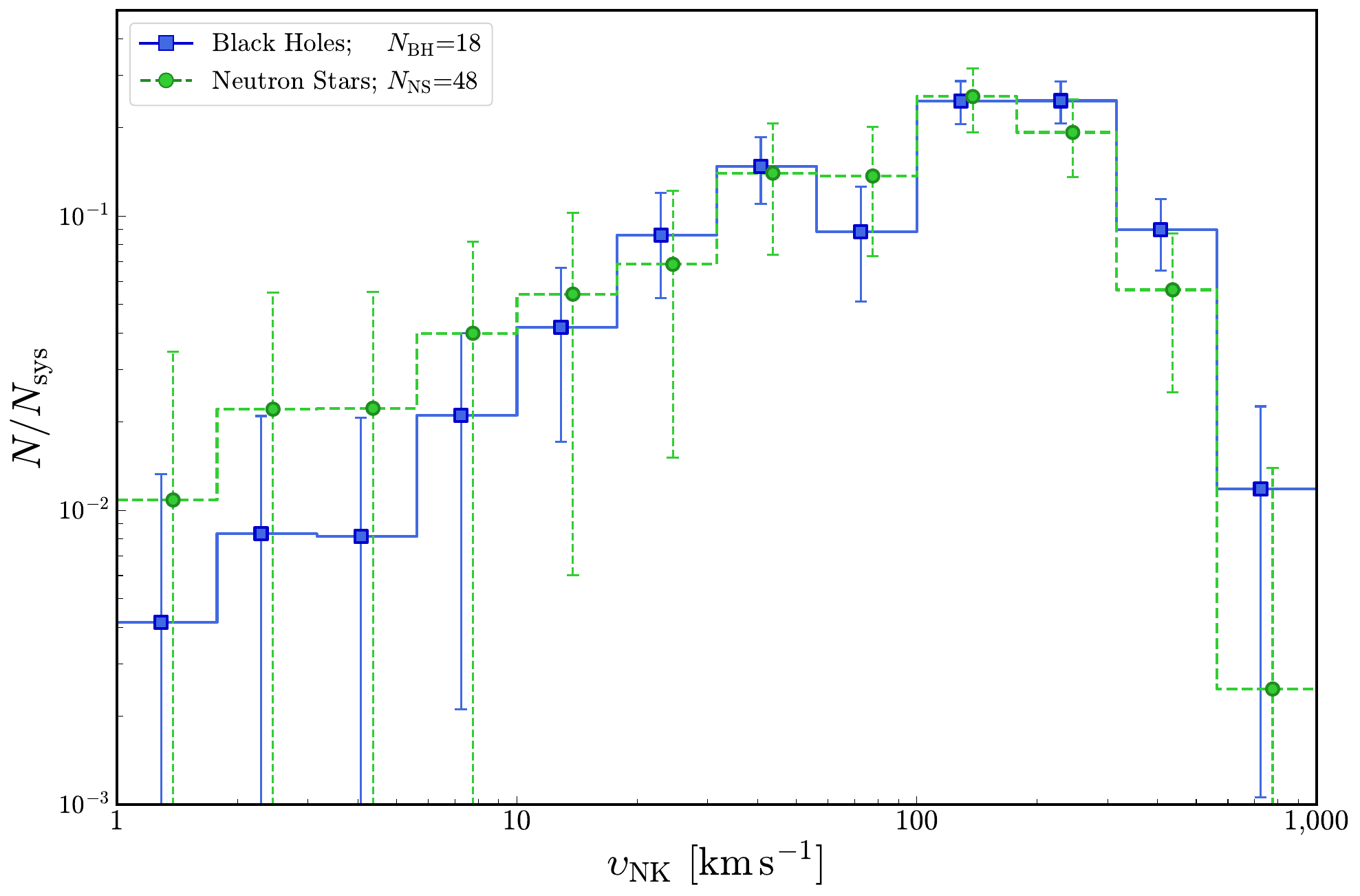}
    \caption{Inferred distribution functions of natal kicks for black holes and for neutron stars. Ensembles of natal kicks are simulated from each of the distributions for individual sources, each iteration corresponding to a possible physical scenario. The histograms here denote the mean and one standard deviation amongst these ensembles (the distributions of y-values for a given bin are generally Gaussian in linear space). The blue solid line indicates the distribution of natal kicks for black-hole compact objects in binaries, and the green dashed line represents neutron stars. \label{Figure:BHNSHistogram}}
\end{figure*}

Our key consequent result is in Figure\,\ref{Figure:BHNSHistogram}, showing the overall natal-kick distributions for compact binaries segregated by compact-object class. These global distributions are extracted through Monte Carlo draws of the individual-source $\vnk$ distributions. Specifically, the distribution of each source is sampled to extract one $\vnk$ value, resulting in a randomised global $\vnk$ histogram comprising $N_{\rm src}$\,=\,66 values. This process was iterated 10 million times, resulting in an ensemble of distributions, which can then be binned for display purposes, and its moments (mean and standard deviation) computed.

\subsection{Statistical comparison between black holes and neutron stars}
\label{Section:BHNSComparison}

Figure\,\ref{Figure:BHNSHistogram} shows no apparent segregation between the black-hole and neutron-star systems, with both distributions showing a broad peak between $\vnk$\,$\approx$\,100--200\,km\,s$^{-1}$. Both classes comprise systems that could have received large ($\vnk$\,$>$\,200\,\kmps) or very small ($\vnk$\,$\lesssim$\,5\,\kmps) kicks. The 68\% confidence intervals for both classes (considering all systems) are $\upsilon_{\rm NK,\,BHs}^{68\%}=$ 0--147\,\kmps, and $\upsilon_{\rm NK,\,NSs}^{68\%}=$ 0--187\,\kmps\ ($\upsilon_{\rm NK,\,BHs}^{\rm MAP}=7$\,\kmps\ \& $\upsilon_{\rm NK,\,NSs}^{\rm MAP}=36$\,\kmps).

In order to quantitatively compare the distributions for black-hole and neutron-star systems, 10 million statistical tests were carried out, in which one possible $\vnk$ value from the kick distribution for each of 66 systems is randomly drawn. These are then subject to two-sample Kolmogorov-Smirnov (KS) testing between the 18 natal kicks associated with black holes and the 48 natal kicks associated with neutron stars, taking $p$-value\,$\geq$\,0.05 as our cut-off threshold for rejecting the null hypothesis that the two samples are drawn from the same underlying population. Results of KS testing are shown in Table \ref{Table:KSTests}; these tests confirmed that there is no statistically significant evidence for an intrinsically different kick distribution (i.e. $p$-value=0.60$^{+0.28}_{-0.20}$); with $p$-value being more than 0.05 in more than 99\% of the tests. \

Given that some of the systems within our study are controversial (e.g., the distance to BW Cir is subject to significant uncertainty; \citealt{Casares2009, Gandhi2019, Plotkin2021}, and the nature of the compact object in LS 5039 remains unclear; \citealt{Casares2005, Yoneda2020, Zeng2024}), we also present the results of KS testing between the black-hole and neutron-star samples with these systems excluded in the same table. The main inference is unchanged - the natal kicks applied to black holes appear entirely consistent with those applied to neutron stars, and our results are not conditional on a few systems whose compact objects may have been incorrectly classified (for further discussion, see Appendix\,\ref{Section:App_WeirdSystems}). 

\begin{table}
\renewcommand{\arraystretch}{1.25}
\begin{tabular}{ l | c } 
Systems & $p$-value \\
\hline\hline
\textbf{All Black Holes} & \multirow{2}{3em}{\textbf{0.60$\mathbf{_{-0.20}^{+0.28}}$}} \\
\textbf{All Neutron Stars} & \\
\hline
BHs (excluding BW Cir \& SS 433) & \multirow{2}{3em}{0.67$_{-0.26}^{+0.20}$} \\ 
NSs (excluding Aql X-1 \& LS 5039) & \\
\end{tabular}
\caption[Results of Kolmogorov-Smirnov testing]{Results of Kolmogorov-Smirnov testing comparing black hole (BH) and neutron star (NS) natal kicks. These are conducted by generating ensembles of natal kicks (one per system, drawn from each systems distribution) and conducting KS tests between those associated with black hole systems vs neutron star systems. This is repeated 10 million times, resulting in a range of $p$-values (presented as MAP \& 68\% HDI). Given the controversial nature of some of the systems, most notably Aql X-1, BW Cir, LS 5039, \& SS 433 (see Appendix\,\ref{Section:App_WeirdSystems} for discussion), in addition to our core KS-tests (using all 66 systems in our sample), we conduct tests excluding the above systems. This demonstrates that the natal kick distributions we find for some of the less-well characterised systems do not significantly impact our results.}
\label{Table:KSTests}
\end{table}

Before moving on, it is worth emphasizing explicitly what our results do and do not establish. No statistical analysis can "prove" that two samples -- such as our NS and BH samples -- must be drawn from a common parent distribution. For example, the $p$-value we obtain from our KS test is specifically the probability of observing a KS-distance between the two samples that is as large as that observed, under the assumption that the samples {\em are} drawn from the same parent distribution. A value as large as the one we find -- $p \simeq 0.6$ -- therefore simply means we cannot rule out this hypothesis based on the present data. It is always possible that, with more and/or better data, this same hypothesis {\em will} eventually be ruled out. However, the physically interesting point here, of course, is that any differences between BH and NS kick distributions appear to be much subtler than previously expected -- to the point that present-day observations are insensitive to them.

\subsection{Hierarchical Bayesian modelling of the
\texorpdfstring{$\upsilon_{\rm NK}$}{v_NK} distributions}
\label{Section:KickDistributions}

We next characterised the empirically-derived $\vnk$ distributions analytically. We choose the 2-parameter Gamma distribution for a phenomenological description, as it provides a good combination of form flexibility and simplicity of modelling. As we shall say, it provides acceptable fits to the empirical distributions. 

The Gamma distribution can be written as 
\begin{equation}
    P\,(\vnk\,|\,\alpha,\theta) = 
    \frac{1}
    {\Gamma(\alpha)\, \theta^{\alpha}}\,
    \vnk^{\alpha - 1}\,e^{-\vnk/\theta}; ~~~~~\vnk\,\ge\,0,
\label{Eq:Gamma}
\end{equation}

\noindent
where $\alpha$ is a shape parameter, $\Gamma$ represents the gamma function, and $\theta$ is a characteristic scale length (in the units of $\vnk$). The first three moments of the Gamma distribution are the mean $\mu$~=~$\langle\vnk\rangle = \alpha\theta$, the variance $\sigma^2_\vnk = \alpha\theta^2$ and the skewness $s_\vnk = 2/\sqrt{\alpha}$. 

Given empirically-based $\vnk$ estimates for any given sample (e.g. BH systems, NS systems, all systems), our goal is to obtain optimal estimates of the parameters $\alpha$ and $\theta$. In order to achieve this, we set up a hierarchical Bayesian model that allows us to generate samples -- using {\sc dynesty} \citep{dynesty} -- from the joint posterior distribution of $\alpha$ and $\theta$ via nested sampling. We then adopt the joint maximum a posteriori (MAP) values as optimal and use the 68\% marginal highest-posterior-density interval to define the parameter uncertainties. Note that our empirically-based $\vnk$ estimates already come in the form of posterior distributions, but these were obtained under the assumption of a flat prior in $\vnk$ between 0 \kmps and 1000 \kmps. In our model, this prior is replaced with the Gamma distribution, and we instead adopt wide, uninformative log-uniform priors on the parameters of that distribution: $\alpha \in [10^{-2}, 10^{2}]$ and $\theta \in [1, 10^{4}]$. 

In order to test whether our fits are statistically acceptable, we carry out posterior predictive checks. This involves creating $j = 1 ... M$ empirical and control mock samples for comparison. The empirical mock samples are generated from the updated posterior $\vnk$ distributions for each of our $N$ systems. Specifically, for each mock sample, we draw $(\alpha_j, \theta_j)$ from the joint posterior and construct an updated $\vnk$ posterior for each system by replacing the original flat prior with the Gamma distribution corresponding to that draw \citep[e.g.][]{Hogg2010, Mandel2019}. We then generate one $\vnk$ value for each system in this mock sample by a random draw from this posterior. We also generate a matching control sample of $N$ values directly from the same Gamma distribution, this time ignoring the individual-source likelihoods. Both samples are compared against the analytic CDF associated with $(\alpha_j, \theta_j)$ via the Kolmogorov-Smirnov distance, $d_{KS}$, i.e. the maximum absolute difference between the empirical and analytic CDFs. Our goodness-of-fit metric is then the fraction of pairs, $f_p$, for which the analytical Gamma distribution was a better match (smaller $d_{KS}$) to the empirically-based mock sample than to the control sample. Since the control sample {\em was} in fact drawn from the Gamma distribution, we expect to find $f_p \simeq 0.5$ if the empirically-based $\vnk$ are fully consistent with being drawn from a Gamma parent distribution. If the Gamma distribution is a poor fit to the empirically-based $\vnk$ values, we expect to find $f_p \simeq 0$. In practice, we find $f_p \simeq 0.37$ for the full (NS+BH) sample, $f_p \simeq 0.41$ for the NS-only sample, and $f_p \simeq 0.48$ for the BH-only sample. Thus the Gamma distribution provides quite a good fit to our empirical distributions.

The global (NS + BH) distribution of natal kicks is well described by a Gamma distribution with $\alpha=1.20^{+0.17}_{-0.26}$ \& $\theta=107^{+26}_{-22}$\,\kmps\ (corresponding to a mean of  $\langle\vnk\rangle\approx128$\,\kmps\, a standard deviation of  $\sigma$\,=\,117\,km\,s$^{-1}$ and a skew $s\approx1.83$); see Table\,\ref{Table:GammaParams}. This simple, observationally motivated prescription is well-suited for implementation in future population synthesis models (to the extent that our observational sample is not too badly affected by selection effects; this is discussed further in the following section). It should also provide a valuable empirical benchmark for theoretical models that predict the intrinsic kick distribution(s).

\begin{table}[h!]
\centering
\renewcommand{\arraystretch}{1.25}
\begin{tabular}{ l | c c | c c } 
 & $\alpha$ & $\theta$  [\kmps] & $\mu$  [\kmps] & $s$ \\
\hline\hline
\textbf{All Systems} & $\mathbf{1.20^{+0.17}_{-0.26}}$ & $\mathbf{107^{+26}_{-22}}$ & $\mathbf{128^{+17}_{-15}}$ & $\mathbf{1.81^{+0.21}_{-0.14}}$\\ 
Neutron Stars & $1.33^{+0.21}_{-0.37}$ & $105^{+32}_{-27}$ & $141^{+21}_{-19}$ & $1.73^{+0.24}_{-0.17}$ \\ 
Black Holes & $1.03^{+0.20}_{-0.50}$ & $97^{+57}_{-38}$ & $102^{+27}_{-26}$ & $1.95^{+0.45}_{-0.29}$ \\ 
\end{tabular}
\caption{Best-fit parameters for the Gamma distribution describing the natal kicks of black holes, neutron stars, and all systems. $\alpha$ and $\theta$ are constants as in Equation\,\ref{Eq:Gamma}; $\mu$ and $s$ represent the mean and the skew of the distribution.}
\label{Table:GammaParams}
\end{table}

\begin{figure}
    \includegraphics[width=0.45\textwidth]{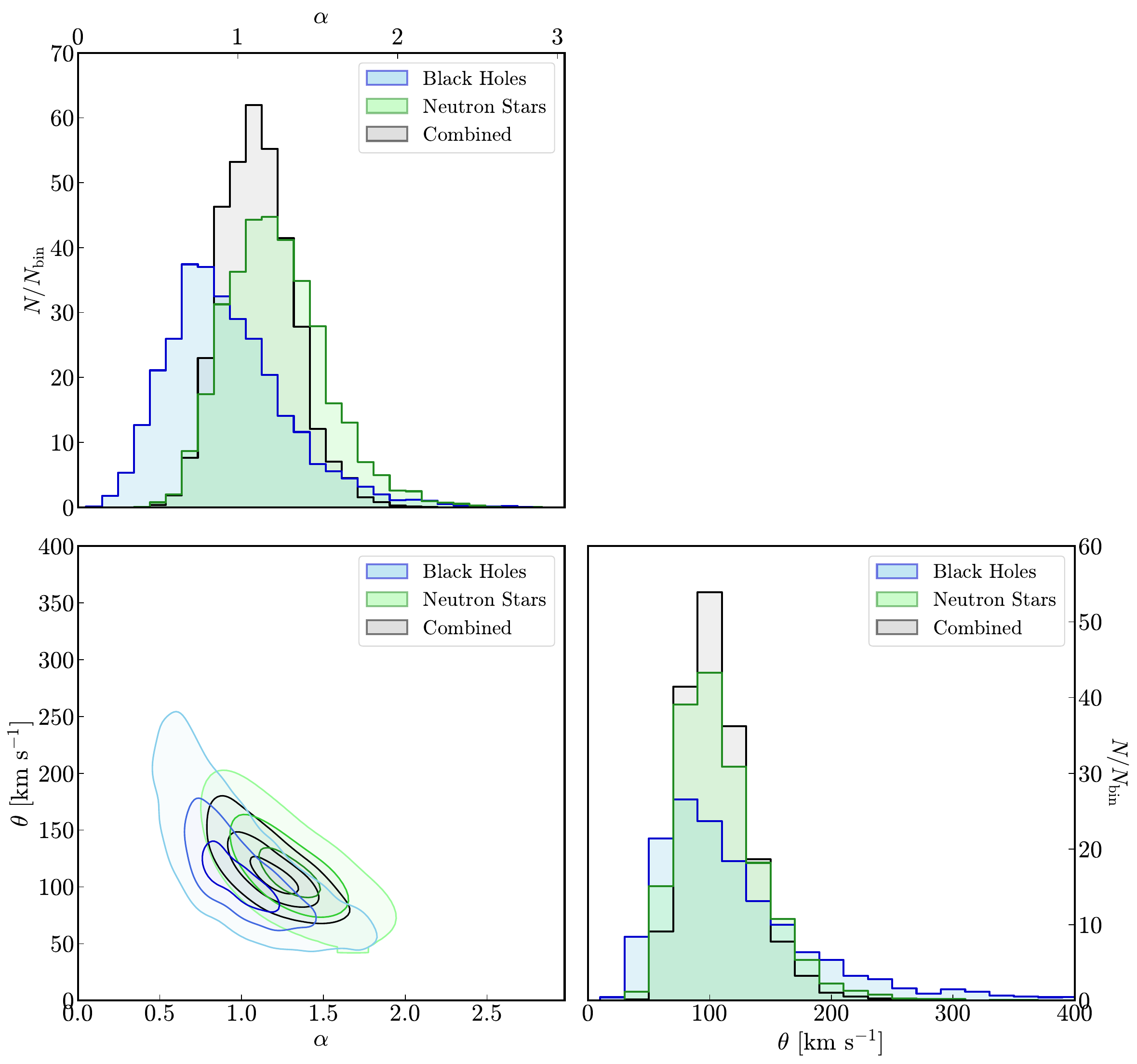}   
    \caption{Corner plot showing the parameters for the Gamma distribution which best describes our data. The parameters are $\alpha$, the shape (unitless), and $\theta$, the scale (units \kmps) - a single Gamma distribution is fit to each possible physical scenario (see Section\,\ref{Section:KickDistributions}), for our sample of black holes (blue contours), neutron stars (green contours), and all systems together (black contours). Though the best fit parameters are subtly different when segregating based on compact-object class, there exists a range of parameters which produce statistically acceptable fits for both black holes and neutron stars}
    \label{Figure:GammaDistributionParams}
\end{figure}

Finally, we note that the strongest black hole kicks are still weaker than the strongest neutron star kicks - BW Cir is subject to the strongest natal kick of all black holes in our sample, and that system is the subject of much debate (see Appendix\,\ref{Section:App_WeirdSystems}), however, as we have discussed previously, our broader results are not contingent on this system alone. It may be the case that future studies and large sample sizes find a more significant difference between neutron star and black hole natal kicks than our current data; however, for the time being, there is no evidence for such a distinction.

\section{Discussion}
\label{Section:Discussion}

Our work has derived the plausible natal kicks for 66 systems and, consequently, provided insight into the distribution of natal kicks for the wider population of compact objects. Here, we discuss some of the significant implications of these results in the wider context, including evidence for the long-disputed high natal kicks necessary for black holes, relationships between natal kick magnitude and various physical parameters, and the potential consequences for supernova physics. This is not intended to be a complete review of every facet of our results, but rather to highlight some of the key takeaways, inspire discussion, and hone in on some specific avenues for further discussion and research.

\subsection{Black-hole systems with natal kicks in excess of 100 \kmps}
\label{Section:StrongBHKicks}

Three of the 18 black-hole systems (17$_{-10}^{+17}$\,\%, for 1$\sigma$ Poisson sampling uncertainties; \citealt{gehrels86}) require kicks $\upsilon_{\rm NK,min}\,\geq\,100$\,\kmps, compared to 17 out of 46 (37$_{-10}^{+13}$\,\%) neutron stars. For an additional 4 black holes, the 95\%\ $\vnk$\ confidence region exceeds 100\,\kmps\ at the low end. In other words, $39_{-17}^{+24}$\% of black holes in our sample have natal kicks $\geq100$\,\kmps\ at more than 95\% confidence; these are: GRS\,1124--684 ($\upsilon_{\rm NK}^{95\%}$=102\,\kmps); H\,1705--250 ($\upsilon_{\rm NK}^{95\%}$=121\,\kmps); XTE\,J1118+480 ($\upsilon_{\rm NK}^{95\%}$=124\,\kmps); GX\,339--4 ($\upsilon_{\rm NK}^{95\%}$=\,142\,\kmps); GRO\,1655--40 ($\upsilon_{\rm NK}^{95\%}$=166\,\kmps); Swift\,J1727.8--1613 ($\upsilon_{\rm NK}^{95\%}$=178\,\kmps); and BW\,Cir ($\upsilon_{\rm NK}^{95\%}$=296\,\kmps). Despite losing appreciable mass during the supernova ($\sim$\,20--55\% of their progenitor mass), the mass-loss kicks of these systems are limited to $\vmlk$\,$\leq$\,40\,\kmps\ (Appendix\,\ref{Section:App_AllResults}), insufficient to explain their large peculiar velocities (spanning $\vpec$\,$\approx$\,118--350\,\kmps). In the case of neutron stars, 20 systems have natal kicks $\geq100$\,\kmps at more than 95\% confidence, comprising $43_{-12}^{+14}$\% of the neutron-star systems studied here. That a similar proportion of neutron stars and black holes are likely to have received substantial natal kicks contradicts the idea that black holes may only form with low kicks.

\subsection{Negligible natal kicks}
\label{Section:ZeroKicks}

At least 19 Galactic X-ray binaries may have formed without a natal kick (i.e. $\upsilon_{\rm NK,min}=0$\,\kmps): 5 black-hole and 14 neutron-star systems --- a similar fraction of each class. The full range of their modelled kicks, however, allow many of these systems to have potentially formed with larger kicks --- of order 10s--100s \kmps.
The mean peculiar velocities of these systems range from 8 to $>200$\,\kmps. 

A larger fraction of high-mass systems may have formed without a natal kick, as was proposed in \citetalias{Zhao2023} - however, the mean natal kicks for many of these systems are still substantial. Only two pulsars, B1259--63 (PSR J1302--6350) and J2339--053, may have been formed without a natal kick - this is primarily due to the fact that their peculiar velocities are uncharacteristically low for pulsars (45\,\kmps \& 64\,\kmps respectively). Furthermore, these statements are ultimately conservative assessments of the simulations, considering only the \textit{minimum} natal kick required and do not necessarily indicate the most probable scenario (i.e. for 9 of these 19 systems, fewer than 5\% of kicks are less than 10\,\kmps). In B1259--63, a spin-orbit misalignment and a significant eccentricity, has been used to infer that the system was likely impacted by a significant kick in the polar direction \citep{Valli2025}; these higher-order measurements are not modelled here (cf. discussions on limitations of our approach in Section\,\ref{Section:Biases} and in Paper~I). 

\subsection{Implications for natal-kick and supernova mechanisms}
\label{Section:KickMechanisms}

A variety of physical mechanisms have been put forward to explain kicks, including binary centre-of-mass recoil due to baryonic mass loss \citep{Blaauw1961,Nelemans1999}; asymmetric stellar collapse \citep{Burrows1996}; anisotropic neutrino emission \citep{Chugai1984,Dorofeev1985,Arras1999,Lai04,Tamborra2014,Janka2024}; and hydrodynamic instabilities leading to asymmetric mass ejection \citep{Blondin2003,Wongwathanarat2013}. Large-scale asymmetries can impart recoil through anisotropic momentum transfer during and after core bounce \citep{Nordhaus2012,Wongwathanarat2013}, while the post-explosion compact object can undergo sustained directional gravitational acceleration due to asymmetrically ejected mass \citep{Janka2017}. Given a total radiated neutrino energy of $\sim$\,3\,$\times$\,10$^{53}$\,erg, even percent-level global asymmetries in the time-integrated neutrino emission are sufficient to impart kick velocities $\vnk$\,$\sim$\,$\mathcal{O}(100)$\,km\,s$^{-1}$ or more to neutron stars \citep{Janka2024}.

\subsubsection{BH Kick moderation by fallback}

For black holes, the magnitudes of natal kicks are thought to be tempered relative to those imparted to neutron stars. Firstly, {\em fallback} of material that fails to escape the explosion and returns onto the newly formed compact object increases its mass and can modify or reduce the impulse associated with initially outflowing ejecta \citep{Janka2013}. \citet{Fryer2012} adopted a fallback-modulated natal-kick prescription in which kicks are moderated by the fraction of matter that falls back onto the proto-neutron star ($f_{\rm fb}$), scaling as 
%\vnk = \left(1-f_{\rm fb}\right)\upsilon_{\rm NS}$, 

\begin{equation}
    \upsilon_{\rm NK}^{\rm BH} = \left(1-f_{\rm fb}\right)\upsilon_{\rm NS}
    \label{eq:fryer12}
\end{equation}

\noindent
where $\upsilon_{\rm NS}$ is drawn from a typical neutron-star kick distribution (e.g., the Maxwellian distribution of \citealt{Hobbs2005}, in which case $\upsilon_{\rm NS}$\,$\equiv$\,$\upsilon_{\rm H05}$). Since the amount of fallback is expected to be significant for many black holes, this prescription predicts natal kicks smaller than those applied to neutron stars, asymptoting to zero for complete fallback.

Explosion outcome and asymmetry are key decisive factors in predicting the magnitudes of kicks. Several recent 3D simulations have identified non-exploding progenitors around 12--14\,\Msun\ although explodability seems to depend non-monotonically on progenitor mass in this regime \citep[e.g.,][]{Burrows2024a,Sykes2025}. In such failed explosions, hydrodynamic kicks are minimal and recoil is dominated by anisotropic neutrino emission. The transient proto-neutron star may acquire neutrino-driven velocities of tens of km\,s$^{-1}$, but subsequent collapse and accretion of the stellar core dilute the final black-hole recoil to only a few km\,s$^{-1}$ \citep{Janka2024}.

Furthermore, failed supernovae, or direct-collapse black-hole formation events, are often assumed to be a major channel through which black holes form. However, the mapping between progenitor mass and explosion outcome is complex, and recent simulations predict multiple black-hole formation channels \citep{Burrows2024a,Burrows2024b}. Black holes formed through failed explosions are expected to receive only weak, neutrino-driven natal kicks of up to a few km\,s$^{-1}$, whereas some formation channels following successful asymmetric explosions or fallback allow for substantially larger hydrodynamic kicks \citep{Burrows2024a}.

Our results stand in contrast to expectations of uniformly tempered kicks in black holes, at least in the mass regime of compact binaries that we have investigated. Despite being limited by small number statistics, there is no obvious observational bias that could result in rapidly moving black holes preferentially falling within our sample. The absolute differences between the baseline theoretical predictions above (with black hole kick expectations aggregating around just a few km\,s$^{-1}$) and the mean distribution that we infer (with $\langle\vnk\rangle$\,$\approx$\,128\,km\,s$^{-1}$) is even more stark.

This indicates the need for a revised understanding of fallback formation channels or the level of asymmetry that nature may permit in supernova explosions which form black holes. A recent proposed mechanism that could deliver strong kicks in black holes is through the actions of large-scale ordered magnetic fields around rapidly rotating collapsars \citep{li2026}. The fields sustain an accretion-powered, asymmetric outflow post-SN, leading to non-uniform escape of material and consequent strong kicks. Simulated explosions in this regime have enormous energy reservoirs ($E_{\rm explosion}$\,$\gtrsim$\,10$^{52}$\,erg) and may naturally associated with gamma-ray-burst progenitors. It remains to be understood whether these engines are able to account for the kicks in the XRB regime. 

Furthermore, both the mass and fraction of matter that may be totally ejected or fallback onto the black hole are dependent on the progenitor mass as well as the final mass, and it is therefore difficult to make inferences on the entire population of black holes based on the (relatively) low mass ones included in this study. 

\subsubsection{Electron-Capture Supernovae}
\label{Section:ECSN}

Whilst most X-ray binaries are considered to be the result of core-collapse supernovae, electron-capture supernovae (ECSN) may be responsible for some neutron stars. These ECSN involve an ultra-stripped progenitor and very little ejecta mass (0.01\textendash0.2\,\Msun), resulting in weaker natal kicks (typically a few kilometres per second; \citealt{Nomoto1984,Podsiadlowski2004,Tauris2013,Igoshev2021}).

According to our \cosmic\ simulations, 16 systems may have formed via ECSN: 9 NS-LM-PSRs, 6 NS-LMXBs, and 1 NS-HMXB (see Table\,\ref{Table:ECSN}). However, only 4 of these systems may be formed via ECSN without a natal kick ($\upsilon_{\rm NK,min}\leq10$\,\kmps): 4U 1700+24; Aql X-1; 1417-4402; and RX J0146.9+6121. The remaining ECSN analogues require stronger natal kicks (averaging over 100\,\kmps). In cases where not all analogues undergo ECSN (i.e. some evolve via core-collapse), one can, perhaps, discard the ECSN analogues requiring substantial natal kicks as implausible. Where ECSN appears as the only viable supernova channel and are always accompanied by non-negligible kicks (as is the case for: J1431-4751; J1622-0315; J1628-3205; and J2039-5617), things become more complex. In these instances, our population-synthesis modelling, combined with the peculiar-velocity boundary constraints from kinematic measurements, directly contradicts hydrodynamical simulations and similar theoretical studies. However, it should be kept in mind that our kick modelling is decoupled from the explosion mechanism prescriptions of \cosmic, so we cannot test any direct causal relationship between the ECSN mechanism and inferred kicks.

\subsection{Testing for physical scalings}
\label{Section:PhysicalScalings}

\begin{figure*}
    \centering
    \includegraphics[width=0.9\linewidth]{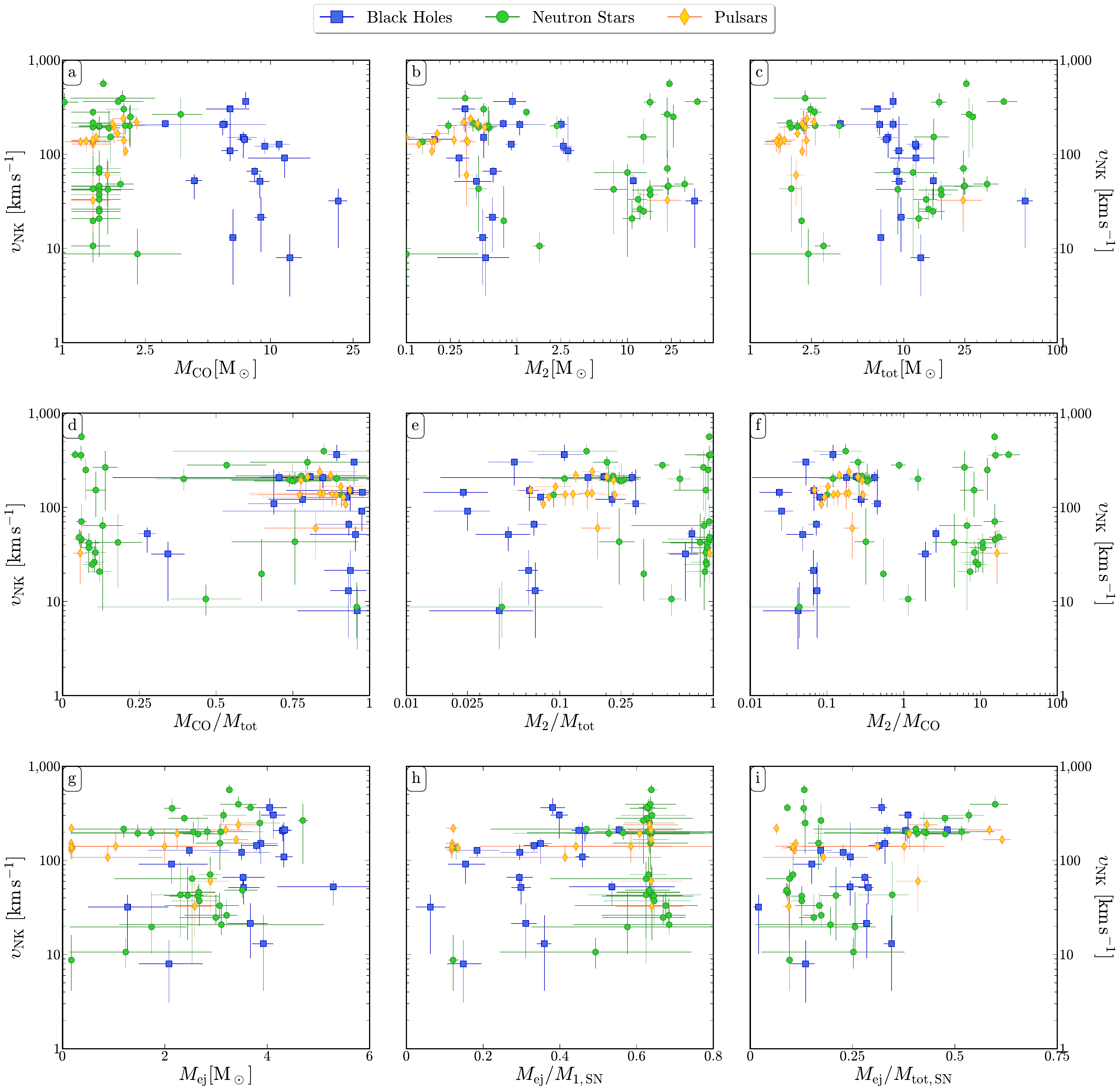}
    \caption{Natal kick magnitude compared to: [a] the (current) mass of the compact object $M_{\rm CO}$ (and nominal observational uncertainties); [b] the mass of the stellar companion $\mtwo$; [c] the total mass of the system $M_{\rm tot}$; [d] the ratio  $M_{\rm CO}$/$M_{\rm tot}$; [e] the ratio $\mtwo/M_{\rm tot}$; [f] the (current) mass ratio, $q$; [g] the mass of material ejected $M_{\rm ej}$ due to the supernova (markers indicate mean ejecta mass, errors indicate 68\% confidence intervals); [h] the fraction of the compact progenitor that is ejected (markers indicate mean, errors indicate 68\% confidence intervals); [i] the fraction of the total progenitor system mass that is ejected (markers indicate mean, errors indicate 68\% confidence intervals). The plotted natal kicks in all panels represent the mean $\vnk$ magnitudes, and the errors indicate the 68\% confidence intervals. System types are distinguished by colour \& marker type: blue squares for black holes; green circles for neutron stars; and orange diamonds for pulsars.}
    \label{Figure:KickScalings}    
\end{figure*}

Figures\,\ref{Figure:KickScalings} show the natal kick constraints for our successfully simulated systems as a function of mass (compact object $M_{\rm CO}$; companion $\mtwo$; total $M_{\rm tot}$; and ejected mass $M_{\rm ej}$). Table\,\ref{Table:ScalingCorrelations} tabulates results of Spearman's rank-order correlation tests, quantifying potential correlations between various system parameters and natal kick magnitude. We present the MAP and 68\% HDI of the $p$-values for each test. Overall, there is limited statistical evidence to identify a meaningful correlation between natal kick and, for example, progenitor $M_{\rm 1,pre-SN}$, total mass $M_{\rm tot}$ or ejected mass $M_{\rm ej}$. 

That the magnitude of the natal kick does not appear to be related to the total system mass might seem surprising, given that there is a known correlation between observed peculiar velocity and $M_{\rm tot}$ \citepalias{Zhao2023}; however, this is to be expected. The systemic velocity that is  attained following a natal kick has a strong dependence on both component masses (particularly the ratio between $\mtwo$\ and $M_{\rm tot}$; this partly accounts for many NS-HMXBs being consistent with a broad range of natal kicks). This further highlights the limitations in using $\vpec$ as a proxy for natal kick velocity, an issue discussed extensively in Paper~I.

\begin{table*}
\centering
\renewcommand{\arraystretch}{1.25}
\begin{tabular}{c | l | c c } 
& $\vnk$ vs. & $\rho_{s}$ &  $p$-value $\left(\rho_s\right)$ \\ [0.5ex] 
 \hline\hline
\multirow{6}{4em}{All Systems} & $M_{\rm 1,pre-SN}$ & $-0.06^{+0.06}_{-0.06}$ & $0.611^{+0.264}_{-0.261}$ \\ 
\cline{2-4}
&        $M_{\rm 1,core,pre-SN}$ & $-0.08^{+0.06}_{-0.06}$ & $0.527^{+0.273}_{-0.256}$ \\ 
\cline{2-4}
&        $\Mc$ & $-0.04^{+0.05}_{-0.05}$ & $0.704^{+0.216}_{-0.218}$ \\ 
\cline{2-4}
&        $M_{\rm ej}$ & $0.12^{+0.08}_{-0.08}$ & $0.382^{+0.32}_{-0.269}$ \\ 
\cline{2-4}
&        $M_{\rm ej}$/$M_{\rm1,pre-SN}$ & $0.07^{+0.08}_{-0.08}$ & $0.545^{+0.304}_{-0.302}$ \\ 
\cline{2-4}
&        $M_{\rm ej}$/$M_{\rm tot,pre-SN}$ & $0.33^{+0.07}_{-0.07}$ & $0.022^{+0.015}_{-0.021}$ \\ 
\hline\hline
\multirow{6}{4em}{Neutron Stars} & $M_{\rm 1,pre-SN}$ & $0.13^{+0.1}_{-0.1}$ & $0.415^{+0.331}_{-0.293}$ \\ 
\cline{2-4}
&        $M_{\rm 1,core,pre-SN}$ & $0.1^{+0.1}_{-0.1}$ & $0.511^{+0.322}_{-0.315}$ \\ 
\cline{2-4}
&        $\Mc$ & $0.17^{+0.09}_{-0.09}$ & $0.307^{+0.267}_{-0.233}$ \\ 
\cline{2-4}
&        $M_{\rm ej}$ & $0.13^{+0.1}_{-0.1}$ & $0.428^{+0.338}_{-0.301}$ \\ 
\cline{2-4}
&        $M_{\rm ej}$/$M_{\rm1,pre-SN}$ & $-0.07^{+0.11}_{-0.11}$ & $0.54^{+0.319}_{-0.324}$ \\ 
\cline{2-4}
&        $M_{\rm ej}$/$M_{\rm tot,pre-SN}$ & $0.3^{+0.09}_{-0.09}$ & $0.083^{+0.072}_{-0.076}$ \\ 
\hline\hline
\multirow{7}{4em}{Black Holes} & $M_{\rm 1,pre-SN}$ & $-0.53^{+0.11}_{-0.11}$ & $0.042^{+0.037}_{-0.039}$ \\ 
\cline{2-4}
&        $M_{\rm 1,core,pre-SN}$ & $-0.53^{+0.11}_{-0.1}$ & $0.044^{+0.039}_{-0.039}$ \\ 
\cline{2-4}
&        $\Mc$ & $-0.52^{+0.1}_{-0.1}$ & $0.048^{+0.041}_{-0.042}$ \\ 
\cline{2-4}
&        $M_{\rm ej}$ & $0.48^{+0.12}_{-0.12}$ & $0.075^{+0.065}_{-0.067}$ \\ 
\cline{2-4}
&        $M_{\rm ej}$/$M_{\rm1,pre-SN}$ & $0.51^{+0.1}_{-0.1}$ & $0.053^{+0.044}_{-0.046}$ \\ 
\cline{2-4}
&        $f_{\rm fb}$ & $-0.51^{+0.11}_{-0.11}$ & $0.049^{+0.04}_{-0.044}$ \\ 
\end{tabular}
\caption{Spearman ($\rho_{s}$) coefficients and corresponding $p$-values when testing correlations between $\vnk$ and compact object mass ($M_{\rm CO}$); compact object progenitor mass ($M_{\rm 1,pre-SN}$); ejected mass ($M_{\rm ej}$); fraction of compact object progenitor mass ejected ($M_{\rm ej}$/$M_{\rm1,pre-SN}$); fraction of total pre-SN system mass ejected ($M_{\rm ej}$/$M_{\rm tot,pre-SN}$); and fallback fraction ($f_{\rm fb}$) describing how much of the matter initially ejected during the supernova fallback onto the newly formed black hole (i.e. if $f_{\rm fb}=1$, $M_{\rm ej}=0$). Uncertainties correspond to the 68\% confidence level.}
\label{Table:ScalingCorrelations}
\end{table*}

There appears to be some evidence of relation between natal kick magnitude and the fraction of total pre-SN system mass ejected ($M_{\rm ej}$/$M_{\rm tot,pre-SN}$), however, this is likely an artifice introduced by population bias rather than evidence of a physical correlation. The fraction of total mass that is ejected directly influences whether the system remains bound - in the absence of a natal kick, a given system can lose no more than half its total mass due to the supernova and remain bound. This limit changes when considering the impact of a given natal kick, being proportional to both the fraction of total mass which is lost and the kick magnitude (see Appendix\,\ref{Section:App_HalfMassLoss}). Given that, by definition, the XRBs in our sample survived the supernova as binaries, both the natal kick magnitude and ejected mass (fraction) must satisfy the following condition:

\begin{equation}
    2\left(1-\frac{M_{\rm ej}}{M_{\rm tot,pre-SN}}\right)>\frac{\vnk^2}{\vorb^2}+\frac{\vnk\cphi\stheta}{\vorb}+1,
\end{equation}
where $M_{\rm ej}$ and $M_{\rm tot,pre-SN}$ are the ejecta mass and total system mass at the instant of supernova, respectively; $\vorb$ is the relative velocity of the compact progenitor at the instant of supernova; and $\vnk$, $\cphi$, and $\stheta$ describe the magnitude and direction of the natal kick. This constraint is likely the origin of the apparent correlation seen in Table\,\ref{Table:ScalingCorrelations}. 

There is some evidence of an anti-correlation between black hole progenitor (\monepre\, \& $\Mc$) and natal kick magnitude ($\vnk$)  (note, $M_{\rm 1,pre-SN}$, $M_{\rm 1,core,pre-SN}$, $M_{\rm BH}$, and $f_{\rm fb}$ are all closely related), however, again this correlation is not consistently significant to make broad claims about a generalised relationship. Whilst this could be readily explained as a consequence of natal kicks arising from asymmetries in ejected matter (and, perhaps, mediated by fallback), we hold off on presenting this as hard evidence, instead suggesting this be the subject of future, more focused studies, with the aid of additional data (including from other types of systems).  

Whilst we find limited evidence for a correlation between remnant and ejected mass and natal kick magnitude, we point out that our population synthesis framework uses the delayed remnant mass function from \citet{Fryer2012}, which naturally incorporates fallback for large progenitors (core mass $M_{\rm core}\geq11$\,\Msun), though is but one of many putative descriptions of remnant mass determination. Further studies exploring the variation of inferred natal kicks with different remnant mass functions may be able to reconcile the apparent contradiction between our results and expectations from theoretical studies.

\subsection{Comparison with previous studies}
\label{Section:PrevDistributions}

\citet{Hobbs2005} studied the proper motion of a sample of 233 pulsars, determining their systemic velocity. They proposed that the velocity of young, predominantly isolated pulsars within their larger sample is well described by a Maxwellian distribution, with rms $\sigma$\,=\,265\,\kmps. As discussed in the introduction (and, extensively, in Paper I), systemic/peculiar velocities have long-since been viewed as an approximate estimate of natal kicks, and the Hobbs Maxwellian has therefore been adopted as the canonical natal-kick distribution in a number of studies concerning compact objects, and built upon in subsequent works. \citet{Igoshev2021} suggested a bimodality in the natal-kick distribution, where the Maxwellian component of rms $\sigma$\,=\,265\,\kmps is relevant for the case of  core-collapse supernovae, proposing that the natal kicks associated with ECSN are drawn from a Maxwellian with a smaller rms $\sigma$\,=\,30\,\kmps. 

\citet{Bray2016} express the natal kick as $\vnk=\alpha\left(M_{\rm ej}/\Mc\right)+\beta$, calibrating $\alpha$ and $\beta$ against the velocities of young, single neutron stars selected from the \citet{Hobbs2005} sample. For their preferred binary-progenitor population, comparison with the observed 2D velocities yielded $\alpha=70^{+15}_{-15}$\,km\,s$^{-1}$ and $\beta=110^{+70}_{-90}$\,km\,s$^{-1}$. \citet{bray2018} subsequently revised the treatment of projecting synthetic 3D velocities into the observed 2D plane and obtained substantially revised parameters, particularly $\beta$, with a preferred 2D fit of $\alpha=100^{+30}_{-20}$\,km\,s$^{-1}$ and $\beta=-170^{+100}_{-100}$\,km\,s$^{-1}$. Although the prescription has subsequently been applied within population-synthesis calculations that include black holes, its empirical calibration is based on neutron-star velocities.  \citet{mandel2020} introduced stochasticity into remnant-mass and natal-kick prescriptions to account for variations in core-collapse outcomes, with kicks scaling with the asymmetrically ejected carbon--oxygen core mass rather than the total ejecta mass.

\citet{Fortin2022} studied a sample of 35 NS HMXBs, determining their peculiar velocities and inferring their associated natal kicks. Their inferred natal kicks were derived by using the observed values of $\vsys$, $M_2$, $\porb$, $e$, and taking $\Mc$=1.4\,\Msun (assumed as constant for all neutron stars) and solving the same natal kick equations we use in our work \citep{Brandt1995a,Kalogera1996}, assuming uniform priors for the unknown parameters ($M_{\rm 1, pre-SN}$; $P_{\rm orb, pre-SN}$; $\vnk$) and isotropic kicks ($\thetank$ and $\phink$ sampled uniformly from a sphere). They use Bayesian analysis to construct likelihoods for the observable properties, and infer posterior-distributions for the above unknowns. Their work proposed neutron star natal kicks are best described by a Gamma distribution with mean $116^{+18}_{-15}$\,\kmps, which is similar to our findings (although we propose that the Gamma distribution is also a valid fit for kicks associated with black holes); however, estimates for individual sources are discrepant with our findings in many cases (see Section\,\ref{Section:App_PrevEstimates} for examples and further discussion). 

\citet{Disberg2025} carried out a statistical analysis of neutron-star proper motions, finding that pulsars with ages below 40\,Myr have velocities consistent with a log-normal distribution with parameters $\mu=5.67\pm0.10$ \& $\sigma=0.59\pm0.08$, and present this as an alternative distribution from which natal kicks may be drawn. In their work, they return to the results of \citet{Hobbs2005}, and amend the analysis therein (by considering the necessary Jacobian transformation, accounting for the use of logarithmic binning), finding that the results from that study are an overestimate. 

\citet{Giacobbo2019} offer two descriptions of natal kicks, both being dependent on system parameters in addition to a random Maxwellian. $\vnk=\upsilon_{\rm H05}\frac{M_{\rm ej}}{\langle M_{\rm ej}\rangle}\frac{\langle\Mc\rangle}{\Mc}$\,\kmps (or $\vnk=\upsilon_{\rm H05}\frac{M_{\rm ej}}{\langle M_{\rm ej}\rangle}$\,\kmps) where: $\upsilon_{\rm H05}$ is a random number drawn from the Maxwellian presented by \citet{Hobbs2005}; $M_{\rm ej}$ is the mass ejected by the supernova (with mean $\langle M_{\rm ej}\rangle=9$\,\Msun); and $\Mc$ is the mass of the remnant compact object (with mean $\langle\Mc\rangle=1.2$\,\Msun). In contrast to the majority of the works discussed above, which focus in neutron stars, these equations are proposed to be suitable for both neutron stars and black holes. In the case of black holes, particularly those with minimal mass loss, natal kicks are expected to be systematically lower than those associated with neutron stars. Whilst these prescriptions are consistent with our findings in a small number of cases (for example, we find the plausible natal kick range for MAXI\,J1305-704 to be $\vnkconf$\,=\,34--62\,\kmps, where equation 1 from \citealt{Giacobbo2019} predicts $\vnkconf$\,=\,15--47\,\kmps), for the majority of the population our natal kicks appear to lie outside the predicted range. However, given that these mass-dependent models are subject to the inherent uncertainties in the binary evolution physics we have implemented (i.e. the remnant mass function adopted from \citealt{Fryer2012}), we cannot rule out these models conclusively. It may be that a different remnant mass function (or some other metaparameter in binary population synthesis) produces results more in line with the expectations from the above prescriptions. The same holds for the \citeauthor{Bray2016} prescriptions, which suggest that natal kicks are drawn from a linear equation, including a dependence on $M_{\rm ej}/\Mc$. Testing these models against the output of our method under different binary and stellar evolution assumptions is an avenue for further research. 

\begin{figure}
    \includegraphics[width=0.95\linewidth]{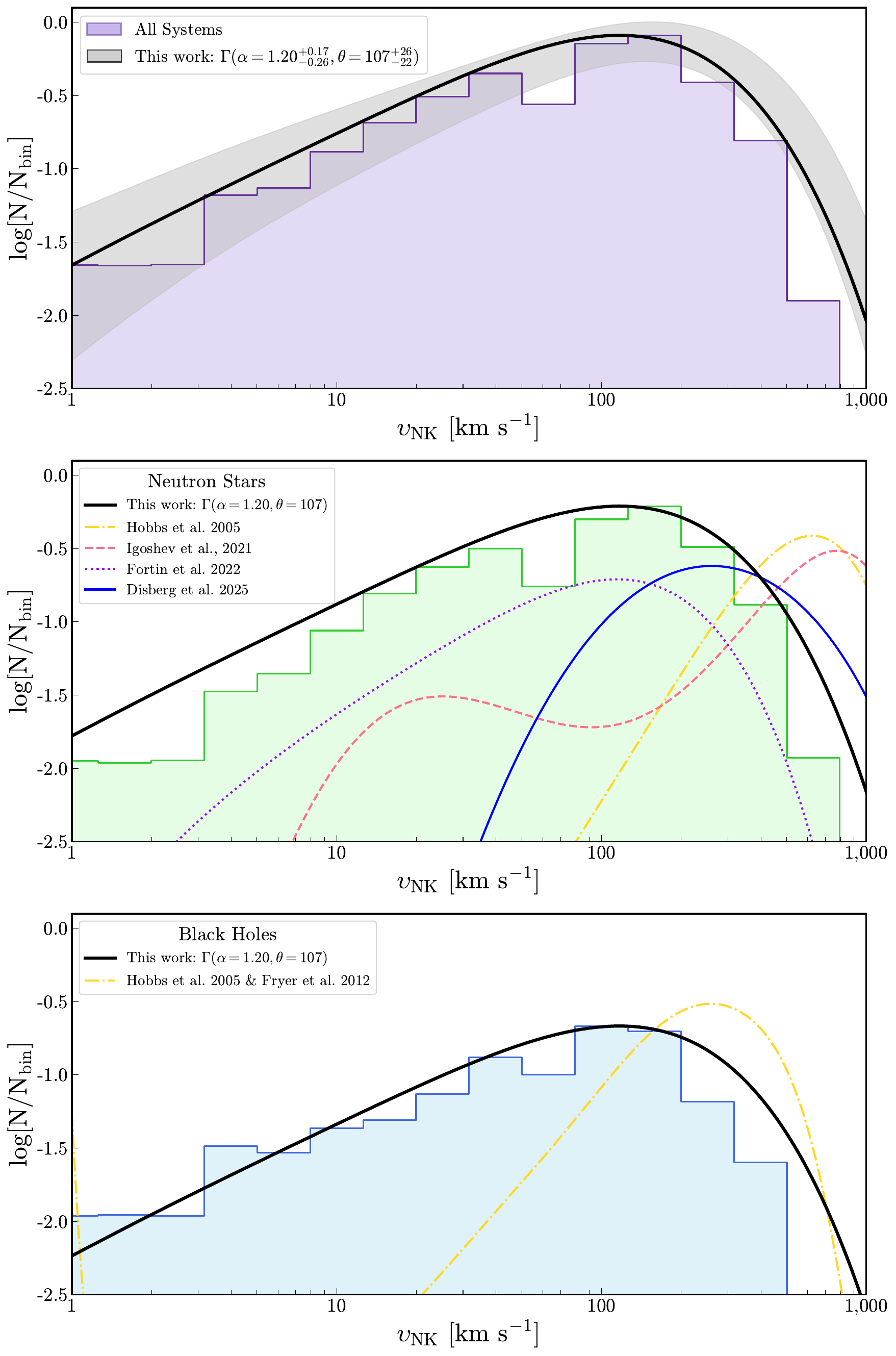} 
    \caption{$\vnk$\ distributions for (a) all sources, (b) neutron stars, and (c) black holes, compared to our Gamma distribution fit, and literature model distribution curves. All distributions are normalised to the size of the relevant population (66 sources total (a), 48 neutron stars (b), 18 black holes (c)).}
    \label{Figure:VNKLitDistributions}
\end{figure}

The above serves as a summary of some of the more commonly cited/implemented descriptions of natal kicks, and is in no way an exhaustive literature review on the subject. In Figure\,\ref{Figure:VNKLitDistributions}, we show the comparison between our proposed kick distribution and other probabilistic distributions put forth by previous studies. The Maxwellian of mean $\sigma=265$\,\kmps\ is not consistent with our data - even when considering fallback processes for black holes (however, given that subsequent research has called this into question, this is, perhaps, unsurprising). The log-normal distribution put forth by \citet{Disberg2025} also fails to describe our results (although we cannot rule out a log-normal fit with different parameters).  The Gamma distribution proposed by \citet{Fortin2022} is also a reasonable fit, consistent with our results (now including black holes) $\sim80$\% of the scenarios (when comparing this distribution to 66 distinct natal kicks, one drawn from each systems' distribution, 1 million times) (as may be expected, given they are both Gamma distributions of similar means). Discussion on natal kick estimates for {\em individual} systems can be found in Appendix\,\ref{Section:App_PrevEstimates}, and comparison between our results and predictions from deterministic models of natal kicks is presented in Appendix\,\ref{Section:App_GBBE}.

\subsection{Irreproducible systems}
\label{Section:Failed}

There are 89 compact object binaries presented in \citetalias{Zhao2023}; of these, 9 are non-interacting systems \& 8 do not have well-constrained companion masses or orbital periods. Two of the black hole XRBs have since been reclassified as likely not hosting compact objects (HD 96670 [\citealt{Naze2025}] \& MWC 656 [\citealt{janssens2023, rivinius2024, Dzib2025}]), and a new black-hole XRB has since been identified (Swift J1727.8-1613 [\citealt{MataSanchez2025}]). This leaves a catalogue of 71 XRBs with sufficiently well-constrained characteristics to be included in our modelling; 66 of these are successfully simulated and these make up our final sample and the basis for this paper's results.

Regarding the remaining 5 objects, these comprise 4 pulsars and 1 NS-HMXB for which constraints on their formation history are indeterminate: J1012+5307; J1023+0038; J1311--3430; J1653--0158; and 4U\,1700--37 (see Table\,\ref{Table:Failed}). The NS-HMXB, 4U\,1700--37, has a dedicated discussion in Paper~I; here, we focus on the remaining 4 pulsar systems,  each of which consists of a neutron star (\moneobs\,=\,1.6--2.4\,\Msun) and a low-mass stellar companion (\mtwoobs\,=\,0.01--0.22\,\Msun), with orbital periods spanning \porbobs\,=\,1--5 hours. These systems proved difficult to replicate through standard population synthesis channels - likely due to their low system masses and low mass ratios ($q$\,=\,0.05--0.15) making it more difficult for them to survive mass-transfer \& common envelope intervals without merging (this is also the case for a handful of other systems for which only a small number of analogues are identified, such as J1431--4715). These systems may be modelled (more) successfully under different physical assumptions (e.g. different common envelope efficiencies or descriptions of stellar winds), however this is beyond the scope of this work. Given that each of these 5 systems have `mid-range' peculiar velocities, we are satisfied that their exclusion is unlikely to drastically impact our results. However, future studies will benefit from increased numbers of systems, perhaps including these in addition to, for example, non-interacting systems \& gravitational-wave sources.

%\subsection{Population and Observational Selection Biases}
\subsection{Selection Biases and Limitations of our Approach}
\label{Section:Biases}

The most important selection bias here is that our work has focused exclusively on compact objects in binary systems, by virtue of the predominant optical selection. So our sample carries with it biases inherent to source {\em survival} (i.e., which binaries survive a natal kick) as well as {\em detection} (which may be related to orbital separation, accretion activity, and/or Galactic location). While our framework allows partial control of the first of these, as part of the \cosmic\ simulation framework, addressing the second is not possible without detailed population-level modelling including corrections for telescope sensitivities, extinction along various sightlines, and duty cycles of X-ray binary outbursts, amongst other effects. We thus cannot, and do not claim to, present a complete and unbiased natal-kick distribution for the population of compact objects in the Milky Way. 

Several other limitations of our work are discussed in Paper I. These include the assumption of an idealised progenitor population drawing uniform samples from a wide swath of $M_1$, $M_2$ and $P_{\rm orb}$ parameter space, in order to isolate the effects of natal kicks without worrying about progenitor priors; the assumptions that are inherent to BSE-like semi-analytic frameworks that optimise on speed instead of physical fidelity; and our inability to simulate a few specific systems (e.g., \fouru) that may have followed rare evolutionary channels (further details on the sources that we failed to model can be found in Appendix\,\ref{Section:App_Failures}). We also adopted canonical assumptions for some key physical parameters that control common envelope efficiency, stellar-wind mass-loss, and magnetic braking (see Paper I for validation tests of variations in these canonical assumptions, and Section\,\ref{Section:KickMechanisms} for specific discussion on supernova channels and remnant mass-functions). 

When generating our population we employ flat priors for initial binary masses, orbital periods, and eccentricities (for details, see Paper I) rather than a more `realistic' Galactic distribution of these parameters. Whilst such choice of priors will not impact any single simulation nor its resultant natal kick, it would change the overall {\em distribution} of kicks due to different weightings of the ZAMS progenitor properties. We conducted a limited test of this alternative approach by adopting a weighting based on varying the initial mass function, binary mass ratio, orbital-period and eccentricity distributions to match constraints from observations, and running our full suite of simulations with these choices \citep[][details can be found in Appendix\,\ref{Section:App_Results}]{Kroupa2001, Sana2012}. Figure\,\ref{Figure:App_WeightedViolins} shows the resultant $\vnk$ distributions based on this alternative prior weighting. 

Qualitatively, there is little difference when compared to our primary methodology without weighting in most cases, though there are instances where larger changes become apparent. While this serves as a limited test of the robustness f our results, a more complete investigation of realistic Galactic priors must be reserved for future work. For further discussion, see Appendix \ref{Section:App_Results}.

An additional restriction is our implicit assumption that XRBs formed exclusively through isolated binary evolution in the Galactic disc, which allows $\vpec$ to be used as a tracer for the kinematic recoil imparted at compact-object formation. This assumption need not hold for systems that formed in, or were subsequently modified by, interactions in  dense stellar environments. Dynamical encounters in globular clusters can assemble compact-object binaries and ultimately also eject them into the Galactic field, resulting in a systemic velocity that is not a good proxy of the supernova impulse \citep[e.g., ][]{Giesler2018,Kremer2018}. Similarly, other plausible origin sites in the halo would also  invalidate the assumption of disc birth location, as might be the case for Gaia-BH3 \citep{Gaia2024, balbinot24, marinpina26}. However, cluster calculations suggest that dynamically ejected XRBs likely constitute only a small fraction of the presently observable Galactic field population \citep{Kremer2018}, making this channel unlikely to dominate in our sample.

Yet another potential pathway to formation of XRBs that we do not consider here is through evolution in higher-order multiples, most notably triple systems. Triple evolution allows for qualitatively different pathways to XRB formation: Kozai--Lidov oscillations can drive initially wide inner binaries to high eccentricities, followed by tidal shrinkage and eventual mass transfer, thereby producing progenitor configurations that would not arise through isolated binary evolution \citep[e.g., ][]{Naoz2016,Shariat2025}. The presence of a tertiary also alters the response of the system to supernova mass loss and kicks, potentially changing or disrupting both the inner and outer orbits \citep{LuNaoz2019}. V404\,Cyg is thought to possess a tertiary at a projected separation of at least $\sim$\,3,500\,AU, and the survival of the triple system strongly constrains the black-hole natal kick to $\lesssim$\,5\,km\,s$^{-1}$ \citep{Burdge2024}. Reassuringly, V404,Cyg is also among the systems for which we infer a relatively small $\vnk$ (though the MAP value is formally above the limit found by \citeauthor{Burdge2024}, we concur that the system's parameters are consistent with no natal kick (i.e. $\upsilon_{\rm NK,min}=0$\,\kmps)), despite our binary-only modelling; this qualitative success can be traced to the slow observed kinematic boundary condition that needs to be satisfied within our synthesis framework. Properly quantifying the contribution of this channel will require explicit triple-population modelling, which is not included in \cosmic\ at present. 

For several, and potentially all, of the limitations listed above, it is important to keep in mind that our work is focusing on comparisons between representative samples of black-hole and neutron-star binaries. These two populations suffered similar observational selection biases in many respects --- X-ray duty-cycle filters, optical follow-up sensitivity limits; though it is arguable whether more time and effort may be spent by the community on characterising black hole candidates, as opposed to neutron stars. The effect of these biases are thus somewhat mitigated when discussing our comparisons of the $\vnk$ distributions between the compact-object types.

\section{Summary}
\label{Section:Summary}

The primary result of this study is that black holes and neutron stars in compact binaries with a stellar companion appear to share similar natal kick distributions (see Figures\,\ref{Figure:BHNSHistogram} \&\,\ref{Figure:VNKLitDistributions}) which are well-described with a Gamma distribution with $\alpha=1.20^{+0.17}_{-0.26}$ \& $\theta=107^{+26}_{-22}$\,\kmps.

In the absence of a comprehensive and established description of the physical drivers of kicks and their expected direction \& velocity, the Gamma distribution of mean $\approx128$\,\kmps\ and scatter $\sigma$\,=\,117\,\kmps\ provides a robust observationally motivated model, suitable for implementation in population synthesis studies and related modelling, in so far as selection effects are similar for sampled black-hole as well as neutron-star systems. 

Studies of supernova mechanisms and the associated natal kicks have propounded the idea that neutron-star natal kicks are primarily attributed to asymmetries in the progenitor baryonic envelope, whereas black-hole natal kicks are dominated by neutrino emission  and are, consequently, smaller. Whilst the specific drivers of these natal kicks and their relative contribution cannot be exactly constrained, the fact that evidence shows black holes can receive kicks consistent with those imparted to neutron stars upon formation points to an incomplete understanding of some supernovae processes. Specifically, there is evidence that the dominant channel for black-hole formation is one that imparts moderate kicks (a few 10s to 200\,\kmps).

It has been suggested that a delayed explosion may be more conducive to larger natal kicks \citep{Fryer2004,Scheck2004} - if this holds true for black holes, this may be responsible for the larger natal kicks they are subject to. However, it is these delayed supernovae that are expected to populate the observed mass gap between 3-5\,\Msun, the validity of which remains contentious.

Whilst hydrodynamic simulations and theoretical studies of black hole formation are valuable, the high-precision kinematic observations of the Galactic population analysed herein have shown that these must be complemented by data. A reckoning regarding the disparate behaviour of natal kicks for black holes and neutron stars now seems unavoidable. Any future work constraining the mechanisms behind natal kicks ought to be consistent with the possibility of equivalent natal kicks for neutron stars and black holes. 

The legitimacy (and precision) of these constraints will be reinforced with further study. Upcoming {\gaia} data releases DR4 and DR5 will improve current astrometric constraints, while also extending the sample of sources for which kinematics is measured. Beyond there are large potential samples of accreting binaries being investigated \citep[e.g., ][]{gandhi2022, zhao26}. The challenge will be the robustly characterise these, and to measure their kinematics. The Vera Rubin Observatory's Legacy Survey of Space and Time (LSST) should extend \gaia's reach deeper by several mag in brightness \citep{johnson19, saunders_lsstastrometry}, though spectroscopy on the faintest candidates will likely need to await the Extremely Large Telescope era \citep{elt}. Furthermore, extending the above analysis to populations of non-interacting compact objects expected with DR4 and DR5 will remove some of the astrophysical bias associated with our focus here on the accreting binary channel alone \citep{janssens2022, janssens2023, elbadry24, chawla2025, nagarajan2025_dr4_dr5}. 

%\vspace{-0.4cm}
\section*{Acknowledgments}
\label{LastPage}
CDB, PG, YZ and CK thank UKRI Science \& Technology Facilities Council for support. PG also acknowledges funding from a Royal Society Senior Leverhulme Trust Fellowship (SRF$\backslash$R1$\backslash$241074). PN is supported by the Thai Government’s Development and
Promotion of Science and Technology Talents Project (DPST). PC is supported by the Leverhulme Trust. PG and KB acknowledge fruitful discussions with participants at the KITP workshop ``Stellar-Mass Black Holes at the Nexus of Optical, X-ray, and Gravitational Wave Surveys'' coordinated by Kareem El-Badry, Jessica Lu, Selma de Mink and Tejaswi Venumadhav, and supported by grant NSF PHY-2309135 to the Kavli Institute for Theoretical Physics (KITP). KB further acknowledges support from the Falco-DeBenedetti Career Development Professorship, \emph{Chandra} Award Numbers GO223031A, GO324037X, and GO425032X, NASA ATP Grant 80NSSC24K0768, and NASA LISA Preparatory Science Program Grant 80NSSC24K0361.

Part of the simulations/calculations were completed using an 36-core Intel Xeon machine with 512 GB RAM at the AI Center of Excellence, Shiv Nadar Institution of Eminence, NCR Delhi. 

We wish to thank all those involved in the development and maintenance of \cosmic\ \citep{cosmic}, and acknowledge the use of {\tt TOPCAT} \citep{topcat} and {\tt NumPy} \citep{numpy}. This work made use of OverCite \citep{Shariat2026}, an in-editor citation tool for \LaTeX. NumPy (Harris et al. 2020); Pandas(McKinney et al. 2010); Matplotlib (Hunter 2007); SciPy (Virtanen et al. 2020b); Astropy (Robitaille et al. 2013).

%\vspace{-0.4cm}
\section*{Data Availability}
No new data are presented in this work. For details on the observational parameter boundary conditions for individual compact object binaries, we refer the reader to \citepalias{Zhao2023} and references therein. \textsc{cosmic} is publicly available on {\tt GitHub} -- refer to \citet{cosmic}. 

\bibliographystyle{mnras}
\bibliography{references} 

\clearpage
\onecolumn

\appendix

\section{Terminology}
\label{Section:App_Terminology}
\begin{figure*}
    \centering
    \includegraphics[width=0.45\linewidth]{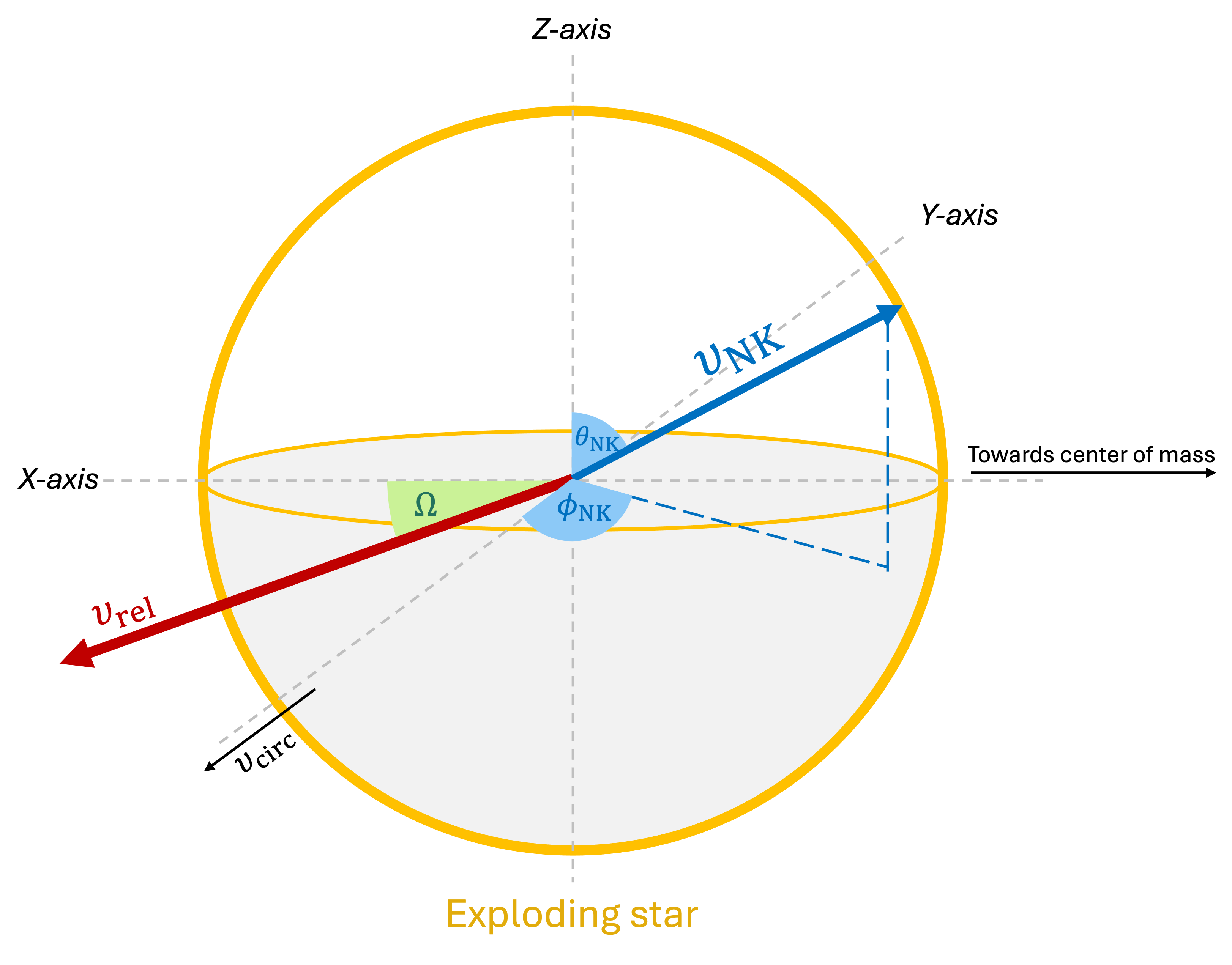}
    \caption{Diagram showing geometry of natal kicks in this study. The X- and Y- axes define the binary orbital plane: X is directed towards the centre of mass and Y is in the direction of orbital motion in the event of a circular orbit. The polar and azimuthal angles are $\thetank$ \& $\phink$, respectively. $\pi-\Omega$ gives the angle between the orbital motion and the centre of mass. Note: our geometry tries to adhere to generic spherical geomtery conventions, but results in slight differences from some previous kick studies.}
    \label{Fig:App_NKConf}
\end{figure*}

This section lists salient notations and terminology used in this work (Table \ref{Table:App_Terminology}) and summarises the geometry used when applying the natal kicks (\ref{Fig:App_NKConf}). These were presented in Paper~I, and are reproduced here to aid the reader.

\begin{table*}[h!]
    \centering
    \caption{Summary of terminology used in this paper. All masses are in units \Msun; orbital periods in days (unless otherwise specified); separations in \Rsun\ (unless otherwise specified); velocities in \kmps; and angles in degrees.}
    \begin{tabular}{lcr}
     Physical Parameter    &  Symbol\\
     \hline\hline
     ZAMS mass (compact object progenitor)  & \monei \\
     ZAMS mass (companion progenitor)       & \mtwoi \\
     ZAMS orbital period                    & \porbi  \\
     ZAMS orbital separation                & \sepi  \\
     ZAMS eccentricity                      & \ei\\
     Mass at onset of SN (compact object progenitor)& \monepre \\
     Mass at onset of SN (companion progenitor)     & \mtwopre \\
     Orbital period at onset of SN                  & \porbpre  \\
     Orbital separation at onset of SN              & \seppre  \\
     Eccentricity at onset of SN    & \epre\\
     Post-SN mass (compact object)  & \monepost \\
     Post-SN mass (companion)       & \mtwopost \\
     Post-SN orbital period         & \porbpost  \\
     Post-SN orbital separation     & \seppost  \\
     Post-SN eccentricity           & \epost\\
     Current mass (compact object)  & \moneobs \\
     Current mass (companion)       & \mtwoobs \\
     Current orbital period         & \porbobs  \\
     Current eccentricity           & \eobs\\
     Compact object mass            & $\Mc$ \\
     Mass transfer rate             & \mdot \\
     Mass ratio (=\,$M_2/M_1$)      & $q$ \\
     Relative velocity              & $\vorb$ \\
     Natal kick magnitude           & $\vnk$ \\
     Blaauw (mass-loss) kick magnitude      & $\vmlk$  \\
     Polar angle of kick (see Figure\,\ref{Fig:App_NKConf})  & $\theta$\ \\
     Azimuthal angle of kick (see Figure\,\ref{Fig:App_NKConf}) & $\phi$\ \\
 The angle between $\vorb$ and centre of mass subtracted from $\pi$ (see Figure\,\ref{Fig:App_NKConf}) & $\Omega$ \\
     \hline
    \end{tabular}
    \label{Table:App_Terminology}
\end{table*}

\clearpage

\section{Summary $\upsilon_{\rm NK}$ for all systems}
\label{Section:App_Results}

\subsection{Compilation of primary $\vnk$ estimates}

Table\,\ref{Table:AllSystems} tabulates the best $\vnk$ estimators for each of the 66 systems, listing the maximum aposteriori probability (MAP) value and the 68\% highest-density interval (HDI) confidence region. The same values are listed for the best-estimates of the mass-loss kicks $\vmlk$. We opt to show the minimum allowed natal ($\upsilon_{\rm NK,min}$) kick to highlight the handful of black holes which \textit{require} substantial natal kicks, and the maximum mass-loss kick ($\upsilon_{\rm MLK,max}$) to highlight the cases where this is insufficient to explain the system's peculiar velocity. The $\vpec$ values from \citetalias{Zhao2023}, corrected for Galactic velocity dispersion $\upsilon_{\rm dis}$), are also listed as the kinematic boundary condition for each system. 

%Table\,\ref{Table:AllSystems} tabulates the $\vnk$ estimates for each of our 66 simulated systems, including MAP values and HDI intervals of the respective probability distributions, together with the observational boundary conditions (i.e., the $\vpec$ values from \citetalias{Zhao2023}, corrected for Galactic velocity dispersion $\upsilon_{\rm dis}$). 

Figure\,\ref{Figure:App_WeightedViolins} displays the full distribution of possible natal kicks for each system as a violin plot, collated in a single panel and arranged in alphabetical order of source name for ease of the reader. These results are based on the methodology detailed in Paper\,I, and briefly outlined in the main text of the present work, wherein possible ZAMS parameter space is uniformly sampled for our core simulations. Our primary natal-kick inferences are thus agnostic to any reality of any particular Galactic population of objects.

\onecolumn

\begin{center}
\begin{longtable}{@{\extracolsep{\fill}}|l|c|cc|ccc|ccc|@{}}
\caption[Natal Kick Constraints (a)]{\textbf{Natal Kick Constraints for our sample sources:} Results for all 66 compact-object binaries in our sample. $\upsilon_{\rm pec}$ is the peculiar velocity of a given system \citep{Zhao2023}; $\left(\upsilon_{\rm pec}^2 -\upsilon_{\rm dis}^2\right)^{0.5}$ indicates the peculiar velocity of the system once a fixed intrinsic stellar velocity dispersion ($\upsilon_{\rm dis}$\,=\,40\,\kmps - see \citet{dashwoodbrown2026i} for further discussion) has been subtracted - it is this velocity that we match our simulations to. The range is based on the nominal errors; where $\upsilon_{\rm pec}<40$\,\kmps, the system's peculiar velocity may be entirely due to the intrinsic velocity dispersion, and the range is quoted as $0-10$\,\kmps. $\vnk$ indicates inferred natal kick velocities based on uniform priors, including the maximum, MAP, and 68\% highest-posterior-density interval for the mass-loss-kick, and the minimum, MAP, and 68\% highest-posterior-density interval for the natal kick magnitudes. All velocities are in \kmps.} \\
\hline
Name & Type & $\upsilon_{\rm pec}$ &  $\left(\upsilon_{\rm pec}^2 -\upsilon_{\rm dis}^2\right)^{0.5}$ & \multicolumn{3}{@{}c@{}}{$\vmlk$} & \multicolumn{3}{@{}c@{}|}{$\vnk$}  \\
 &  & &  & MAP & 68\% & Max & MAP & 68\% & Min \\
    \hline
    \hline
1A 0535+262 & NS-HMXB & 45$\pm$4 & 18--22 & 19 & 17--22 & 34 & 42 & 14--85 & 0 \\ \hline
        1E 1145.1-6141 & NS-HMXB & 56$\pm$10 & 33--47 & 28 & 22--36 & 57 & 152 & 79--239 & 0 \\ \hline
        2A 1822-371 & NS-LMXB & 254$\pm$33 & 218--283 & 64 & 45--70 & 140 & 301 & 253--350 & 178 \\ \hline
        2FGL J1019.0-5856 & NS-HMXB & 31$\pm$3 & 0--10 & 3 & 3--4 & 8 & 46 & 37--56 & 28 \\ \hline
        2S 0114+650 & NS-HMXB & 22$\pm$2 & 0--10 & 5 & 4--5 & 8 & 42 & 34--53 & 13 \\ \hline
        2S 0921-630 & NS-LMXB & 38$\pm$7 & 0--10 & 3 & 3--3 & 5 & 9 & 4--16 & 0 \\ \hline
        2S 1145-619 & NS-HMXB & 16$\pm$4 & 0--10 & 5 & 4--6 & 19 & 26 & 20--39 & 2 \\ \hline
        4U 1254-69 & NS-LMXB & 155$\pm$18 & 133--167 & 41 & 27--60 & 78 & 196 & 166--230 & 127 \\ \hline
        4U 1538-52 & NS-HMXB & 77$\pm$10 & 58--75 & 32 & 30--44 & 59 & 357 & 310--448 & 102 \\ \hline
        4U 1543-475 & BH-LMXB & 99$\pm$15 & 77--104 & 17 & 11--42 & 98 & 122 & 101--147 & 8 \\ \hline
        4U 1636-536 & NS-LMXB & 165$\pm$13 & 148--172 & 40 & 7--46 & 66 & 194 & 173--240 & 150 \\ \hline
        4U 1700+24 & NS-LMXB & 39$\pm$9 & 0--10 & 0 & 0--1 & 10 & 11 & 7--15 & 0 \\ \hline
        4U 1735-444 & NS-LMXB & 60$\pm$31 & 22--68 & 31 & 26--38 & 69 & 43 & 15--97 & 0 \\ \hline
        4U 2206+543 & NS-HMXB & 28$\pm$4 & 0--10 & 3 & 3--4 & 8 & 46 & 37--55 & 30 \\ \hline
        A 0620-00 & BH-LMXB & 44$\pm$7 & 15--20 & 18 & 14--19 & 33 & 13 & 4--26 & 0 \\ \hline
        Aql X-1 & NS-LMXB & 42$\pm$12 & 10--18 & 17 & 8--20 & 21 & 20 & 10--46 & 0 \\ \hline
        B1259-63 & NS-HM-PSR & 25$\pm$8 & 0--10 & 4 & 3--5 & 13 & 33 & 15--58 & 0 \\ \hline
        BW Cir & BH-LMXB & 352$\pm$99 & 252--448 & 18 & 12--39 & 129 & 363 & 309--461 & 267 \\ \hline
        Cen X-3 & NS-HMXB & 102$\pm$4 & 91--98 & 46 & 41--47 & 50 & 561 & 518--638 & 442 \\ \hline
        Cen X-4 & NS-LMXB & 419$\pm$69 & 349--485 & 34 & 30--43 & 64 & 394 & 378--475 & 353 \\ \hline
        Cyg X-1 & BH-HMXB & 22$\pm$3 & 0--10 & 11 & 3--11 & 14 & 32 & 10--43 & 0 \\ \hline
        Cyg X-2 & NS-LMXB & 164$\pm$8 & 151--167 & 46 & 38--62 & 124 & 191 & 171--290 & 62 \\ \hline
        GRO 1655-40 & BH-LMXB & 162$\pm$6 & 151--162 & 32 & 20--43 & 129 & 208 & 184--240 & 42 \\ \hline
        GRS 1124-684 & BH-LMXB & 119$\pm$15 & 97--126 & 7 & 5--15 & 23 & 128 & 110--133 & 89 \\ \hline
        GRS 1915+105 & BH-LMXB & 32$\pm$15 & 0--10 & 5 & 3--8 & 51 & 8 & 3--14 & 0 \\ \hline
        GX 1+4 & NS-LMXB & 189$\pm$8 & 177--193 & 43 & 35--55 & 92 & 281 & 250--303 & 215 \\ \hline
        GX 301-2 & NS-HMXB & 58$\pm$4 & 39--45 & 18 & 17--21 & 28 & 362 & 336--407 & 282 \\ \hline
        GX 339-4 & BH-LMXB & 166$\pm$35 & 127--195 & 30 & 21--43 & 128 & 207 & 159--231 & 52 \\ \hline
        H 1705-250 & BH-LMXB & 221$\pm$109 & 110--325 & 20 & 11--23 & 29 & 303 & 170--331 & 95 \\ \hline
        HD 259440 & NS-HMXB & 10$\pm$3 & 0--10 & 5 & 3--5 & 15 & 37 & 20--45 & 0 \\ \hline
        Her X-1 & NS-LMXB & 126$\pm$11 & 109--129 & 63 & 49--87 & 118 & 200 & 150--254 & 14 \\ \hline
        IGR J00370+6122 & NS-HMXB & 23$\pm$8 & 0--10 & 3 & 3--6 & 19 & 64 & 8--79 & 0 \\ \hline
        IGR J08408-4503 & NS-HMXB & 39$\pm$5 & 0--10 & 4 & 3--4 & 7 & 48 & 40--58 & 33 \\ \hline
        IGR J17544-2619 & NS-HMXB & 44$\pm$4 & 16--19 & 14 & 13--16 & 26 & 71 & 61--109 & 42 \\ \hline
        J0348+0432 & NS-LM-PSR & 62$\pm$12 & 38--57 & 23 & 4--26 & 33 & 108 & 95--145 & 42 \\ \hline
        J1048+2339 & NS-LM-PSR & 158$\pm$35 & 118--187 & 36 & 32--42 & 86 & 239 & 208--281 & 111 \\ \hline
        J1306-4035 & NS-LM-PSR & 121$\pm$17 & 98--130 & 39 & 26--45 & 83 & 193 & 131--221 & 92 \\ \hline
        J1417-4402 & NS-LM-PSR & 99$\pm$19 & 73--107 & 35 & 27--53 & 94 & 140 & 94--219 & 1 \\ \hline
        J1431-4715 & NS-LM-PSR & 123$\pm$16 & 101--131 & 4 & 3--5 & 5 & 128 & 115--138 & 108 \\ \hline
        J1622-0315 & NS-LM-PSR & 144$\pm$10 & 129--149 & 5 & 3--5 & 5 & 152 & 141--162 & 129 \\ \hline
        J1628-3205 & NS-LM-PSR & 114$\pm$28 & 80--133 & 6 & 4--6 & 6 & 136 & 104--147 & 84 \\ \hline
        J1723-2837 & NS-LM-PSR & 110$\pm$11 & 93--113 & 6 & 5--7 & 35 & 136 & 120--141 & 109 \\ \hline
        J1816+4510 & NS-LM-PSR & 165$\pm$26 & 135--186 & 35 & 28--38 & 40 & 166 & 157--179 & 134 \\ \hline
        J2039-5617 & NS-LM-PSR & 113$\pm$28 & 80--132 & 6 & 5--6 & 7 & 138 & 105--148 & 84 \\ \hline
        J2129-0429 & NS-LM-PSR & 201$\pm$24 & 174--221 & 42 & 34--51 & 74 & 211 & 182--227 & 128 \\ \hline
        J2215+5135 & NS-LM-PSR & 118$\pm$17 & 94--127 & 5 & 5--7 & 22 & 219 & 185--248 & 152 \\ \hline
        J2339-0533 & NS-LM-PSR & 64$\pm$13 & 40--60 & 38 & 33--43 & 69 & 60 & 28--89 & 0 \\ \hline
        LS 5039 & NS-HMXB & 88$\pm$3 & 76--82 & 73 & 60--77 & 92 & 265 & 92--399 & 0 \\ \hline
        LS I +61 303 & NS-HMXB & 9$\pm$4 & 0--10 & 5 & 4--6 & 17 & 33 & 24--45 & 0 \\ \hline
        MAXI J1305-704 & BH-LMXB & 58$\pm$22 & 26--58 & 11 & 10--16 & 29 & 52 & 34--62 & 7 \\ \hline
        MAXI J1820+070 & BH-LMXB & 72$\pm$16 & 47--74 & 10 & 8--14 & 29 & 66 & 50--73 & 29 \\ \hline
        MXB 1659-298 & NS-LMXB & 260$\pm$204 & 55--458 & 43 & 30--62 & 147 & 201 & 124--321 & 0 \\ \hline
        RX J0146.9+6121 & NS-HMXB & 12$\pm$4 & 0--10 & 5 & 4--8 & 21 & 21 & 16--32 & 0 \\ \hline
        SS 433 & BH-HMXB? & 44$\pm$16 & 12--26 & 35 & 7--41 & 44 & 52 & 33--61 & 11 \\ \hline
        Sco X-1 & NS-LMXB & 168$\pm$10 & 153--172 & 6 & 5--41 & 63 & 215 & 174--233 & 113 \\ \hline
        Ser X-1 & NS-LMXB & 110$\pm$31 & 73--132 & 6 & 4--7 & 36 & 136 & 100--153 & 75 \\ \hline
        Swift J1727.8-1613 & BH-LMXB & 207$\pm$7 & 196--210 & 39 & 33--54 & 71 & 211 & 188--234 & 154 \\ \hline
        Swift J1753.5-0127 & BH-LMXB & 110$\pm$57 & 50--156 & 11 & 11--20 & 42 & 151 & 91--173 & 17 \\ \hline
        V404 Cyg & BH-LMXB & 45$\pm$3 & 20--23 & 21 & 11--24 & 42 & 21 & 9--35 & 0 \\ \hline
        V4641 Sgr & BH-HMXB & 92$\pm$7 & 77--89 & 85 & 26--93 & 111 & 109 & 84--253 & 0 \\ \hline
        Vela X-1 & NS-HMXB & 59$\pm$7 & 38--48 & 28 & 24--45 & 56 & 249 & 118--335 & 0 \\ \hline
        X Persi & NS-HMXB & 18$\pm$8 & 0--10 & 5 & 4--6 & 17 & 25 & 21--43 & 5 \\ \hline
        XSS J12270-4859 & NS-LM-PSR & 131$\pm$21 & 105--145 & 6 & 5--32 & 62 & 141 & 122--204 & 118 \\ \hline
        XTE J1118+480 & BH-LMXB & 143$\pm$11 & 126--148 & 5 & 4--12 & 42 & 144 & 127--152 & 104 \\ \hline
        XTE J1550-564 & BH-LMXB & 79$\pm$32 & 40--96 & 1 & 1--3 & 17 & 91 & 56--98 & 36 \\ \hline
        XTE J1814-338 & NS-LMXB & 171$\pm$51 & 117--216 & 42 & 31--54 & 94 & 202 & 163--232 & 113 \\ \hline
\end{longtable}
\label{Table:AllSystems}
\end{center}

\section{Exploring more realistic progenitor parameter priors}

We also conducted one test with a different assuming a difference weighting of initial binary parameters, in order to qualitatively assess the impact of varying the ZAMS sampling. In this case, we assume the Galactic initial mass function (IMF) is as described in \citet{Kroupa2001}, and that initial mass ratios are drawn $\left[0,1\right]$ where $p(q)=q^{-0.1}$, as in \citet{Sana2012}. Initial orbital periods and eccentricities are also assumed to follow the prescriptions laid out by \citeauthor{Sana2012}, as shown in Table\,\ref{Table:IMF}. 

\begin{table*}[h]
\centering
\renewcommand{\arraystretch}{1.25}
\begin{tabular}{ l | c | c } 
Parameter & Probability & Range \\
\hline\hline
\multirow{5}{12em}{Primary Mass ($M_{\rm 1,i}$) $\left[\rm{M}_{\odot}\right]$}
& $f_{\rm p}\left(M_1\right) = M^{0.3}$ &  $M<0.8$ \\
& $f_{\rm p}\left(M_1\right) = M^{2.7}$ &  $0.08\leq M<0.5$ \\
& $f_{\rm p}\left(M_1\right) = M^{2.7}$ &  $0.5\leq M<1$ \\
& $f_{\rm p}\left(M_1\right) = M^{2.3}$ &  $1\leq M<8$ \\
& $f_{\rm p}\left(M_1\right) = M^{1.15}$ &  $M\geq8$ \\
\hline
Mass Ratio ($q_{\rm i}$) & $f_{\rm p}\left(q_{i}\right) = q^{-0.1}$ & $q\in\left[0.0,1\right]$\\
\hline
Orbital Period ($P_{\rm orb,i}$) [days] & $f_{\rm p}\left(\log P_{\rm orb,\,i}\right) = \left(\log P_{\rm orb}\right)^{-0.55}$ & $\log P_{\rm orb}\in\left[0.15,5.5\right]$ \\
\hline
Orbital Eccentricity ($e_{\rm i}$) & $f_{\rm p}\left(e_i\right) = e^{-0.45}$ & $e\in\left[0.0,0.9\right]$ \\
\end{tabular}
\caption{Descriptions of initial binary parameters used to weight natal kick distributions. The initial mass function (for the primary mass, $M_1$) comes from \citet{Kroupa01}, the remaining distributions come from \citet{Sana2012}. }
\label{Table:IMF}
\end{table*}

\begin{figure*}
    \centering
    \includegraphics[width=0.89\textwidth]{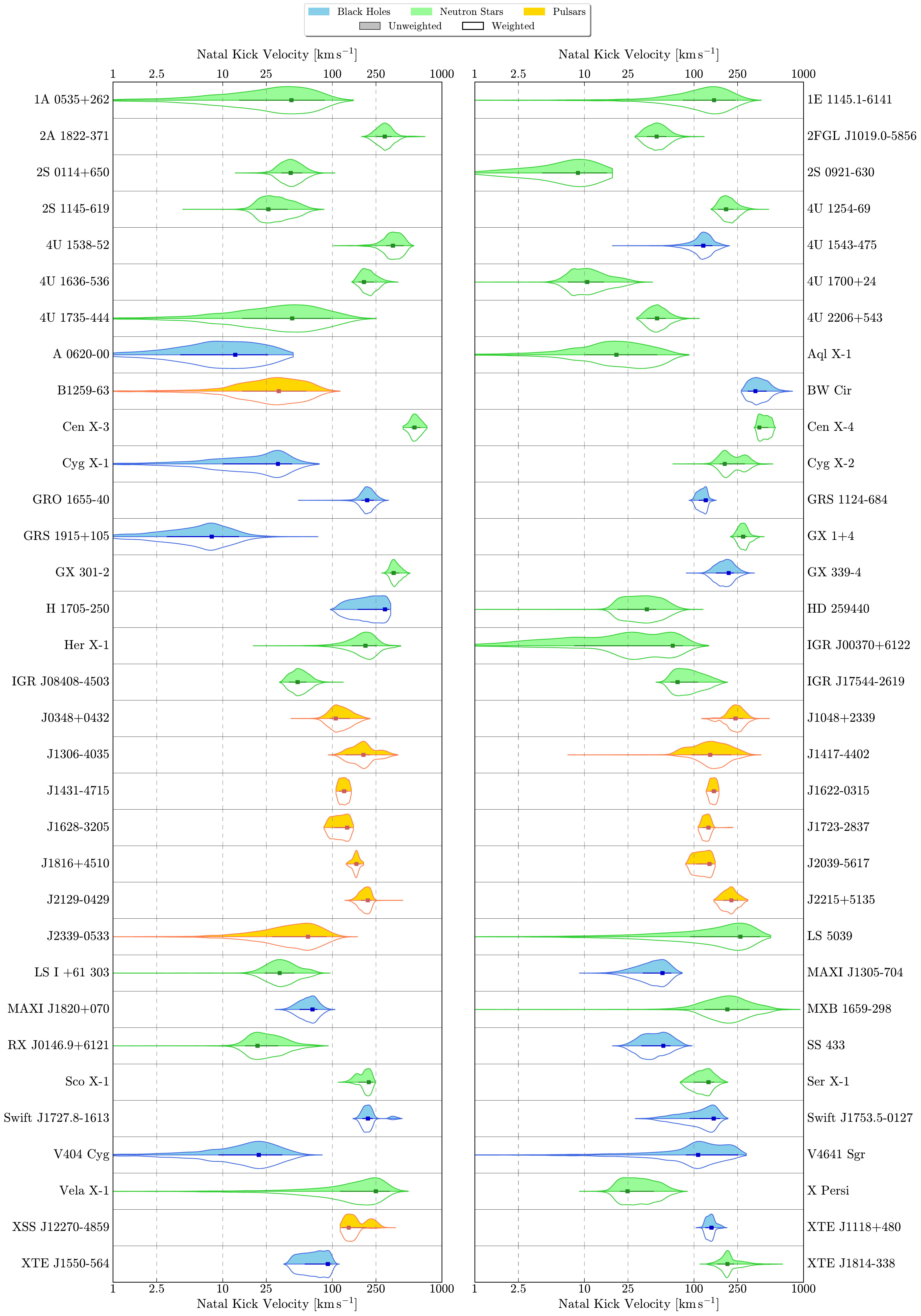}
    \caption{Violin plots showing distribution of all viable natal kick magnitudes for each system in this study. The upper (filled) violins show the raw distributions, the lower (unfilled) violins show the distributions weighted according to their initial binary parameters (see Table \ref{Table:IMF}). MAP values for the raw distributions are indicated by small square markers; internal horizontal lines indicate the HDI 68\% confidence interval of the unweighted distribution. Compact-object types are distinguished by their colour: blue for black holes, green for neutron stars, orange for pulsars. }
    \label{Figure:App_WeightedViolins}
\end{figure*}

Treating these initial binary prescriptions as probabilities, we assign a weighting for each simulated binary according to their relative `likelihood', and use these to produce the unfilled violin plots in Figure \ref{Figure:App_WeightedViolins}. This offers additional insight into the possible natal kicks for any given system, by accounting for the fact that our methodology uses broad priors which are not motivated by any observations or Galactic formation physics - however, we do not account for this in the core analysis of our work as aspects of stellar populations such as the IMF are not concretely known, adding additional sources of errors/uncertainties into our work. 

The primary focus of this work is the comparison between black holes and neutron stars, and as the use of broad \& physically unmotivated priors applies to both populations, these systematics are consistent across the two populations and therefore of limited significance in our work. Furthermore, given that the minimum and maximum natal kick values remain unchanged regardless of the shape of any initial binary priors, our inferences regarding the necessity of significant natal kicks for some black holes are unaffected. Indeed, examination of Figure \ref{Figure:App_WeightedViolins} shows that, for many systems, the distribution of natal kicks is similar when weighting according to the initial binary characteristics. The most significant differences appear to correspond to distinct formation channels. For example, for the pulsar XSS\,J12270--4859, our work shows a bimodality in the natal kicks, with peaks at roughly 125\,\kmps and 225\,\kmps. However when weighting according to initial binary characteristics, the relative contribution of the higher magnitude natal kick channel reduces to be negligible, as it corresponds to a population of binaries with higher initial mass-ratios.   
%\clearpage

\clearpage

\section{Literature Comparisons}\label{Section:App_PrevEstimates}

In this section, we comment on comparisons between our $\vnk$ estimators for individual objects, and those published in the literature. 

\begin{figure}
    \centering
    \includegraphics[width=0.85\linewidth]{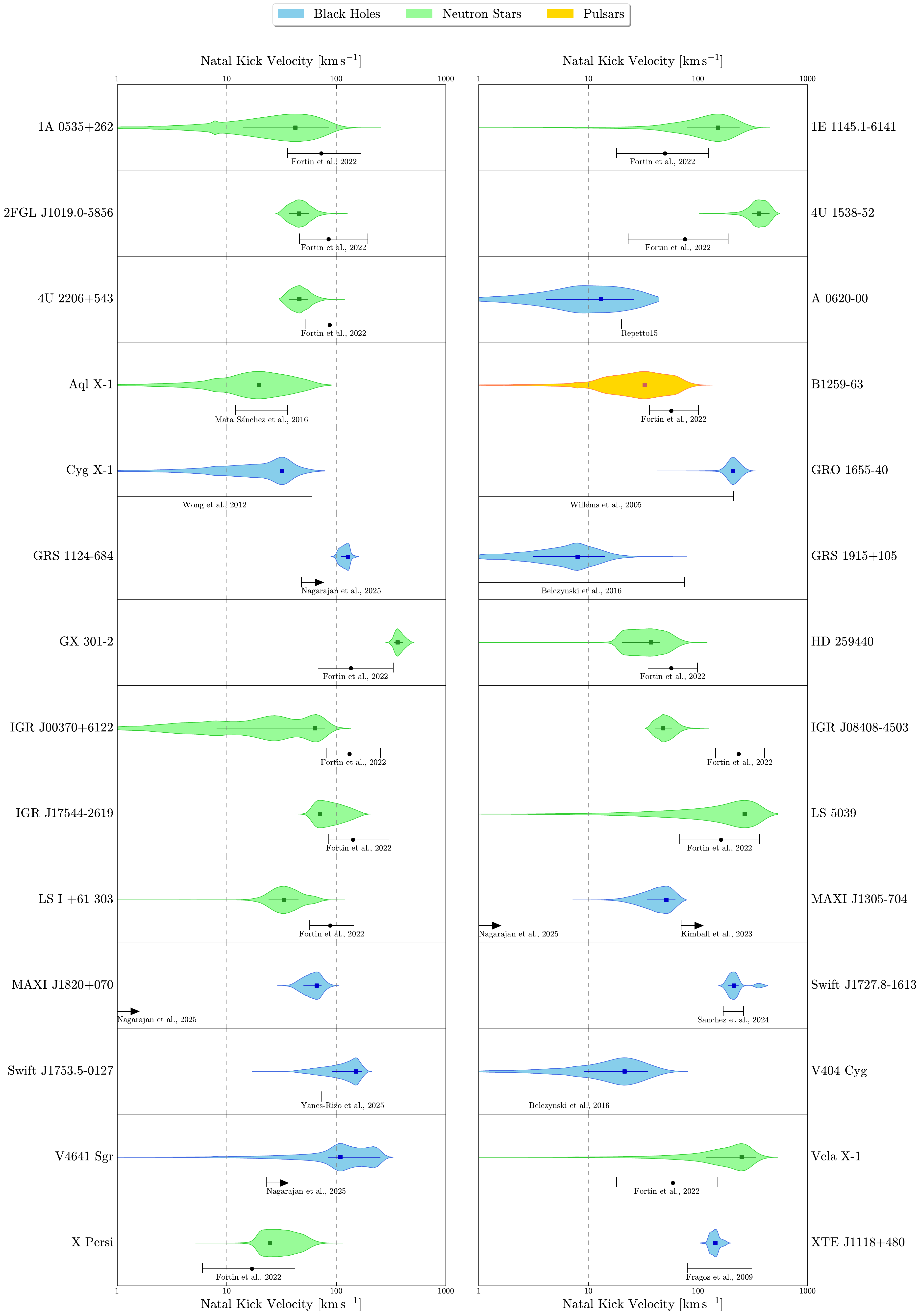}
    \caption{The distributions of plausible natal kicks for 28 systems in this study for which previous estimates on their natal kicks are available compared to those estimates. Violin plots show distribution of all viable natal kick magnitudes determined by this study; internal horizontal line indicates 68\% HDI, and MAP values are indicated as a small square. Violin plots are shaded according to their system type (blue for black holes, green for neutron stars, and yellow for neutron star pulsars). The natal kicks determined by previous studies are annotated in black, alongside the relevant citation - horizontal lines indicated ranges; small black circles indicate mean value (where applicable); and arrows denote lower limits. Comparisons between our results and those from the literature are discussed in \ref{Section:App_PrevEstimates} - they generally arise from; updated observational constraints; inclusion of pre-SN evolution in our study; and/or inclusion of an intrinsic Galactic dispersion velocity.}
    \label{Figure:App_LitSystems}
\end{figure}

The section begins with the executive summary plot in Fig.\,\ref{Figure:App_LitSystems}, showing our $\vnk$ distributions as a violin plot, with published values and/or limits indicated. The respective tabulated values are also found in Table\,\ref{Table:LiteratureValues}. In some cases where no previous natal kick estimates based on evolutionary modelling have been published, we indicate this and discuss the comparison against published $\vpec$ values (or proxies). 

Following this plot, a succinct discussion follows for each source in its dedicated sub-section. In each case, brief commentary is presented on likely causes of any discrepancies with respect to literature values. 

\begin{table}[h!]
\renewcommand{\arraystretch}{1.25}
\centering
\begin{tabular}{|l | c | c | c | l|}
\hline
Name & Type & $\vnk$ & $\upsilon_{\rm NK, lit.}$ & Literature Reference \\
\hline
\hline
1A 0535+262 & NS-HMXB & 42$^{+213}_{-42}$ & 36--167 & \citet{Fortin2022} \\
\hline
1E 1145.1-6141 & NS-HMXB & 152$^{+300}_{-152}$ & 18--125 & \citet{Fortin2022} \\
\hline
2FGL J1019.0-5856 & NS-HMXB & 46$^{+80}_{-18}$ & 46--194 & \citet{Fortin2022} \\
\hline
4U 1538-52 & NS-HMXB & 357$^{+197}_{-255}$ & 23--188 & \citet{Fortin2022} \\
\hline
4U 2206+543 & NS-HMXB & 46$^{+73}_{-16}$ & 52--172 & \citet{Fortin2022} \\
\hline
A 0620-00 & BH-LMXB & 13$^{+31}_{-13}$ & 20--43 & \citet{Repetto2015} \\
\hline
Aql X-1 & NS-LMXB & 20$^{+70}_{-20}$ & 12--36 & \citet{MataSanchez2017} \\
\hline
B1259-63 & NS-HM-PSR & 33$^{+102}_{-33}$ & 36--101 & \citet{Fortin2022} \\
\hline
\multirow{2}{*}{Cyg X-1} & \multirow{2}{*}{BH-HMXB} & \multirow{2}{*}{32$^{+47}_{-32}$} & 0$\leq$ & \citet{Nagarajan2025} \\
\cline{4-5}
 & & & 0--60 & \citet{Wong12} \\
\hline
\multirow{2}{*}{GRO J1655-40} & \multirow{2}{*}{BH-LMXB} & \multirow{2}{*}{208$^{+127}_{-166}$} & 62$\leq$ & \citet{Nagarajan2025} \\
\cline{4-5}
 & & & 0--210 & \citet{Willems2005} \\
\hline
GRS 1124-684 & BH-LMXB & 128$^{+32}_{-39}$ & 48$\leq$ & \citet{Nagarajan2025} \\
\hline
GRS 1915+105 & BH-LMXB & 8$^{+71}_{-8}$ & 0--75 & \citet{Belczynski16} \\
\hline
GX 301-2 & NS-HMXB & 362$^{+150}_{-80}$ & 68--330 & \citet{Fortin2022} \\
\hline
HD 259440 & NS-HMXB & 37$^{+84}_{-37}$ & 35--99 & \citet{Fortin2022} \\
\hline
IGR J00370+6122 & NS-HMXB & 64$^{+72}_{-64}$ & 81--253 & \citet{Fortin2022} \\
\hline
IGR J08408-4503 & NS-HMXB & 48$^{+78}_{-15}$ & 144--405 & \citet{Fortin2022} \\
\hline
IGR J17544-2619 & NS-HMXB & 71$^{+134}_{-29}$ & 85--303 & \citet{Fortin2022} \\
\hline
LS 5039 & NS-HMXB & 265$^{+267}_{-265}$ & 68--364 & \citet{Fortin2022} \\
\hline
LS I +61 303 & NS-HMXB & 33$^{+87}_{-33}$ & 57--145 & \citet{Fortin2022} \\
\hline
\multirow{2}{*}{MAXI J1305-704} & \multirow{2}{*}{BH-LMXB} & \multirow{2}{*}{52$^{+26}_{-45}$} & 70$\leq$ & \citet{Kimball2023} \\
\cline{4-5}
 & & & 0$\leq$ & \citet{Nagarajan2025} \\
\hline
MAXI J1820+070 & BH-LMXB & 66$^{+40}_{-37}$ & 0$\leq$ & \citet{Nagarajan2025} \\
\hline
\multirow{2}{*}{Swift J1727.8-1613} & \multirow{2}{*}{BH-LMXB} & \multirow{2}{*}{211$^{+223}_{-57}$} & 77$\leq$ & \citet{Nagarajan2025} \\
\cline{4-5}
 & & & 170--260 & \citet{MataSanchez2025} \\
\hline
Swift J1753.5-0127 & BH-LMXB & 151$^{+59}_{-134}$ & 73--179 & \citet{YanesRizo2025} \\
\hline
V404 Cyg & BH-LMXB & 21$^{+60}_{-21}$ & 0--45 & \citet{Belczynski16} \\
\hline
V4641 Sgr & BH-HMXB & 109$^{+221}_{-109}$ & 23$\leq$ & \citet{Nagarajan2025} \\
\hline
Vela X-1 & NS-HMXB & 249$^{+286}_{-249}$ & 18--151 & \citet{Fortin2022} \\
\hline
X Persi & NS-HMXB & 25$^{+90}_{-20}$ & 6--42 & \citet{Fortin2022} \\
\hline
\multirow{2}{*}{XTE J1118+480} & \multirow{2}{*}{BH-LMXB} & \multirow{2}{*}{144$^{+56}_{-40}$} & 80--310 & \citet{Fragos2009} \\
\cline{4-5}
 & & & 36$\leq$ & \citet{Nagarajan2025} \\
\hline

\end{tabular}
\caption{Natal kick constraints for 28 compact objects for which previous literature estimates are available. Natal kicks from this study ($\vnk$) are presented with their MAP values, along with the minimum and maximum. We present the plausible range of natal kicks (or their minimum) derived by previous studies ($\upsilon_{\rm NK, lit.}$), alongside the associated citation. All velocities are in \kmps.}
\label{Table:LiteratureValues}
\end{table}

\clearpage

\clearpage

\subsection{V404\,Cyg \& GRS\,1915+105}
Our results are consistent with \citet{Belczynski16} - the MAP and HDI values lie within the previously expected range, the minimum natal kick value (0) is the same in both their and our studies (for both V404 \& GRS1915). 

Our derived $\vnk$ distributions formally allow a small number of kicks above the ranges presented by \citeauthor{Belczynski16} for both sources. Different formation pathways could subdue these further, e.g., a triple evolution scenario has been proposed for V404\,Cyg by \citet{Burdge2024}, which would limit the maximum allowed kick further. %- i.e. our maximum natal kicks for V404\,Cyg and GRS\,1915+105 are 81\,\kmps and 79\,\kmps, respectively, as compared to respective values of 45\,\kmps\ and 75\,\kmps\ from their work.

\subsection{GRO\,J1655--40, XTE\,J1118+480, and Cyg\,X--1}
In a series of works, \citeauthor{Willems2005,Fragos2009,Wong12} present natal kick estimates for GRO\,J1655--50, XTE\,J1118+480, and Cyg\,X--1, respectively. The methodology for each paper is broadly the same (hence why we are grouping them in this section). 

\subsubsection{GRO\,J1655--40}
This source is one of the early suggested `runaway' black hole systems \citep{mirabel02} based upon astrometric measurements from the {\em Hubble Space Telescope (HST)}. Evolutionary modelling of post-SN mass-transfer tracks led to estimates of either between $\vnk$\,=\,30--160\,\kmps\ or $\vnk$\,$<$\,210\,\kmps, depending on differing treatments of the source inclinations and masses \citep{Willems2005}, including contribution from a symmetric mass-loss kick \citep{Nelemans1999}. Uncertainties in evolutionary modelling, and on estimates based on supernova characteristics from metal pollution of the donor, have been discussed by \citet{Belczynski16} who suggested that the source likely received a moderate kick across the wide allowed range of 0\,$<$\,$\vnk$\,$<$\,210\,\kmps. 

Our estimated $\vnk=208^{+127}_{-166}$\,km\,s$^{-1}$ lies at the upper end of this allowed range, with a full (min-max) range of 42--335\,\kmps. This difference can be largely attributed to new observational measurements of the source kinematics. \gaia\ DR3 reports proper motion values of $(\mu_\alpha^*$, $\mu_\delta)$\,=\,(--4.33\,$\pm$\,0.06, --7.46\,$\pm$\,0.05) mas\,y$^{-1}$, approximately 80\% larger than the early published values of (--3.3\,$\pm$\,0.5, --4.0\,$\pm$\,0.4) mas\,y$^{-1}$ based on limited {\em HST} epochs \citep{mirabel02} and also about an order of magnitude more precise along each coordinate axis. The systemic radial velocity adopted by \citetalias{Zhao2023} is $\gamma$\,=\,--167\,$\pm$\,0.6\,\kmps\ \citep{GonzalezHernandez2008}, also significantly higher in magnitude than prior measurements \citep{bailyn95, orosz97, shahbaz99}. These values by themselves would result in a $\vpec$ boost of about 40\%, together with a substantial reduction in the allowed $\vpec$\ range, relative to the early estimates, and thus accounting for our $\vnk$\ distribution excluding the lower end of the range from \citet{Willems2005}. Moreover, intervening years have seen improvements in population  modelling, including RLOF prescriptions, now incorporated within \cosmic. We also simulate the complete evolution of the source from ZAMS, not restricted to the post-SN timeframe as in \citet{Willems2005}, all of which could contribute some additional differences to the inferred kick values.

\subsubsection{XTE\,J1118+480}
This source has been extensively discussed in Paper I, including at least two possible evolutionary channels allowed within {\tt COSMIC} that might successfully produce it. 

Our $\vnk$ range is consistent with previous estimates from \citet{Fragos2009}, albeit a bit tighter, which may be attributed to reduced kinematic measurement gathered in the interim (specifically, a more tightly constrained proper motion from \gaia, smaller distance uncertainties and incorporating a directly systemic RV measurement, all discussed in \citetalias{Zhao2023}). 

There are also subtle methodological differences between the respective adopted approaches. For instance, \citeauthor{Fragos2009} fit the donor temperature and indirectly incorporate a donor mass lower-limit by consideration of evidence for CNO processing highlighted by \citet{haswell2002}. They also imposed an upper-limit of 20\,\Msun\ on the pre-SN core of the black-hole progenitor. Our population-wide approach does not incorporate such fine-tuning; despite this, it is encouraging to see the match on the final $\vnk$ inferences. It is also noteworthy that `Channel B' for this source (a scenario in which the binary avoided interaction prior to the supernova, see Paper I) prefers a massive donor that allows for CNO processing, and is also consistent with the presence of high metallicity gas as observed by \citet{GonzalezHernandez2006,GonzalezHernandez2008}, who used this as strong evidence for a natal kick for a progenitor born in the Galactic thin disc.

\subsubsection{Cyg X--1}
Our results are almost synonymous with previous estimates derived by \citet{Wong12} - the MAP and HDI values lie within the previously expected range, the minimum natal kick value (0\,\kmps) is the same in both their and our studies. %It is noteworthy that this study considered the companion mass to be $19.16\pm1.90$\,\Msun, as presented by \citet{Orosz2011}. Subsequent studies have put the mass of the stellar companion as much higher, closer to $40$\,\Msun \cite{MillerJones2021}

\subsection{MAXI\,J1305--704}

\citet{Kimball2023} determine the magnitude of the natal kick imparted to MAXI J1305--704 using a very similar method to our own. First, its trajectory (as determined by {\em Gaia}) is integrated back through the Galactic potential, with its peculiar velocity sampled when it crosses the Galactic plane, finding a birth peculiar velocity (assuming formation in the Galactic plane) of $74^{+19}_{-11}$\,\kmps. Periods off mass transfer are simulated to determine the characteristics at the onset of Roche lobe overflow that would reproduce the present day black hole mass, companion mass, companion temperature, and orbital period (within 2$\sigma$). Further binary populations are simulated from ZAMS to supernova, with natal kicks and mass loss sampled uniformly, and the resultant post-supernova orbits and systemic velocities calculated. Systems that match the parameters of viable binaries at onset of Roche lobe overflow are credible, and the resultant posterior on the natal kick velocity finds $\upsilon_{\rm NK}\geq70$\,\kmps at 95\% confidence. 

Our constraints for this system are slightly lower, with $\upsilon_{\rm NK}^{68\%}=34-62$\,\kmps - around 15\% of our simulations require a natal kick greater than 70\,\kmps. This is likely, at least in part, due to the different values of systemic velocity which are matched to in our respective methods. Where \citeauthor{Kimball2023} matches to the peculiar velocity determined through backward integration of the system's current orbit, we take the present day peculiar velocity (58$\pm$22\,\kmps) and subtract an assumed intrinsic stellar velocity dispersion, meaning our simulations are required to satisfy $\upsilon_{\rm sys,\,post-SN}=26-58$\,\kmps, compared to 63-93\,\kmps used by \citeauthor{Kimball2023}. 

\subsection{A\,0620--00}

This is the nearest dynamically confirmed black-hole XRB \citep{Cantrell2010}. \citet{Repetto2012} determining the minimum natal kick applied to A\,0620--00 as $\upsilon_{\rm NK,\,min}$=24--43\,\kmps. Their method also involved reproducing the evolutionary history of A\,0620--00 (and the other 7 systems in their study), integrating back to the onset of Roche-lobe overflow between the black hole and its companion and determining the masses and orbital separation at that time. They conduct Monte Carlo simulations with priors for the compact progenitor mass, ejected mass, pre-SN orbital separation, and natal kick magnitude ($M_{\rm 1,\,SN}$, $M_{\rm ej}$, $a_{\rm SN}$, \& $\vnk$), and identify scenarios in which the post-SN orbital separation is sufficiently tight to ensure RLO onset whilst the companion is on the main sequence. They use the Galactic scale height (0.11-0.13\,kpc) to determine the minimum birth peculiar velocity ($\upsilon_{\rm\bot,\,min}$, where the natal kick is directly exactly perpendicular to the Galactic plane), and discard any scenarios in which the post-SN systemic velocity is less than this. 

Though their minimum natal kick lies within our viable natal kick, our study finds the system can be consistent with no natal kick ($\vpec=0$\,\kmps). This likely arises from their use of flat priors for the pre-SN characteristics of the system. For example, they find a minimum ejecta mass of $M_{\rm ej}\geq0.1$\,\Msun, whereas our population synthesis points to much larger ejecta masses ($M_{\rm ej}\geq3$\,\Msun) - this means we determine a larger `mass-loss' kick, reducing the necessary natal kick strength required to achieve a given systemic velocity. 

\subsection{Aql\,X--1 \& Swift\,J1727.8--1613}
In two distinct works, \citet{MataSanchez2017} and \citet{MataSanchez2025} present natal kick estimates for Aql\,X--1 and Swift\,J1727.8--1613, respectively. The methods employed by both papers are the same - based off the observed systemic velocity (observed at $104\pm3$\,\kmps in that study) in relation to the Galactic radial velocity curve \cite{Clemens1985,Nakanishi2003} (effectively equating $\vnk$ with $\vpec$).

\subsubsection{Aql\,X--1}
\citet{MataSanchez2017} propose that Aql\,X--1 recieved a natal kick of 12-36\,\kmps, given that the Galactic radial velocity curve suggests an object in the location of Aql\,X--1 ought to have a velocity 68--92\,\kmps. Our natal kicks do encompass that range, spanning 0--90\,\kmps, with a mean of 20\,\kmps. The broader constraints are the result of updated peculiar velocity (30--54\,\kmps compared to 9--39\,\kmps, calculated from the formula presented by \citet{Reid2009} rather than subtraction), and thorough treatment of binary evolution and the relationship between $\vnk$ and $\vpec$. For example, we find this system can be consistant with no natal kick, as the peculiar velocity may be attributable to the `mass-loss' kick alone - something that is not considered in the previous paper. 

Finally, we note that there appears to be ambiguity regarding the \gaia\ counterpart of \aqlx. Based on independent constraints from the radio, the impact on our analysis may be minimal, though this remains to be verified. This is discussed in Section\,\ref{Section:App_WeirdSystems}.

\subsubsection{Swift\,J1727.8--1613}
The natal kicks quoted by \citet{MataSanchez2025} are included in our natal kick distributions for this source, however our range is broader, once again due to accounting for contributions to the peculiar velocity beyond a natal kick - for instance, the `mass-loss' kick for this system ranges 13--71\,\kmps. Observing a systemic velocity of $201\pm7$\,\kmps, they apply the methodology presented by \citet{atri2019} to determine the systemic velocity at birth (assuming formation within the Galactic plane, finding $\upsilon_{\rm pec,\,z=0}=210^{+40}_{-50}$\,\kmps, and defining this as the natal kick.

The MAP and HDI values from our natal kicks distributions are consistent with the above estimates, though we note that our distribution is multi-modal, with a small number of systems consistent with significantly higher natal kicks ($\geq300$\,\kmps), corresponding to an evolutionary channel in which the companion star is more massive at the instant of supernova, shrinking to the observed value after periods of significant mass transfer. This, again, highlights the influence of binary/stellar evolution on constraining natal kicks.  

\subsection{Nagarajan et al., 2025}
Previous constraints for these systems were presented by \citet{Nagarajan2025}. The authors determine the space velocities of 12 black holes (from distances, proper motions, and radial velocity measurements of the luminous companion) and compare these to the local stellar population. The minimum natal kick for each system is computed as the difference between its total space velocity and the 68\% velocity threshold of a local comparison sample — i.e., $\upsilon_{\rm NK,\,min} = \max(0, \upsilon_{\rm BH} - \upsilon_{\rm stellar}^{68\%})$. 

For the majority of the 8 systems which appear in both their and our study, their minimum values are lower or the same as ours. The one exception is GRO\,J1655-40, where they quote a minimum natal kick $\upsilon_{\rm NK,\,min}=62$\,\kmps, where we find $\upsilon_{\rm NK,\,min}=42$\,\kmps (though we find $\upsilon_{\rm NK}\geq150$\,\kmps at 95\% confidence). Our results are more precise than those determined by this study, due to the combined constraints of peculiar velocity and detailed evolutionary modelling.

\subsection{Swift\,J1753.5--0127}
\citet{YanesRizo2025} conduct a study focused on refining the characteristics of the black hole XRB, Swift\,J1753.5--0127. This includes updated measurements of the distance to the system, using photometry to constrain the distance to $3.9\pm0.7$\,kpc, corresponding to a scale height of $z=0.8\pm0.2$\,kpc. Considering the anti-correlation between orbital period and Galactic scale height reported by \citet{Gandhi2020}, they use this as evidence for a substantial natal kick. By following the method outlined by \citet{atri2019}, they determine a birth velocity of $\upsilon_{\rm pec,\,z=0}=110^{+69}_{-37}$\,\kmps, and report this as the natal kick velocity. Our work is based on the compilation of measurements in \citetalias{Zhao2023}, which yield a higher current $\vpec$, but a near-identical $\upsilon_{\rm pec}^{z=0}$ at Galactic plane crossing to \citet{YanesRizo2025}. Despite this, our natal kicks are consistent with the inference of a relatively strong kick by \citeauthor{YanesRizo2025}, albeit a broader range due to the inclusion of other factors in constraining the natal kick magnitude (such as the orbital configuration/evolution). %\poshak{(What 'other factors' specifically? If referring to the following paragraph, I would just join them up instead of making the following on as 'aside').} 

As an aside, it may seem counter-intuitive that applying {\em additional} constraints results in {\em broader} natal kick distributions, as supposed to making them more precise. This is because the factors used to calculate natal kicks are all interconnected, and can all act to moderate or enhance the influence of each other. For example, the peculiar velocity of a system is just as much related to its pre-supernova properties and ejected mass as it is to the natal kick velocity - the `mass-loss' kick may counteract the momentum imparted by a given natal kick, reducing the resultant peculiar velocity, or it may serve to increase the post-SN peculiar velocity depending on the orbital configuration and kick direction. 

\subsection{Other sources in common with \citet{Fortin2022}}

\begin{figure}
    \centering
    \includegraphics[width=0.9\linewidth]{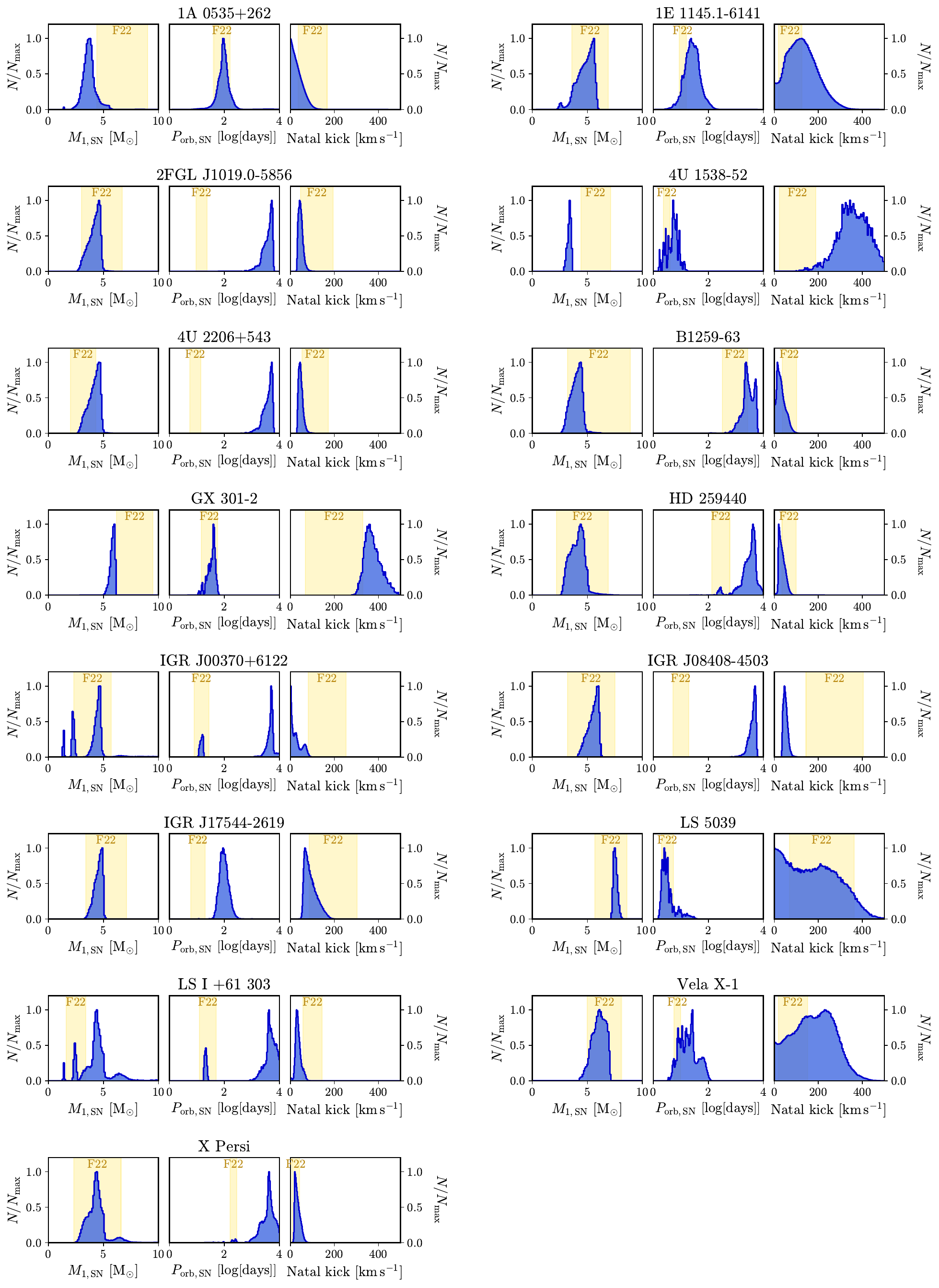}
    \caption{Distribution of compact progenitor masses ($M_{\rm 1,\,SN}$); progenitor system orbital periods ($P_{\rm orb,\,SN}$; and natal kicks $\vnk$ as determined by our study (blue histograms) compared to the ranges quoted by \citeauthor{Fortin2022} for the same system.}
    \label{Figure:App_FortinSystems}
\end{figure}

As discussed in Section \ref{Section:KickDistributions}, \citet{Fortin2022} used Bayesian analysis to infer natal kicks for 35 NS-HMXBs, finding distributions for individual systems and using that to construct a distribution for the population as a whole. 15 systems appear in both their catalogue and ours; whilst there is some overlap in the proposed natal kicks across the two studies, there are also significant differences in many cases. This is due to a number of differences in our analyses, primarily: a) the use of the currently observed properties as the immediate post-SN properties (i.e. \porbobs = $P_{\rm orb, post-SN}$), neglecting any potential change in the systems properties between the supernova and the current epoch (such as orbital period changes; circularisation; mass-transfer etc.) and b) using uniform priors for the unknown pre-SN quantities, such as compact object progenitor mass. In contrast, our work uses pre-SN systems which are physically motivated by thorough population synthesis, thereby excluding progenitor systems which are viable \textit{on paper} though unlikely/impossible to form when considering detailed binary evolution. Similarly, our population synthesis framework also accounts for any changes in the system after the formative supernova. In short, whilst the values proposed by \citeauthor{Fortin2022} are hypothetically plausible, they lack justification and the differences between our results highlight the sensitivity of kick estimates to `unknown' (i.e. un-observable) parameters, which must be adopted based on other criteria (e.g. considering binary evolution). Figure \ref{Figure:App_FortinSystems} shows the distribution of compact progenitor mass ($M_{\rm 1,\,SN}$); progenitor system orbital period ($P_{\rm orb,\,SN}$; and natal kick $\vnk$ (blue histograms) compared to the ranges quoted by \citeauthor{Fortin2022} - the differences in the progenitor systems leads to the differences in derived natal kicks. 

For example, in the case of IGR\,J08408-4503, there is zero overlap between our kicks and values presented by \citet{Fortin2022}. Whilst their proposed neutron star progenitor mass is similar to those identified in our simulations ($\approx3-7\,$\Msun), they identify an orbital period of $P_{\rm orb\,pre-SN}$=5.2--19.5\,days at the time of supernova. In contrast, our simulations point to a larger pre-SN orbital period, a few thousand days, as such short periods in binaries with two high mass stars (recalling a companion mass of $>30$\,\Msun) struggle to survive without merging. Given the intrinsic role of the progenitor system properties in determining its post-SN characteristics, and therefore in inferring the viable natal kicks, it is unsurprising that different priors are suggestive of different natal kicks. 

\clearpage

\section{Comparison between our work and mass-dependent natal kick prescriptions}\label{Section:App_GBBE}
In addition to the purely probablistic natal kick distributions proposed by this work and previous studies, there have been a number of prescriptions proposed which include system parameters such as the mass ejected by the supernova \& the remnant compact object. These are outlined in Section\,\ref{Section:PrevDistributions}, and here we expand upon the comparison between our work and three alternative prescriptions: the two main equations presented in \citet{Giacobbo2019} and the equation presented by \citet{Bray2016}, updated in \citet{Richards2023}. 

\begin{figure*}
    \centering
    \includegraphics[width=0.8\textwidth]{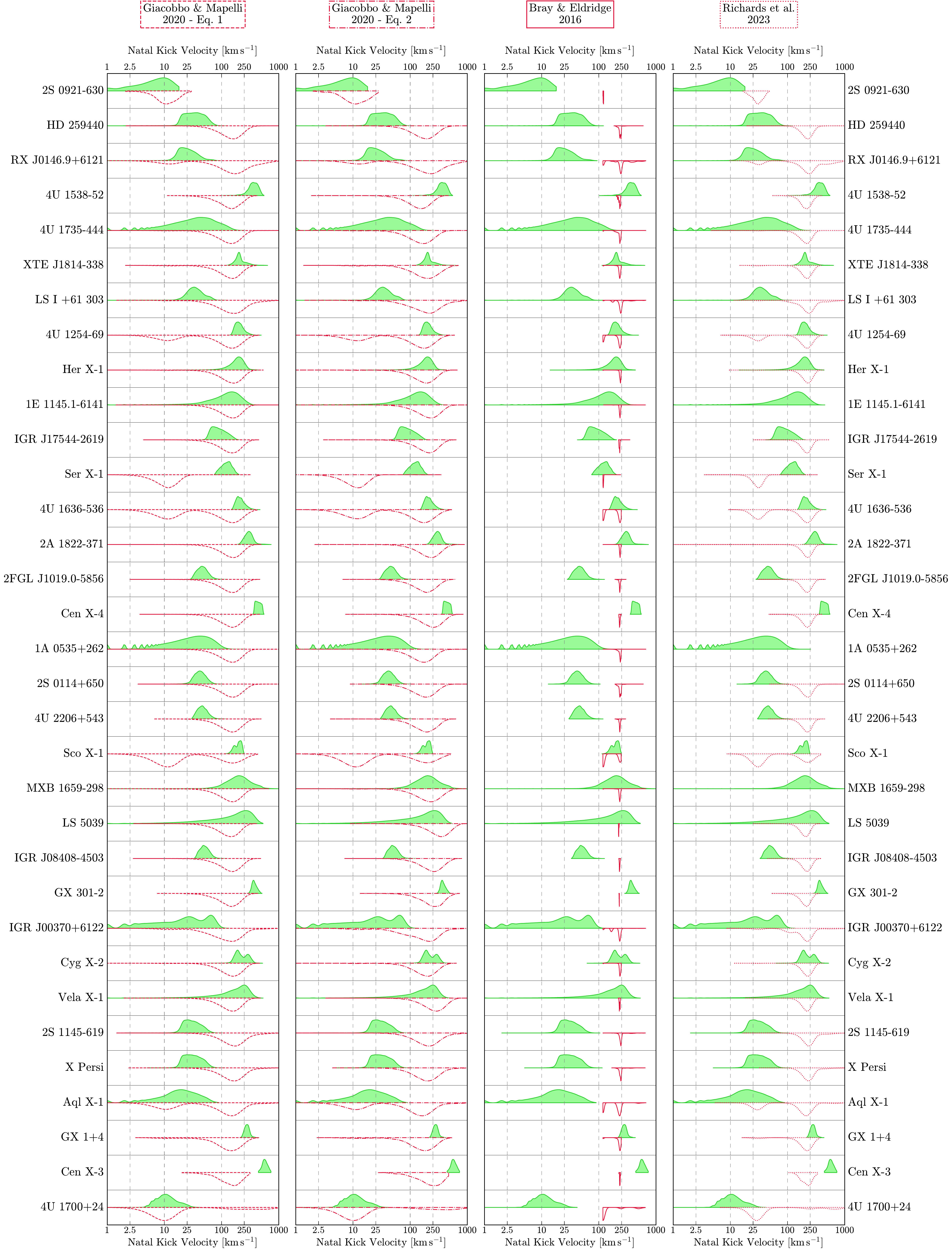}
    \caption{Distribution of natal kicks determined by this study (filled violins) compared to expected natal kicks as described by \citet{Giacobbo2019}; \citet{Bray2016}; and \citet{Richards2023} -- unfilled violins -- for all 33 neutron-star systems in our sample.}
    \label{Fig:App_MassDependentA}
\end{figure*}

\begin{figure*}
    \centering
    \includegraphics[width=0.8\textwidth]{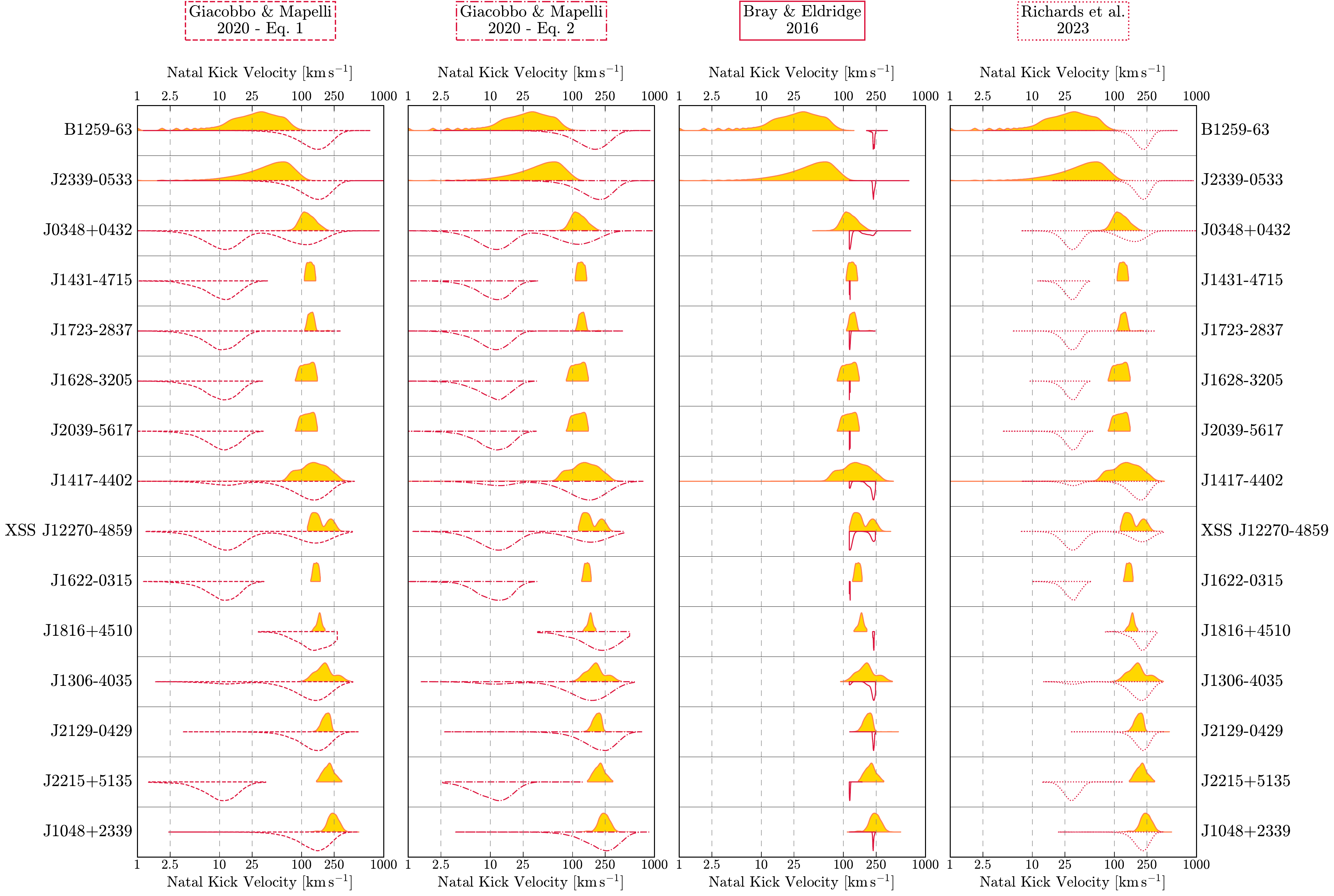}
    \caption{Distribution of natal kicks determined by this study (filled violins) compared to expected natal kicks as described by \citet{Giacobbo2019}; \citet{Bray2016}; and \citet{Richards2023} -- unfilled violins -- for all 15 pulsar systems in our sample.}
    \label{Fig:App_MassDependentB}
\end{figure*}

\begin{figure*}
    \centering
    \includegraphics[width=0.8\textwidth]{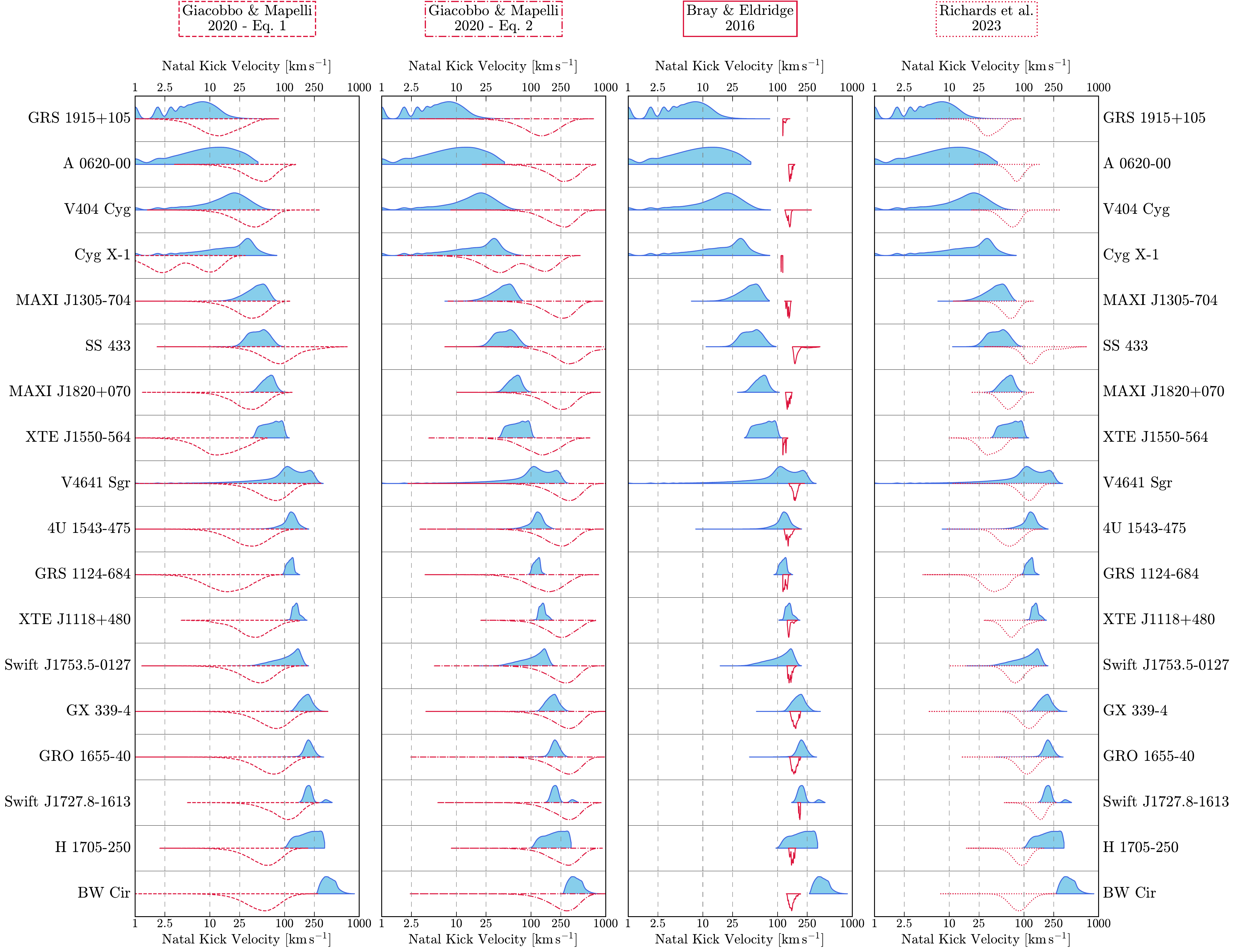}
    \caption{Distribution of natal kicks determined by this study (filled violins) compared to expected natal kicks as described by \citet{Giacobbo2019}; \citet{Bray2016}; and \citet{Richards2023} -- unfilled violins -- for all 18 black-hole systems in our sample.}
    \label{Fig:App_MassDependentC}
\end{figure*}

\citeauthor{Giacobbo2019} propose that the natal kicks for both neutron stars and black holes may be described by either of the following: Here, $\upsilon_{\rm NK,\,GM20_1} = \upsilon_{\rm H05}\frac{M_{\rm ej}}{\langle M_{\rm ej}\rangle}\frac{\langle\Mc\rangle}{\Mc}$ and $\upsilon_{\rm NK,\,GM20_2} = \upsilon_{\rm H05}\frac{M_{\rm ej}}{\langle M_{\rm ej}\rangle}$, where: $\upsilon_{\rm H05}$ is random number extracted from a Maxwellian distribution with one-dimensional rms $\sigma=265$\,\kmps; $M_{\rm ej}$ is the ejected mass; $\Mc$ is the mass of the remnant compact object; and $\langle M_{\rm ej}\rangle$ \& $\langle\Mc\rangle$ are the 9\,\Msun and 1.4\,\Msun respectively, (representing the mean of the two values). In Figures\,\ref{Fig:App_MassDependentA},\ref{Fig:App_MassDependentB}, \& \ref{Fig:App_MassDependentC}, one can see that the distribution of natal kicks as predicted by \citeauthor{Giacobbo2019} is similar for many systems - this is because the driving factor in these equations is the factor $\upsilon_{\rm H05}$, drawn randomly from the Maxwellian presented by \citet{Hobbs2005}. This factor is of order 10s-100\,\kmps for all systems, whereas the factors $\frac{M_{\rm ej}}{\langle M_{\rm ej}\rangle}$ and $\frac{\langle\Mc\rangle}{\Mc}$ are typically of order unity (in our sample ranging 0.01--1.5). It is worth noting that this Maxwellian is implemented based on the work of \citet{Hobbs2005}, in order to reproduce the observed velocity distribution of pulsars - this work has since been revisited the velocity distribution corrected, such that that dataset is now best described by a lognormal distribution \cite{Disberg2024}. It is possible that implementing random variables from this distribution may result in improved predictions, however this is beyond the scope of this work. 

\citeauthor{Bray2016} put forth that natal kicks are determined by a simple linear equation: $\upsilon_{\rm NK,\,BE16}=\alpha\frac{M_{\rm ej}}{\Mc}+\beta$ where $\alpha=70$\,\kmps and $\beta=110$\,\kmps. This prescription does not include stochasticity, as is realistically expected; hence the plotted predicted distributions appear artificially narrow. Whilst some of the natal kicks predicted by this equation are consistent with our study, there is insufficient overlap for our work considered to be compatible with this prescription. The limited range of ejecta and remnant masses for some of the systems in our sample results in much more precise estimates in natal kick velocity. In \citet{Richards2023} this formalism is updated - $\alpha=115^{+40}_{-55}$\,\kmps \& $\beta=15^{+10}_{-15}$\,\kmps are identified as regions of parameter space which best satisfy the constraints used in their study (namely: binary neutron star (BNS) merger rate; the period–eccentricity distribution of Galactic BNSs; velocity distribution of isolated pulsars; and likelihood for low ejecta mass supernovae to produce low-velocity kicks. 
\citet{Richards2023} employ the same linear equation, with updated coefficients, calibrated against BPASS version 2.2.1 models. Using these values with other BPS codes should be approached with caution. %So we should acknowledge that we are aware of this, and that we are essentially testing BPASS-calibrated Richards/Bray prescription on COSMIC evolutionary outputs, not validating their kick parameters, per se.

In Table\,\ref{Table:App_MassDependent} and Figures\,\ref{Fig:App_MassDependentA}, \ref{Fig:App_MassDependentB}, \& \ref{Fig:App_MassDependentC} we present the natal kicks for each system as determined by our study, alongside those predicted by the above (based on the remnant and ejecta masses determined through out simulations). There are some systems for which previous predictions are consistent with our results, however the majority of systems in our study are not well-described by such equations. Whilst comparisons between these global prescriptions to individual sources show discrepancy, this is not a verdict on the correctness of these studies - each has its own assumptions built in, and the dependence on latent variables will inevitable introduce some error. 
%\poshak{(Incomplete? Need statements ensuring that while these comparisons show that translating from their global prescriptions to individual sources (using our compulation of observed CO masses and simulated results on ejected masses) does not always show a match, this is not a verdict on the correctness or incorrectness of their studies. Each has its own assumptions built in, and translating from global to latent variables will inevitably introduce some scatter anyway. Such comparisons will nevertheless be useful to examine evolutionary pathways assumptions under different models. etc.}

\clearpage

\clearpage

\renewcommand{\arraystretch}{1.25}
\begin{longtable}{| l | c c c c | c c c c |@{}}
\caption{Natal kick estimates from this study compared to the predictions
from previously proposed natal kick distributions, alongside the relevant variables (ejected mass $M_{\rm ej}$, and compact remnant mass $\Mc$). Here, $\upsilon_{\rm NK,\,GM20_1} = \upsilon_{\rm H05}\frac{M_{\rm ej}}{\langle M_{\rm ej}\rangle}\frac{\langle\Mc\rangle}{\Mc}$ and $\upsilon_{\rm NK,\,GM20_2} = \upsilon_{\rm H05}\frac{M_{\rm ej}}{\langle M_{\rm ej}\rangle}$ as in \citet{Giacobbo2019}, where $\upsilon_{\rm H05}$ is random number extracted from a Maxwellian distribution with one-dimensional rms $\sigma=265$\,\kmps. $\upsilon_{\rm NK,\,BE16}=\alpha\frac{M_{\rm ej}}{\Mc}+\beta$ as in \citet{Bray2016}, where $\alpha=70$\,\kmps and $\beta=110$\,\kmps. $\upsilon_{\rm NK,\,R23}=\alpha\frac{M_{\rm ej}}{\Mc}+\beta$ as in \citet{Bray2016}, where $\alpha$ is drawn from a normal distribution of mean $\mu=115$\,\kmps\ and scale $\sigma=25$\,\kmps, and $\beta$ is drawn from a normal distribution of mean $\mu=15$\,\kmps\ and scale $\sigma=5$\,\kmps (these coefficients are intended to be {\em representative} of those found by \citet{Richards2023}, though their study does not present these as Gaussian). All values show MAP \% 68\% HDI. All masses are in units of $\rm{M_{\odot}}$, and all velocities in \kmps.\label{Table:App_MassDependent}
} \\
\hline
\multirow{2}{8em}{Name} & \multicolumn{4}{@{}c@{}|}{This Study} & \multicolumn{4}{@{}c@{}|}{Previous Kick Prescriptions}  \\
 & $M_{\rm ej}$ & $\Mc$ & $M_{\rm ej}$/$\Mc$ & $\vnk$ &  $\upsilon_{\rm NK,\,GM20_1}$ & $\upsilon_{\rm NK,\,GM20_2}$ & $\upsilon_{\rm NK,\,BE16}$ & $\upsilon_{\rm NK,\,RE23}$ \\
\hline
\hline
        1A 0535+262 & 42$_{-28}^{+43}$ & 2.46$_{-0.49}^{+0.27}$ & 1.28$_{-0.05}^{+0.1}$ & 1.83$_{-0.12}^{+0.11}$ & 158$_{-67}^{+75}$ & 178$_{-77}^{+93}$ & 238$_{-8}^{+8}$ & 224$_{-51}^{+53}$ \\ \hline
        1E 1145.1-6141 & 152$_{-73}^{+87}$ & 3.54$_{-0.74}^{+0.12}$ & 2.09$_{-0.48}^{+0.04}$ & 1.7$_{-0.01}^{+0.13}$ & 160$_{-64}^{+74}$ & 224$_{-90}^{+125}$ & 229$_{-1}^{+9}$ & 227$_{-48}^{+47}$ \\ \hline
        2A 1822-371 & 301$_{-48}^{+49}$ & 3.06$_{-0.39}^{+0.45}$ & 1.84$_{-0.37}^{+0.13}$ & 1.79$_{-0.07}^{+0.06}$ & 163$_{-67}^{+70}$ & 227$_{-88}^{+123}$ & 236$_{-5}^{+4}$ & 226$_{-42}^{+51}$ \\ \hline
        2FGL J1019.0-5856 & 46$_{-9}^{+11}$ & 2.92$_{-0.45}^{+0.16}$ & 1.61$_{-0.17}^{+0.19}$ & 1.74$_{-0.05}^{+0.1}$ & 148$_{-53}^{+85}$ & 190$_{-72}^{+109}$ & 232$_{-3}^{+7}$ & 224$_{-47}^{+46}$ \\ \hline
        2S 0114+650 & 42$_{-8}^{+11}$ & 2.87$_{-0.68}^{+0.29}$ & 1.28$_{-0.04}^{+0.27}$ & 1.77$_{-0.05}^{+0.19}$ & 165$_{-71}^{+70}$ & 188$_{-75}^{+115}$ & 234$_{-4}^{+13}$ & 226$_{-50}^{+52}$ \\ \hline
        2S 0921-630 & 9$_{-5}^{+7}$ & 0.19$_{-0.03}^{+0.01}$ & 1.24$_{-0}^{+0}$ & 0.15$_{-0.02}^{+0.01}$ & 13$_{-6}^{+6}$ & 12$_{-4}^{+7}$ & 121$_{-2}^{+0}$ & 31$_{-6}^{+5}$ \\ \hline
        2S 1145-619 & 26$_{-6}^{+13}$ & 2.91$_{-0.57}^{+0.41}$ & 1.51$_{-0.22}^{+0.08}$ & 1.89$_{-0.17}^{+0.06}$ & 165$_{-71}^{+102}$ & 211$_{-93}^{+134}$ & 244$_{-13}^{+3}$ & 239$_{-64}^{+67}$ \\ \hline
        4U 1254-69 & 196$_{-30}^{+34}$ & 2.39$_{-0.73}^{+0.7}$ & 1.26$_{-0.06}^{+0.1}$ & 1.78$_{-0.42}^{+0.17}$ & 155$_{-114}^{+179}$ & 176$_{-132}^{+217}$ & 234$_{-29}^{+12}$ & 221$_{-100}^{+111}$ \\ \hline
        4U 1538-52 & 357$_{-47}^{+90}$ & 2.2$_{-0.14}^{+0.15}$ & 1.27$_{-0.02}^{+0.01}$ & 1.76$_{-0.11}^{+0.09}$ & 148$_{-58}^{+71}$ & 165$_{-68}^{+71}$ & 233$_{-8}^{+6}$ & 222$_{-58}^{+33}$ \\ \hline
        4U 1543-475 & 122$_{-20}^{+26}$ & 3.54$_{-0.35}^{+0.79}$ & 8.57$_{-1.58}^{+1.79}$ & 0.41$_{-0.15}^{+0.14}$ & 33$_{-13}^{+29}$ & 257$_{-103}^{+136}$ & 139$_{-11}^{+9}$ & 65$_{-19}^{+24}$ \\ \hline
        4U 1636-536 & 194$_{-21}^{+46}$ & 0.16$_{-0.03}^{+2.23}$ & 1.25$_{-0.05}^{+0.09}$ & 1.79$_{-1.67}^{+-0.02}$ & 155$_{-143}^{+63}$ & 174$_{-162}^{+70}$ & 235$_{-117}^{+-1}$ & 222$_{-191}^{+39}$ \\ \hline
        4U 1700+24 & 11$_{-4}^{+4}$ & 0.14$_{-0.02}^{+0.02}$ & 1.24$_{-0.04}^{+0.04}$ & 0.12$_{-0.01}^{+0.02}$ & 10$_{-5}^{+9}$ & 10$_{-6}^{+9}$ & 118$_{-0}^{+1}$ & 29$_{-8}^{+10}$ \\ \hline
        4U 1735-444 & 43$_{-28}^{+54}$ & 2.47$_{-0.53}^{+0.51}$ & 1.28$_{-0.06}^{+0.16}$ & 1.76$_{-0.14}^{+0.18}$ & 149$_{-63}^{+79}$ & 169$_{-73}^{+102}$ & 233$_{-10}^{+13}$ & 219$_{-56}^{+53}$ \\ \hline
        4U 2206+543 & 46$_{-9}^{+9}$ & 2.92$_{-0.47}^{+0.16}$ & 1.6$_{-0.17}^{+0.2}$ & 1.75$_{-0.05}^{+0.09}$ & 162$_{-68}^{+67}$ & 199$_{-84}^{+91}$ & 232$_{-4}^{+6}$ & 221$_{-42}^{+52}$ \\ \hline
        A 0620-00 & 13$_{-9}^{+13}$ & 3.97$_{-0.23}^{+0.18}$ & 7.57$_{-0.9}^{+0.34}$ & 0.51$_{-0.04}^{+0.09}$ & 49$_{-19}^{+26}$ & 310$_{-127}^{+121}$ & 145$_{-0}^{+9}$ & 79$_{-13}^{+23}$ \\ \hline
        Aql X-1 & 20$_{-10}^{+26}$ & 2.17$_{-2.04}^{+0.06}$ & 1.26$_{-0.04}^{+0.04}$ & 1.74$_{-1.63}^{+0}$ & 152$_{-139}^{+84}$ & 160$_{-147}^{+100}$ & 232$_{-114}^{+0}$ & 213$_{-181}^{+60}$ \\ \hline
        B1259-63 & 33$_{-17}^{+26}$ & 2.8$_{-0.49}^{+0.19}$ & 1.29$_{-0.02}^{+0.28}$ & 1.73$_{-0.03}^{+0.13}$ & 159$_{-64}^{+75}$ & 196$_{-85}^{+88}$ & 231$_{-2}^{+9}$ & 226$_{-48}^{+49}$ \\ \hline
        BW Cir & 363$_{-54}^{+98}$ & 4.18$_{-0.37}^{+0.29}$ & 6.36$_{-0.92}^{+1.54}$ & 0.64$_{-0.2}^{+0.12}$ & 54$_{-23}^{+33}$ & 304$_{-124}^{+138}$ & 155$_{-14}^{+8}$ & 84$_{-21}^{+30}$ \\ \hline
        Cen X-3 & 561$_{-43}^{+77}$ & 3.24$_{-0.03}^{+0.11}$ & 1.87$_{-0.01}^{+0.02}$ & 1.74$_{-0}^{+0.04}$ & 157$_{-61}^{+65}$ & 239$_{-99}^{+88}$ & 232$_{-0}^{+3}$ & 234$_{-53}^{+39}$ \\ \hline
        Cen X-4 & 394$_{-16}^{+81}$ & 3.39$_{-0.25}^{+0.46}$ & 1.91$_{-0.13}^{+0.24}$ & 1.71$_{-0.03}^{+0.07}$ & 161$_{-67}^{+67}$ & 263$_{-111}^{+111}$ & 230$_{-2}^{+5}$ & 221$_{-43}^{+47}$ \\ \hline
        Cyg X-1 & 32$_{-22}^{+11}$ & 0.5$_{-0}^{+1.47}$ & 18.57$_{-1.61}^{+0.97}$ & 0.11$_{-0.09}^{+0}$ & 2$_{-0}^{+10}$ & 38$_{-10}^{+138}$ & 118$_{-6}^{+0}$ & 26$_{-9}^{+7}$ \\ \hline
        Cyg X-2 & 191$_{-20}^{+100}$ & 2.59$_{-0.29}^{+0.43}$ & 1.46$_{-0.19}^{+0.11}$ & 1.77$_{-0.07}^{+0.1}$ & 159$_{-63}^{+80}$ & 205$_{-88}^{+97}$ & 234$_{-5}^{+7}$ & 231$_{-51}^{+46}$ \\ \hline
        GRO 1655-40 & 208$_{-24}^{+32}$ & 4.34$_{-0.19}^{+0.17}$ & 4.88$_{-0.55}^{+1.47}$ & 0.8$_{-0.17}^{+0.17}$ & 73$_{-31}^{+40}$ & 331$_{-135}^{+144}$ & 166$_{-12}^{+12}$ & 110$_{-26}^{+34}$ \\ \hline
        GRS 1124-684 & 128$_{-18}^{+5}$ & 3.23$_{-1.37}^{+0.17}$ & 12.86$_{-2.95}^{+0.34}$ & 0.12$_{-0.01}^{+0.18}$ & 18$_{-9}^{+15}$ & 166$_{-68}^{+124}$ & 118$_{-1}^{+13}$ & 36$_{-8}^{+20}$ \\ \hline
        GRS 1915+105 & 8$_{-5}^{+6}$ & 1.5$_{-0.1}^{+0.95}$ & 13.29$_{-2.52}^{+0.58}$ & 0.11$_{-0}^{+0.11}$ & 12$_{-5}^{+12}$ & 139$_{-59}^{+91}$ & 118$_{-0}^{+8}$ & 33$_{-7}^{+15}$ \\ \hline
        GX 1+4 & 281$_{-31}^{+23}$ & 2.57$_{-0.31}^{+0.37}$ & 1.29$_{-0.07}^{+0.22}$ & 1.78$_{-0.06}^{+0.1}$ & 156$_{-63}^{+89}$ & 186$_{-83}^{+95}$ & 235$_{-5}^{+7}$ & 221$_{-45}^{+60}$ \\ \hline
        GX 301-2 & 362$_{-27}^{+45}$ & 3.75$_{-0.13}^{+0.09}$ & 2.26$_{-0.1}^{+0.06}$ & 1.69$_{-0.02}^{+0}$ & 152$_{-59}^{+71}$ & 272$_{-97}^{+145}$ & 228$_{-1}^{+0}$ & 220$_{-47}^{+38}$ \\ \hline
        GX 339-4 & 207$_{-48}^{+24}$ & 4.33$_{-0.16}^{+0.16}$ & 4.8$_{-1}^{+0.92}$ & 0.94$_{-0.27}^{+0.12}$ & 77$_{-32}^{+44}$ & 318$_{-123}^{+154}$ & 176$_{-19}^{+8}$ & 119$_{-29}^{+35}$ \\ \hline
        H 1705-250 & 303$_{-132}^{+28}$ & 4.16$_{-0.26}^{+0.22}$ & 5.27$_{-0.15}^{+1.29}$ & 0.63$_{-0.01}^{+0.22}$ & 60$_{-25}^{+29}$ & 307$_{-122}^{+142}$ & 154$_{-0}^{+16}$ & 94$_{-21}^{+24}$ \\ \hline
        HD 259440 & 37$_{-17}^{+8}$ & 2.9$_{-0.69}^{+0.25}$ & 1.28$_{-0.03}^{+0.29}$ & 1.76$_{-0.04}^{+0.2}$ & 163$_{-70}^{+71}$ & 192$_{-79}^{+106}$ & 245$_{-15}^{+2}$ & 225$_{-49}^{+55}$ \\ \hline
        Her X-1 & 200$_{-50}^{+54}$ & 2.46$_{-0.16}^{+0.51}$ & 1.3$_{-0.06}^{+0.25}$ & 1.86$_{-0.11}^{+0.04}$ & 161$_{-65}^{+77}$ & 193$_{-78}^{+104}$ & 239$_{-7}^{+3}$ & 230$_{-47}^{+52}$ \\ \hline
        IGR J00370+6122 & 64$_{-56}^{+15}$ & 2.91$_{-0.91}^{+0.57}$ & 1.27$_{-0.12}^{+0.53}$ & 1.75$_{-0.14}^{+0.57}$ & 152$_{-87}^{+89}$ & 216$_{-135}^{+124}$ & 233$_{-10}^{+40}$ & 225$_{-97}^{+76}$ \\ \hline
        IGR J08408-4503 & 48$_{-8}^{+10}$ & 3.76$_{-0.42}^{+0.14}$ & 2.22$_{-0.26}^{+0.06}$ & 1.71$_{-0.02}^{+0.06}$ & 157$_{-61}^{+74}$ & 265$_{-109}^{+118}$ & 229$_{-1}^{+4}$ & 221$_{-42}^{+49}$ \\ \hline
        IGR J17544-2619 & 71$_{-10}^{+39}$ & 2.97$_{-0.2}^{+0.2}$ & 1.84$_{-0.24}^{+0.03}$ & 1.74$_{-0.07}^{+0.03}$ & 152$_{-58}^{+76}$ & 222$_{-93}^{+94}$ & 232$_{-5}^{+2}$ & 217$_{-39}^{+50}$ \\ \hline
        J0348+0432 & 108$_{-13}^{+38}$ & 0.17$_{-0.04}^{+1.28}$ & 1.26$_{-0.04}^{+0.03}$ & 0.13$_{-0.02}^{+1.01}$ & 12$_{-4}^{+111}$ & 12$_{-4}^{+114}$ & 119$_{-2}^{+71}$ & 31$_{-6}^{+139}$ \\ \hline
        J1048+2339 & 239$_{-31}^{+42}$ & 3.8$_{-0.59}^{+0.15}$ & 1.95$_{-0.11}^{+0.33}$ & 1.71$_{-0.04}^{+0.05}$ & 157$_{-59}^{+81}$ & 255$_{-101}^{+132}$ & 230$_{-3}^{+4}$ & 221$_{-42}^{+49}$ \\ \hline
        J1306-4035 & 193$_{-62}^{+29}$ & 2.3$_{-0.69}^{+0.87}$ & 1.27$_{-0.07}^{+0.2}$ & 1.78$_{-0.18}^{+0.16}$ & 146$_{-63}^{+100}$ & 173$_{-80}^{+156}$ & 234$_{-12}^{+12}$ & 213$_{-54}^{+64}$ \\ \hline
        J1417-4402 & 140$_{-46}^{+79}$ & 2.28$_{-0.9}^{+0.94}$ & 1.27$_{-0.06}^{+0.07}$ & 1.74$_{-0.39}^{+0.17}$ & 140$_{-71}^{+101}$ & 162$_{-90}^{+141}$ & 232$_{-28}^{+11}$ & 209$_{-66}^{+73}$ \\ \hline
        J1431-4715 & 128$_{-13}^{+10}$ & 0.19$_{-0.04}^{+0}$ & 1.25$_{-0.04}^{+0.02}$ & 0.15$_{-0.03}^{+0}$ & 12$_{-4}^{+6}$ & 12$_{-5}^{+6}$ & 121$_{-2}^{+0}$ & 33$_{-7}^{+5}$ \\ \hline
        J1622-0315 & 152$_{-10}^{+10}$ & 0.19$_{-0.05}^{+0}$ & 1.24$_{-0.03}^{+0.03}$ & 0.15$_{-0.03}^{+0}$ & 11$_{-4}^{+8}$ & 11$_{-4}^{+7}$ & 121$_{-2}^{+0}$ & 31$_{-6}^{+7}$ \\ \hline
        J1628-3205 & 136$_{-32}^{+11}$ & 0.14$_{-0.01}^{+0.04}$ & 1.24$_{-0.03}^{+0.03}$ & 0.11$_{-0}^{+0.03}$ & 12$_{-5}^{+5}$ & 12$_{-5}^{+5}$ & 118$_{-0}^{+2}$ & 31$_{-6}^{+6}$ \\ \hline
        J1723-2837 & 136$_{-15}^{+6}$ & 0.16$_{-0.02}^{+0.02}$ & 1.24$_{-0.03}^{+0.03}$ & 0.13$_{-0.01}^{+0.02}$ & 12$_{-5}^{+4}$ & 12$_{-5}^{+6}$ & 119$_{-1}^{+1}$ & 31$_{-6}^{+7}$ \\ \hline
        J1816+4510 & 166$_{-9}^{+13}$ & 3.51$_{-0.32}^{+0.14}$ & 1.97$_{-0.12}^{+0.09}$ & 1.78$_{-0.04}^{+0.01}$ & 141$_{-34}^{+87}$ & 248$_{-122}^{+87}$ & 235$_{-3}^{+1}$ & 230$_{-41}^{+60}$ \\ \hline
        J2039-5617 & 138$_{-33}^{+10}$ & 0.14$_{-0.01}^{+0.06}$ & 1.25$_{-0.04}^{+0.02}$ & 0.11$_{-0.01}^{+0.04}$ & 11$_{-4}^{+6}$ & 12$_{-5}^{+7}$ & 118$_{-1}^{+3}$ & 31$_{-6}^{+7}$ \\ \hline
        J2129-0429 & 211$_{-29}^{+16}$ & 3.08$_{-0.14}^{+0.49}$ & 1.83$_{-0.08}^{+0.25}$ & 1.75$_{-0.06}^{+0.07}$ & 161$_{-62}^{+78}$ & 243$_{-101}^{+105}$ & 233$_{-4}^{+5}$ & 224$_{-42}^{+50}$ \\ \hline
        J2215+5135 & 219$_{-34}^{+29}$ & 0.17$_{-0.02}^{+0.02}$ & 1.24$_{-0}^{+0}$ & 0.13$_{-0.02}^{+0.02}$ & 11$_{-4}^{+5}$ & 12$_{-4}^{+7}$ & 119$_{-1}^{+1}$ & 33$_{-8}^{+5}$ \\ \hline
        J2339-0533 & 60$_{-32}^{+29}$ & 2.99$_{-0.52}^{+0.22}$ & 1.87$_{-0.38}^{+0.1}$ & 1.74$_{-0.06}^{+0.1}$ & 162$_{-66}^{+71}$ & 219$_{-89}^{+104}$ & 232$_{-4}^{+7}$ & 227$_{-46}^{+48}$ \\ \hline
        LS 5039 & 265$_{-173}^{+134}$ & 4.67$_{-0.08}^{+0.12}$ & 2.91$_{-0.14}^{+0.03}$ & 1.61$_{-0.03}^{+0.02}$ & 148$_{-58}^{+69}$ & 360$_{-144}^{+162}$ & 223$_{-2}^{+1}$ & 210$_{-44}^{+39}$ \\ \hline
        LS I +61 303 & 33$_{-9}^{+12}$ & 2.93$_{-1.01}^{+0.44}$ & 1.5$_{-0.25}^{+0.05}$ & 1.89$_{-0.46}^{+0.12}$ & 169$_{-88}^{+100}$ & 214$_{-114}^{+137}$ & 243$_{-32}^{+9}$ & 238$_{-82}^{+91}$ \\ \hline
        MAXI J1305-704 & 52$_{-17}^{+11}$ & 3.69$_{-0.38}^{+0.21}$ & 7.45$_{-0.32}^{+1.18}$ & 0.47$_{-0.1}^{+0.04}$ & 38$_{-15}^{+20}$ & 279$_{-116}^{+118}$ & 143$_{-7}^{+3}$ & 69$_{-16}^{+15}$ \\ \hline
        MAXI J1820+070 & 66$_{-16}^{+7}$ & 3.59$_{-0.53}^{+0.29}$ & 9.36$_{-1.55}^{+0.57}$ & 0.35$_{-0.06}^{+0.12}$ & 35$_{-14}^{+22}$ & 254$_{-93}^{+136}$ & 135$_{-4}^{+8}$ & 63$_{-13}^{+19}$ \\ \hline
        MXB 1659-298 & 201$_{-77}^{+120}$ & 3.01$_{-0.32}^{+1.18}$ & 1.29$_{-0.05}^{+0.72}$ & 1.73$_{-0.05}^{+0.12}$ & 153$_{-60}^{+80}$ & 230$_{-101}^{+134}$ & 231$_{-4}^{+8}$ & 223$_{-49}^{+48}$ \\ \hline
        RX J0146.9+6121 & 21$_{-5}^{+11}$ & 2.92$_{-2.79}^{+0.26}$ & 1.27$_{-0.03}^{+0.26}$ & 1.9$_{-1.77}^{+0.05}$ & 171$_{-102}^{+240}$ & 225$_{-136}^{+288}$ & 243$_{-20}^{+217}$ & 240$_{-93}^{+313}$ \\ \hline
        SS 433 & 52$_{-19}^{+9}$ & 4.32$_{-0.23}^{+0.23}$ & 5.2$_{-0.88}^{+0.17}$ & 0.88$_{-0.09}^{+0.23}$ & 84$_{-37}^{+58}$ & 343$_{-152}^{+197}$ & 171$_{-6}^{+16}$ & 125$_{-32}^{+46}$ \\ \hline
        Sco X-1 & 215$_{-41}^{+18}$ & 0.16$_{-0.03}^{+2.38}$ & 1.24$_{-0.06}^{+0.18}$ & 0.13$_{-0.02}^{+1.62}$ & 11$_{-4}^{+153}$ & 12$_{-3}^{+194}$ & 119$_{-1}^{+113}$ & 30$_{-4}^{+208}$ \\ \hline
        Ser X-1 & 136$_{-36}^{+17}$ & 0.16$_{-0.03}^{+0.01}$ & 1.24$_{-0.03}^{+0.03}$ & 0.13$_{-0.02}^{+0.01}$ & 11$_{-4}^{+6}$ & 12$_{-5}^{+6}$ & 119$_{-1}^{+1}$ & 31$_{-6}^{+7}$ \\ \hline
        Swift J1727.8-1613 & 211$_{-23}^{+23}$ & 4.34$_{-0.12}^{+0.13}$ & 3.44$_{-0.1}^{+0.22}$ & 1.27$_{-0.05}^{+0.05}$ & 112$_{-43}^{+56}$ & 333$_{-132}^{+150}$ & 199$_{-3}^{+4}$ & 168$_{-34}^{+33}$ \\ \hline
        Swift J1753.5-0127 & 151$_{-60}^{+22}$ & 3.99$_{-0.41}^{+0.24}$ & 6.71$_{-0.22}^{+1.93}$ & 0.47$_{-0.09}^{+0.13}$ & 48$_{-20}^{+27}$ & 302$_{-127}^{+129}$ & 143$_{-4}^{+12}$ & 76$_{-16}^{+24}$ \\ \hline
        V404 Cyg & 21$_{-12}^{+14}$ & 3.87$_{-0.49}^{+0.24}$ & 7.43$_{-0.58}^{+1.33}$ & 0.51$_{-0.13}^{+0.06}$ & 41$_{-18}^{+21}$ & 262$_{-106}^{+127}$ & 146$_{-9}^{+4}$ & 68$_{-15}^{+21}$ \\ \hline
        V4641 Sgr & 109$_{-25}^{+144}$ & 4.37$_{-0.18}^{+0.15}$ & 4.86$_{-0.78}^{+0.84}$ & 0.84$_{-0.1}^{+0.22}$ & 75$_{-30}^{+43}$ & 317$_{-119}^{+162}$ & 173$_{-11}^{+11}$ & 120$_{-31}^{+29}$ \\ \hline
        Vela X-1 & 249$_{-131}^{+86}$ & 4.18$_{-0.56}^{+0.19}$ & 2.25$_{-0.3}^{+0.21}$ & 1.69$_{-0.01}^{+0.11}$ & 158$_{-63}^{+72}$ & 291$_{-117}^{+138}$ & 228$_{-1}^{+8}$ & 220$_{-43}^{+47}$ \\ \hline
        X Persi & 25$_{-4}^{+18}$ & 2.89$_{-0.55}^{+0.41}$ & 1.51$_{-0.25}^{+0.06}$ & 1.9$_{-0.17}^{+0.06}$ & 169$_{-72}^{+88}$ & 211$_{-96}^{+110}$ & 244$_{-13}^{+3}$ & 236$_{-58}^{+59}$ \\ \hline
        XSS J12270-4859 & 141$_{-19}^{+63}$ & 0.17$_{-0.03}^{+1.99}$ & 1.25$_{-0.06}^{+0.06}$ & 0.14$_{-0.03}^{+1.54}$ & 12$_{-5}^{+133}$ & 13$_{-4}^{+172}$ & 120$_{-1}^{+109}$ & 32$_{-6}^{+174}$ \\ \hline
        XTE J1118+480 & 144$_{-17}^{+8}$ & 3.7$_{-0.31}^{+0.31}$ & 8.27$_{-0.78}^{+0.62}$ & 0.45$_{-0.05}^{+0.09}$ & 45$_{-19}^{+28}$ & 282$_{-106}^{+123}$ & 141$_{-4}^{+6}$ & 66$_{-14}^{+22}$ \\ \hline
        XTE J1550-564 & 91$_{-35}^{+7}$ & 1.51$_{-0.09}^{+1.23}$ & 10.11$_{-0.18}^{+2.89}$ & 0.12$_{-0}^{+0.16}$ & 13$_{-6}^{+13}$ & 151$_{-67}^{+95}$ & 118$_{-0}^{+11}$ & 32$_{-6}^{+18}$ \\ \hline
        XTE J1814-338 & 202$_{-39}^{+30}$ & 2.37$_{-0.22}^{+1.28}$ & 1.31$_{-0.09}^{+0.6}$ & 1.75$_{-0.05}^{+0.11}$ & 146$_{-56}^{+83}$ & 200$_{-78}^{+154}$ & 232$_{-4}^{+7}$ & 218$_{-38}^{+61}$ \\ \hline
\end{longtable}

\clearpage

\section{Systems with poorly-constrained/controversial characteristics} \label{Section:App_WeirdSystems}
\begin{table*}[h!]
\centering
\renewcommand{\arraystretch}{1.25}
\begin{tabular}{| l | c | c c c c c |} 
\hline
Name & Type & $\Mc$ & $M_2$ & $\porb$ & $\upsilon_{\mathrm{pec}}$ & $\upsilon^{68\%}_{\mathrm{NK}}$ \\
 &  & $\left[{\rm M}_{\odot}\right]$ & $\left[{\rm M}_{\odot}\right]$ & [days] & $\left[{\rm km\,s}^{-1}\right]$ & $\left[{\rm km\,s}^{-1}\right]$ \\
\hline
\hline
Aql X-1 & NS-LMXB & $1.4\pm0.$ & $0.76\pm0.0$ & 0.79 & $42\pm12$ & 10--46\\
BW Cir & BH-LMXB & $7.6\pm0.$ & $0.9\pm0.3$ & 2.54 & $352\pm99$ & 309--461\\
LS 5039 & NS-HMXB & $3.7\pm1.0$ & $22.9\pm2.9$ & 3.91 & $88\pm3$ & 92--399\\
SS 433 & BH-HMXB & $4.3\pm0.4$ & $11.3\pm0.06$ & 13.08$\pm2.62$ & $44\pm16$ & 33--61\\
\hline
\end{tabular}
\caption{Characteristics of systems which are subject to uncertainty.}
\label{Table:Weird}
\end{table*}

\subsection{Aql X--1}
Here, we highlight an important caveat regarding the validity of \gaia\ results on \aqlx\ that came to our attention during the final stages of our work. It appears that the commonly adopted \gaia\ counterpart, Gaia DR3 4264296556603631872 (G4264), is unlikely to be a clean counterpart to \aqlx. \citet{Chevalier1999} found that the genuine optical counterpart (their star `$e$') is a faint (quiescent $V\approx 21.6$) star to the west of a brighter interloper. They resolved star $e$ by subtracting the point-spread-function of the interloper; and indeed, star $e$ was clearly resolved in subsequent work by \citet{Hynes12} and by  \citet{MataSanchez2017}. G4264's position, however, is east to star $e$ and has a \gaia\ G magnitude of 18.9. Because the pair is separated by only $\approx$ 0\farcs5, \gaia\ photometry may nevertheless contain flux from both stars, with the degree of blending depending on scan angle and the highly frequent and variable brightness of 
\aqlx\ during outbursts. There is no other \gaia\ DR3 source in the immediate vicinity that could plausibly represent a clean counterpart. 

\aqlx\ has also been observed by radio VLBI in 2009 with VLBA and EVN \citep{Miller-Jones2009, Tudose2009}, and in 2013 with e-EVN \citep{Tudose2013}, yielding a very precise radio localisation $\approx$0\farcs5 east of the \gaia\ position. Using the $\approx 3.6\,\mathrm{yr}$ timeline between the VLBA and e-EVN observation, we estimated linear proper-motion components of $\mu_{\alpha\ast} = -2.63\,\mathrm{mas\ yr^{-1}}$ and $\mu_\delta=-3.50\,\mathrm{mas\ yr^{-1}}$. Using the same distance and systemic radial velocity as in \citetalias{Zhao2023} together with the radio proper motion, results in a similar $\vpec$ estimate to prior values ($\upsilon_{\rm pec}^{\rm radio}$\,=\,43.9$^{+9.4}_{-4.2}$\,$\mathrm{km~s^{-1}}$, compared to $\upsilon_{\rm pec}^{\rm Z23}$\,=\,42.3$^{+13.2}_{-12.0}$\,$\mathrm{km~s^{-1}}$ in \citetalias{Zhao2023}) which, in turn, gives almost identical $\vnk$ estimates. 

While the match between radio and optical proper-motion estimates is encouraging, the radio data also need to be treated with caution, in case variable jet-blob speeds dominate the apparent motion. Thus, both the radio and the optical kinematic bulk-motion estimates of \aqlx\ remain subject to future revision. Previous \gaia\ counterpart analyses (\citealt{ODoherty2023}, \citetalias{Zhao2023}, \citealt{dashwoodbrown2026i}) are similarly subject to the same caveats. 

\subsection{BW Cir}
BW Cir (GS 1354-64) is a LMXB with a well-identified donor and a dynamically-confirmed BH \citep{Casares2004}. However, there has been a significant disagreement on its distance. On one hand, its 2.5-day orbit and X-ray binary nature suggests a Roche-lobe-filling donor, and this G-type donor must be $\gtrsim 25\,\mathrm{kpc}$ to show the apparent optical magnitude (after correction of accretion disc contribution; \citealt{Casares2004}); its {\it Gaia} parallax ($\approx 1.3\pm 0.5\,\mathrm{mas}$), on the other hand, gives much closer yet uncertain estimates between $\approx 1.4-7.0\,\mathrm{kpc}$ \citep{Gandhi2019, Plotkin2021}. In this study, we matched the simulated binary $\vsys$ to the $\vpec$ computed by \citetalias{Zhao2023}, who adopted the more distant estimate. This gives the elevated $\vpec\gtrsim 300\,\mathrm{km~s^{-1}}$. Following the same formulation as in \citetalias{Zhao2023}, one can get $\vpec=140.1^{+20.4}_{-22.3}\,\mathrm{km~s^{-1}}$, and hence somewhat more conservative $\vnk$ values.

\begin{figure*}
    \centering
    \includegraphics[width=0.9\linewidth]{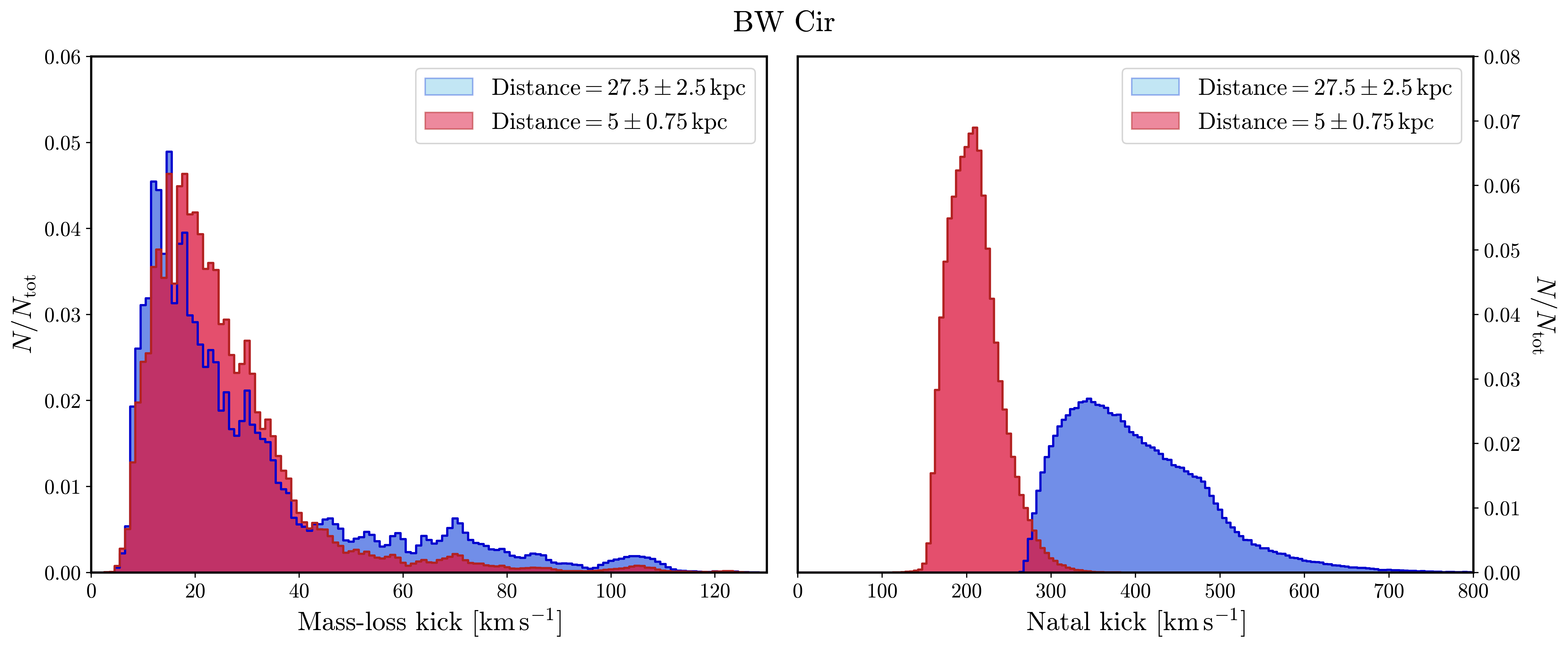}
    \caption{Distribution of mass-loss kicks and natal kicks associated with BW Cir under different distance assumptions. The mass-loss kicks remain consistent with one another, indicative of similar progenitor pathways and characteristics. The natal kick distribution is shifted to lower values when assuming a closer distance and therefore a lower peculiar velocity - though still of the same order of magnitude, and above theoretically expected values for black hole natal kicks.}
    \label{Figure:App_BWCir}
\end{figure*}

For the primary analysis of this study, we match BW Cir analogues according to a peculiar velocity of 352$\pm$99\,\kmps. In this case, we find $\upsilon_{\rm NK}^{68\%}$=309--461\,\kmps (minimum $\vnk$ of 267\,\kmps). However, if we repeat our analysis using the distance from \citeauthor{Plotkin2021} and, consequently, a lower $\vpec$, the allowed $\vnk$ distribution shifts to slightly lower values (140--250\,\kmps, minimum 121\,\kmps; see Figure\,\ref{Figure:App_BWCir}) --- though still well-above the majority of theoretical expectations for BHs. 

\subsection{LS 5039}
LS 5039 is a HMXB whose compact-object nature has not been robustly confirmed. \citet{Casares2005} derived a $\Mc=3.7^{+1.3}_{-1.0}\,\mathrm{M_\odot}$ for an inclination of $\approx 25^\circ$. Later work by found somewhat smaller $\Mc$ values, favouring a NS nature. More recently, \citet{Yoneda2020} found evidence of hard X-ray pulsation with a period $\approx 9\,\mathrm{s}$. \citet{Volkov2021} performed a separate analysis and found that the detection was not statistically significant, which was later revisited by \citet{Makishima2023}, who recovered the pulsation using archival X-ray observations. We note that even if we constrain the remnant mass to $\Mc\leq3$\,\Msun (i.e. assuming its a neutron star) we find the natal kick estimates remain unchanged (0\,\kmps), as is somewhat unsurprising given the non-linear relationship between remnant mass and natal kick magnitude. 

\subsection{SS 433}
SS 433 is a HMXB containing a likely BH and a supergiant donor. The system is persistently accreting at a supercritical level and is surrounded by an accretion disc, dense disc wind, jets, and circumbinary material. These components can produce spectral features similar to those of the donor, making a clean spectroscopic measurement of the motions of either binary component challenging. As a result, the compact-object nature and mass remain uncertain. For example, \citet{Hillwig2008} estimated a $\Mc=4.3\pm0.8\,\mathrm{M_\odot}$ using donor absorption-line radial velocities together with an adopted compact-object radial-velocity amplitude, while \citet{Kubota2010} subsequently argued that irradiation of the donor can distort its measured radial-velocity amplitude, resulting in a lower value of $\Mc=2.5^{+0.7}_{-0.6}\,\mathrm{M_\odot}$. More recent results generally favour a higher $\Mc$ and a BH nature. \citet{Bowler2018} modelled the split stationary emission lines as arising from a circumbinary ring and obtained $\Mc=15\pm 2\,\mathrm{M_\odot}$. The circumbinary component is further favoured by \citet{Dodin26}, who showed that the He II $\lambda4686$ emission exhibits no eclipses by the donor, and hence originates from an extended region substantially larger than the accretion disc. \citet{Cherepashchuk2023} measured long-term orbital-period evolution in SS 433 and constrained the mass ratio using an angular-momentum-loss model, favouring $\Mc\gtrsim 8\,\mathrm{M_\odot}$ for a donor mass of $\sim 10\,\mathrm{M_\odot}$. 

%This casts doubt on spectroscopic component-mass estimates that adopt the He II radial-velocity amplitude for the compact object, including the assumptions underlying some of the earlier low-mass determinations, but does not itself provide a revised compact-object mass. \citet{Bowler2026} subsequently argued that the He II constraint can be discarded while retaining the circumbinary-disc determination, and continued to favour a BH mass of $\sim15,\mathrm{M_\odot}$.
%Most recently, \citet{Dodin26} confirmed that the He II $\lambda4686$ emission is from a region beyond the accretion disc, which should therefore 

With these caveats in mind, we retained these sources within our final sample. Irrespective, excluding or retaining a single object has no material impact on our main inferences. In Table \ref{Table:KSTests} we show the results of KS-testing between the BHs and NSs in our sample when excluding each of these systems (and various combinations thereof), highlighting that, for example, whilst it is the source with the highest $\vnk$ range of the BHs, our results are not reliant on its inclusion, or that of any others. We are therefore satisfied that the uncertainties in some of these systems' characteristics, whilst worthy of consideration, and not significantly impacting our work.

\begin{table*}
\centering
\renewcommand{\arraystretch}{1.25}
\begin{tabular}{| l | c |} 
\hline
\multirow{2}{3em}{Systems} & Results of KS testing \\
 & $p$-value \\
\hline\hline
\textbf{All Black Holes} & \multirow{2}{3em}{\textbf{0.60$\mathbf{_{-0.20}^{+0.28}}$}} \\
\textbf{All Neutron Stars} & \\
\hline\hline
BHs (excluding BW Cir) & \multirow{2}{3em}{ 0.34$_{-0.19}^{+0.22}$} \\ 
NSs & \\
\hline
BHs (excluding SS 433) & \multirow{2}{3em}{ 0.77$_{-0.25}^{+0.18}$} \\ 
NSs & \\
\hline
BHs & \multirow{2}{3em}{0.47$_{-0.22}^{+0.26}$} \\ 
NSs (excluding Aql X-1) & \\
\hline
BHs & \multirow{2}{3em}{ 0.61$_{-0.19}^{+0.29}$} \\ 
NSs (excluding LS 5039) & \\
\hline
BHs (excluding BW Cir) & \multirow{2}{3em}{0.47$_{-0.20}^{+0.24}$} \\ 
NSs (excluding Aql X-1) & \\
\hline
BHs (excluding BW Cir) & \multirow{2}{3em}{ 0.32$_{-0.19}^{+0.23}$} \\ 
NSs (excluding LS 5039) & \\
\hline
BHs (excluding BW Cir \& SS 433) & \multirow{2}{3em}{0.52$_{-0.29}^{+0.12}$} \\ 
NSs & \\
\hline
BHs (excluding SS 433) & \multirow{2}{3em}{0.64$_{-0.16}^{+0.31}$} \\ 
NSs (excluding Aql X-1) & \\
\hline
BHs (excluding SS 433) & \multirow{2}{3em}{ 0.77$_{-0.18}^{+0.23}$} \\ 
NSs (excluding LS 5039) & \\
\hline
BHs & \multirow{2}{3em}{0.60$_{-0.34}^{+0.14}$} \\ 
NSs (excluding Aql X-1 \& LS 5039) & \\
\hline
BHs (excluding BW Cir \& SS 433) & \multirow{2}{3em}{0.64$_{-0.16}^{+0.31}$} \\ 
NSs (excluding Aql X-1) & \\
\hline
BHs (excluding BW Cir \& SS 433) & \multirow{2}{3em}{0.46$_{-0.20}^{+0.24}$} \\ 
NSs (excluding LS 5039) & \\
\hline
BHs (excluding BW Cir) & \multirow{2}{3em}{0.60$_{-0.34}^{+0.14}$} \\ 
NSs (excluding Aql X-1 \& LS 5039) & \\
\hline
BHs (excluding SS 433) & \multirow{2}{3em}{0.68$_{-0.14}^{+0.33}$} \\ 
NSs (excluding Aql X-1 \& LS 5039) & \\
\hline
BHs (excluding BW Cir \& SS 433) & \multirow{2}{3em}{0.67$_{-0.26}^{+0.20}$} \\ 
NSs (excluding Aql X-1 \& LS 5039) & \\
\hline
\end{tabular}
\caption[Results of Kolmogorov-Smirnov testing]{Results of Kolmogorov-Smirnov testing comparing black hole (BH) and neutron star (NS) natal kicks (see Table\,\ref{Table:KSTests}) repeated 10 million times, resulting in a range of $p$-values (presented as MAP \& 68\% HDI). These tests are carried out excluding various combinations of Aql X-1, BW Cir, LS 5039, \& SS 433 due to their somewhat uncertain parameters, as discussed above. The results of the KS-testing do not differ substancially when including vs excluding these systems, and we are therefore confident that our results are not dependent on some of the less-well characterised systems.}
\label{Table:KSTestsWeird}
\end{table*}

\clearpage

\section{Sources that failed to be reproduced}\label{Section:App_Failures}
Here we briefly summarise the 5 systems for which we have all the relevant information (i.e. well-constrained masses, orbital periods, and peculiar velocities) but we are unable to reproduce with our modelling. These systems prove difficult to replicate under the default assumptions for binary evolution, primarily due to their low companion masses and tight orbital periods, which result in mergers (or, in the case of 4U\,1700-37, the high companion mass - see Paper I for in depth discussion). 

This may be, in part, due to fundamental limitations of \cosmic.
Very low-mass donor stars that have undergone significant mass transfer become degenerate, as well as fully convective - BPS codes often employ relations and analytical approximations based on main-sequence or solar type stars; these relations are typically not applicable in the very low mass regime. For example, the mass-radius relation changes for stars $<0.1$\,\Msun, significantly impacting its evolution and interactions with the compact object - however this is often not modelled fully. Furthermore, BSE models treat all stars as in thermal equilibrium, however, a star may be driven out of this equilibrium due to mass loss if the thermal and mass-transfer time-scales are comparable, leading to the unphysical scenarios in lieu of the systems we expect based on observations. Additionally, irradiation-induced mass transfer which may occur in very short period binaries is not fully incorporated into \cosmic's code.

\begin{table*}[h!]
\centering
\renewcommand{\arraystretch}{1.25}
\begin{tabular}{| l | c | c c c c |} 
\hline
Name & Type & $\Mc$ & $M_2$ & $\porb$ & $\upsilon_{\mathrm{pec}}$ \\
 &  & $\left[{\rm M}_{\odot}\right]$ & $\left[{\rm M}_{\odot}\right]$ & [days] & $\left[{\rm km\,s}^{-1}\right]$ \\
\hline
\hline
J1012+5307 & NS-LM-PSR & $1.64\pm0.22$ & $0.16\pm0.02$ & 0.6 & $84\pm12$ \\
J1023+0038 & NS-LM-PSR & $1.65\pm0.16$ & $0.22\pm0.03$ & 0.2 & $131\pm6$ \\
J1311--3430 & NS-LM-PSR & $2.15\pm0.11$ & $0.01\pm0.00$ & 0.07 & $108\pm15$ \\
J1653--0158 & NS-LM-PSR & $2.17\pm0.15$ & $0.01\pm0.00$ & 0.05 & $177\pm9$ \\
4U 1700--37 & NS-HMXB & $1.96\pm0.19$ & $46\pm5$ & 3.41 & $71\pm11$ \\
\hline
\end{tabular}
\caption{Characteristics of systems that could not be successfully modelled.}
\label{Table:Failed}
\end{table*}

\clearpage

\section{Systems which may have undergone ECSN}\label{Section:App_ECSN}
As discussed in Section\,\ref{Section:ECSN}, some neutron stars may form via electron capture supernovae (ECSN). This occurs when a 8-11\,\Msun star has a helium core mass $>$1.38\,\Msun, and never develops an iron core. The collapse is triggered by the sudden drop of electron pressure, resulting from electron capture onto $^{24}$Mg and $^{20}$Ne \citep[e.g.][]{Miyaji1980,Nomoto1984,Podsiadlowski2004,Ivanova2008}. The range of helium core masses for which ECSN occur is debated \citep[e.g.][]{Hurley02,Podsiadlowski2004,Belczynski08,Andrews2015} - \cosmic follows \citet{Podsiadlowski2004}, where the helium core mass ranges 1.4--2.5\,\Msun. 

ECSN are thought to be associated with different, generally much smaller, natal kicks. This is due to a more rapid collapse and more spherically symmetric explosion (due to limited time for perturbations and instabilities to grow). In Table\,\ref{Table:ECSN} we highlight the systems which may have undergone ECSN according to our binary population synthesis, and quote the natal kick ranges associated with the different supernova channels.

\begin{table*}[h!]
\centering
\renewcommand{\arraystretch}{1.25}
\begin{tabular}{| l | l | c | c c c |} 
\hline
Name & Supernova & $N/N_{\rm tot}$ (\%) & \multicolumn{3}{@{}c@{}|}{$\vnk$} \\
 & & & Min & MAP & 68\%\\
\hline\hline
\multirow{3}{8em}{4U 1254-69} & CCSN & 91 & 127 & 179 & 158-207  \\ 
\cline{2-6}
& ECSN & 8 & 171 & 224 & 193-250  \\ 
\cline{2-6}
& AIC & 1 & 176 & 222 & 197-246  \\ 
\hline
\multirow{3}{8em}{4U 1636-536} & CCSN & 87 & 150 & 186 & 164-210  \\ 
\cline{2-6}
& ECSN & 11 & 177 & 226 & 204-259  \\ 
\cline{2-6}
& AIC & 2 & 191 & 221 & 197-237  \\ 
\hline
\multirow{3}{8em}{4U 1700+24} & CCSN & 59 & 0 & 10 & 9-13  \\ 
\cline{2-6}
& ECSN & 10 & 4 & 8 & 5-17  \\ 
\cline{2-6}
& AIC & 31 & 3 & 8 & 5-13  \\ 
\hline
\multirow{3}{8em}{Aql X-1} & CCSN & 91 & 0 & 14 & 0-18  \\ 
\cline{2-6}
& ECSN & 8 & 0 & 41 & 32-55  \\ 
\cline{2-6}
& AIC & 2 & 0 & 19 & 6-28  \\ 
\hline
\multirow{3}{8em}{J0348+0432} & CCSN & 74 & 42 & 105 & 89-123  \\ 
\cline{2-6}
& ECSN & 25 & 47 & 111 & 87-147 \\ 
\cline{2-6}
& AIC & 1 & 53 & 122 & 111-161  \\ 
\hline
\multirow{3}{8em}{J1417-4402} & CCSN & 92 & 8 & 115 & 66-160  \\ 
\cline{2-6}
& ECSN & 7 & 18 & 161 & 133-185  \\ 
\cline{2-6}
& AIC & 1 & 148 & 173 & 162-193  \\ 
\hline
J1431-4751 & ECSN & 100 & 108 & 117 & 110-133  \\ 
\hline
J1622-0315 & ECSN & 100 & 129 & 142 & 136-157  \\ 
\hline
J1628-3205 & ECSN & 100 & 84 & 98 & 91-132  \\ 
\hline
\multirow{3}{8em}{J1723-2837} & CCSN & 43 & 121 & 209 & 203-220  \\ 
\cline{2-6}
& ECSN & 55 & 109 & 129 & 118-138  \\ 
\cline{2-6}
& AIC & 2 & 136 & 148 & 137-150  \\ 
\hline
J2039-5617 & ECSN & 100 & 84 & 99 & 90-131  \\ 
\hline
\multirow{3}{8em}{J2215+5135} & CCSN & 95 & 152 & 217 & 181-238 \\ 
\cline{2-6}
& ECSN & 5 &  160 & 250 & 189-269 \\ 
\hline
\multirow{2}{8em}{RX J0146.9+6121} & CCSN & 93 & 3 & 19 & 16-30  \\ 
\cline{2-6}
& ECSN & 7 & 11 & 17 & 13-32  \\ 
\hline
\multirow{2}{8em}{Sco X-1} & CCSN & 79 & 113 & 165 & 151-179  \\ 
\cline{2-6}
& ECSN & 21 & 179 & 219 & 194-224  \\ 
\hline
\multirow{3}{8em}{Ser X-1} & CCSN & 30 & 95 & 140 & 129-153  \\ 
\cline{2-6}
& ECSN & 69 & 75 & 115 & 93-144  \\ 
\cline{2-6}
& AIC & 1 & 106 & 162 & 130-187  \\ 
\hline
\multirow{2}{8em}{XSS J12270-4859} & CCSN & 95 & 154 & 225 & 195-244  \\ 
\cline{2-6}
& ECSN & 5 & 118 & 132 & 123-155  \\ 
\hline
\end{tabular}
\caption{Natal kick constraints for 16 compact objects segregated by supernova type: core-collapse supernova (CCSN), electron capture supernova (ECSN), and accretion-induced collapse (AIC), including the percentage of analogues which are subject to each supernova ($N/N_{\rm tot}$). The minimum, MAP, and 68\% highest-posterior-density interval natal kick magnitudes are shown. All velocities are in \kmps.}
\label{Table:ECSN}
\end{table*}

\clearpage

\section{Scenario in which a system can lose more than half its total mass and survive}\label{Section:App_HalfMassLoss}
As mentioned in Section \ref{Section:PhysicalScalings}, the fraction of total system mass lose due to the supernova is directly related to whether the binary remains bound. Typically, for a system to avoid disruption, the ejected mass is limited to half the total system mass. However, a finely tuned natal kick can allow a binary to survive losing a larger fraction of its initial mass. Here, we briefly demonstrate this by considering the equation which determines the post-SN orbital separation as a function of system parameters and natal kick - the variables employed in this derivation are as follows: 

\noindent $M_1$ and $M_{\rm CO}$ are the masses of the compact progenitor and the newly formed compact object, respectively; $M_2$ is the mass of the stellar companion; $a_{\rm SN}$ and $a_{\rm post-SN}$ are the orbital semi-major axes at the instant of supernova and directly following the supernova, respectively; $r_{\rm SN}$ is the orbital separation at instant of supernova (related to the semi-major axis by $r_{\rm SN}=a_{\rm SN}\left(1-e\cos M\right)$, where $e$ and $M$ are the eccentricity and mean anomaly); $\upsilon_{\rm rel}$ is the relative velocity of the binary components at instant of supernova; $\vnk$ is the natal kick magnitude; and $\theta$ and $\phi$ are the polar and azimuthal angles describing the direction of the natal kick. Subscripts $X_{\rm S}$ and $X_{\rm C}$ denote the sine and cosine the angle $X$, and G is the gravitational constant. 
\\\\
\noindent Orbital separation is given by $a_{\rm post-SN}^{-1} = 2r_{\rm SN}^{-1} - \left(\rm{G}\left(M_{\rm CO}+M_2\right)\right)^{-1}\left[\upsilon_{\rm NK}^2+\upsilon_{\rm rel}^2+2\upsilon_{\rm NK}\upsilon_{\rm rel}\cphi\stheta\right]$ \citep{Brandt1995a,Kalogera1996}. Given that $\upsilon_{\rm rel}^2={\rm G}\left(M_1+M_2\right)r_{\rm SN}^{-1}$ one can rearrange this to find:

\begin{align*}
    a_{\rm post-SN}^{-1} = &\,\,\frac{2}{r_{\rm SN}} - \frac{1}{{\rm G}\left(M_{\rm CO}+M_2\right)}\left[\upsilon_{\rm NK}^2+\upsilon_{\rm rel}^2+2\upsilon_{\rm NK}\upsilon_{\rm rel}\cphi\stheta\right] \\
     = &\,\,\frac{2}{r_{\rm SN}} - \frac{1}{{\rm G}\left(M_{\rm CO}+M_2\right)}\left[\upsilon_{\rm NK}^2+\frac{{\rm G}\left(M_1+M_2\right)}{r_{\rm SN}}+2\upsilon_{\rm NK}\frac{\sqrt{{\rm G}\left(M_1+M_2\right)}}{\sqrt{r_{\rm SN}}}\cphi\stheta\right] \\
     = &\,\,\frac{2}{r_{\rm SN}} - \frac{{\rm G}\left(M_1+M_2\right)}{{\rm G}\left(M_{\rm CO}+M_2\right)}\left[\frac{\upsilon_{\rm NK}^2}{{\rm G}\left(M_1+M_2\right)}+\frac{1}{r_{\rm SN}}+\frac{2\upsilon_{\rm NK}\cphi\stheta}{\sqrt{{\rm G}\left(M_1+M_2\right)}\sqrt{r_{\rm SN}}}\right] \\
     = &\,\,\frac{2}{r_{\rm SN}} - \frac{M_1+M_2}{M_{\rm CO}+M_2}\left[\frac{\upsilon_{\rm NK}^2}{{\rm G}\left(M_1+M_2\right)}+\frac{1}{r_{\rm SN}}+\frac{2\upsilon_{\rm NK}\cphi\stheta}{\sqrt{{\rm G}\left(M_1+M_2\right)}\sqrt{r_{\rm SN}}}\right] \\
     = &\,\,\frac{2}{r_{\rm SN}} - \frac{M_1+M_2}{M_{\rm CO}+M_2}\left[\frac{\upsilon_{\rm NK}^2}{\upsilon_{\rm rel}^2 r_{\rm SN}}+\frac{1}{r_{\rm SN}}+\frac{2\upsilon_{\rm NK}\cphi\stheta}{\upsilon_{\rm rel} r_{\rm SN}}\right] \\
     = &\,\,\frac{2}{r_{\rm SN}} - \frac{M_1+M_2}{r_{\rm SN}\left(M_{\rm CO}+M_2\right)}\left[\frac{\upsilon_{\rm NK}^2}{\upsilon_{\rm rel}^2}+1+\frac{2\upsilon_{\rm NK}\cphi\stheta}{\upsilon_{\rm rel}}\right] \\
     \frac{r_{\rm SN}}{a_{\rm post-SN}} = &\,\,2  - \frac{M_1+M_2}{M_{\rm CO}+M_2}\left[\frac{\upsilon_{\rm NK}^2}{\upsilon_{\rm rel}^2}+1+\frac{2\upsilon_{\rm NK}\cphi\stheta}{\upsilon_{\rm rel}}\right] \\
\end{align*}

\noindent For a system to remain bound, $a_{\rm post-SN}>0$, therefore $\frac{M_1+M_2}{M_{\rm CO}+M_2}\left[\frac{\upsilon_{\rm NK}^2}{\upsilon_{\rm rel}^2}+1+\frac{2\upsilon_{\rm NK}\cphi\stheta}{\upsilon_{\rm rel}}\right]<2$.

\noindent If $\upsilon_{\rm NK}=0$ (i.e. no natal kick), $\frac{M_1+M_2}{M_{\rm CO}+M_2}$ must be less than 2, meaning the system cannot lose more than its total mass - as expected due to the virial theorem. If there {\em is} a natal kick, we can find the conditions necessary for $\frac{M_1+M_2}{M_{\rm CO}+M_2}>2$ (i.e. special case in which systems can lose more than half their initial mass).

\begin{align*}
     2 > &\,\,\frac{M_1+M_2}{M_{\rm CO}+M_2}\left[\frac{\upsilon_{\rm NK}^2}{\upsilon_{\rm rel}^2}+1+\frac{2\upsilon_{\rm NK}\cphi\stheta}{\upsilon_{\rm rel}}\right] \\
     \frac{2\left(M_{\rm CO}+M_2\right)}{M_1+M_2} > &\,\,\frac{\upsilon_{\rm NK}^2}{\upsilon_{\rm rel}^2}+1+\frac{2\upsilon_{\rm NK}\cphi\stheta}{\upsilon_{\rm rel}} \\
     1 > &\,\,\frac{\upsilon_{\rm NK}^2}{\upsilon_{\rm rel}^2}+1+\frac{2\upsilon_{\rm NK}\cphi\stheta}{\upsilon_{\rm rel}} \\
     0 > &\,\,\frac{\upsilon_{\rm NK}^2}{\upsilon_{\rm rel}^2}+\frac{2\upsilon_{\rm NK}\cphi\stheta}{\upsilon_{\rm rel}} \\
     0 > &\,\,\upsilon_{\rm NK}^2+2\upsilon_{\rm NK}\upsilon_{\rm rel}\cphi\stheta \\
     0 > &\,\,\upsilon_{\rm NK}\left(\upsilon_{\rm NK}+2\upsilon_{\rm rel}\cphi\stheta\right) \\
\end{align*}

\noindent If a system loses more than half its total mass during the supernova, the natal kick magnitude must satisfy $\upsilon_{\rm NK}<-2\upsilon_{\rm rel}\cphi\stheta$.

\clearpage

\section{$\upsilon_{\rm NK}$ distributions for all systems}
\label{Section:App_AllResults}
The following pages show the $\vnk$ distributions for all 66 successfully simulated systems. Each page presents a single object, with the $\vnk$ and $\vmlk$ distributions shown as histograms on the bottom, and a corner plot at the top spanning the key parameter-space regions of the  progenitor in terms of the ZAMS component masses (\monei, \mtwoi) in units of Solar masses, orbital period (\porbi) per 100\,days, eccentricity (\ei), metallicity ($Z$) in units of Solar and the natal kick magnitude ($\vnk$).

\begin{figure*}
    \centering
    \includegraphics[width=0.75\textwidth]{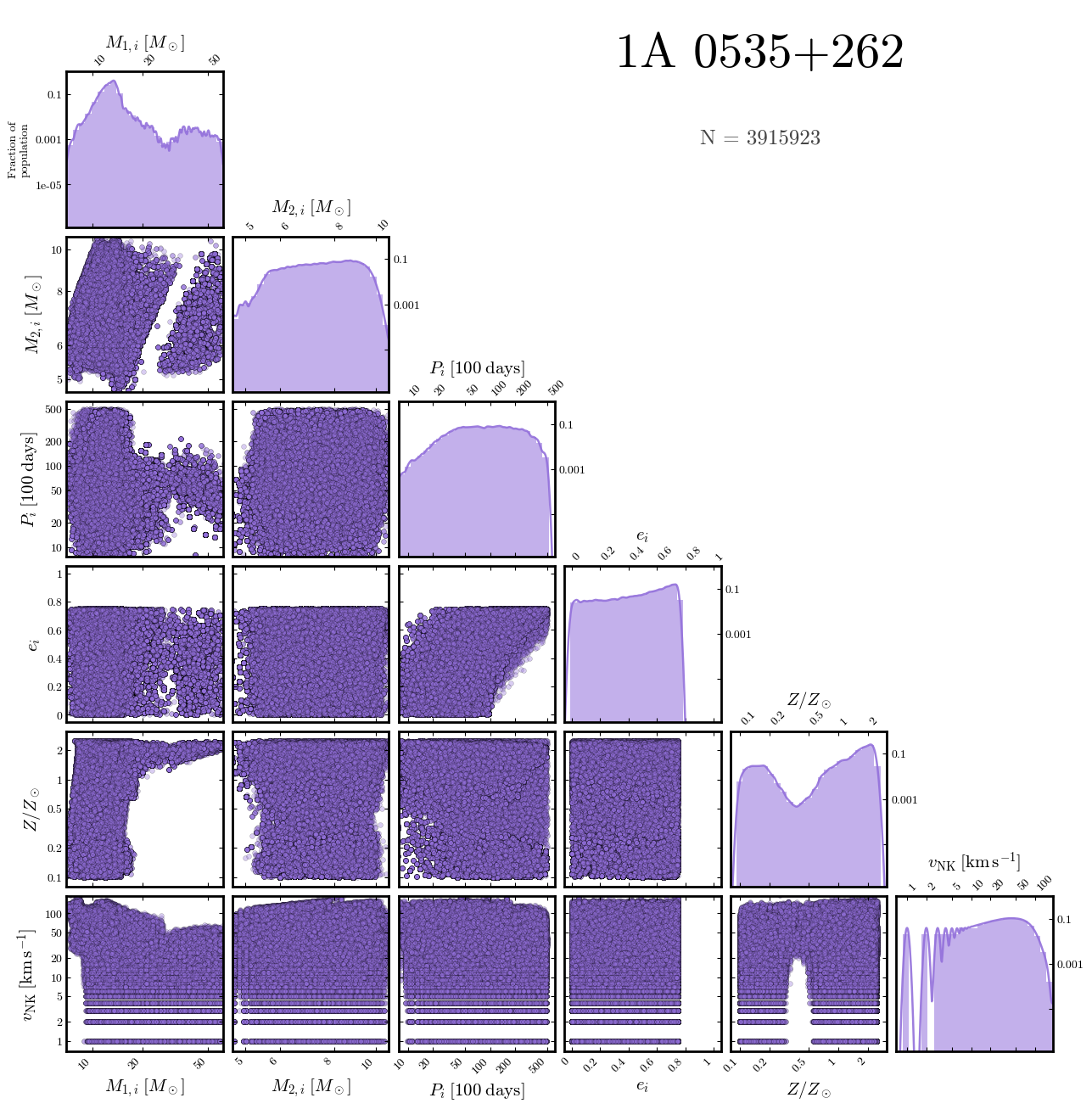}
    \caption{Corner plot highlighting key initial characteristics and acceptable natal kicks for simulated system analogues to 1A 0535+262.}
    \label{Fig:A0535_CornerPlot}
\end{figure*}

\begin{figure*}
    \centering
    \includegraphics[width=0.75\textwidth]{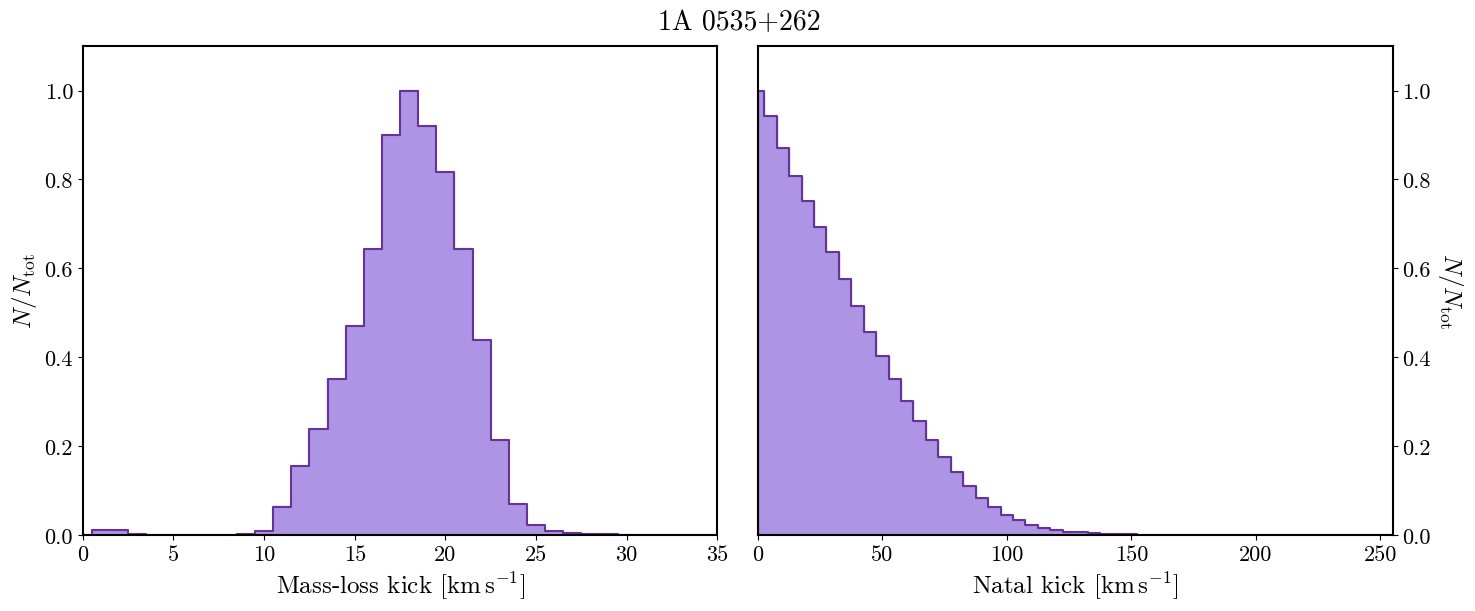}
    \caption{Distribution of mass-loss kicks ($\vmlk$) magnitude (left) and natal kick ($\vnk$) magnitude (right) for simulated system analogues to 1A 0535+262.}
    \label{Fig:A0535_KickHist}
\end{figure*}
\clearpage

\begin{figure*}
    \centering
    \includegraphics[width=0.75\textwidth]{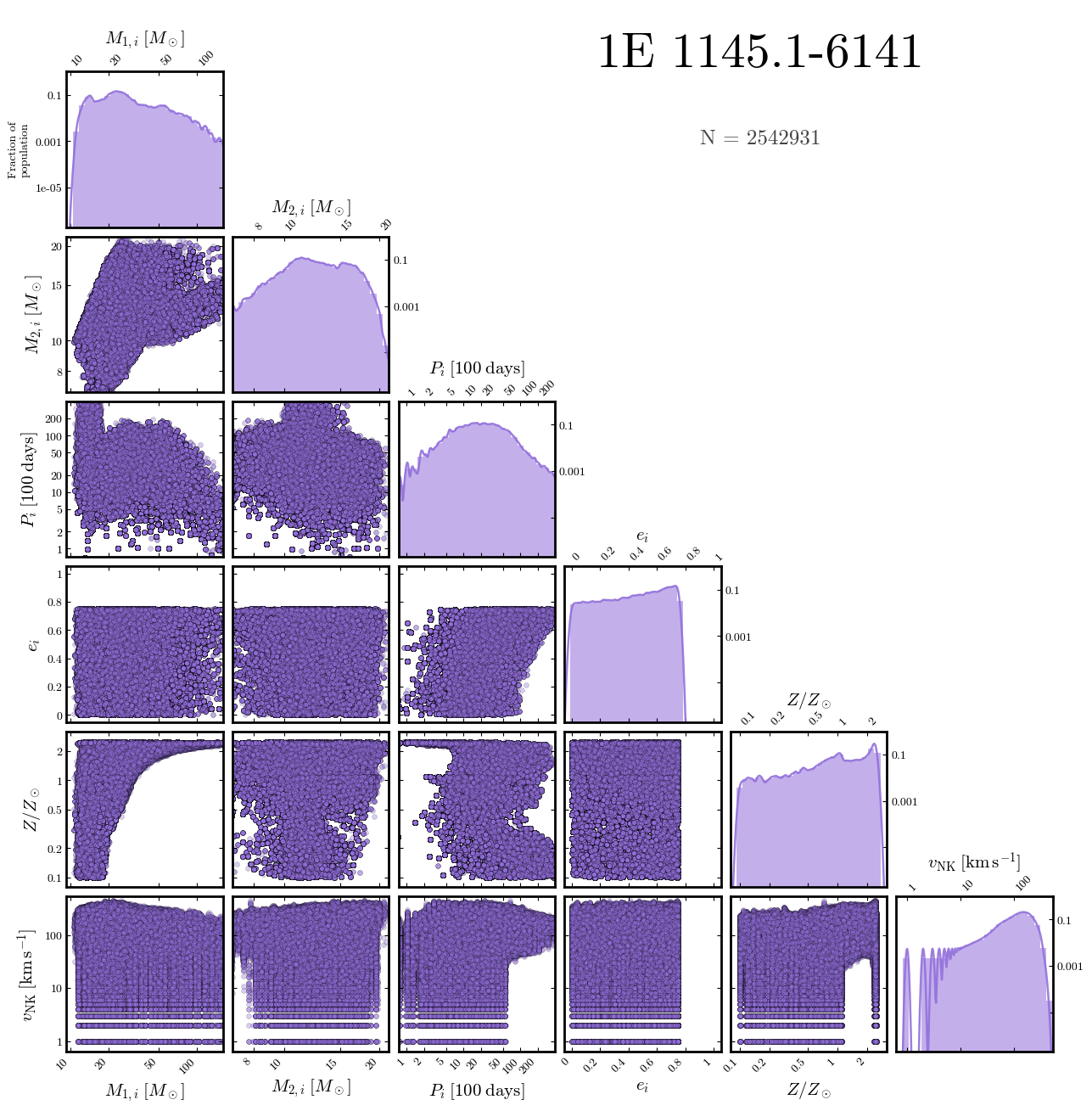}
    \caption{Corner plot highlighting key initial characteristics and acceptable natal kicks for simulated system analogues to 1E 1145.1-6141.}
    \label{Fig:E1145_CornerPlot}
\end{figure*}
\begin{figure*}
    \centering
    \includegraphics[width=0.75\textwidth]{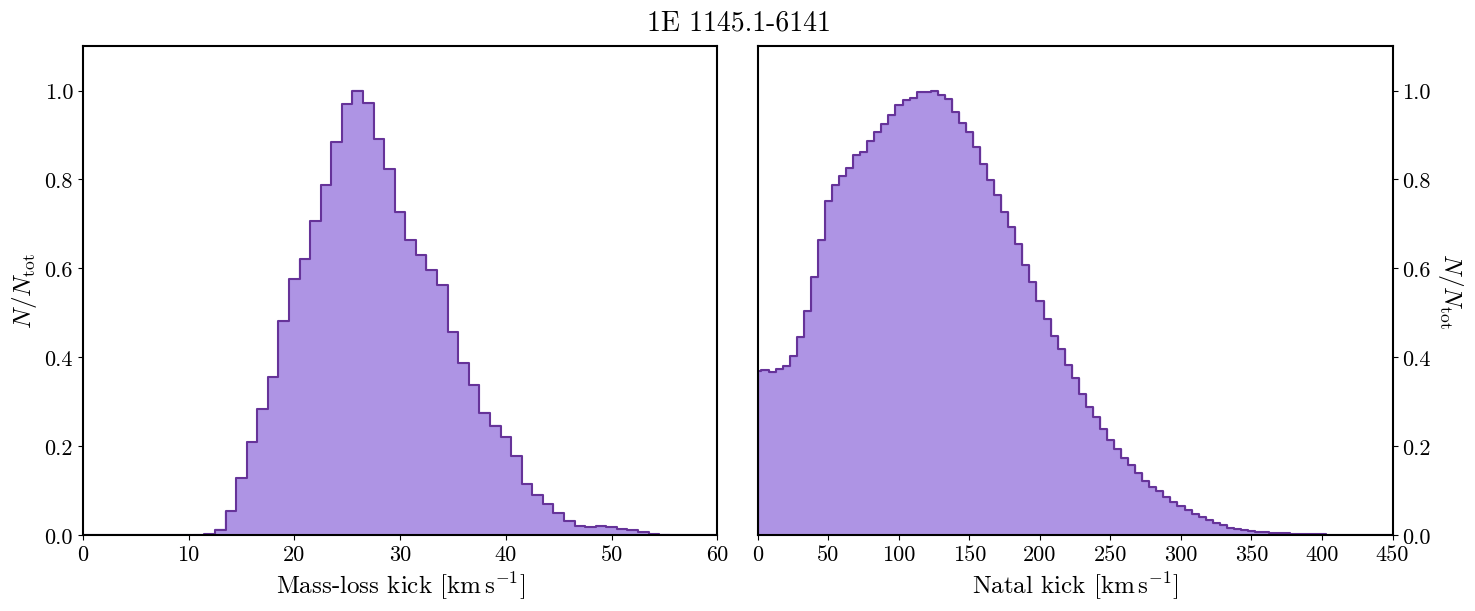}
    \caption{Distribution of mass-loss kicks ($\vmlk$) magnitude (left) and natal kick ($\vnk$) magnitude (right) for simulated system analogues to 1E 1145.1-6141.}
    \label{Fig:E1145_KickHist}
\end{figure*}
\clearpage

\begin{figure*}
    \centering
    \includegraphics[width=0.75\textwidth]{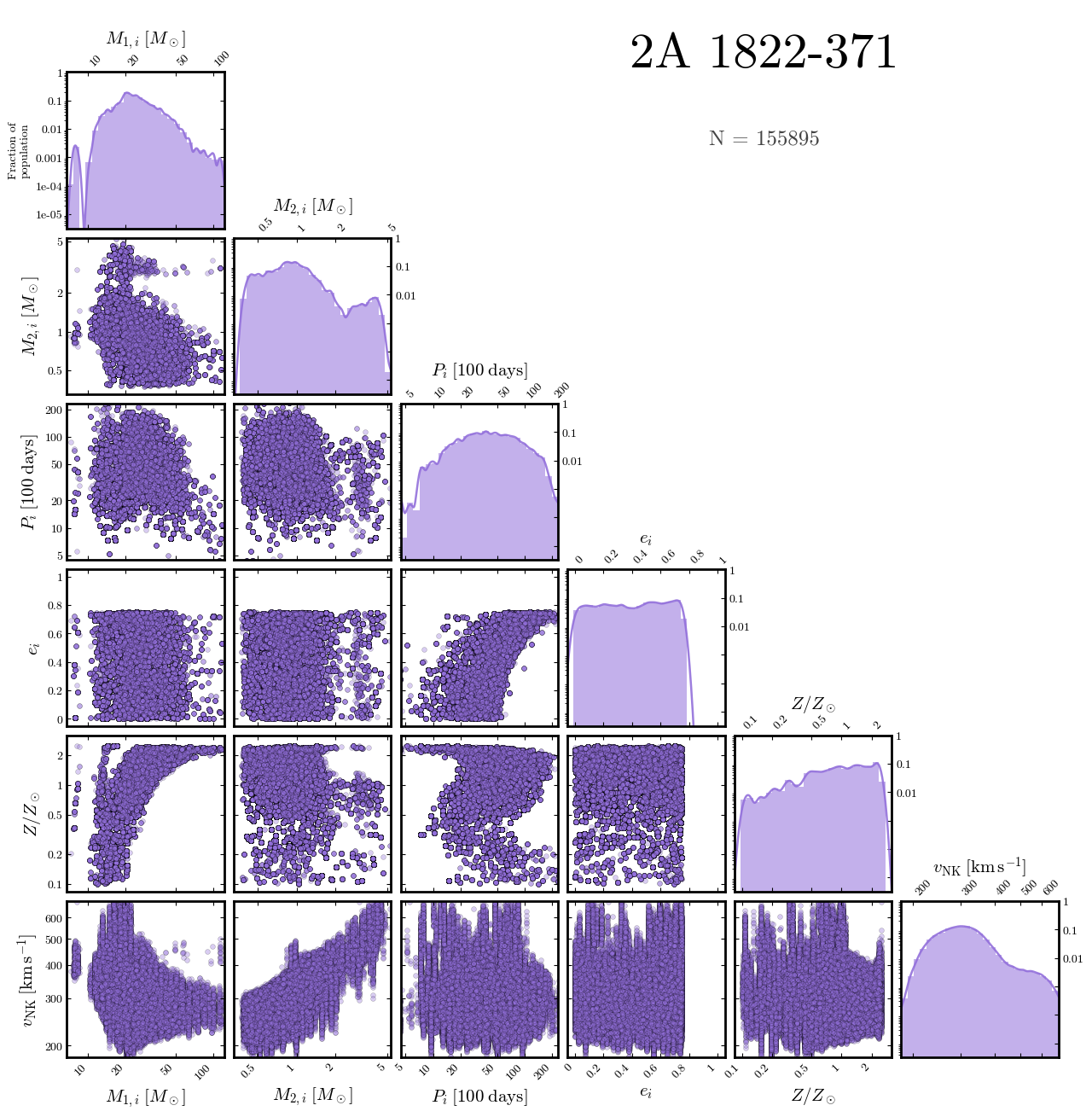}
    \caption{Corner plot highlighting key initial characteristics and acceptable natal kicks for simulated system analogues to 2A 1822-371.}
    \label{Fig:A1822_CornerPlot}
\end{figure*}
\begin{figure*}
    \centering
    \includegraphics[width=0.75\textwidth]{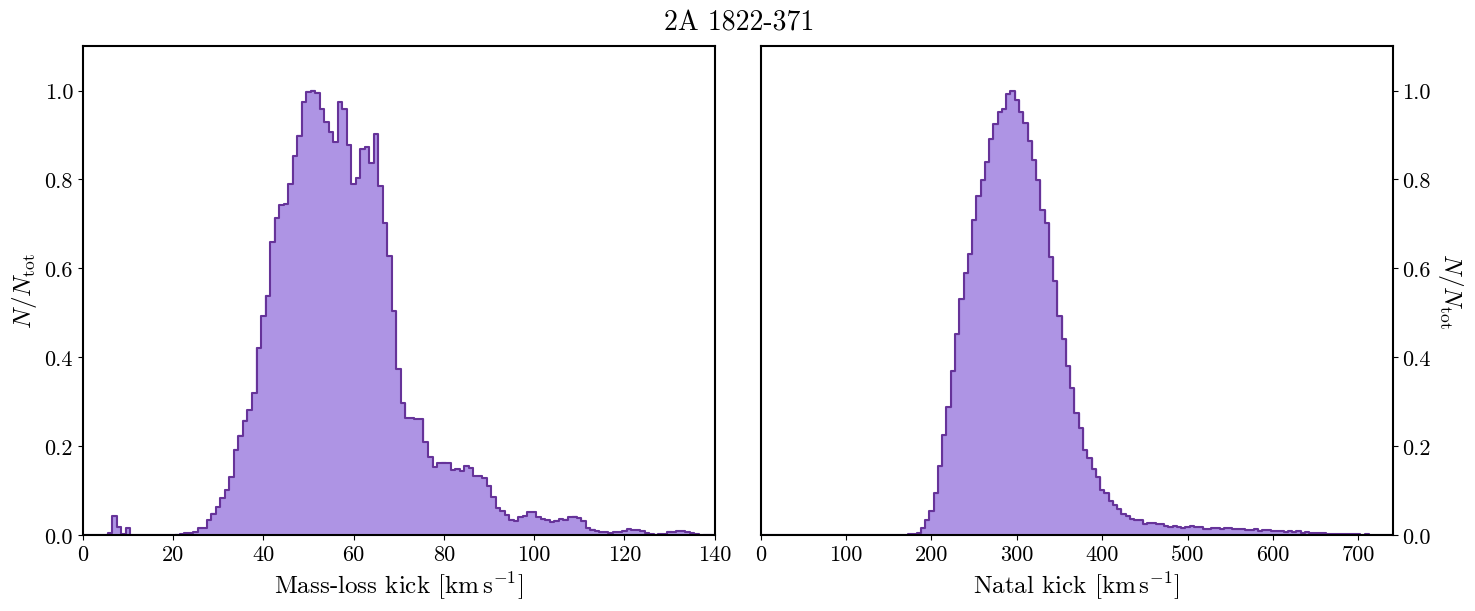}
    \caption{Distribution of mass-loss kicks ($\vmlk$) magnitude (left) and natal kick ($\vnk$) magnitude (right) for simulated system analogues to 2A 1822-371.}
    \label{Fig:A1822_KickHist}
\end{figure*}
\clearpage

\begin{figure*}
    \centering
    \includegraphics[width=0.75\textwidth]{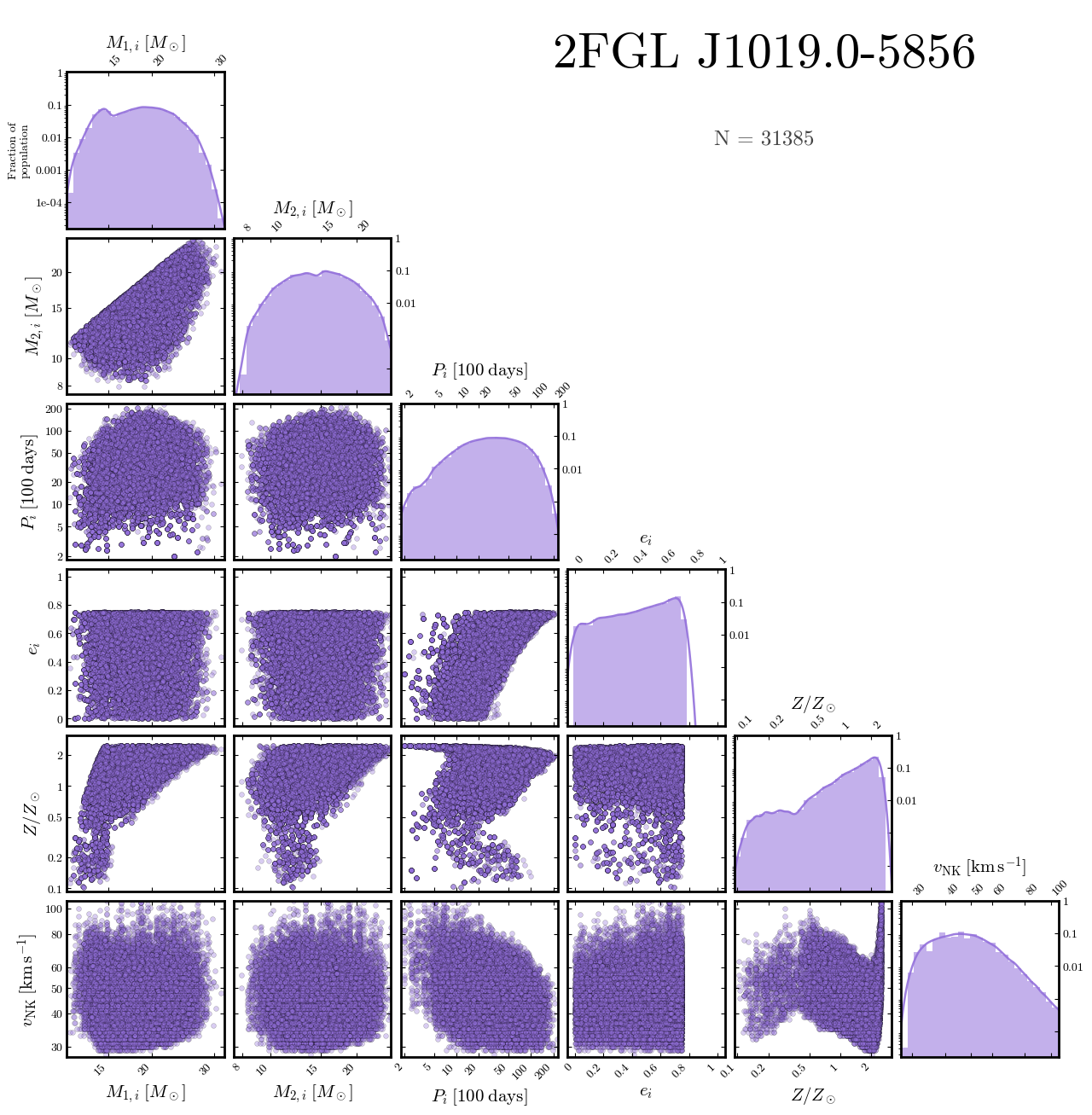}
    \caption{Corner plot highlighting key initial characteristics and acceptable natal kicks for simulated system analogues to 2FGL J1019.0-5856.}
    \label{Fig:J1019_CornerPlot}
\end{figure*}
\begin{figure*}
    \centering
    \includegraphics[width=0.75\textwidth]{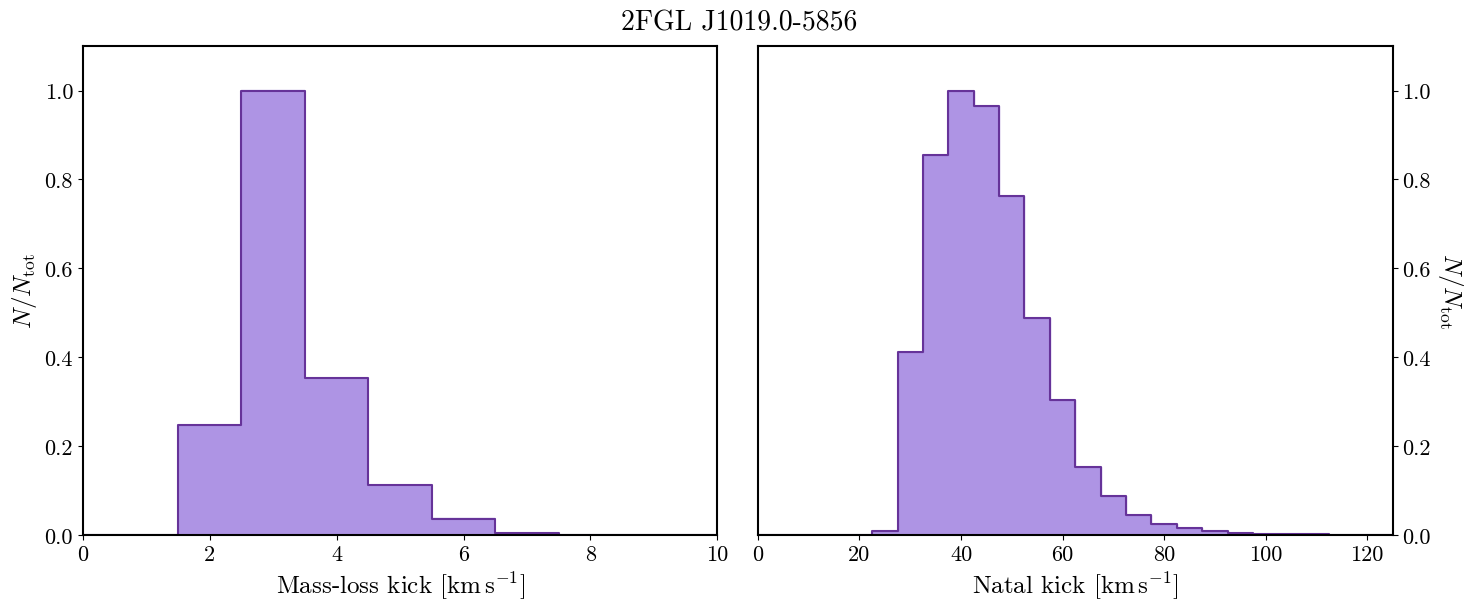}
    \caption{Distribution of mass-loss kicks ($\vmlk$) magnitude (left) and natal kick ($\vnk$) magnitude (right) for simulated system analogues to 2FGL J1019.0-5856.}
    \label{Fig:J1019_KickHist}
\end{figure*}
\clearpage

\begin{figure*}
    \centering
    \includegraphics[width=0.75\textwidth]{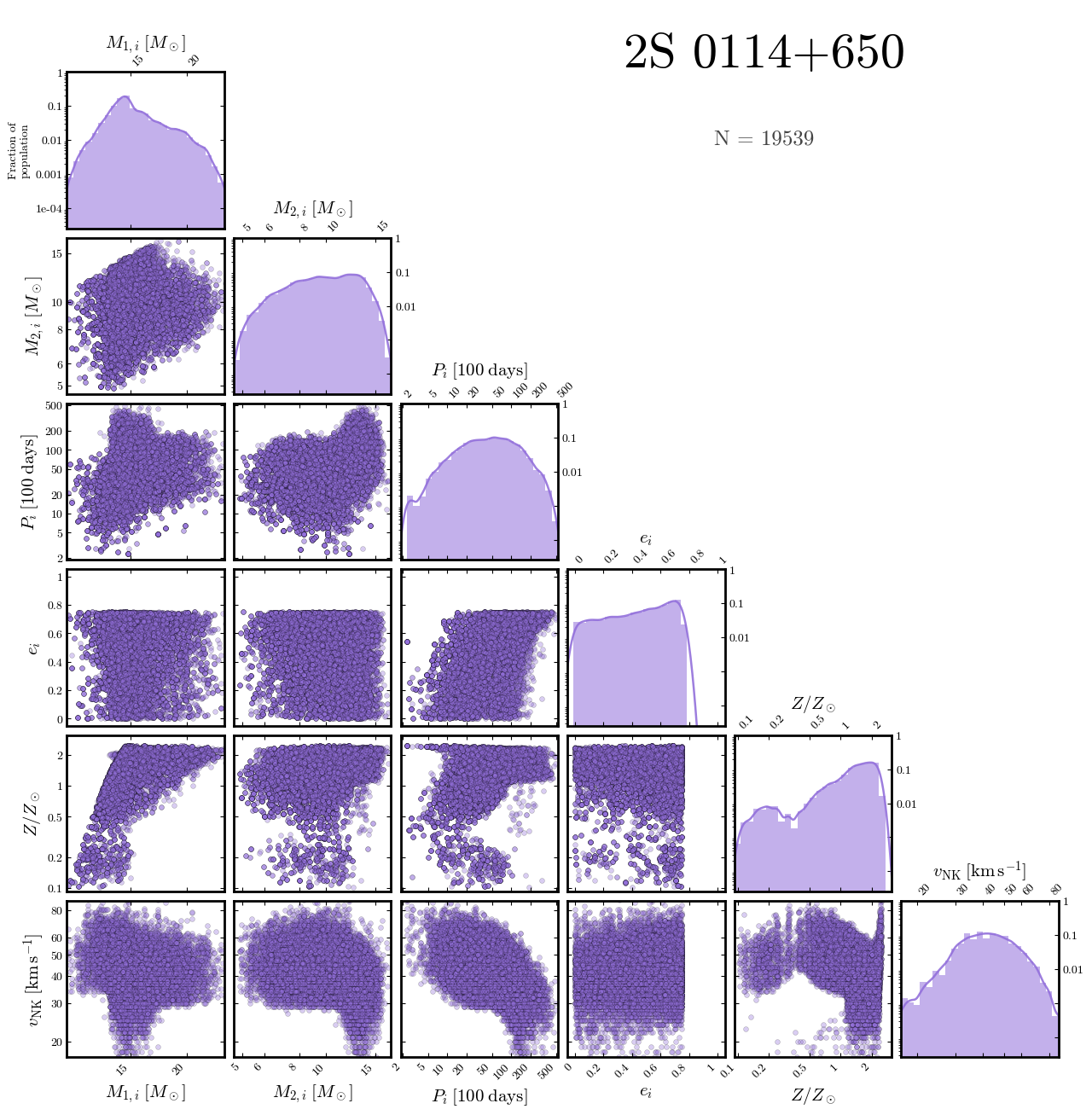}
    \caption{Corner plot highlighting key initial characteristics and acceptable natal kicks for simulated system analogues to 2S 0114+650.}
    \label{Fig:S0114_CornerPlot}
\end{figure*}
\begin{figure*}
    \centering
    \includegraphics[width=0.75\textwidth]{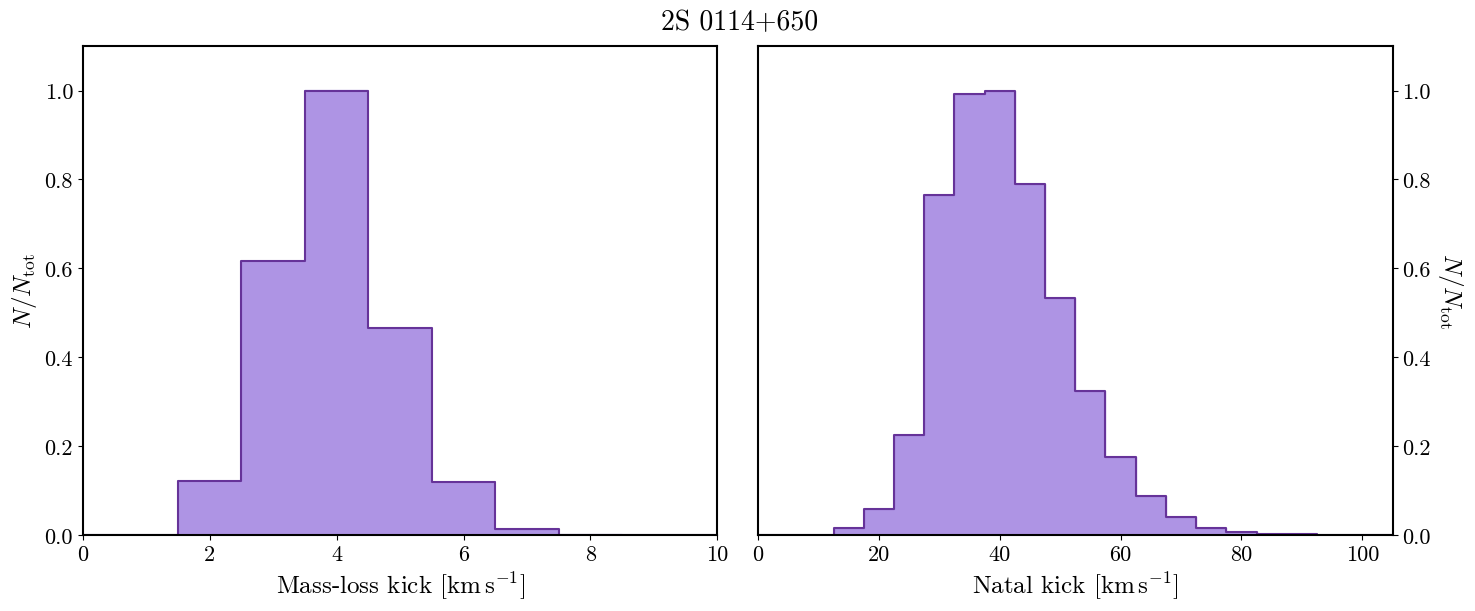}
    \caption{Distribution of mass-loss kicks ($\vmlk$) magnitude (left) and natal kick ($\vnk$) magnitude (right) for simulated system analogues to 2S 0114+650.}
    \label{Fig:S0114_KickHist}
\end{figure*}
\clearpage

\begin{figure*}
    \centering
    \includegraphics[width=0.75\textwidth]{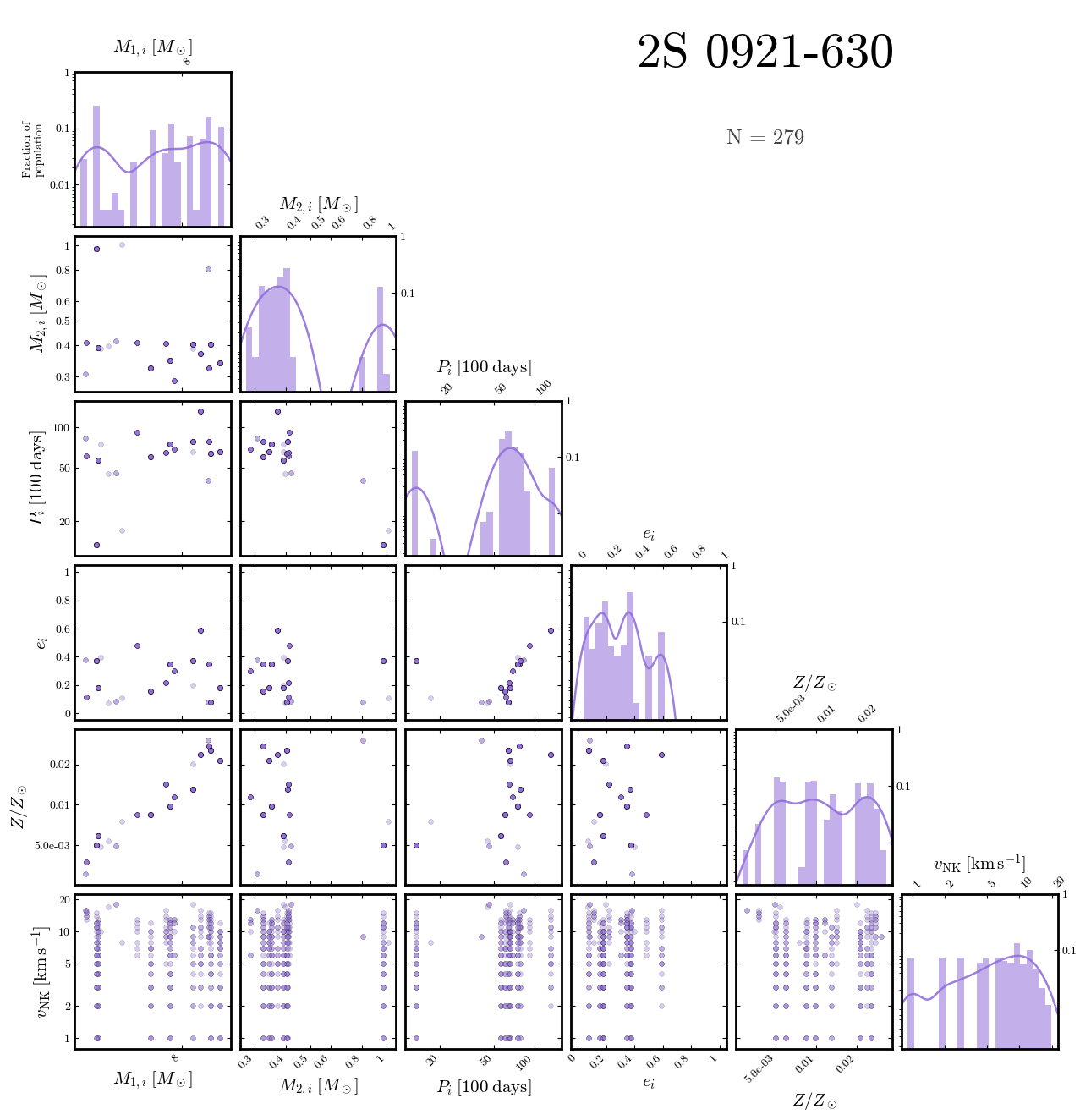}
    \caption{Corner plot highlighting key initial characteristics and acceptable natal kicks for simulated system analogues to 2S 0921-630.}
    \label{Fig:S0921_CornerPlot}
\end{figure*}
\begin{figure*}
    \centering
    \includegraphics[width=0.75\textwidth]{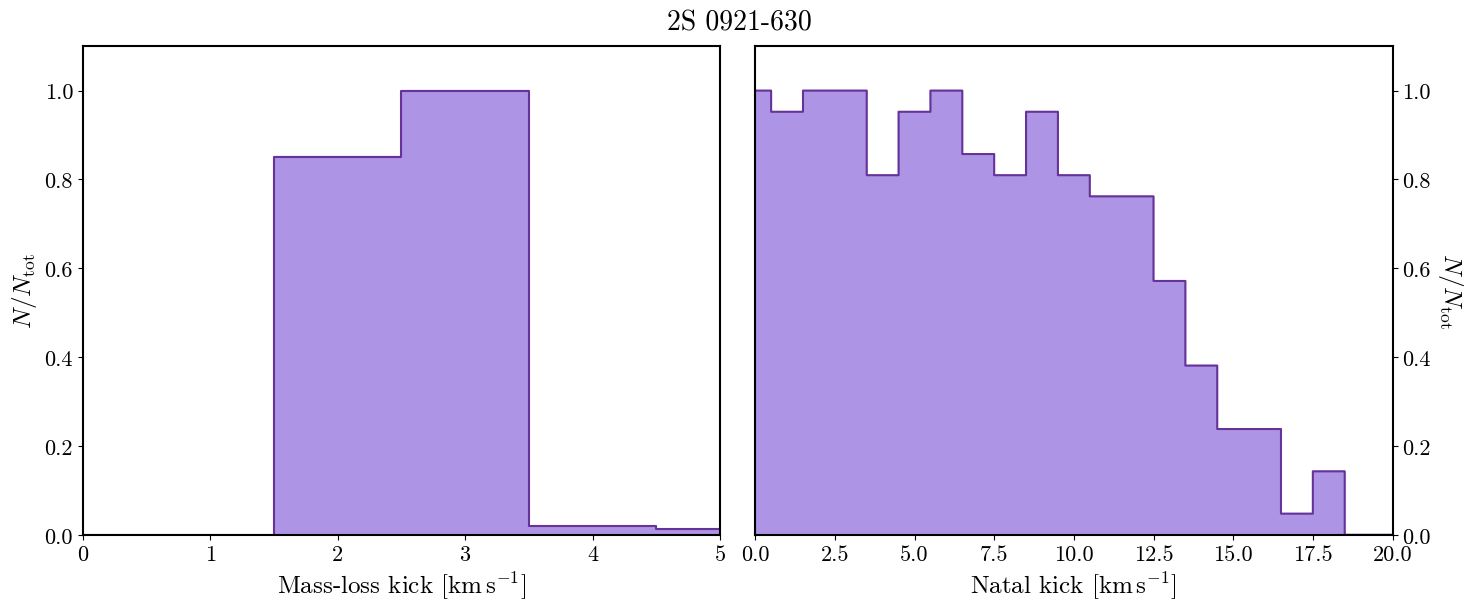}
    \caption{Distribution of mass-loss kicks ($\vmlk$) magnitude (left) and natal kick ($\vnk$) magnitude (right) for simulated system analogues to 2S 0921-630.}
    \label{Fig:S0921_KickHist}
\end{figure*}
\clearpage

\begin{figure*}
    \centering
    \includegraphics[width=0.75\textwidth]{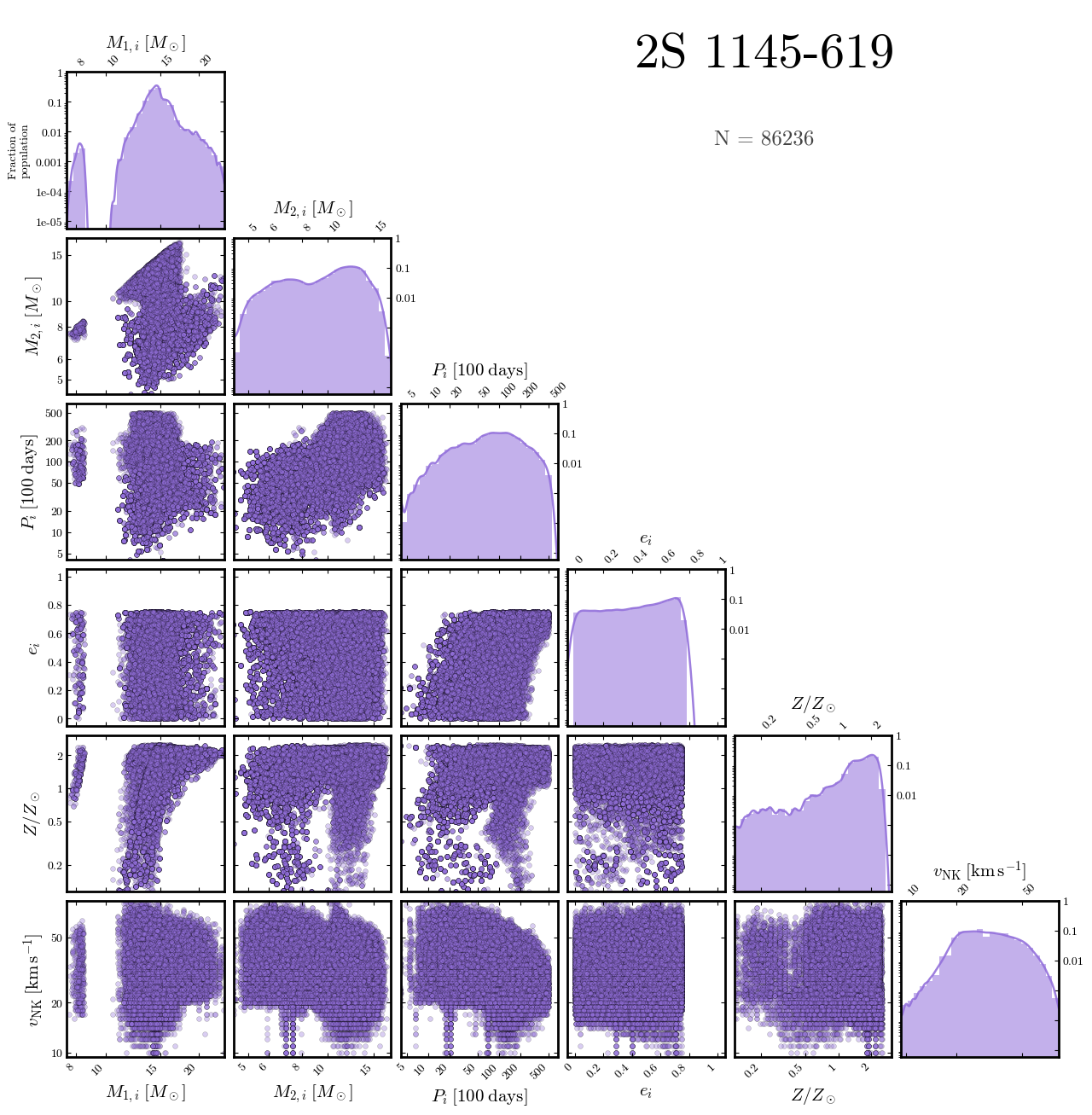}
    \caption{Corner plot highlighting key initial characteristics and acceptable natal kicks for simulated system analogues to 2S 1145-619.}
    \label{Fig:S1145_CornerPlot}
\end{figure*}
\begin{figure*}
    \centering
    \includegraphics[width=0.75\textwidth]{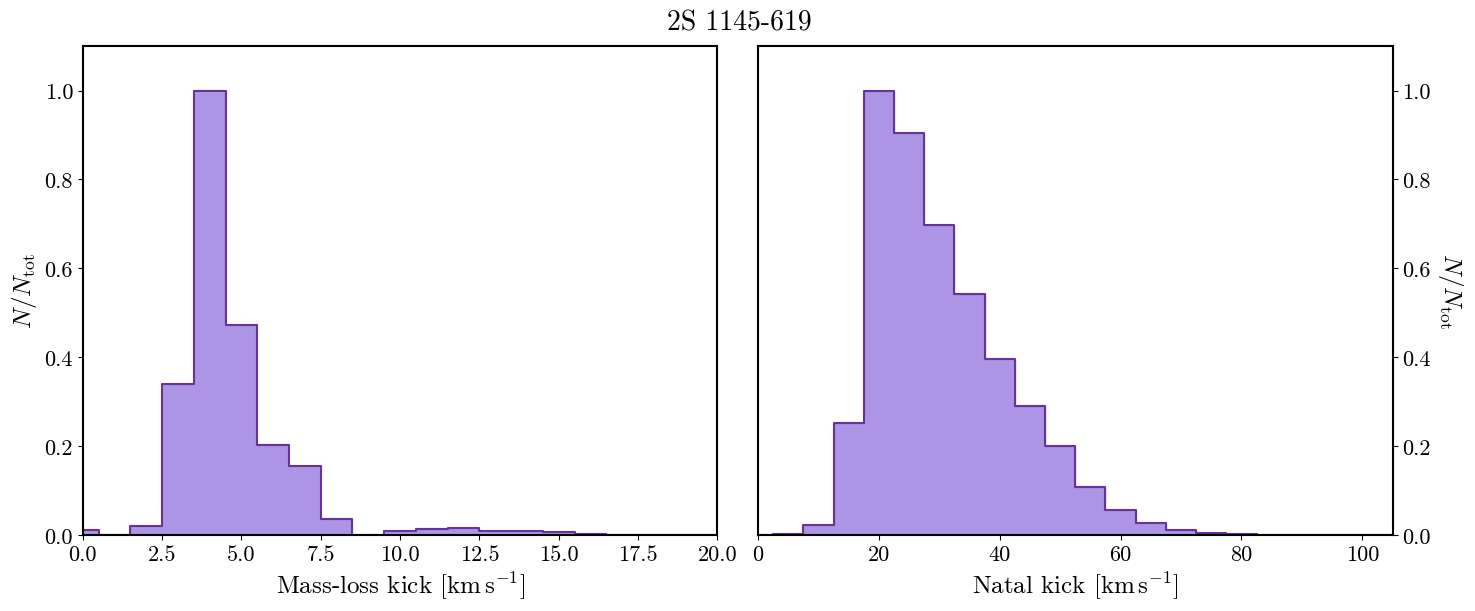}
    \caption{Distribution of mass-loss kicks ($\vmlk$) magnitude (left) and natal kick ($\vnk$) magnitude (right) for simulated system analogues to 2S 1145-619.}
    \label{Fig:S1145_KickHist}
\end{figure*}
\clearpage

\begin{figure*}
    \centering
    \includegraphics[width=0.75\textwidth]{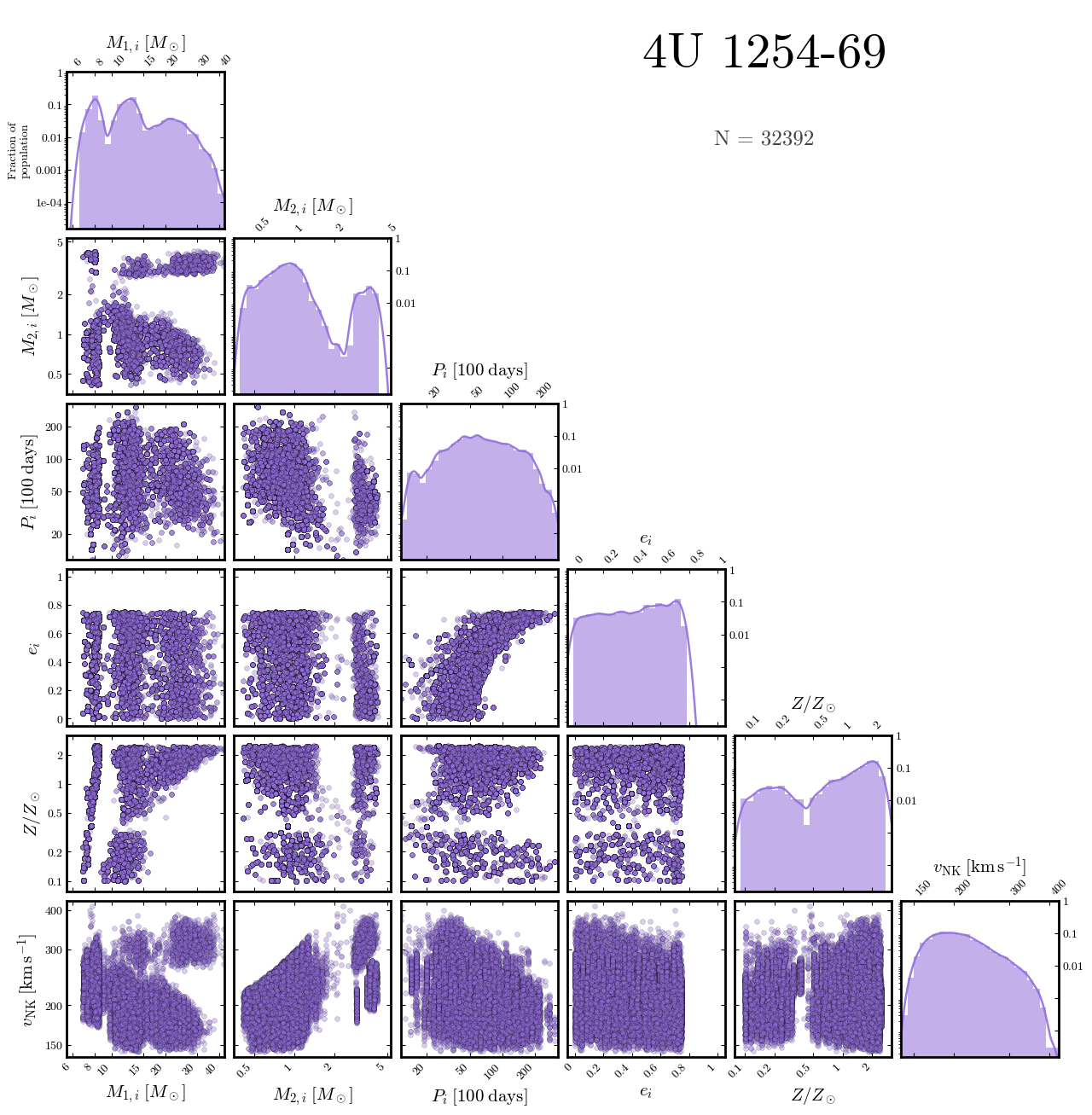}
    \caption{Corner plot highlighting key initial characteristics and acceptable natal kicks for simulated system analogues to 4U 1254-69.}
    \label{Fig:U1254_CornerPlot}
\end{figure*}
\begin{figure*}
    \centering
    \includegraphics[width=0.75\textwidth]{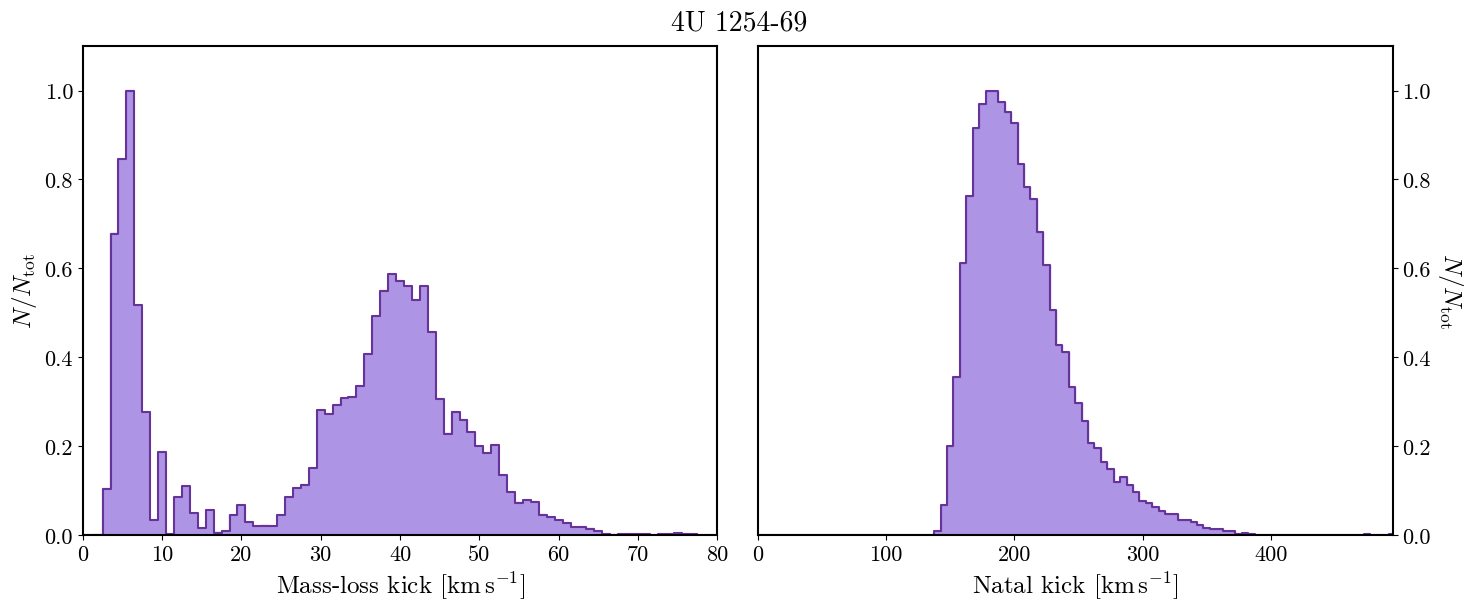}
    \caption{Distribution of mass-loss kicks ($\vmlk$) magnitude (left) and natal kick ($\vnk$) magnitude (right) for simulated system analogues to 4U 1254-69.}
    \label{Fig:U1254_KickHist}
\end{figure*}
\clearpage

\begin{figure*}
    \centering
    \includegraphics[width=0.75\textwidth]{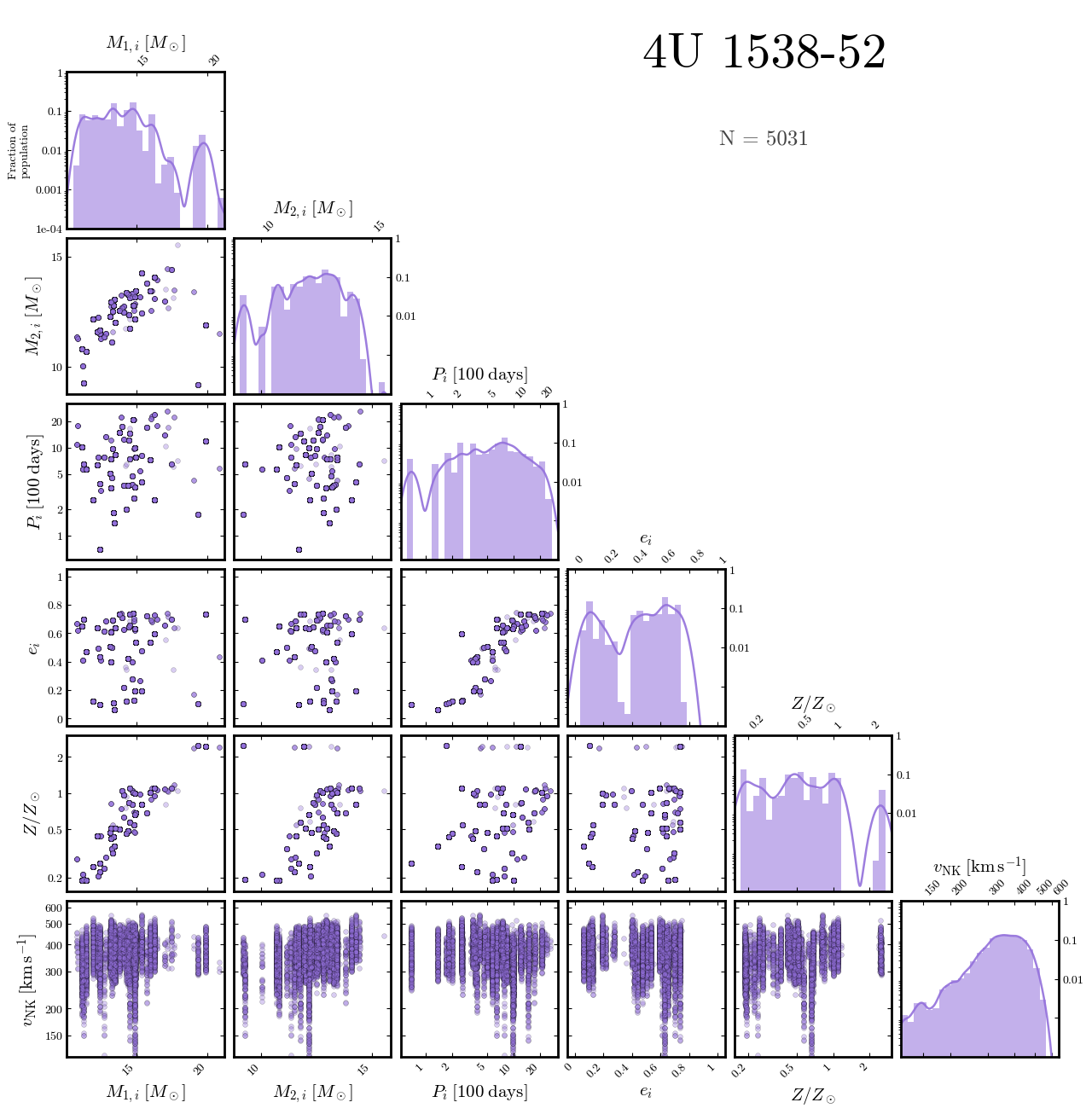}
    \caption{Corner plot highlighting key initial characteristics and acceptable natal kicks for simulated system analogues to 4U 1538-52.}
    \label{Fig:U1538_CornerPlot}
\end{figure*}
\begin{figure*}
    \centering
    \includegraphics[width=0.75\textwidth]{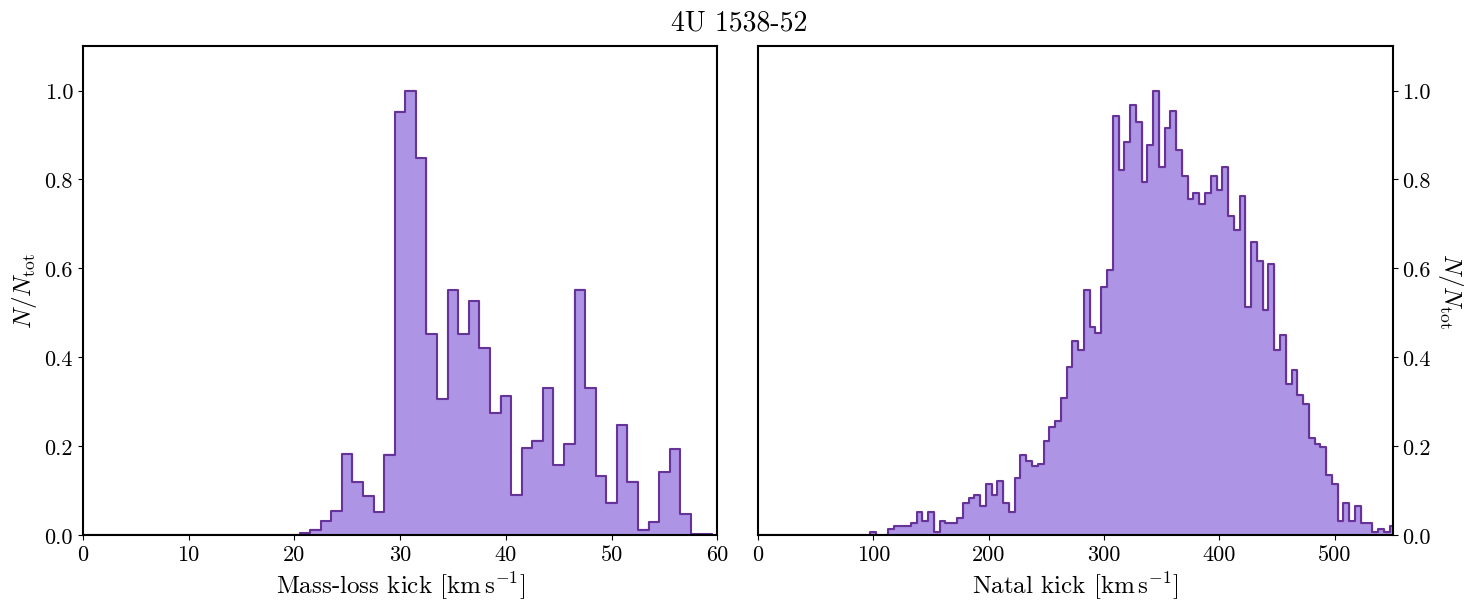}
    \caption{Distribution of mass-loss kicks ($\vmlk$) magnitude (left) and natal kick ($\vnk$) magnitude (right) for simulated system analogues to 4U 1538-52.}
    \label{Fig:U1538_KickHist}
\end{figure*}
\clearpage

\begin{figure*}
    \centering
    \includegraphics[width=0.75\textwidth]{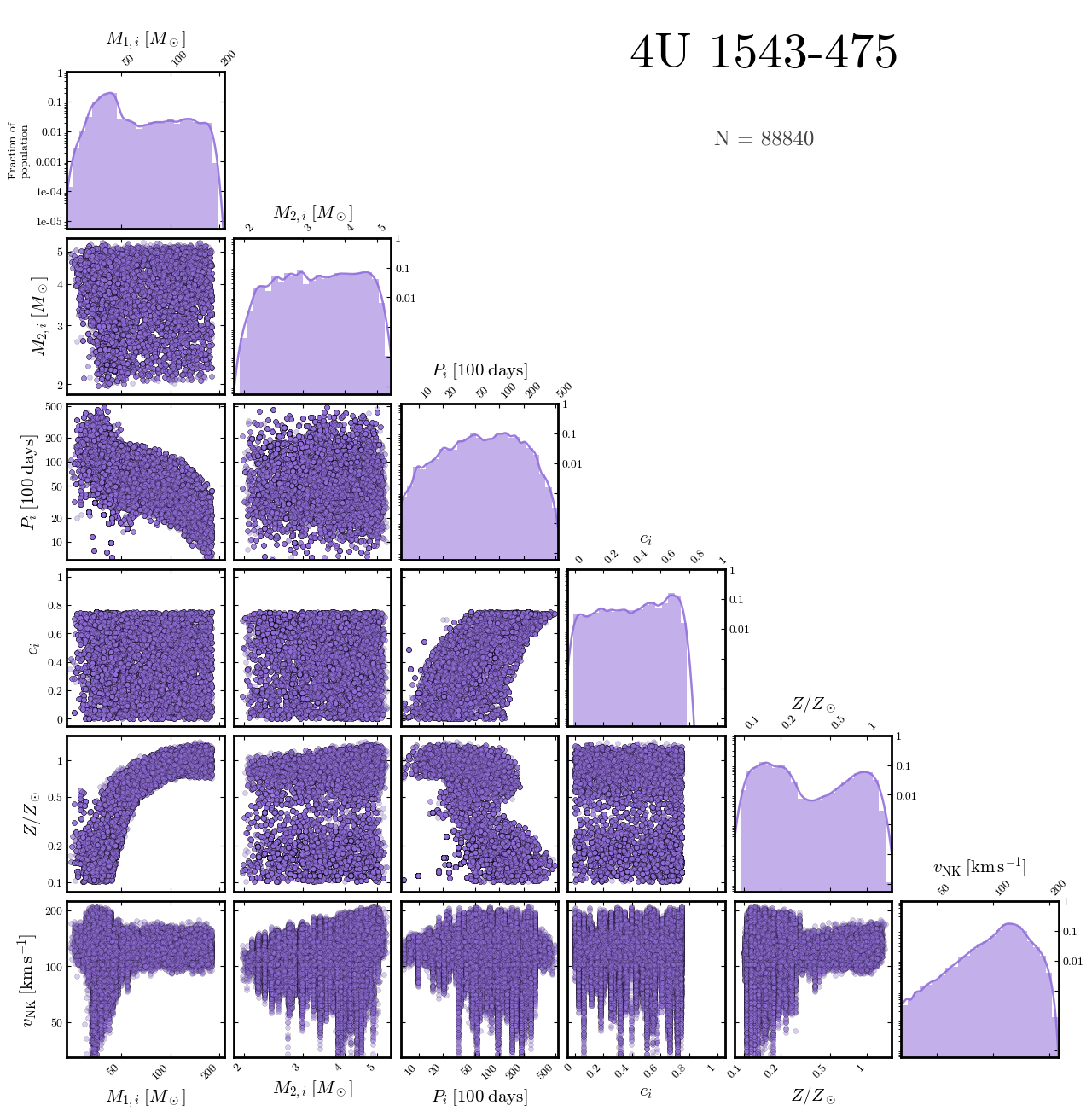}
    \caption{Corner plot highlighting key initial characteristics and acceptable natal kicks for simulated system analogues to 4U 1543-475.}
    \label{Fig:U1543_CornerPlot}
\end{figure*}
\begin{figure*}
    \centering
    \includegraphics[width=0.75\textwidth]{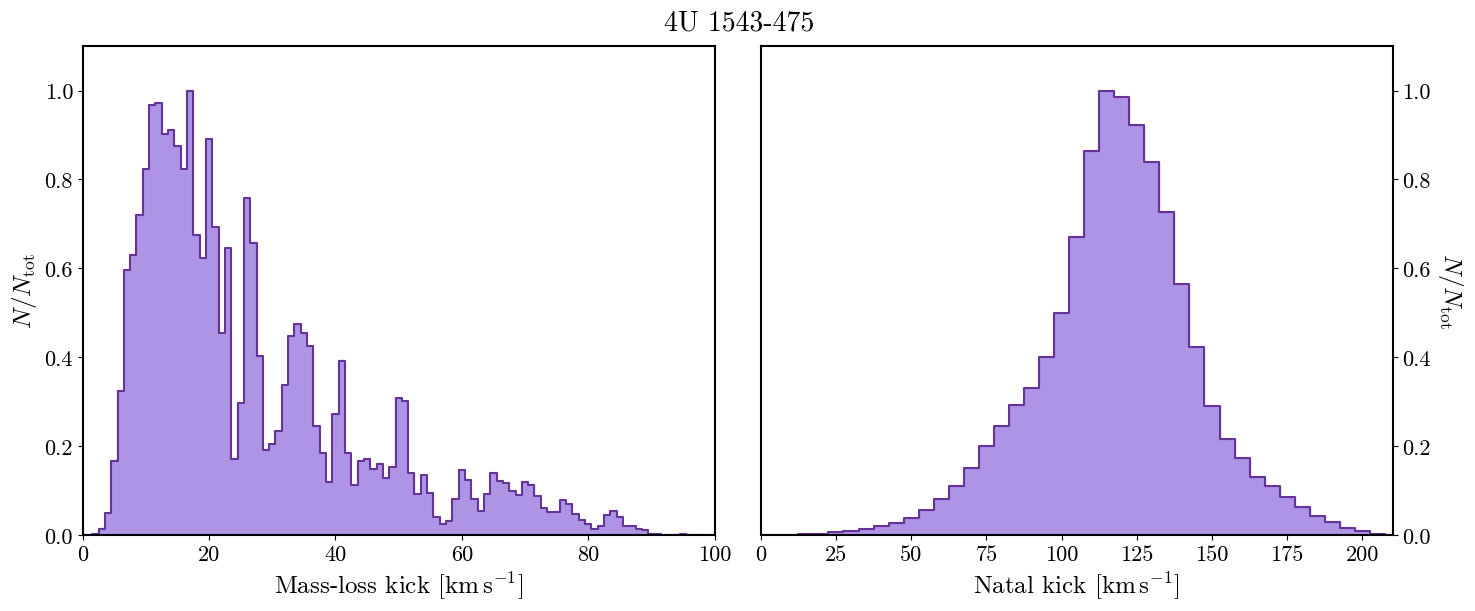}
    \caption{Distribution of mass-loss kicks ($\vmlk$) magnitude (left) and natal kick ($\vnk$) magnitude (right) for simulated system analogues to 4U 1543-475.}
    \label{Fig:U1543_KickHist}
\end{figure*}
\clearpage

\begin{figure*}
    \centering
    \includegraphics[width=0.75\textwidth]{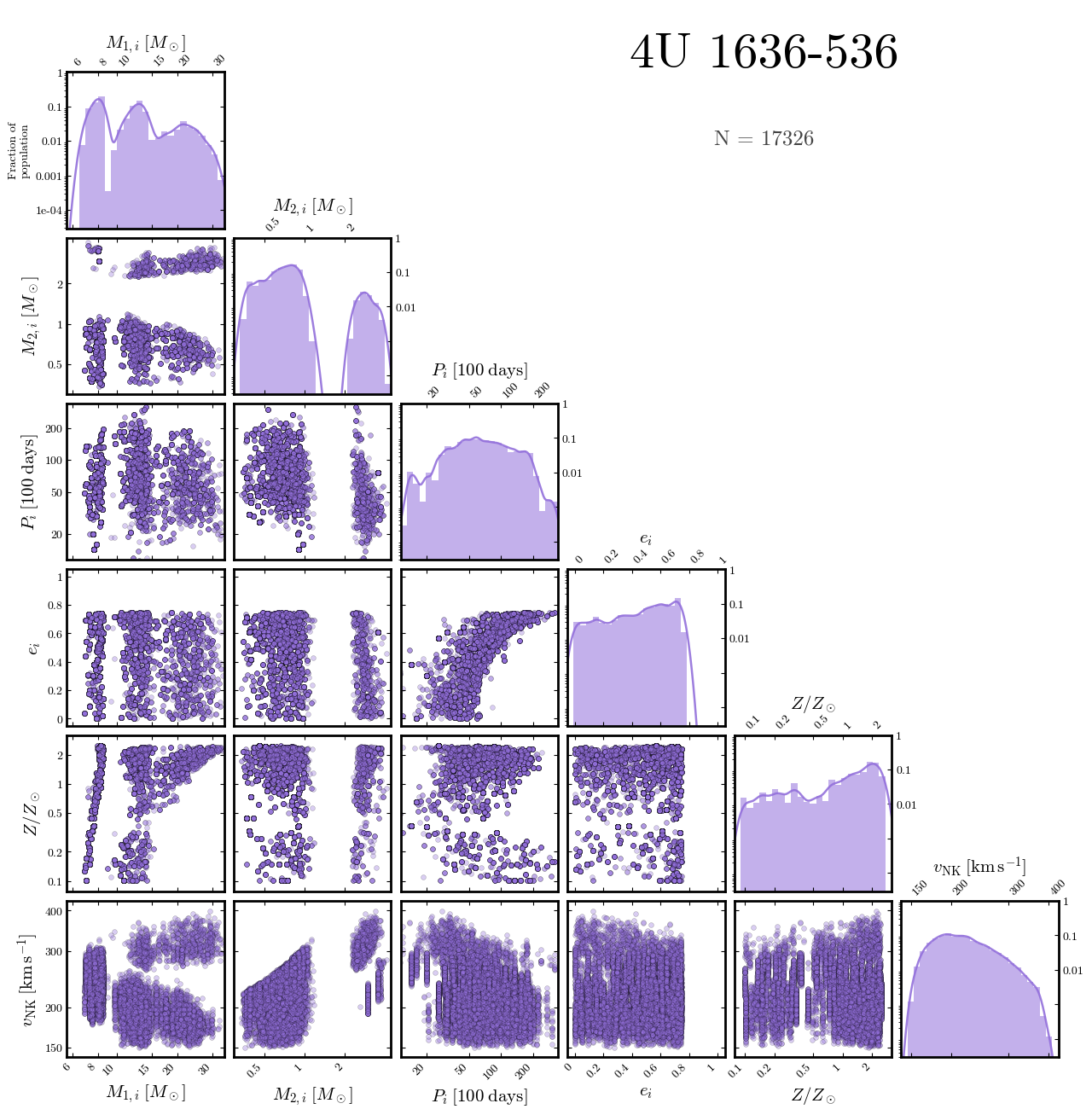}
    \caption{Corner plot highlighting key initial characteristics and acceptable natal kicks for simulated system analogues to 4U 1636-536.}
    \label{Fig:U1636_CornerPlot}
\end{figure*}
\begin{figure*}
    \centering
    \includegraphics[width=0.75\textwidth]{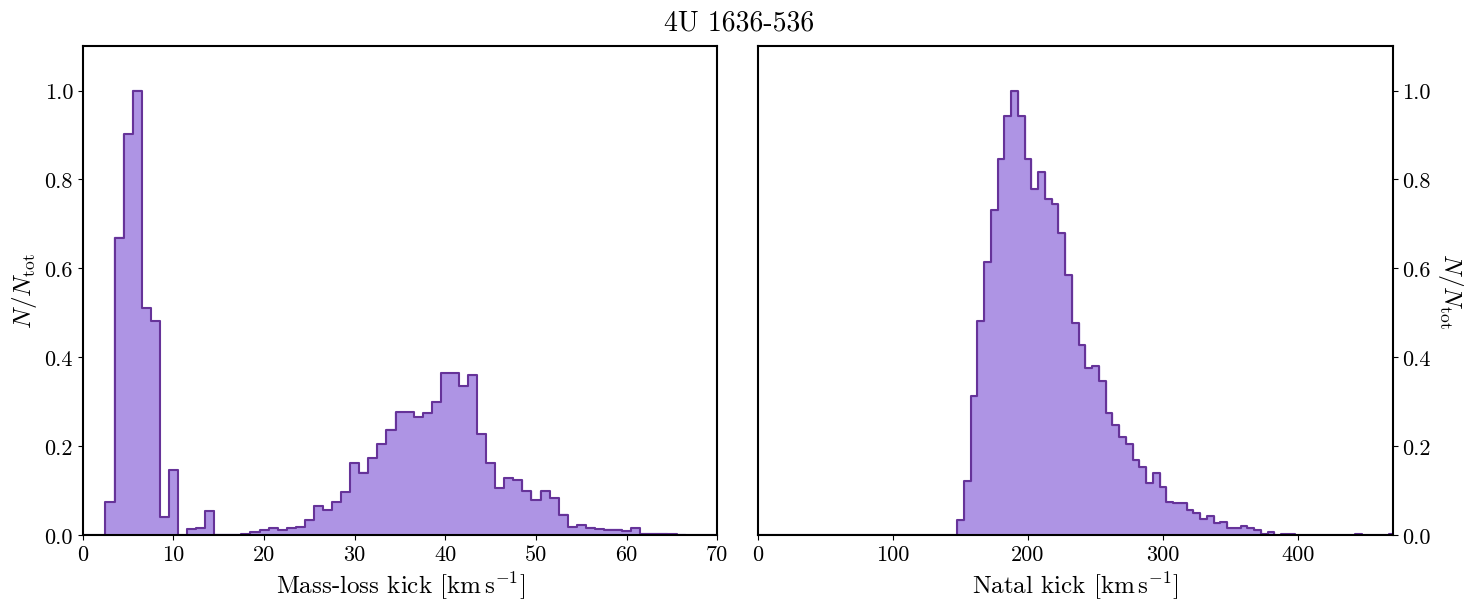}
    \caption{Distribution of mass-loss kicks ($\vmlk$) magnitude (left) and natal kick ($\vnk$) magnitude (right) for simulated system analogues to 4U 1636-536.}
    \label{Fig:U1636_KickHist}
\end{figure*}
\clearpage

\begin{figure*}
    \centering
    \includegraphics[width=0.75\textwidth]{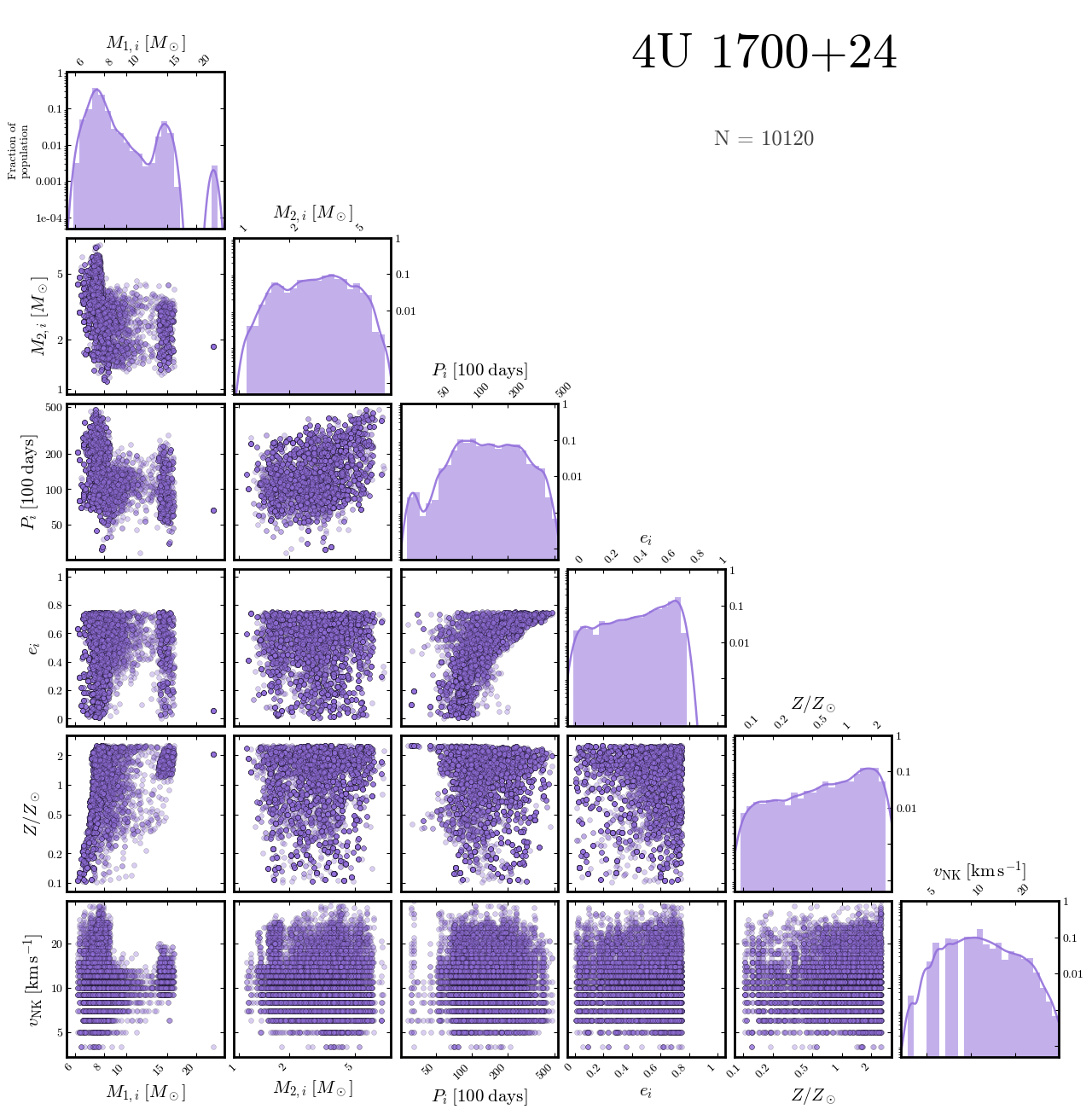}
    \caption{Corner plot highlighting key initial characteristics and acceptable natal kicks for simulated system analogues to 4U 1700+24.}
    \label{Fig:U170024_CornerPlot}
\end{figure*}
\begin{figure*}
    \centering
    \includegraphics[width=0.75\textwidth]{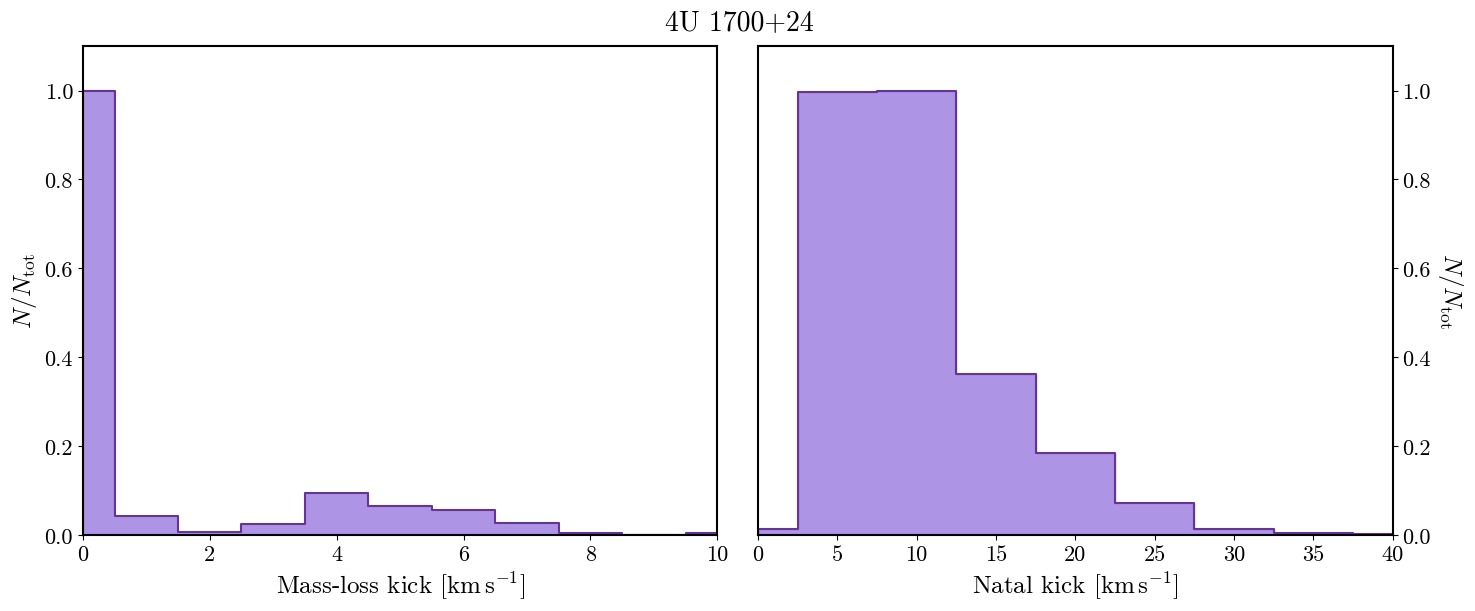}
    \caption{Distribution of mass-loss kicks ($\vmlk$) magnitude (left) and natal kick ($\vnk$) magnitude (right) for simulated system analogues to 4U 1700+24.}
    \label{Fig:U170024_KickHist}
\end{figure*}
\clearpage

\begin{figure*}
    \centering
    \includegraphics[width=0.75\textwidth]{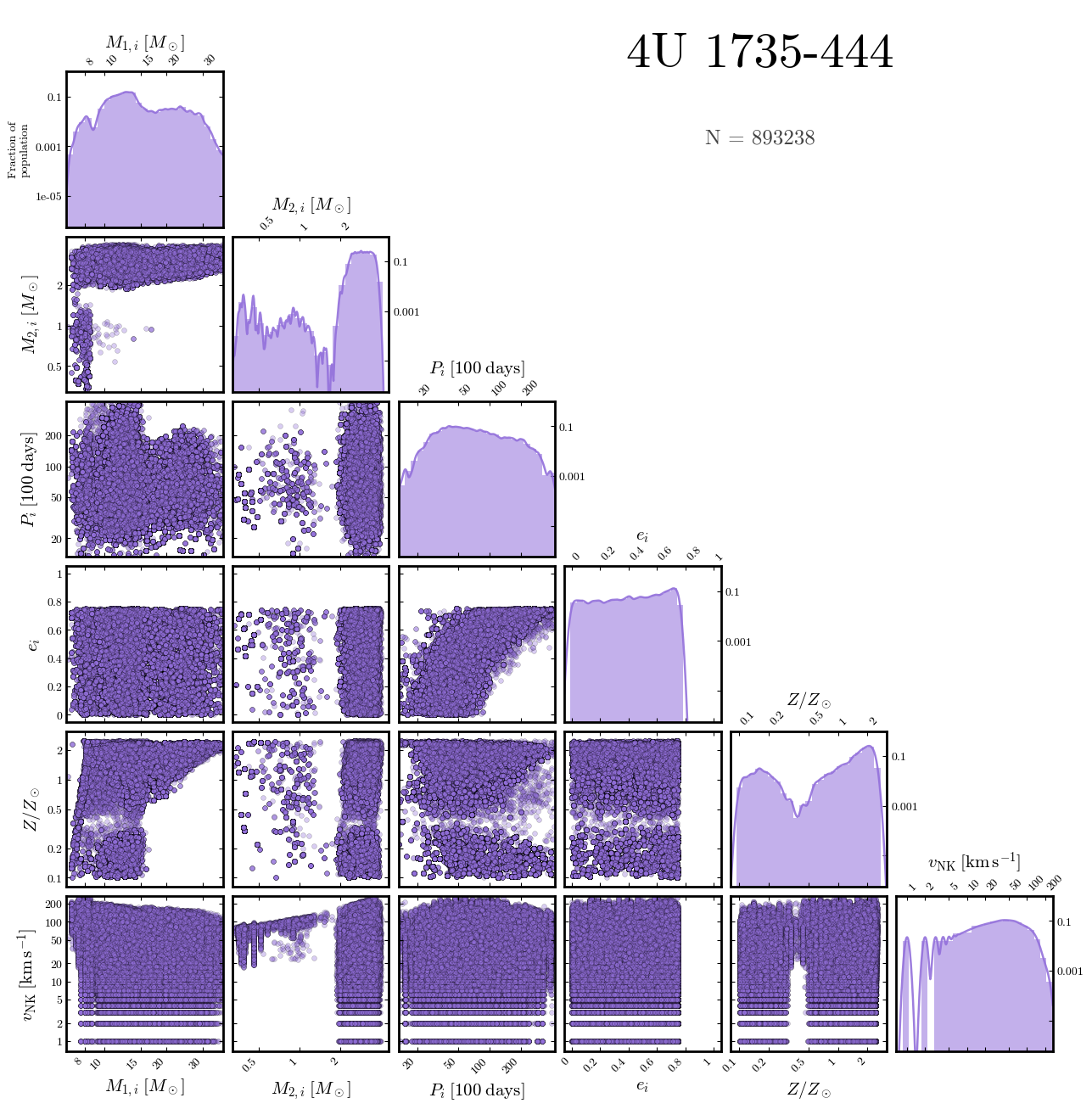}
    \caption{Corner plot highlighting key initial characteristics and acceptable natal kicks for simulated system analogues to 4U 1735-444.}
    \label{Fig:U1735_CornerPlot}
\end{figure*}
\begin{figure*}
    \centering
    \includegraphics[width=0.75\textwidth]{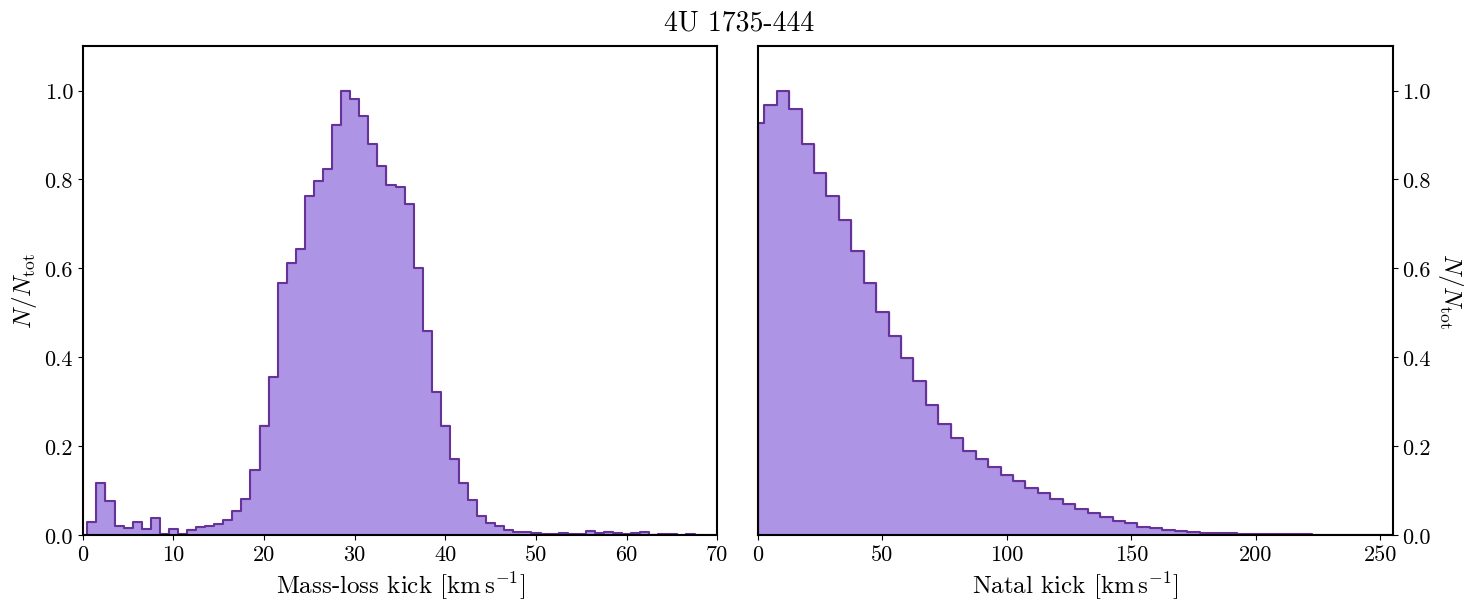}
    \caption{Distribution of mass-loss kicks ($\vmlk$) magnitude (left) and natal kick ($\vnk$) magnitude (right) for simulated system analogues to 4U 1735-444.}
    \label{Fig:U1735_KickHist}
\end{figure*}
\clearpage

\begin{figure*}
    \centering
    \includegraphics[width=0.75\textwidth]{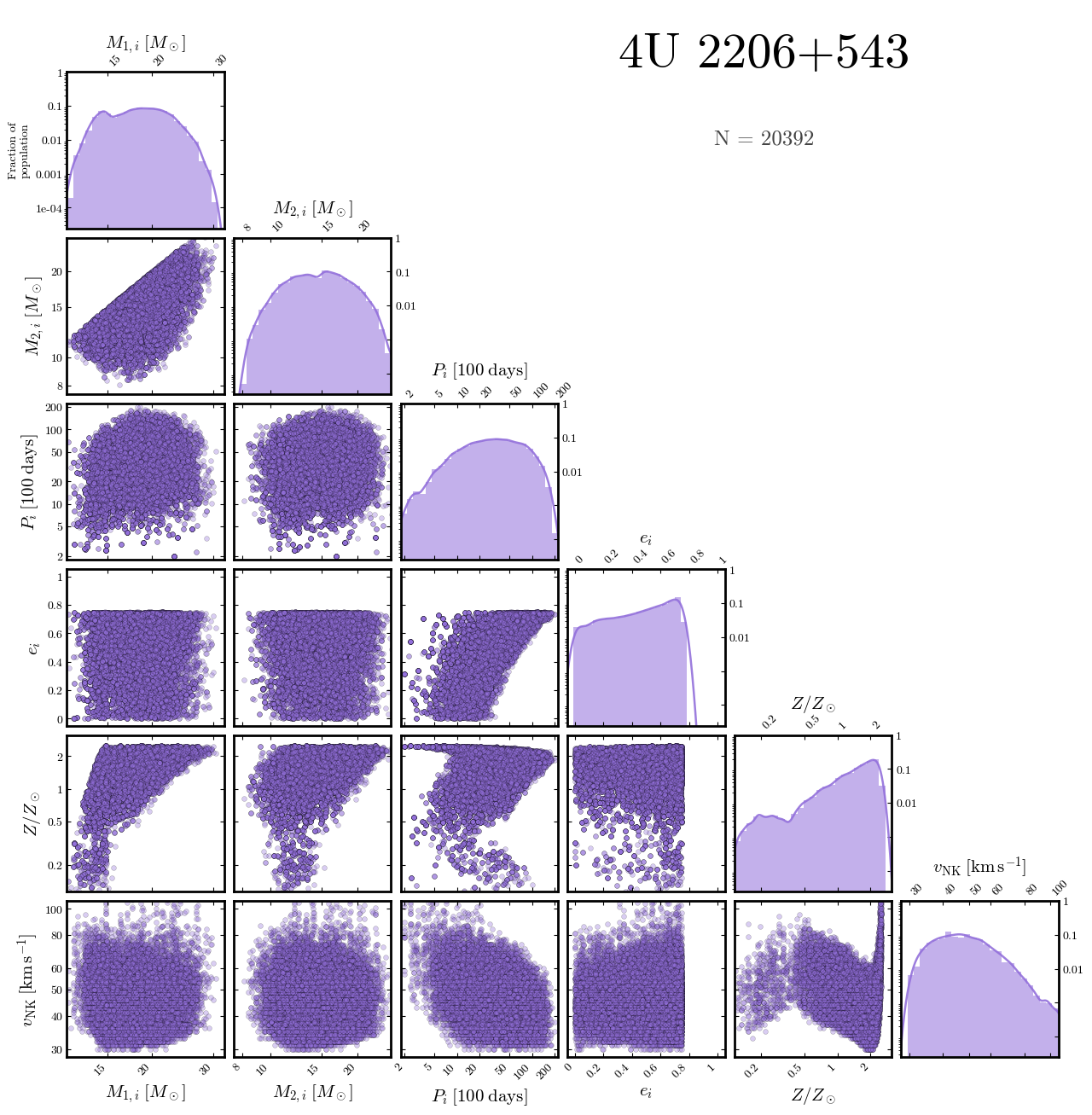}
    \caption{Corner plot highlighting key initial characteristics and acceptable natal kicks for simulated system analogues to 4U 2206+543.}
    \label{Fig:U2206_CornerPlot}
\end{figure*}
\begin{figure*}
    \centering
    \includegraphics[width=0.75\textwidth]{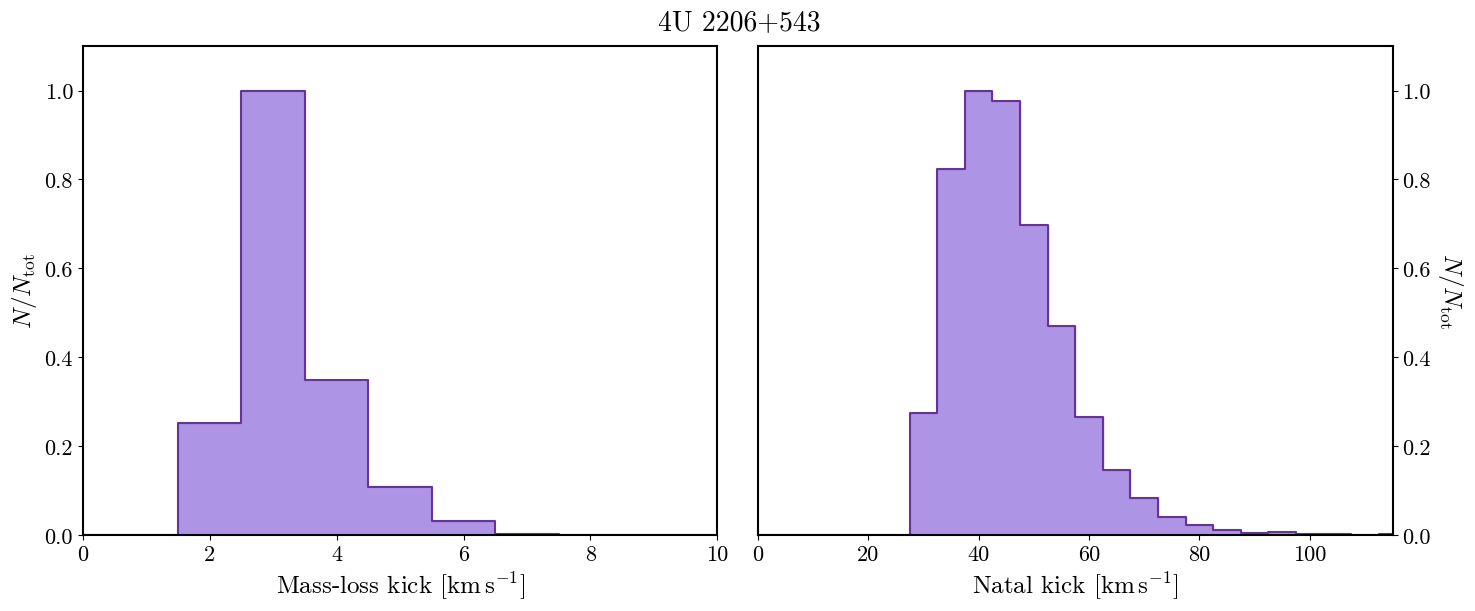}
    \caption{Distribution of mass-loss kicks ($\vmlk$) magnitude (left) and natal kick ($\vnk$) magnitude (right) for simulated system analogues to 4U 2206+543.}
    \label{Fig:U2206_KickHist}
\end{figure*}
\clearpage

\begin{figure*}
    \centering
    \includegraphics[width=0.75\textwidth]{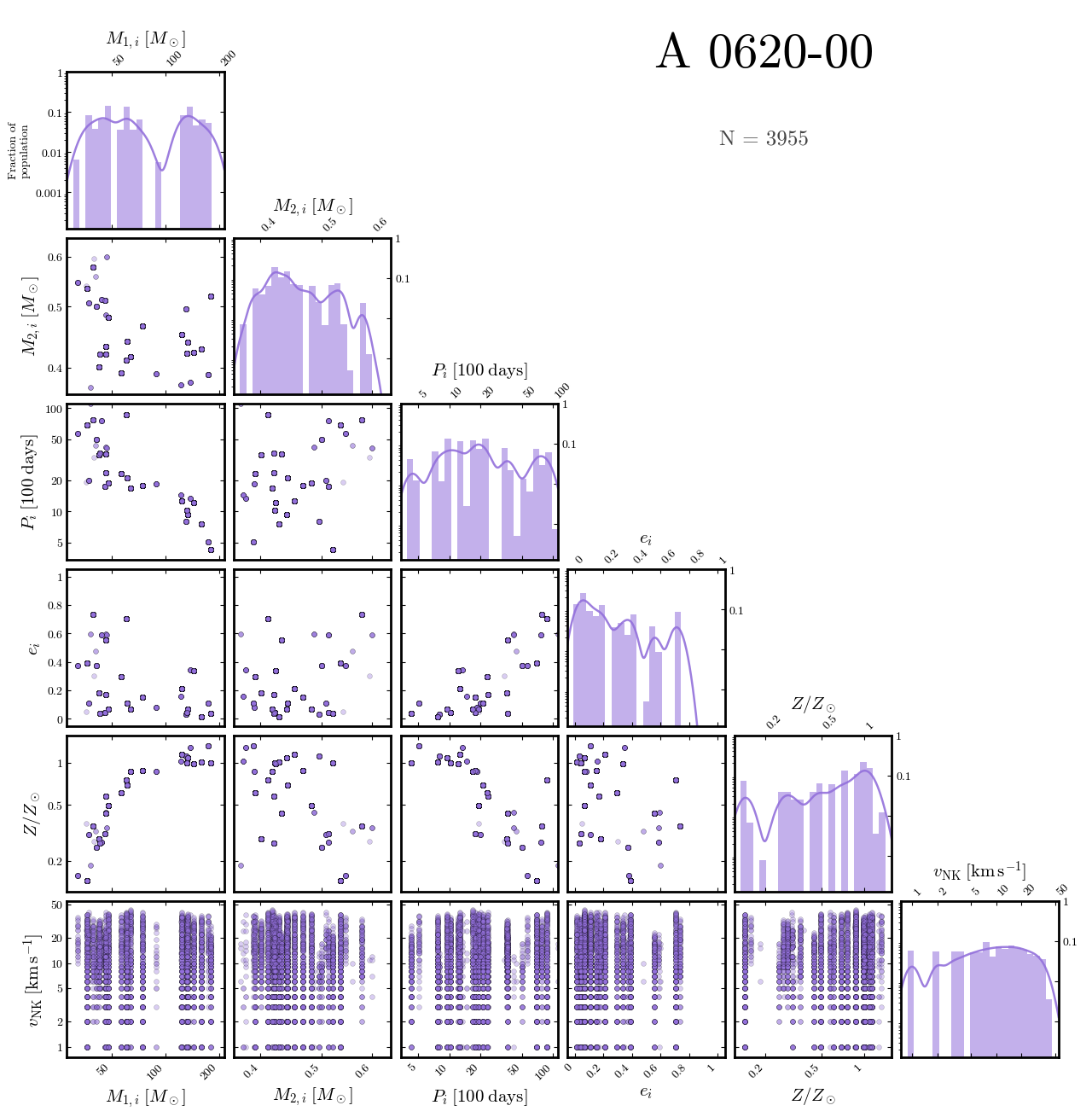}
    \caption{Corner plot highlighting key initial characteristics and acceptable natal kicks for simulated system analogues to A 0620-00.}
    \label{Fig:A0620_CornerPlot}
\end{figure*}
\begin{figure*}
    \centering
    \includegraphics[width=0.75\textwidth]{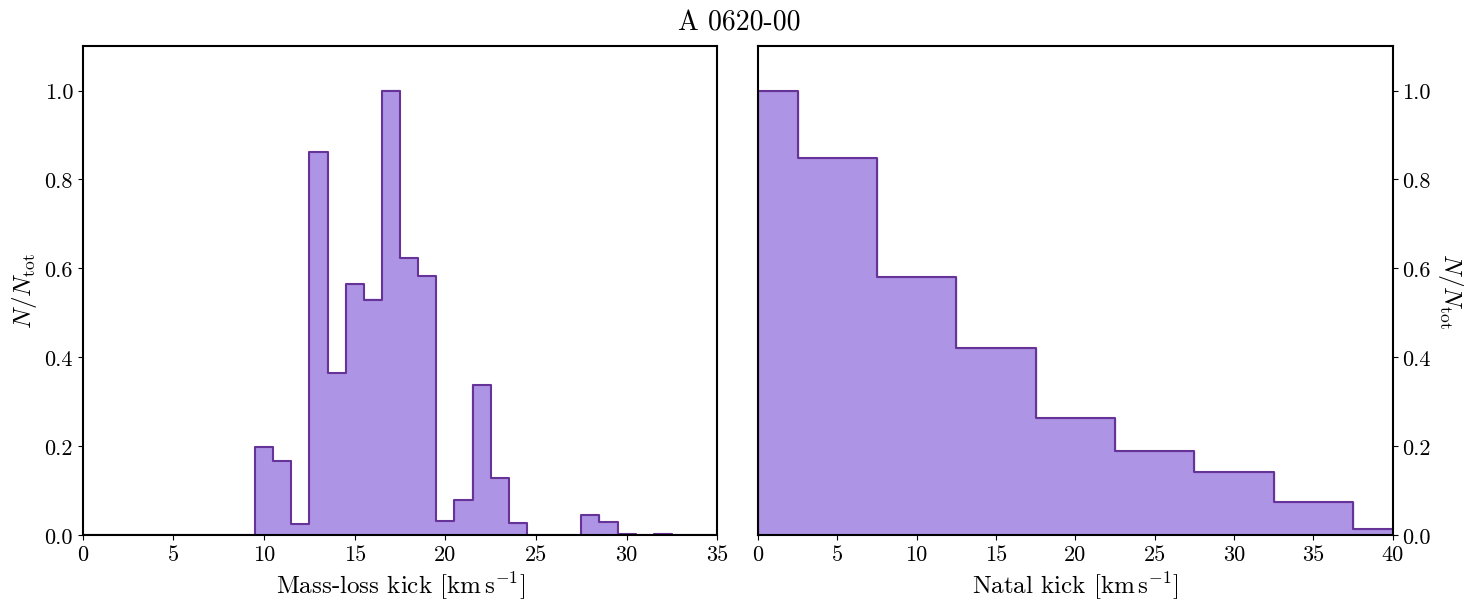}
    \caption{Distribution of mass-loss kicks ($\vmlk$) magnitude (left) and natal kick ($\vnk$) magnitude (right) for simulated system analogues to A 0620-00.}
    \label{Fig:A0620_KickHist}
\end{figure*}
\clearpage

\begin{figure*}
    \centering
    \includegraphics[width=0.75\textwidth]{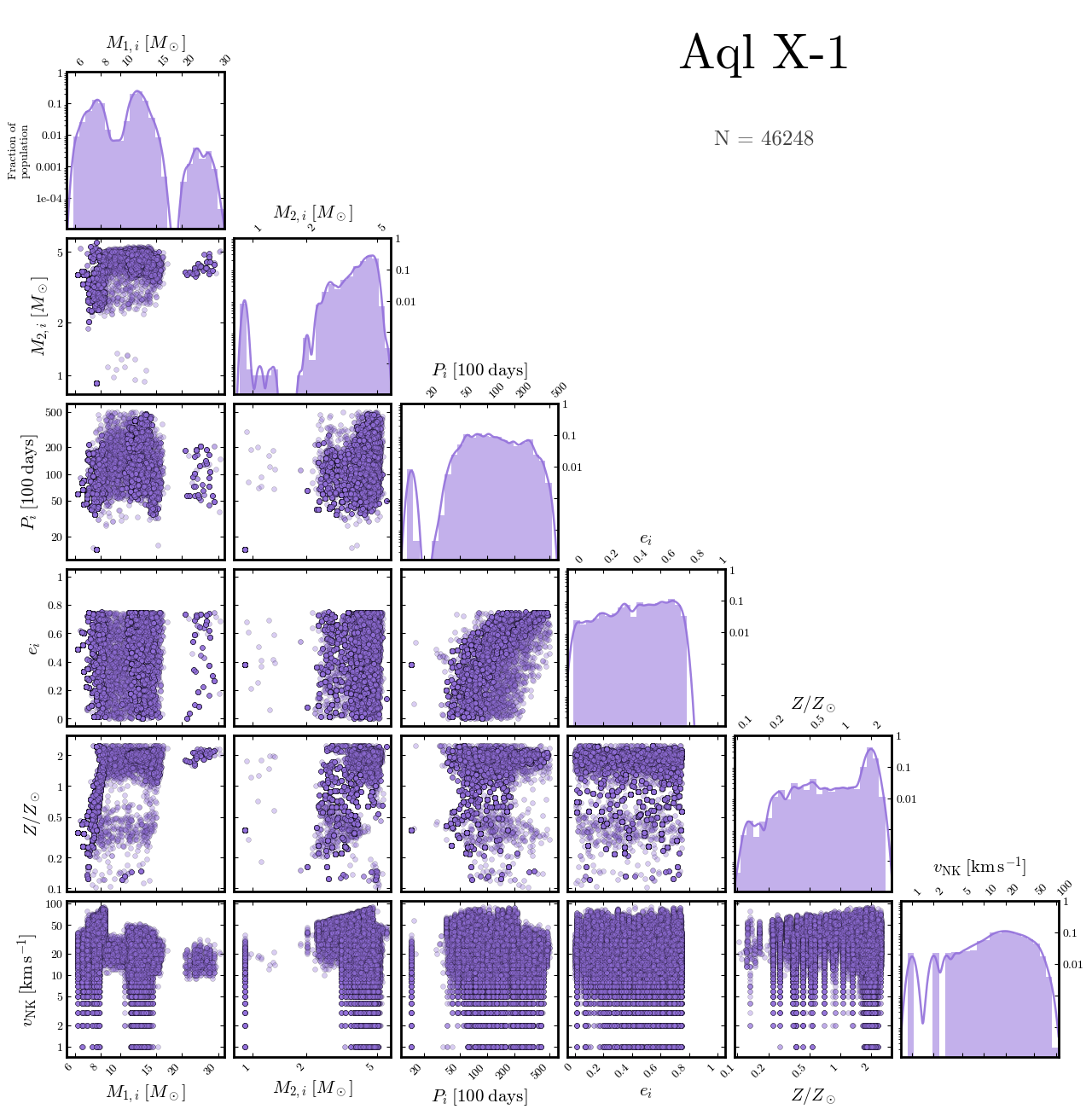}
    \caption{Corner plot highlighting key initial characteristics and acceptable natal kicks for simulated system analogues to Aql X-1.}
    \label{Fig:AqlX1_CornerPlot}
\end{figure*}
\begin{figure*}
    \centering
    \includegraphics[width=0.75\textwidth]{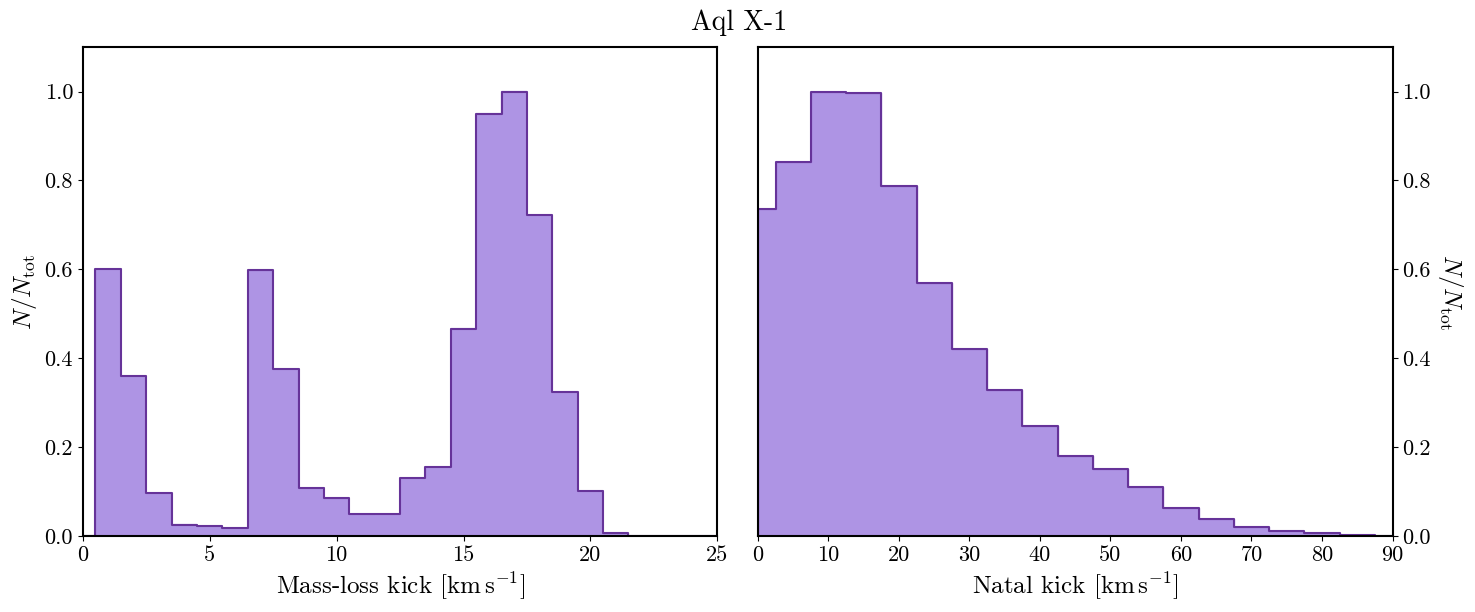}
    \caption{Distribution of mass-loss kicks ($\vmlk$) magnitude (left) and natal kick ($\vnk$) magnitude (right) for simulated system analogues to Aql X-1.}
    \label{Fig:AqlX1_KickHist}
\end{figure*}
\clearpage

\begin{figure*}
    \centering
    \includegraphics[width=0.75\textwidth]{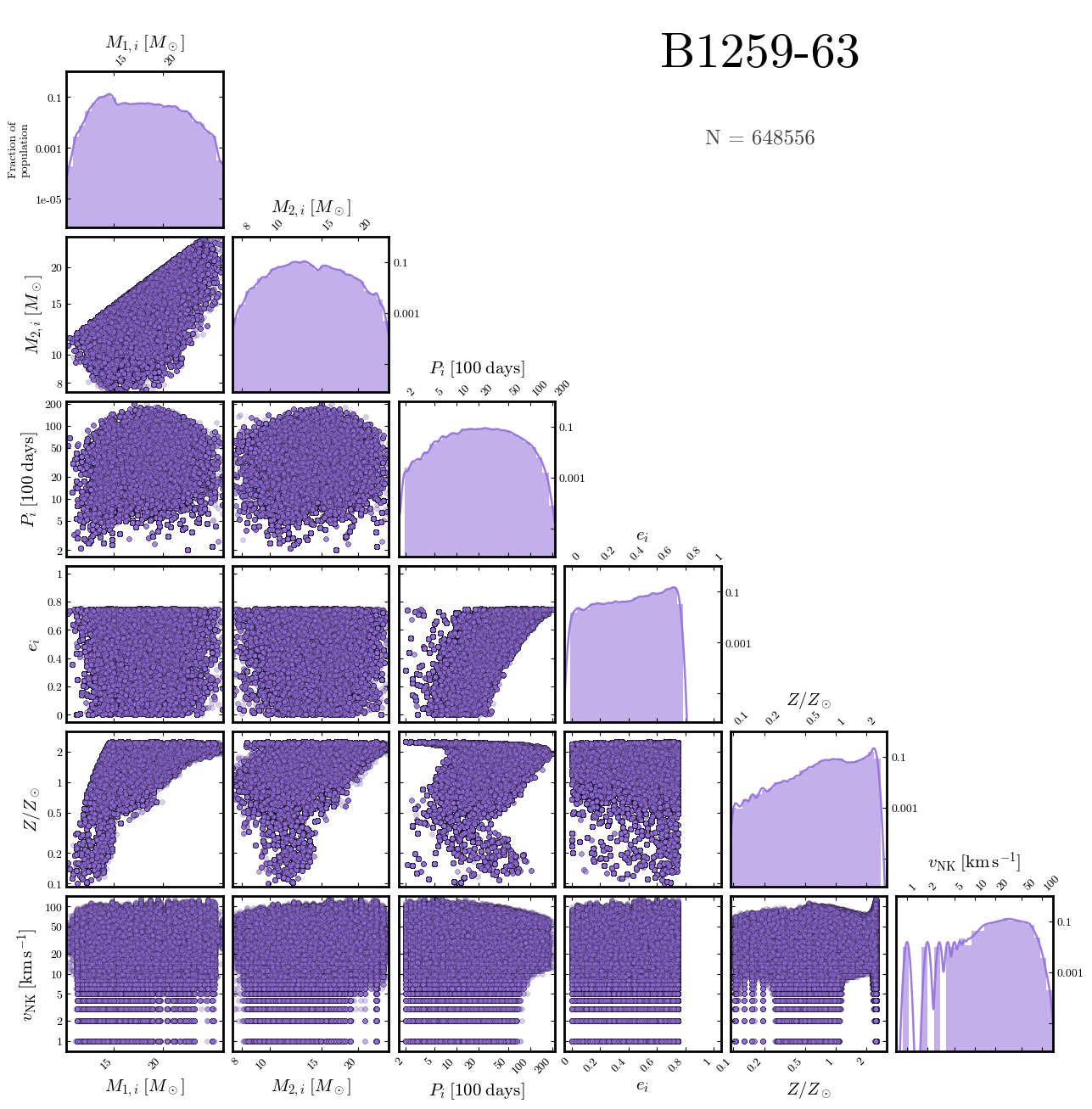}
    \caption{Corner plot highlighting key initial characteristics and acceptable natal kicks for simulated system analogues to B1259-63.}
    \label{Fig:B1259_CornerPlot}
\end{figure*}
\begin{figure*}
    \centering
    \includegraphics[width=0.75\textwidth]{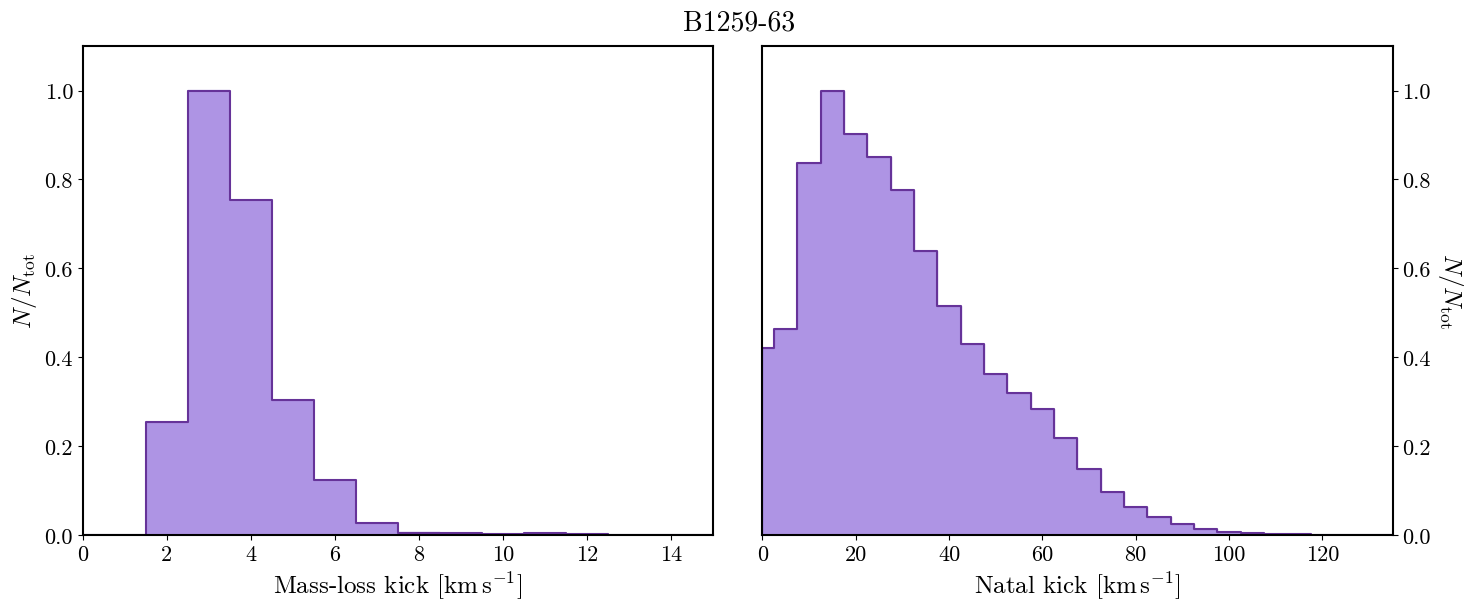}
    \caption{Distribution of mass-loss kicks ($\vmlk$) magnitude (left) and natal kick ($\vnk$) magnitude (right) for simulated system analogues to B1259-63.}
    \label{Fig:B1259_KickHist}
\end{figure*}
\clearpage

\begin{figure*}
    \centering
    \includegraphics[width=0.75\textwidth]{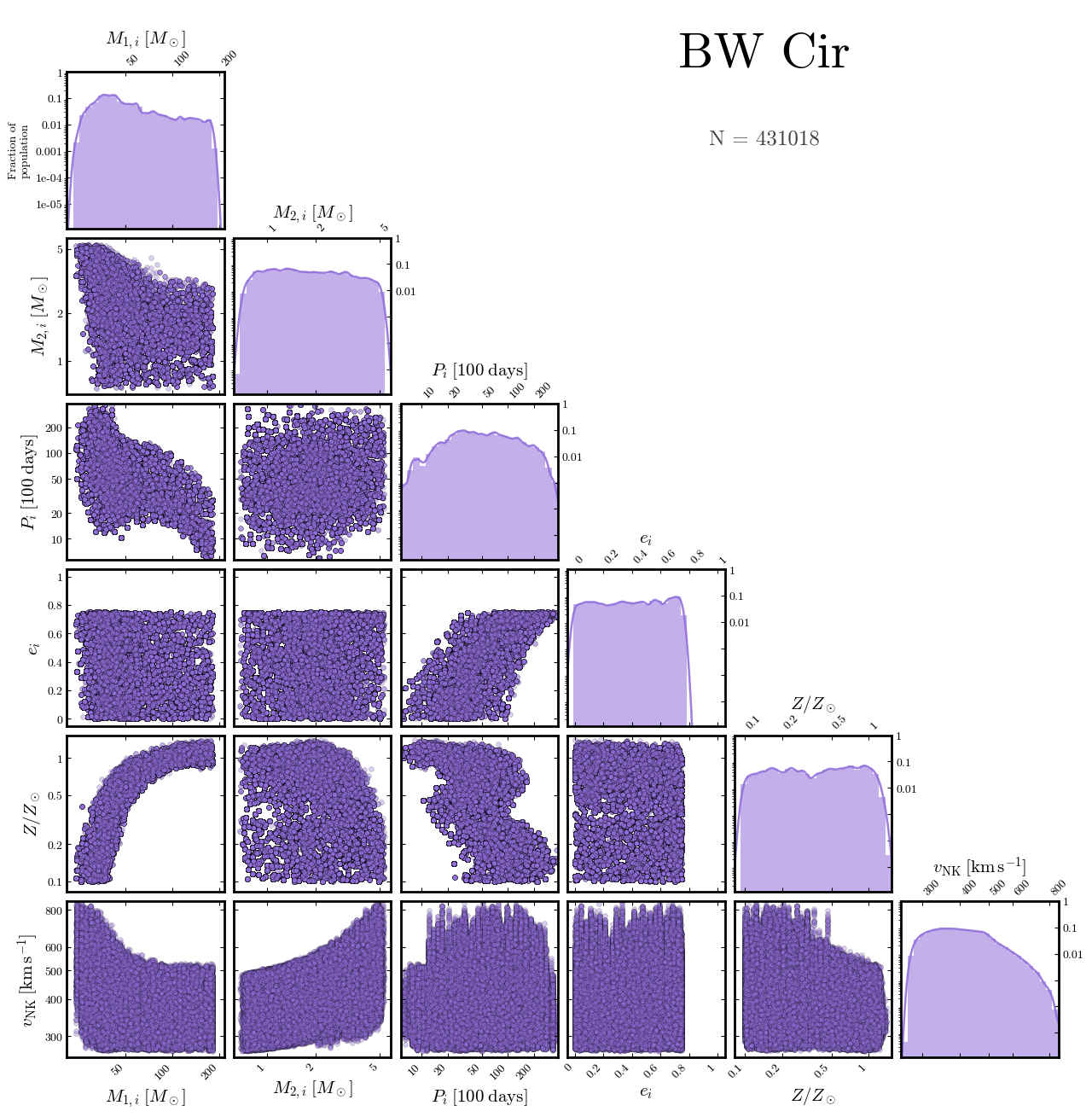}
    \caption{Corner plot highlighting key initial characteristics and acceptable natal kicks for simulated system analogues to BW Cir.}
    \label{Fig:BWCir_CornerPlot}
\end{figure*}
\begin{figure*}
    \centering
    \includegraphics[width=0.75\textwidth]{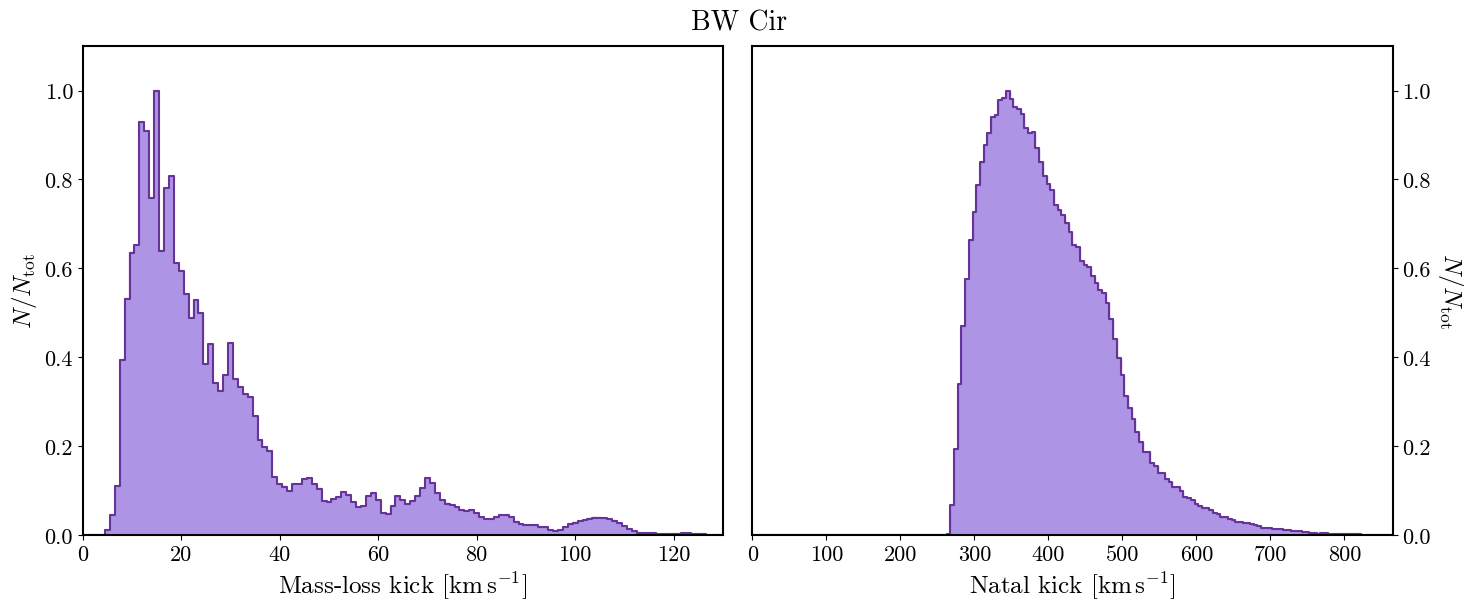}
    \caption{Distribution of mass-loss kicks ($\vmlk$) magnitude (left) and natal kick ($\vnk$) magnitude (right) for simulated system analogues to BW Cir.}
    \label{Fig:BWCir_KickHist}
\end{figure*}
\clearpage

\begin{figure*}
    \centering
    \includegraphics[width=0.75\textwidth]{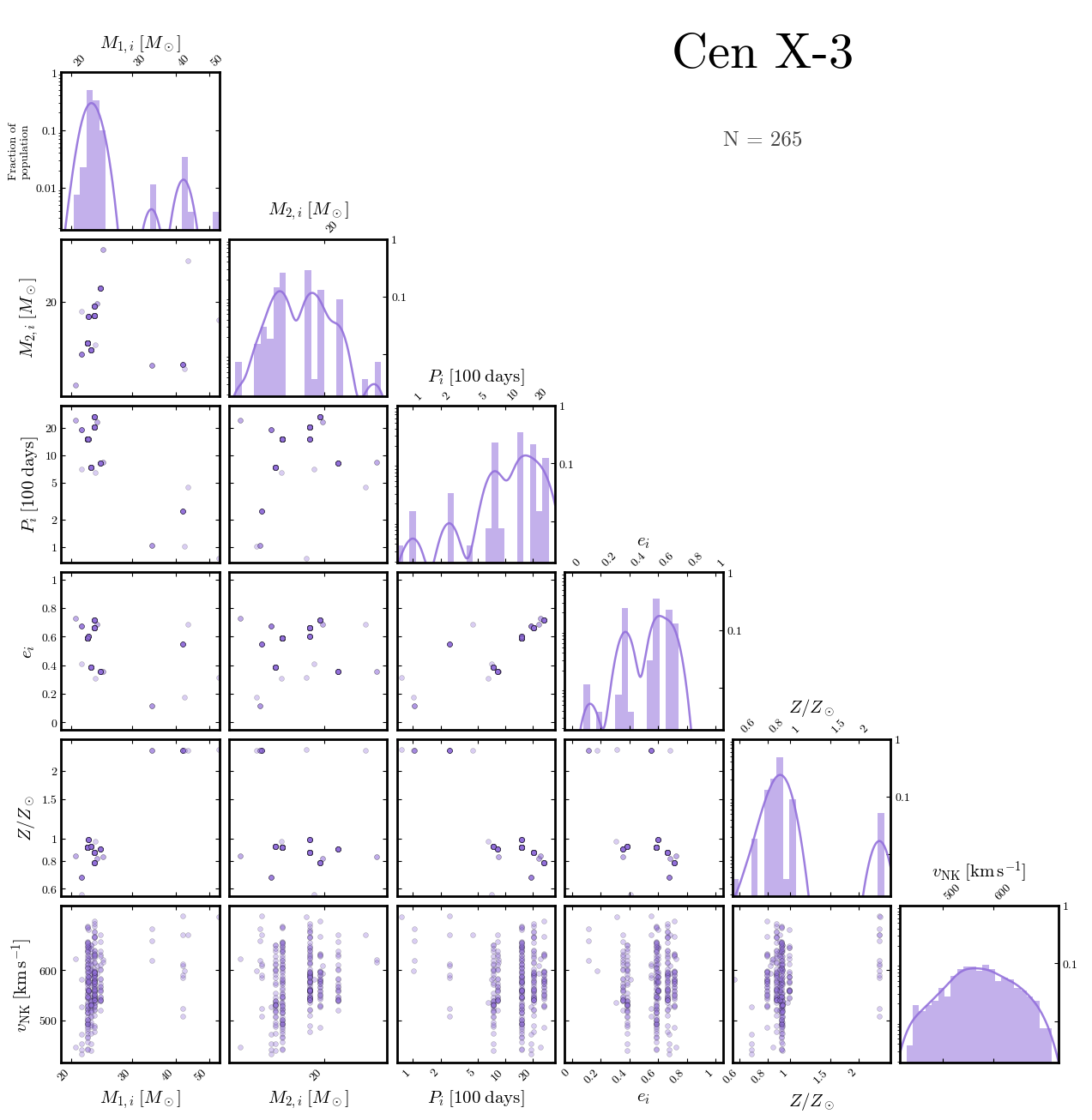}
    \caption{Corner plot highlighting key initial characteristics and acceptable natal kicks for simulated system analogues to Cen X-3.}
    \label{Fig:CenX3_CornerPlot}
\end{figure*}
\begin{figure*}
    \centering
    \includegraphics[width=0.75\textwidth]{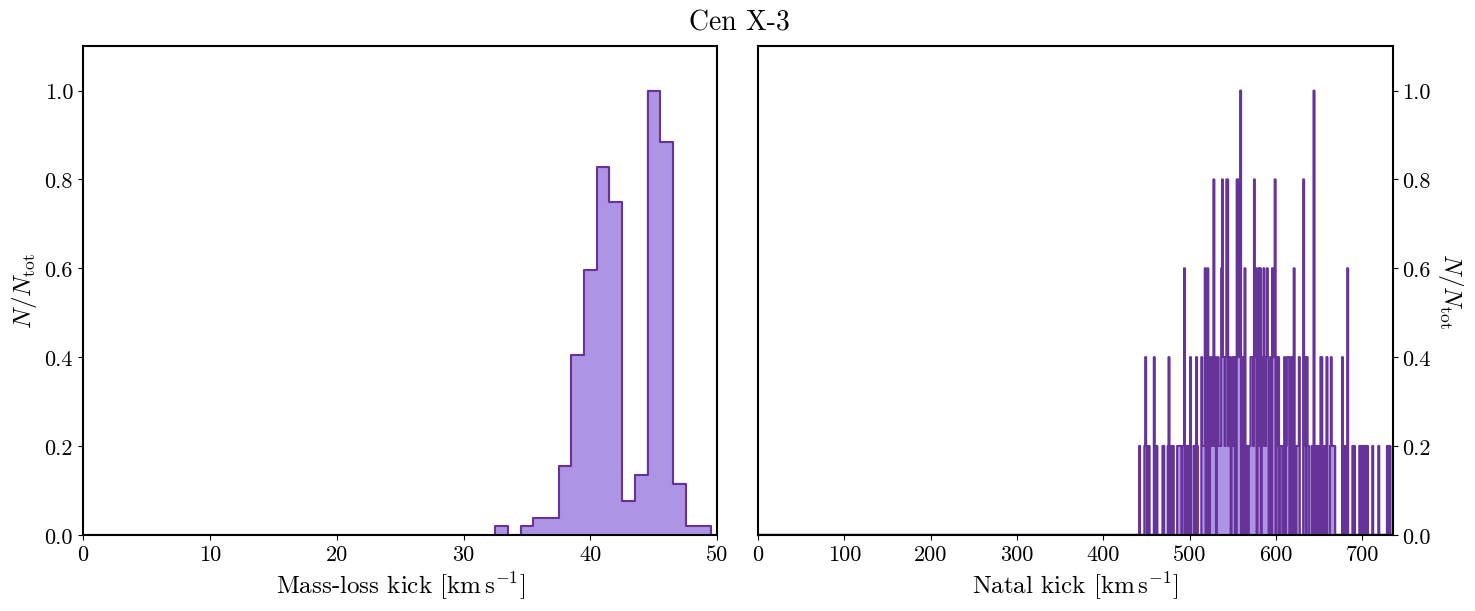}
    \caption{Distribution of mass-loss kicks ($\vmlk$) magnitude (left) and natal kick ($\vnk$) magnitude (right) for simulated system analogues to Cen X-3.}
    \label{Fig:CenX3_KickHist}
\end{figure*}
\clearpage

\begin{figure*}
    \centering
    \includegraphics[width=0.75\textwidth]{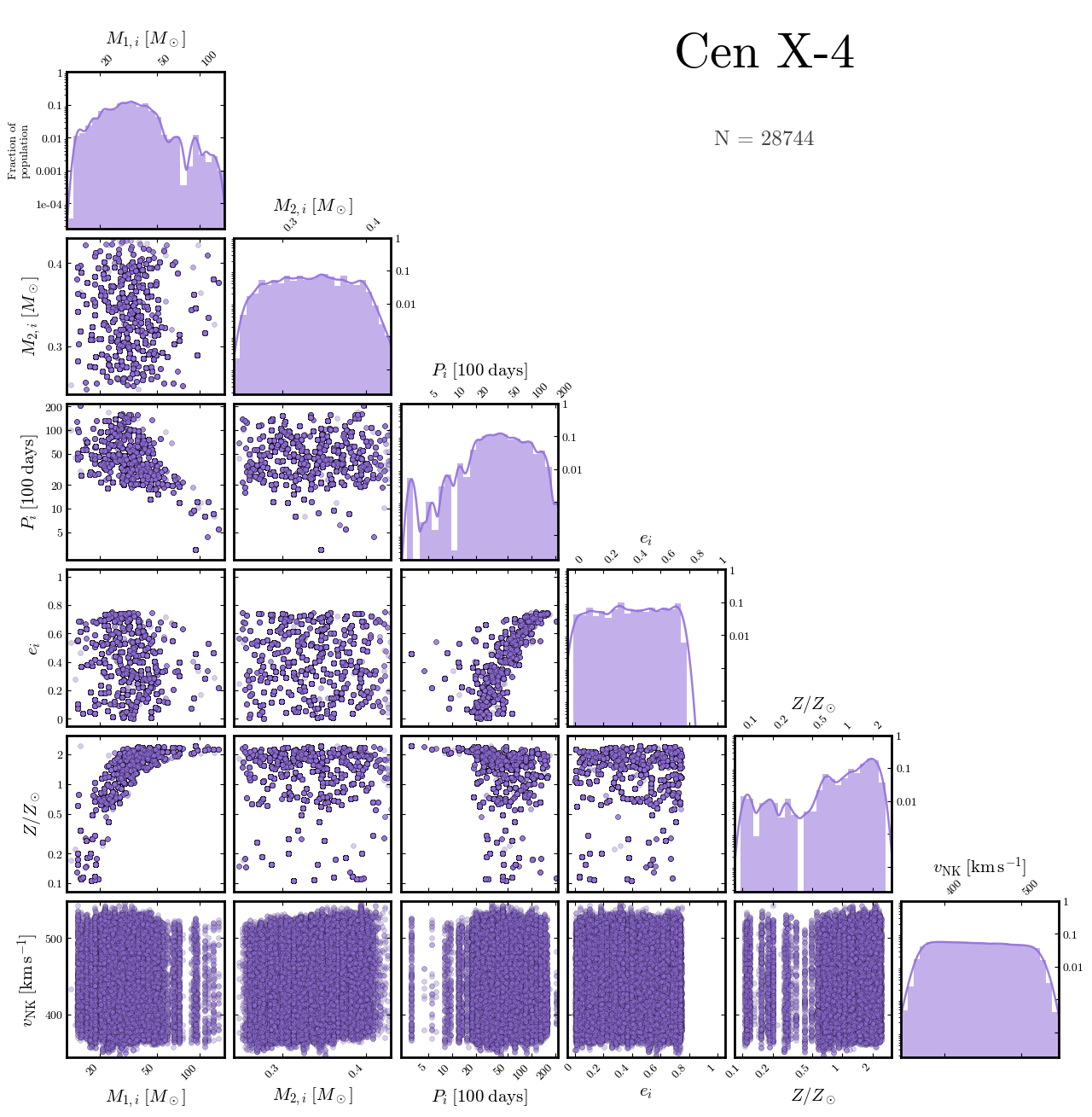}
    \caption{Corner plot highlighting key initial characteristics and acceptable natal kicks for simulated system analogues to Cen X-4.}
    \label{Fig:CenX4_CornerPlot}
\end{figure*}
\begin{figure*}
    \centering
    \includegraphics[width=0.75\textwidth]{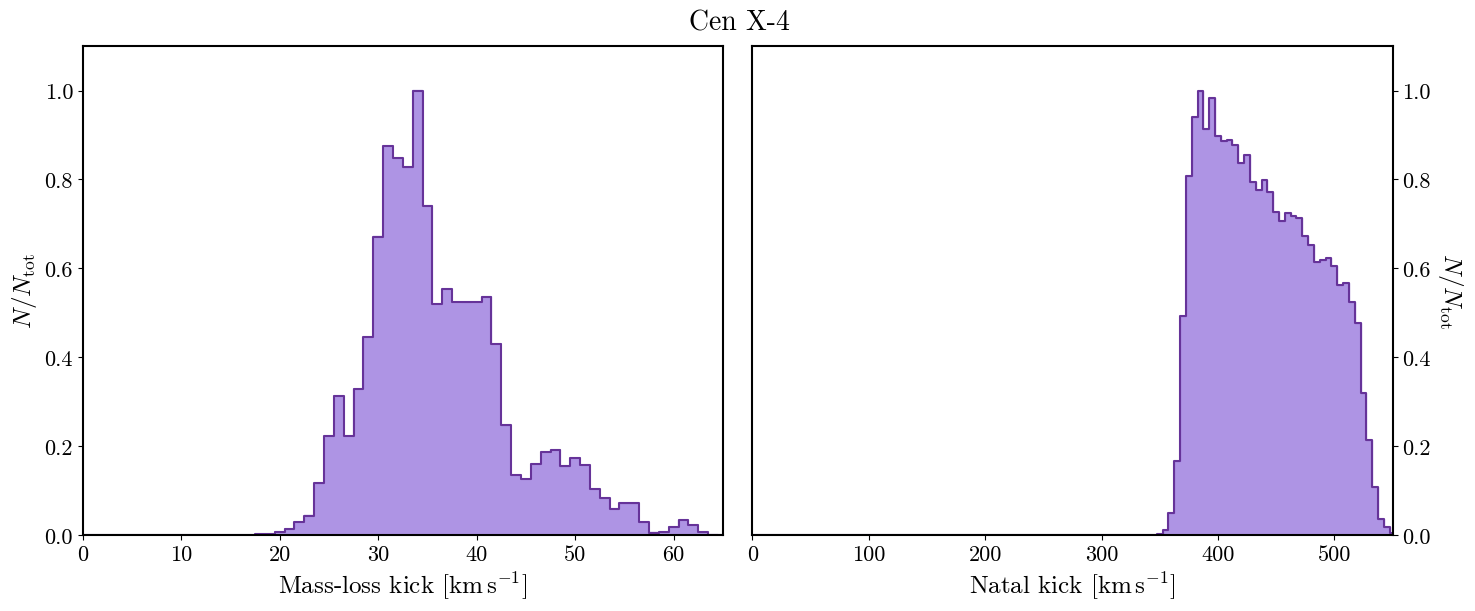}
    \caption{Distribution of mass-loss kicks ($\vmlk$) magnitude (left) and natal kick ($\vnk$) magnitude (right) for simulated system analogues to Cen X-4.}
    \label{Fig:CenX4_KickHist}
\end{figure*}
\clearpage

\begin{figure*}
    \centering
    \includegraphics[width=0.75\textwidth]{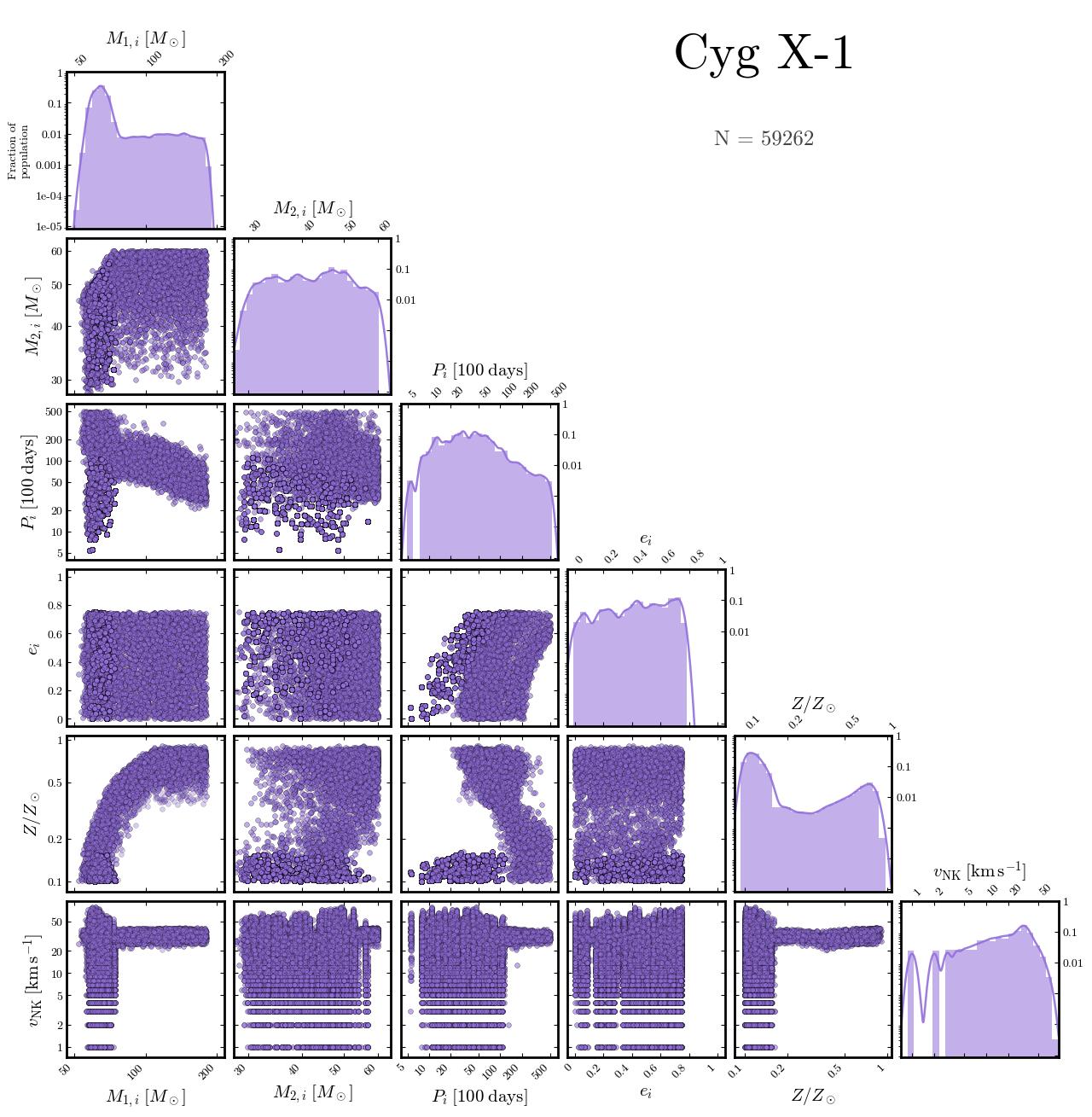}
    \caption{Corner plot highlighting key initial characteristics and acceptable natal kicks for simulated system analogues to Cyg X-1.}
    \label{Fig:CygX1_CornerPlot}
\end{figure*}
\begin{figure*}
    \centering
    \includegraphics[width=0.75\textwidth]{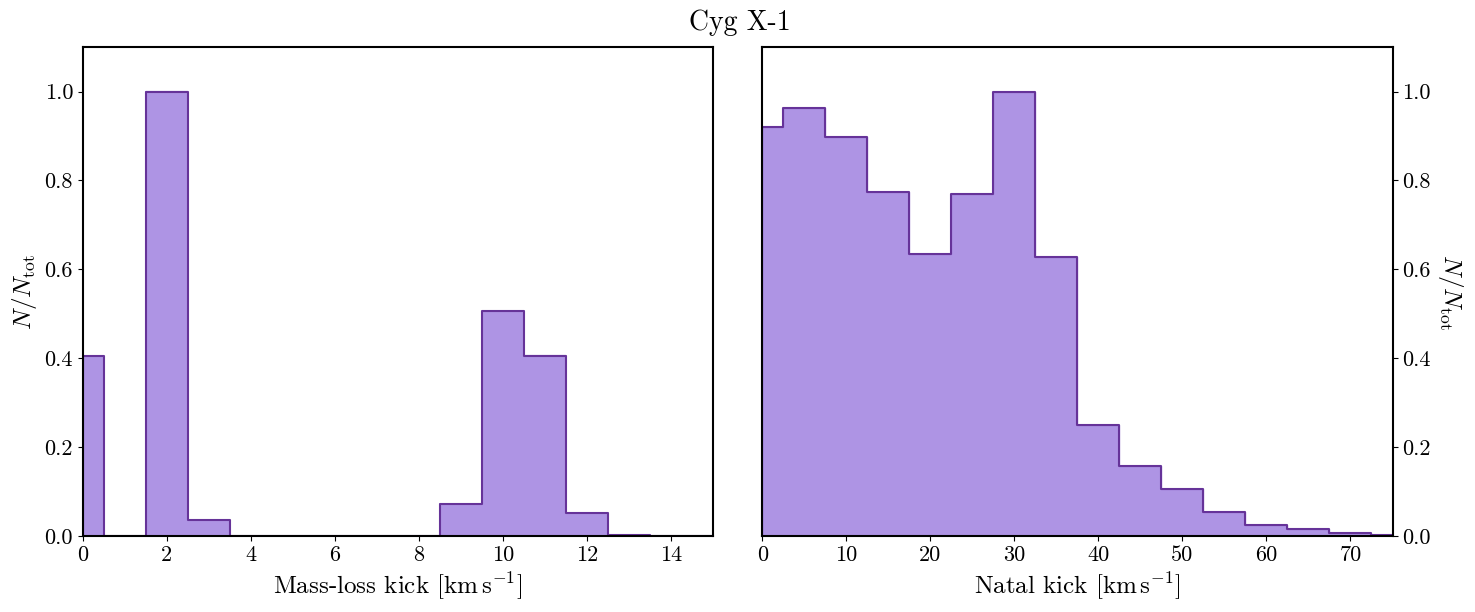}
    \caption{Distribution of mass-loss kicks ($\vmlk$) magnitude (left) and natal kick ($\vnk$) magnitude (right) for simulated system analogues to Cyg X-1.}
    \label{Fig:CygX1_KickHist}
\end{figure*}
\clearpage

\begin{figure*}
    \centering
    \includegraphics[width=0.75\textwidth]{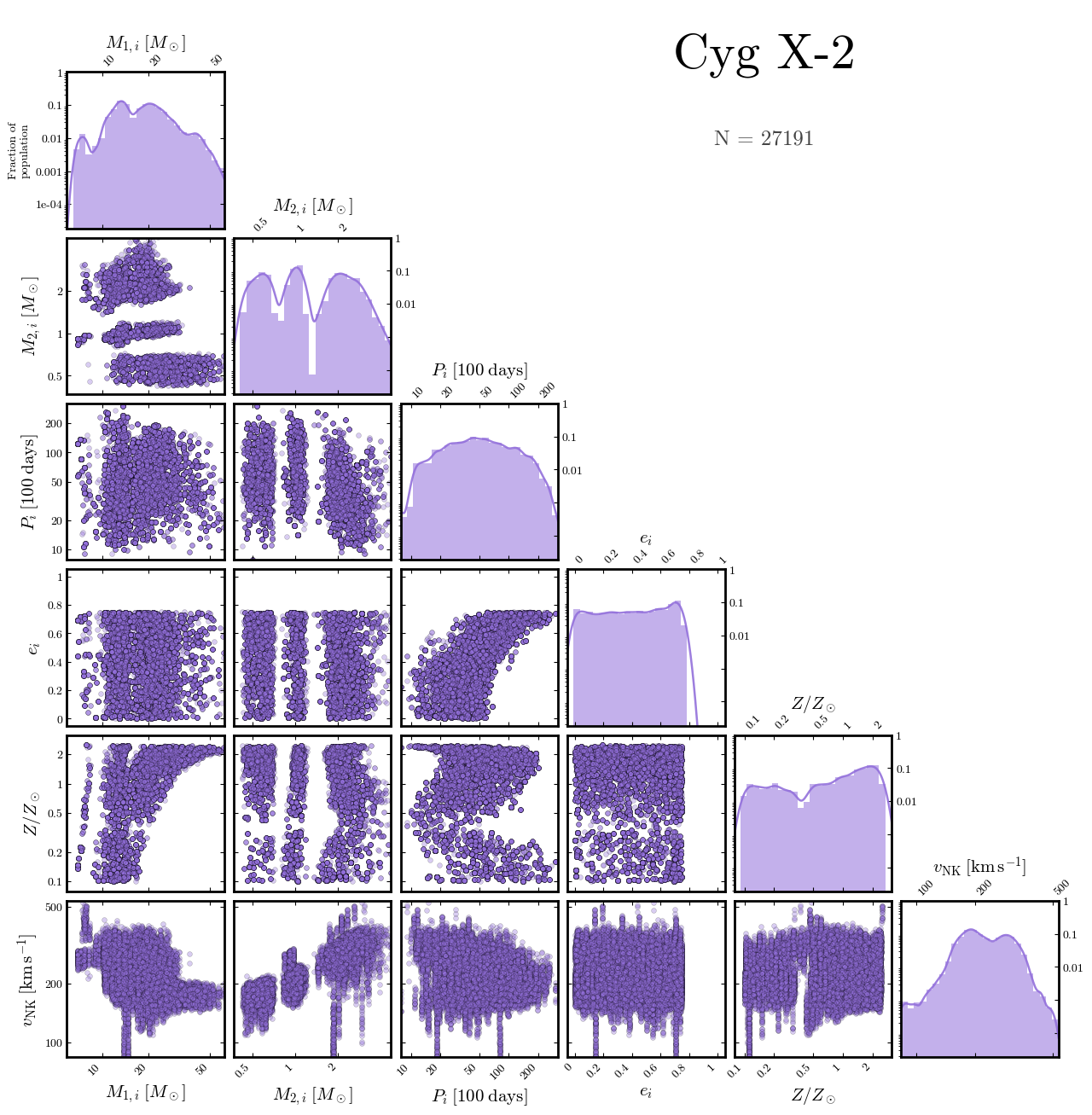}
    \caption{Corner plot highlighting key initial characteristics and acceptable natal kicks for simulated system analogues to Cyg X-2.}
    \label{Fig:CygX2_CornerPlot}
\end{figure*}
\begin{figure*}
    \centering
    \includegraphics[width=0.75\textwidth]{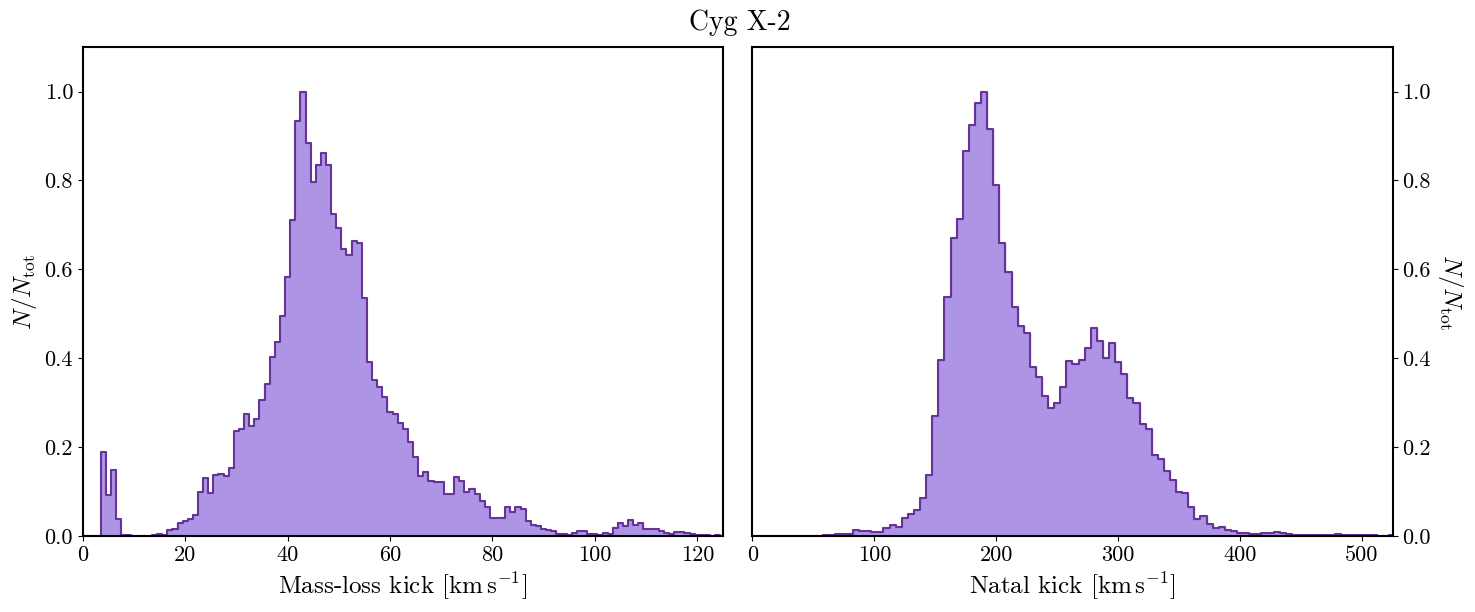}
    \caption{Distribution of mass-loss kicks ($\vmlk$) magnitude (left) and natal kick ($\vnk$) magnitude (right) for simulated system analogues to Cyg X-2.}
    \label{Fig:CygX2_KickHist}
\end{figure*}
\clearpage

\begin{figure*}
    \centering
    \includegraphics[width=0.75\textwidth]{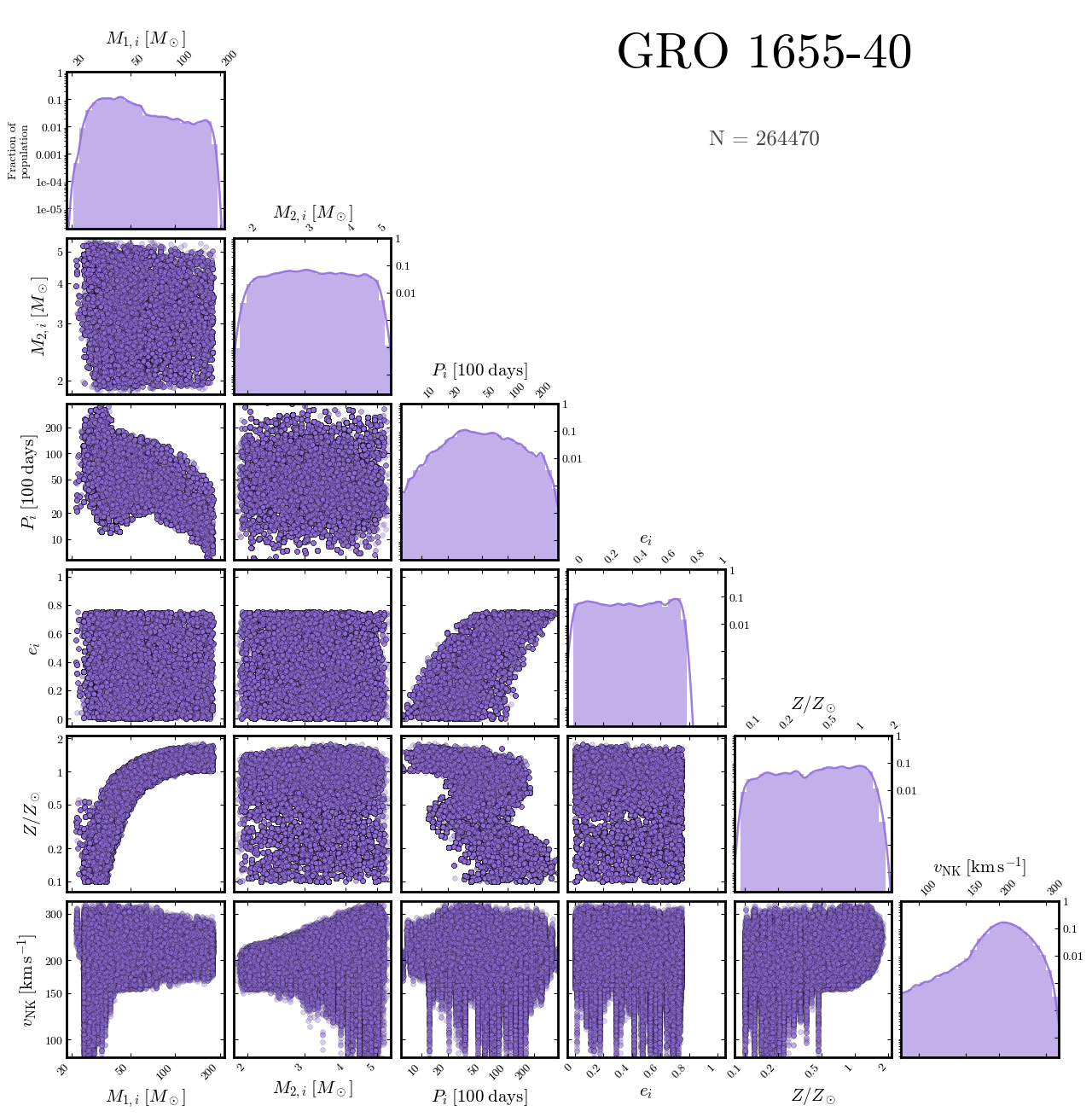}
    \caption{Corner plot highlighting key initial characteristics and acceptable natal kicks for simulated system analogues to GRO 1655-40.}
    \label{Fig:GRO1655_CornerPlot}
\end{figure*}
\begin{figure*}
    \centering
    \includegraphics[width=0.75\textwidth]{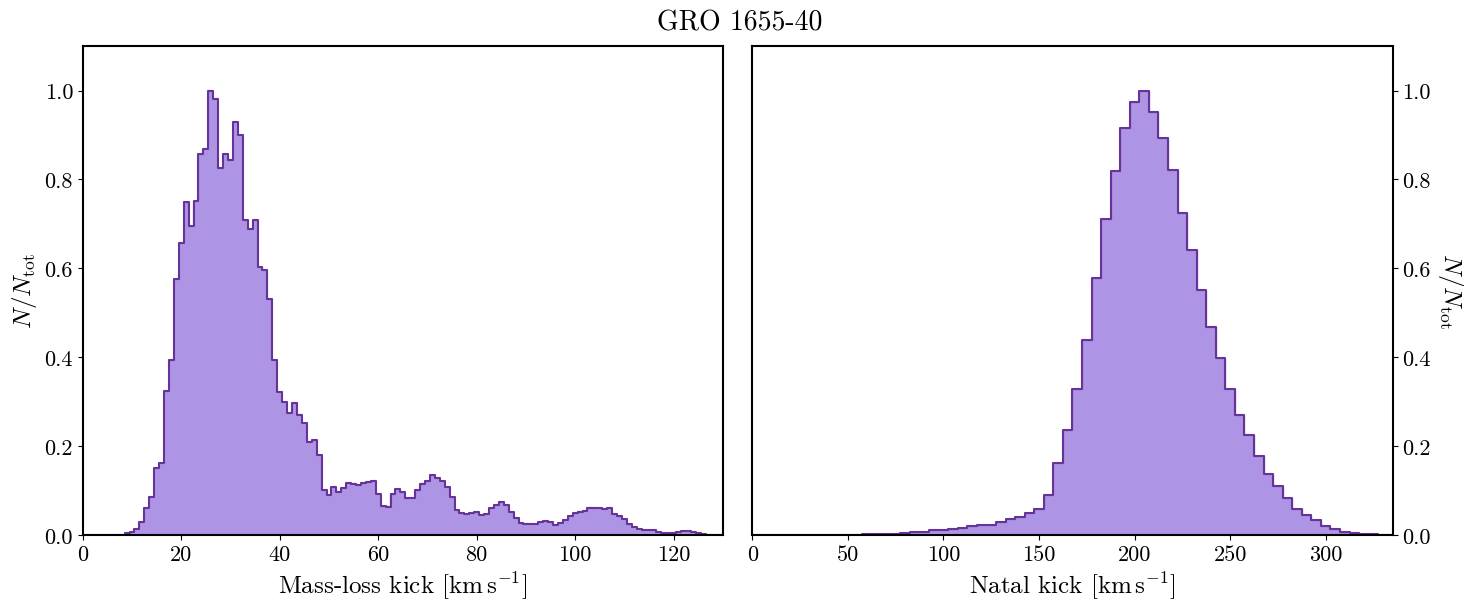}
    \caption{Distribution of mass-loss kicks ($\vmlk$) magnitude (left) and natal kick ($\vnk$) magnitude (right) for simulated system analogues to GRO 1655-40.}
    \label{Fig:GRO1655_KickHist}
\end{figure*}
\clearpage

\begin{figure*}
    \centering
    \includegraphics[width=0.75\textwidth]{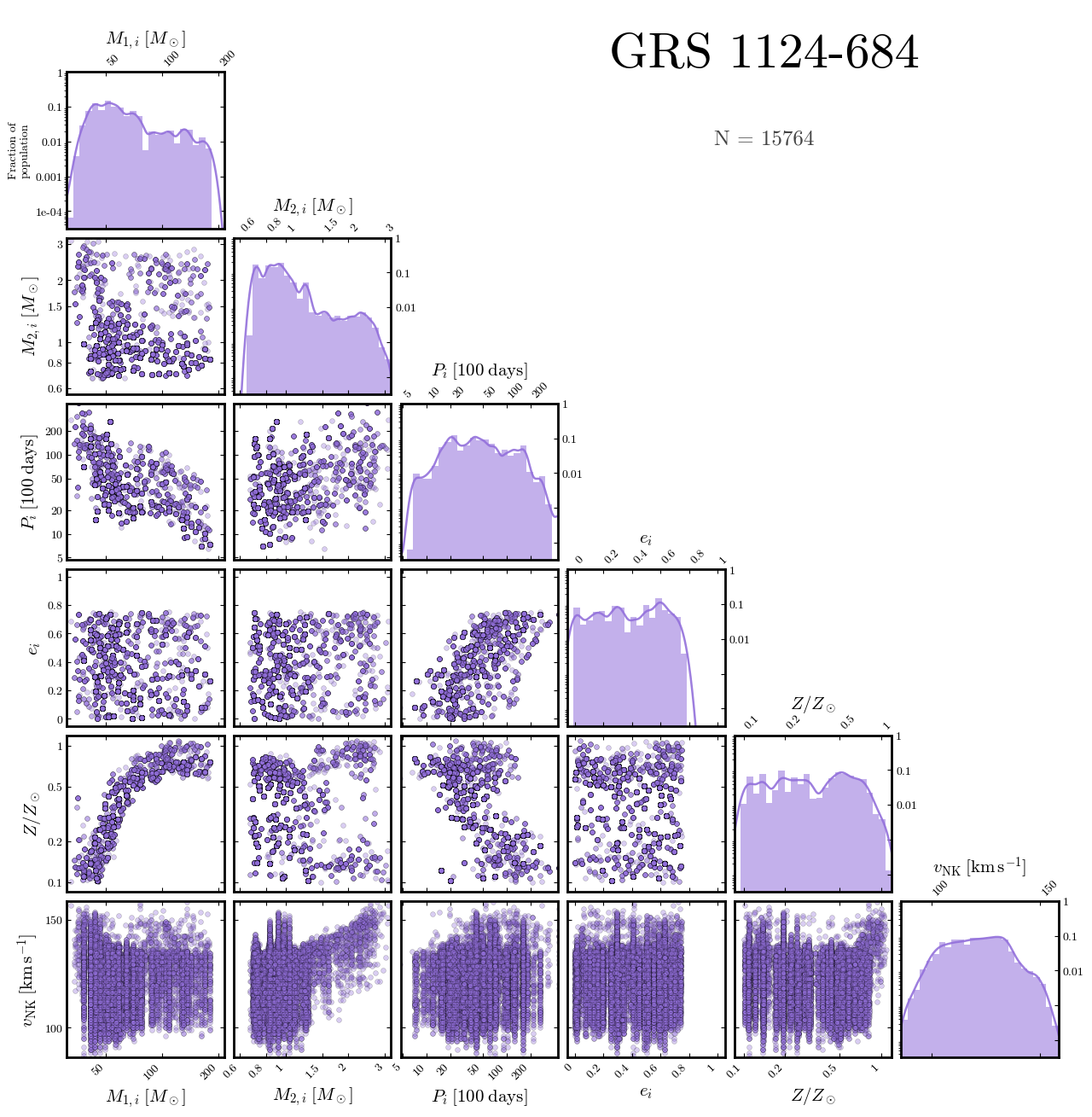}
    \caption{Corner plot highlighting key initial characteristics and acceptable natal kicks for simulated system analogues to GRS 1124-684.}
    \label{Fig:GRS1124_CornerPlot}
\end{figure*}
\begin{figure*}
    \centering
    \includegraphics[width=0.75\textwidth]{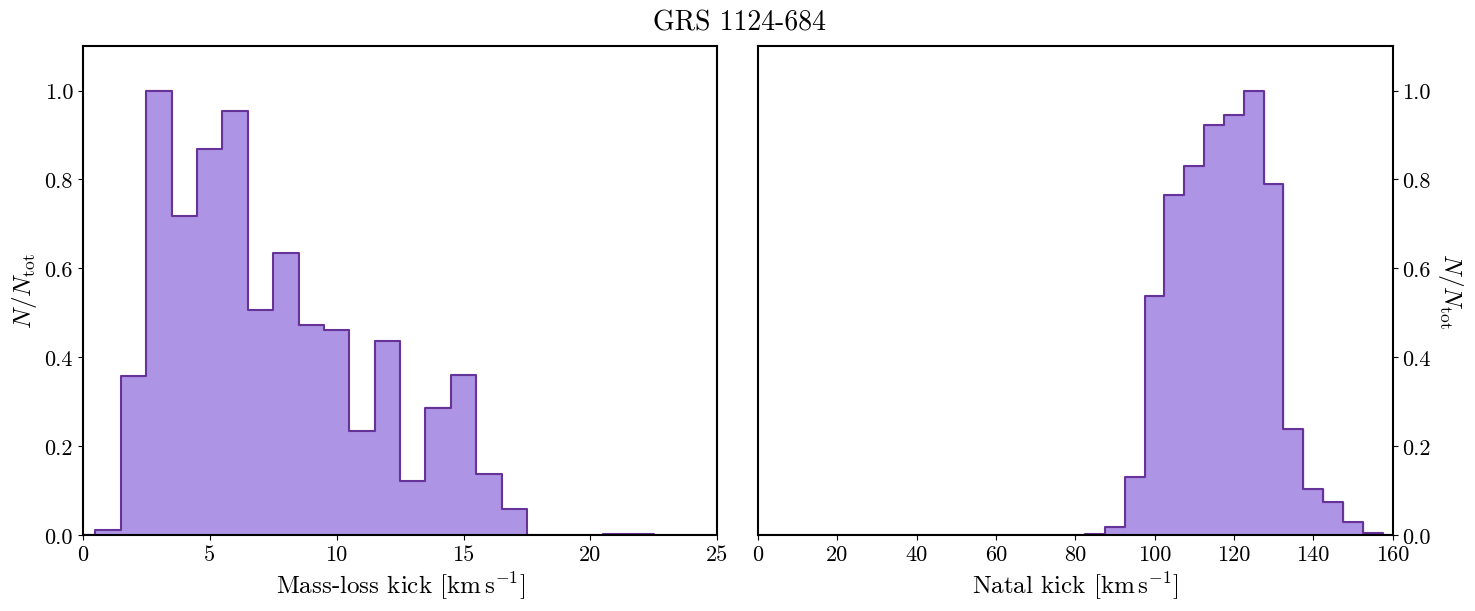}
    \caption{Distribution of mass-loss kicks ($\vmlk$) magnitude (left) and natal kick ($\vnk$) magnitude (right) for simulated system analogues to GRS 1124-684.}
    \label{Fig:GRS1124_KickHist}
\end{figure*}
\clearpage

\begin{figure*}
    \centering
    \includegraphics[width=0.75\textwidth]{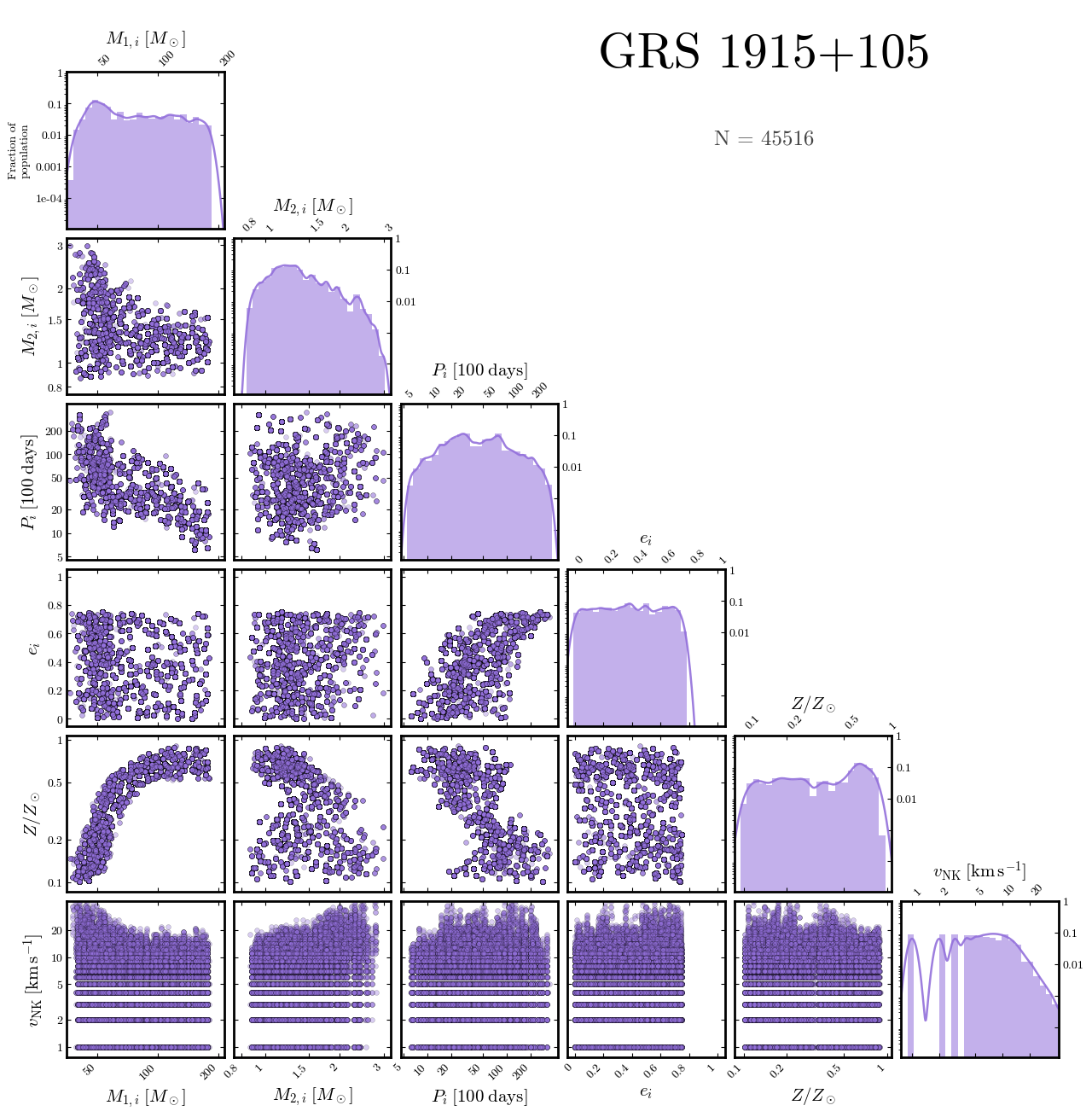}
    \caption{Corner plot highlighting key initial characteristics and acceptable natal kicks for simulated system analogues to GRS 1915+105.}
    \label{Fig:GRS1915_CornerPlot}
\end{figure*}
\begin{figure*}
    \centering
    \includegraphics[width=0.75\textwidth]{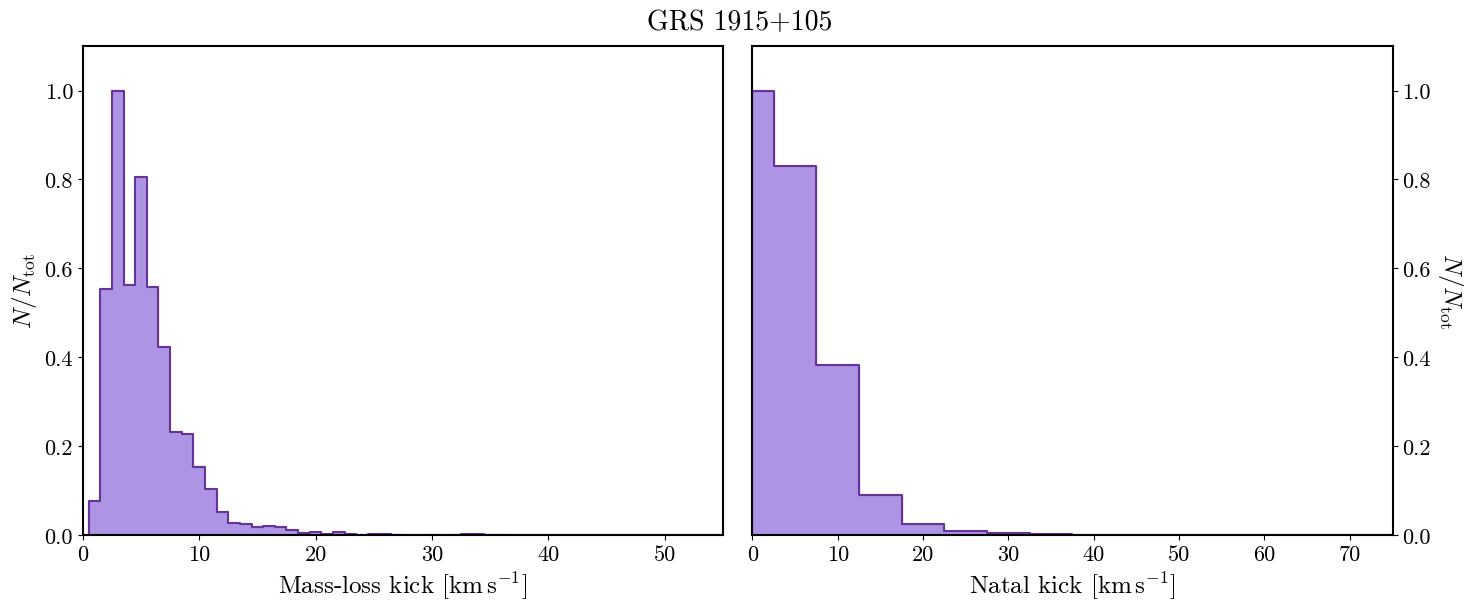}
    \caption{Distribution of mass-loss kicks ($\vmlk$) magnitude (left) and natal kick ($\vnk$) magnitude (right) for simulated system analogues to GRS 1915+105.}
    \label{Fig:GRS1915_KickHist}
\end{figure*}
\clearpage

\begin{figure*}
    \centering
    \includegraphics[width=0.75\textwidth]{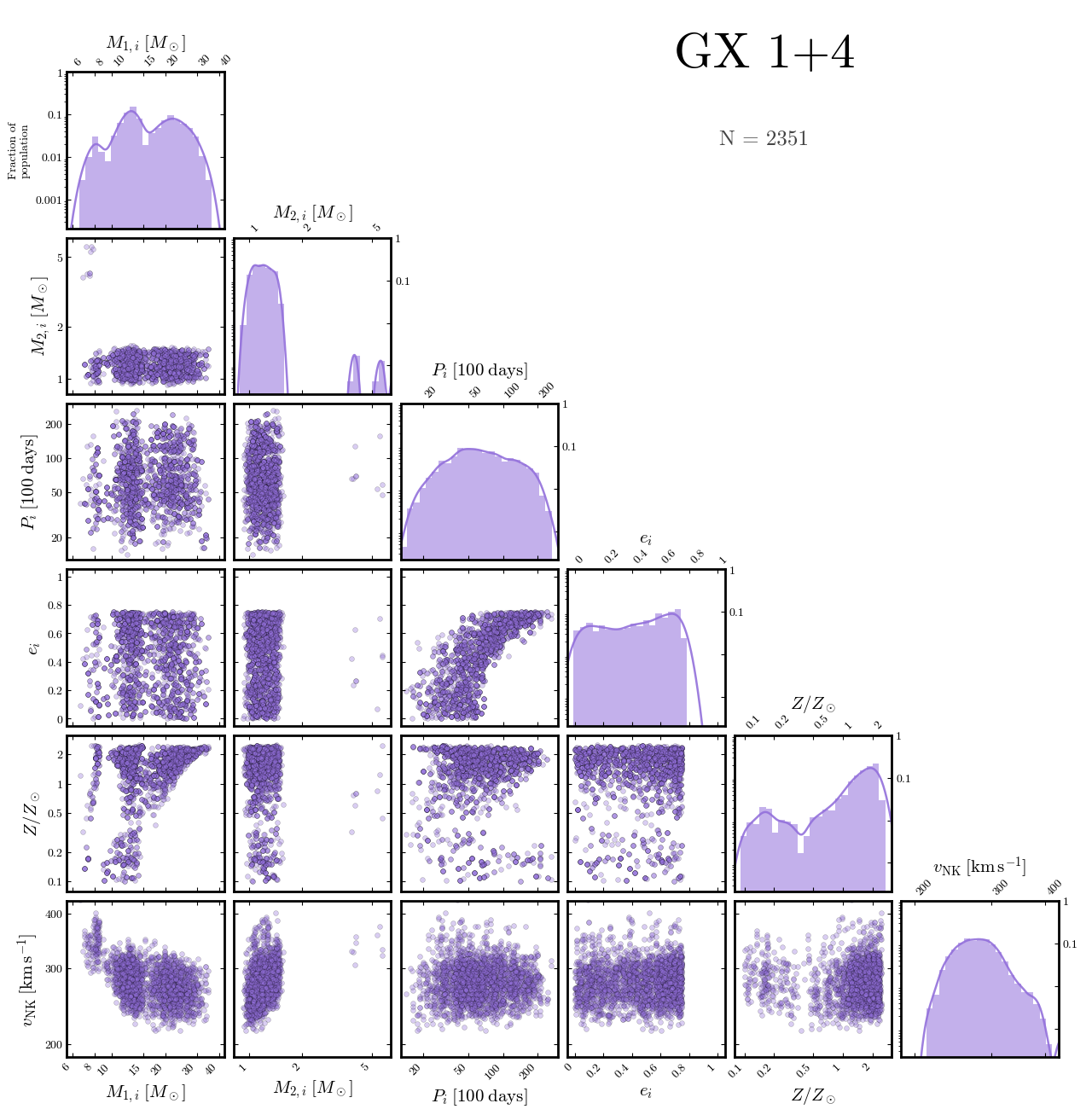}
    \caption{Corner plot highlighting key initial characteristics and acceptable natal kicks for simulated system analogues to GX 1+4.}
    \label{Fig:GX14_CornerPlot}
\end{figure*}
\begin{figure*}
    \centering
    \includegraphics[width=0.75\textwidth]{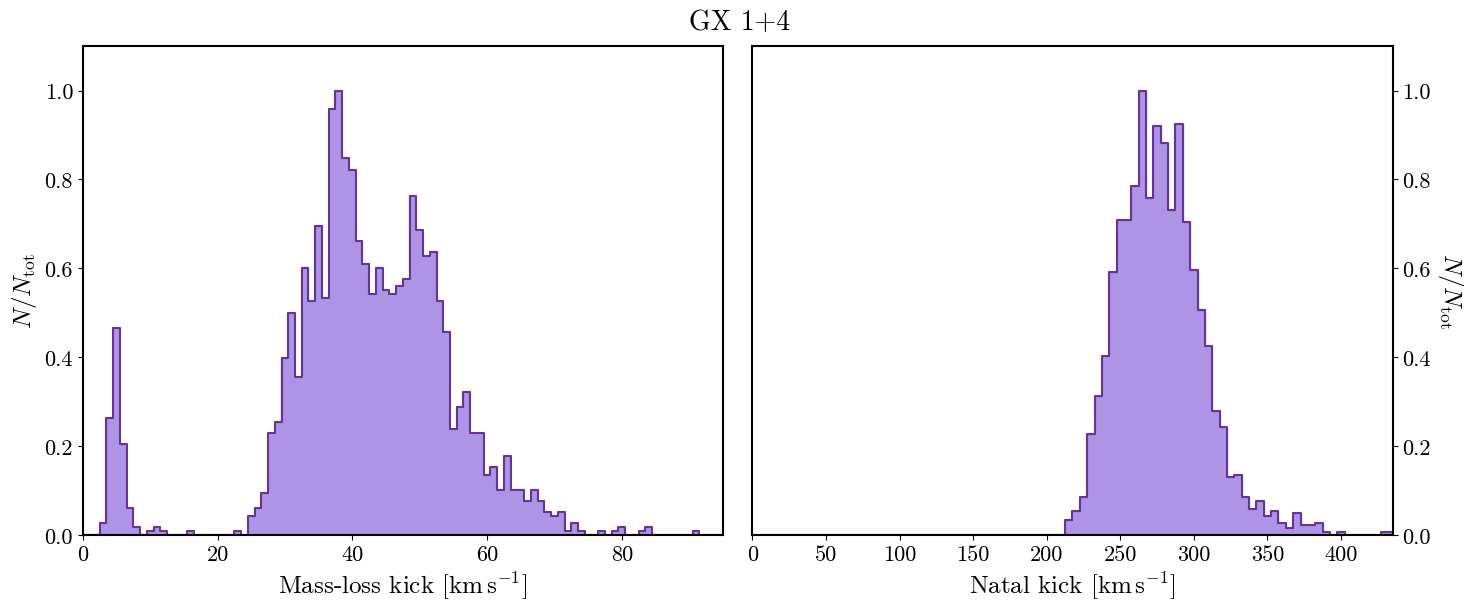}
    \caption{Distribution of mass-loss kicks ($\vmlk$) magnitude (left) and natal kick ($\vnk$) magnitude (right) for simulated system analogues to GX 1+4.}
    \label{Fig:GX14_KickHist}
\end{figure*}
\clearpage

\begin{figure*}
    \centering
    \includegraphics[width=0.75\textwidth]{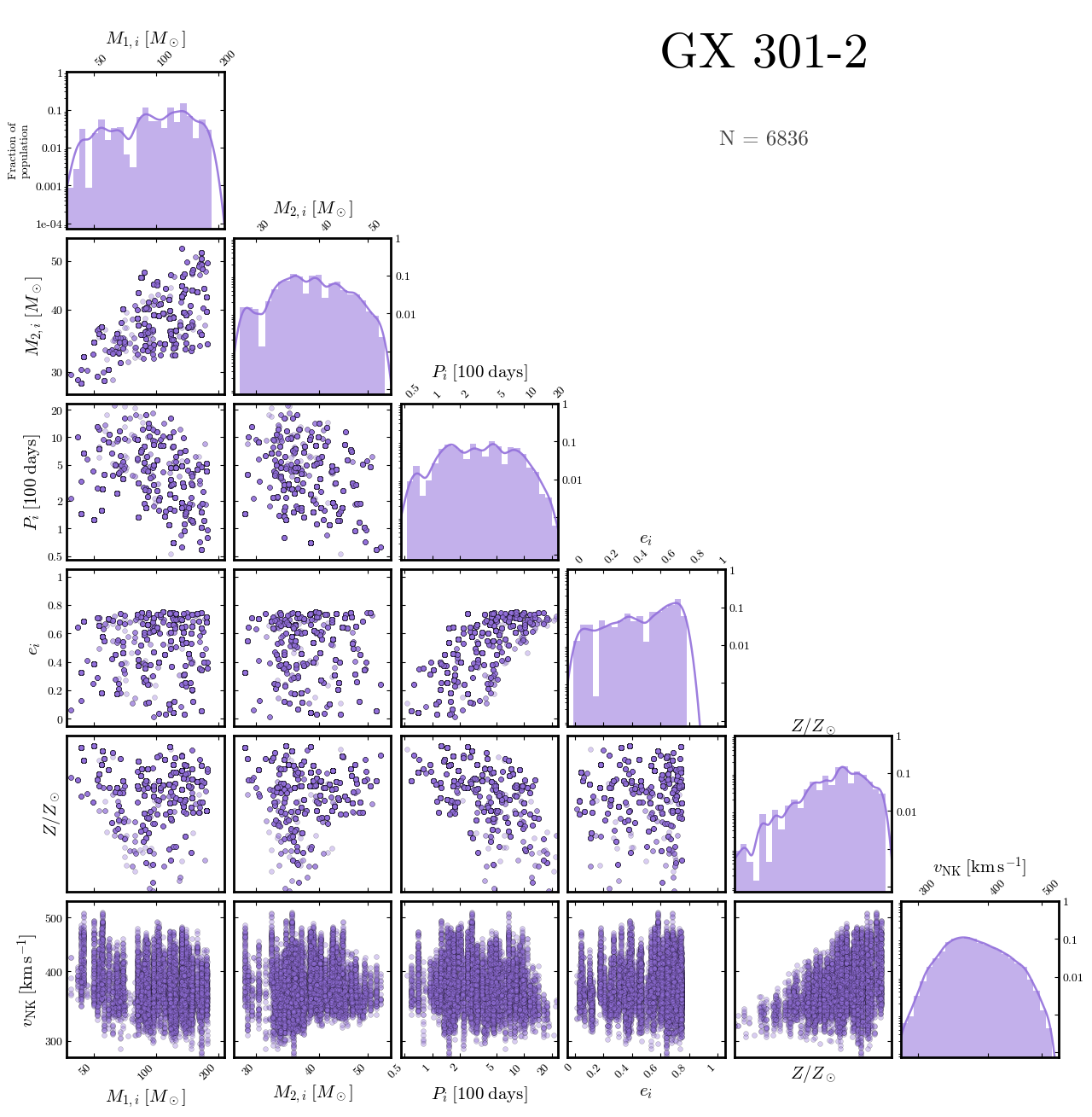}
    \caption{Corner plot highlighting key initial characteristics and acceptable natal kicks for simulated system analogues to GX 301-2.}
    \label{Fig:GX301_CornerPlot}
\end{figure*}
\begin{figure*}
    \centering
    \includegraphics[width=0.75\textwidth]{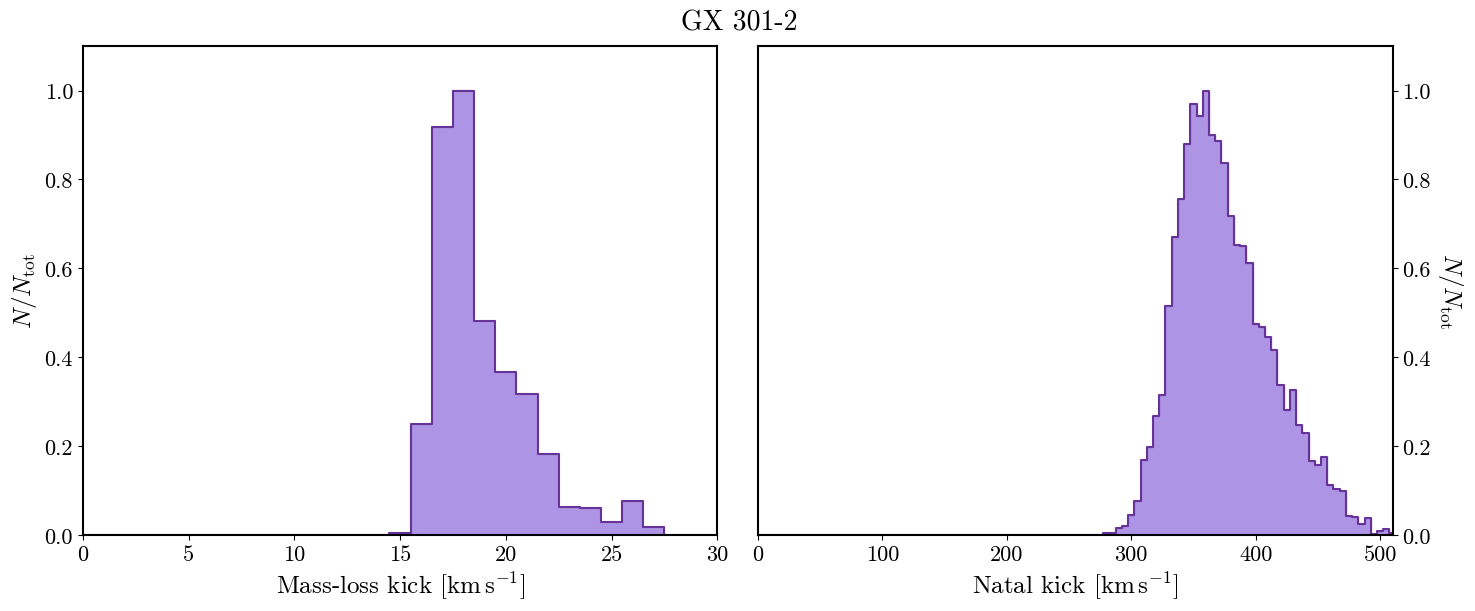}
    \caption{Distribution of mass-loss kicks ($\vmlk$) magnitude (left) and natal kick ($\vnk$) magnitude (right) for simulated system analogues to GX 301-2.}
    \label{Fig:GX301_KickHist}
\end{figure*}
\clearpage

\begin{figure*}
    \centering
    \includegraphics[width=0.75\textwidth]{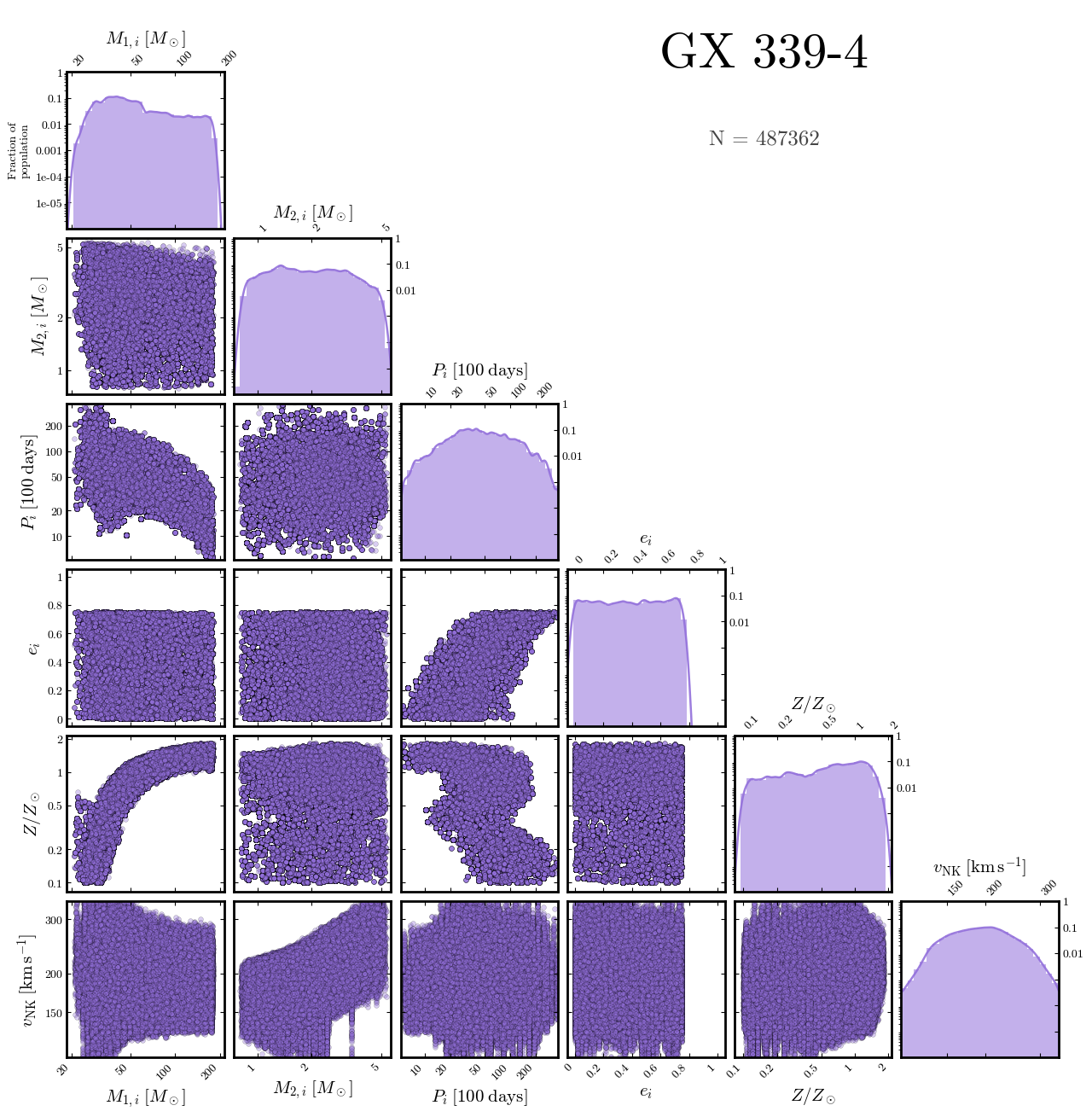}
    \caption{Corner plot highlighting key initial characteristics and acceptable natal kicks for simulated system analogues to GX 339-4.}
    \label{Fig:GX339_CornerPlot}
\end{figure*}
\begin{figure*}
    \centering
    \includegraphics[width=0.75\textwidth]{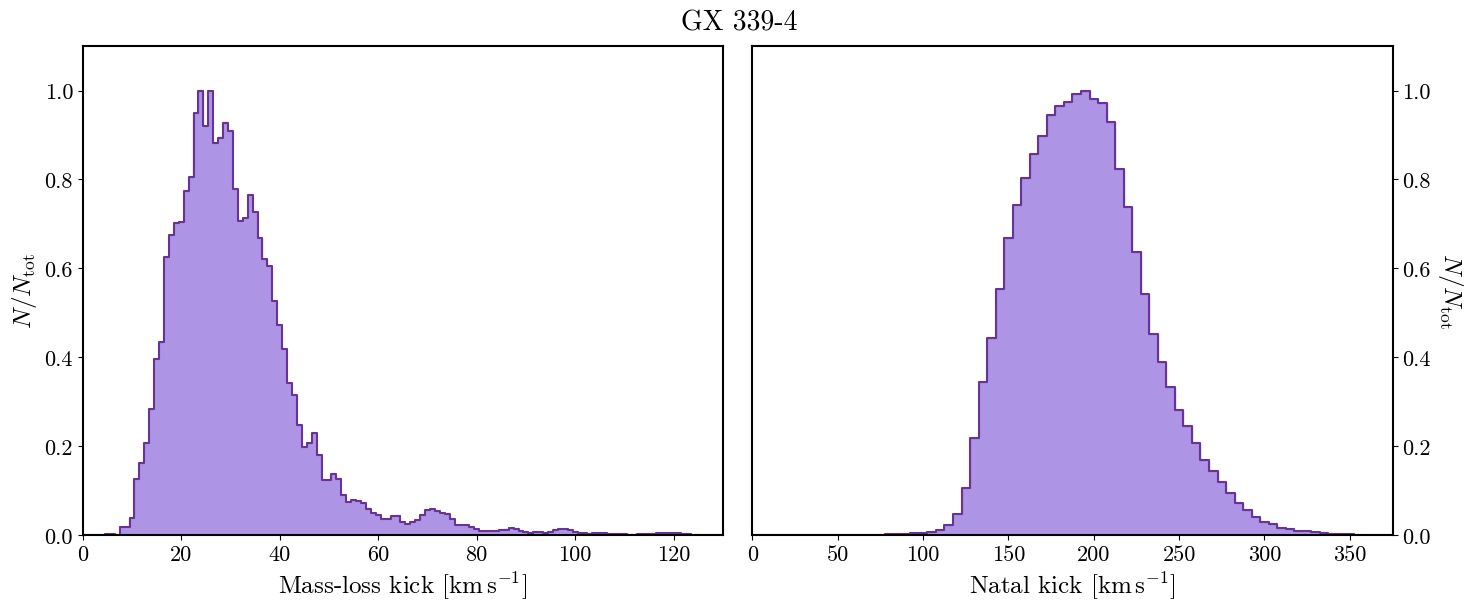}
    \caption{Distribution of mass-loss kicks ($\vmlk$) magnitude (left) and natal kick ($\vnk$) magnitude (right) for simulated system analogues to GX 339-4.}
    \label{Fig:GX339_KickHist}
\end{figure*}
\clearpage

\begin{figure*}
    \centering
    \includegraphics[width=0.75\textwidth]{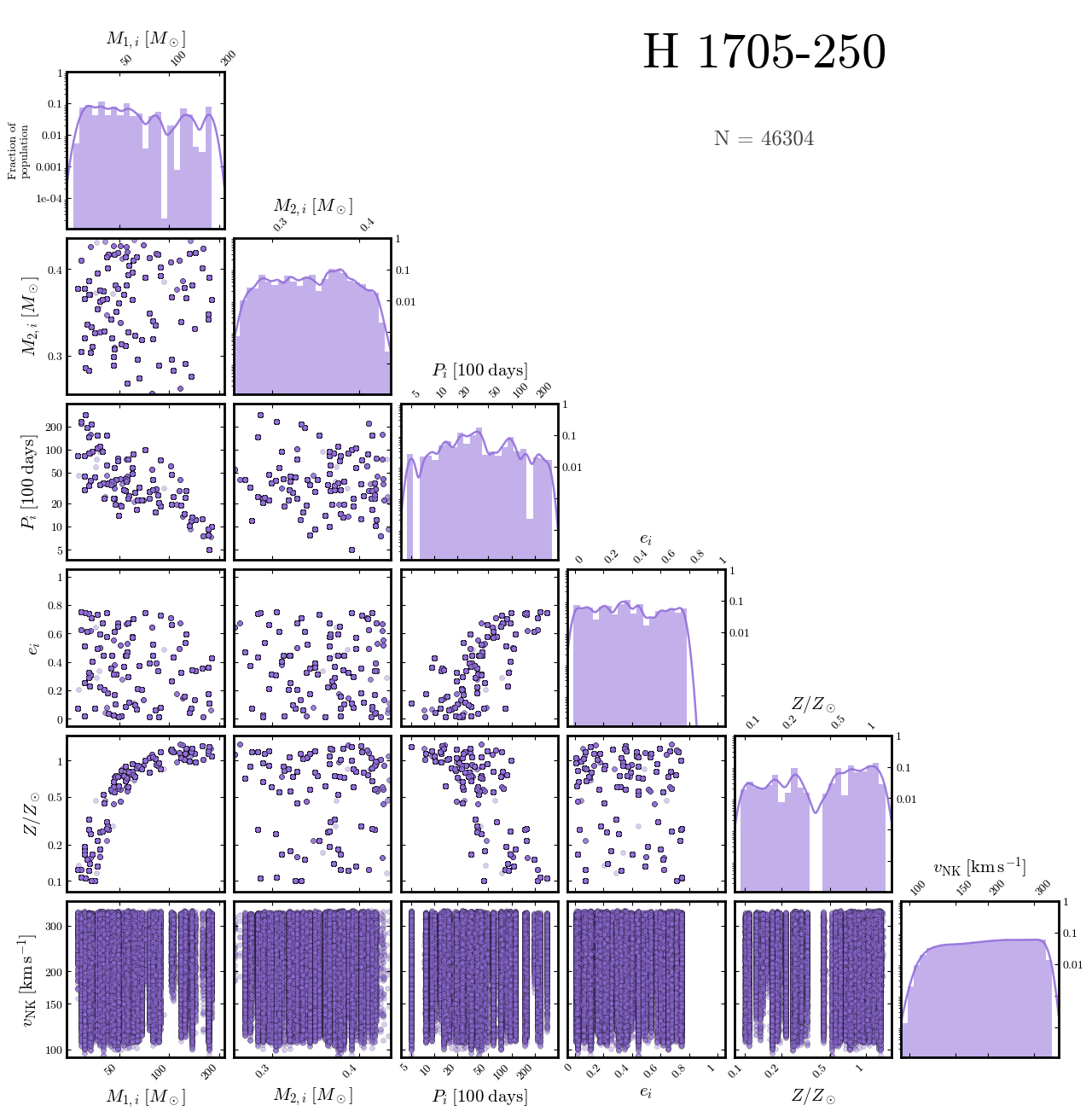}
    \caption{Corner plot highlighting key initial characteristics and acceptable natal kicks for simulated system analogues to H 1705-250.}
    \label{Fig:H1705_CornerPlot}
\end{figure*}
\begin{figure*}
    \centering
    \includegraphics[width=0.75\textwidth]{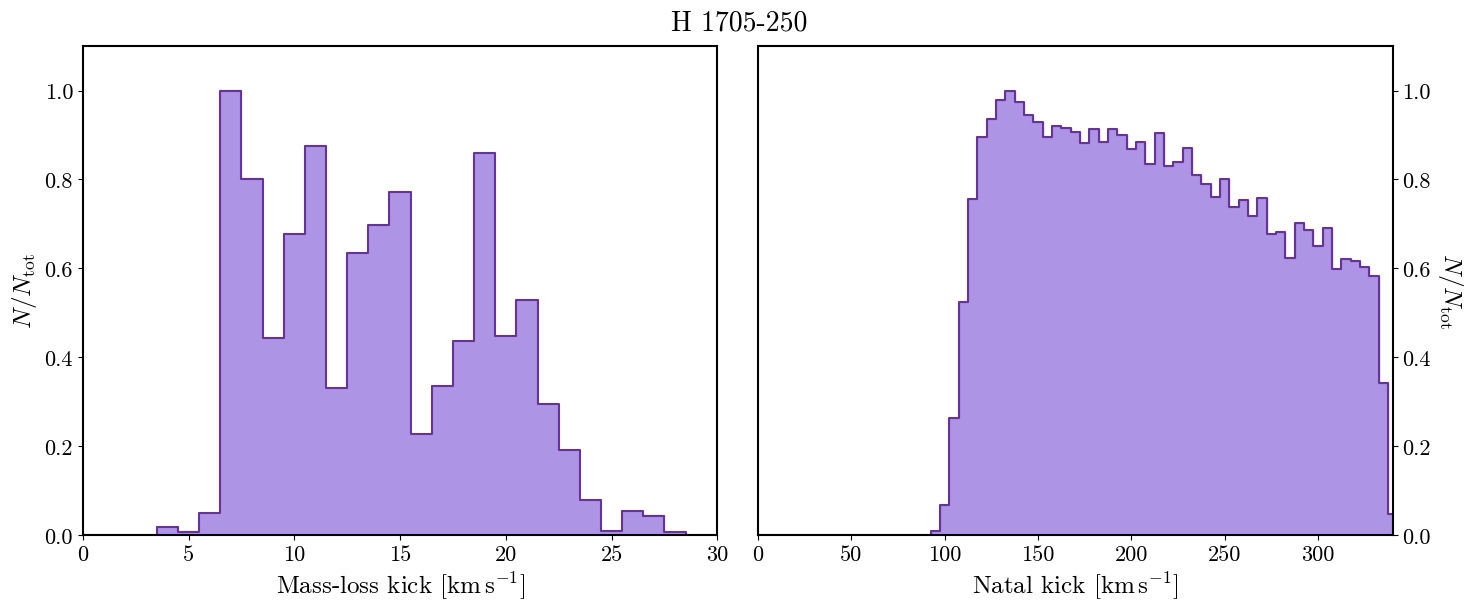}
    \caption{Distribution of mass-loss kicks ($\vmlk$) magnitude (left) and natal kick ($\vnk$) magnitude (right) for simulated system analogues to H 1705-250.}
    \label{Fig:H1705_KickHist}
\end{figure*}
\clearpage

\begin{figure*}
    \centering
    \includegraphics[width=0.75\textwidth]{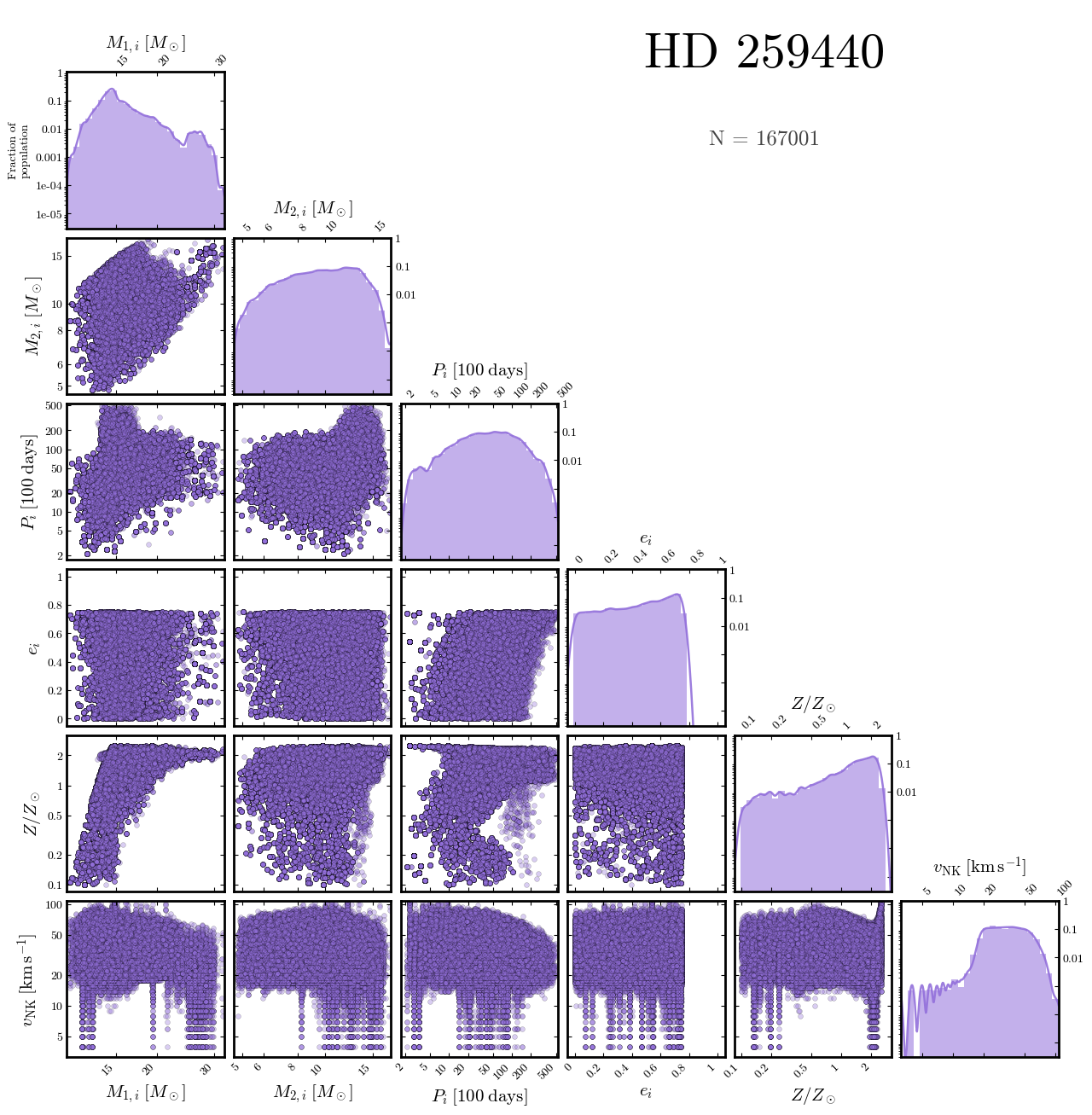}
    \caption{Corner plot highlighting key initial characteristics and acceptable natal kicks for simulated system analogues to HD 259440.}
    \label{Fig:HD259_CornerPlot}
\end{figure*}
\begin{figure*}
    \centering
    \includegraphics[width=0.75\textwidth]{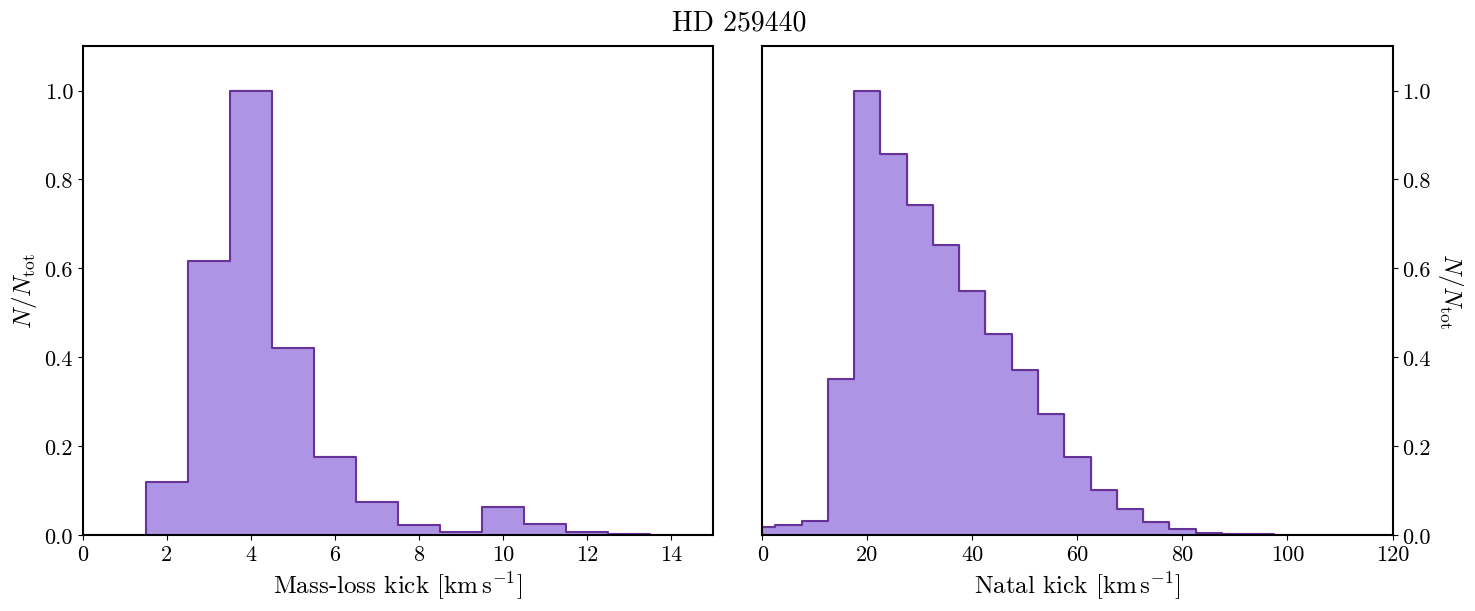}
    \caption{Distribution of mass-loss kicks ($\vmlk$) magnitude (left) and natal kick ($\vnk$) magnitude (right) for simulated system analogues to HD 259440.}
    \label{Fig:HD259_KickHist}
\end{figure*}
\clearpage

\begin{figure*}
    \centering
    \includegraphics[width=0.75\textwidth]{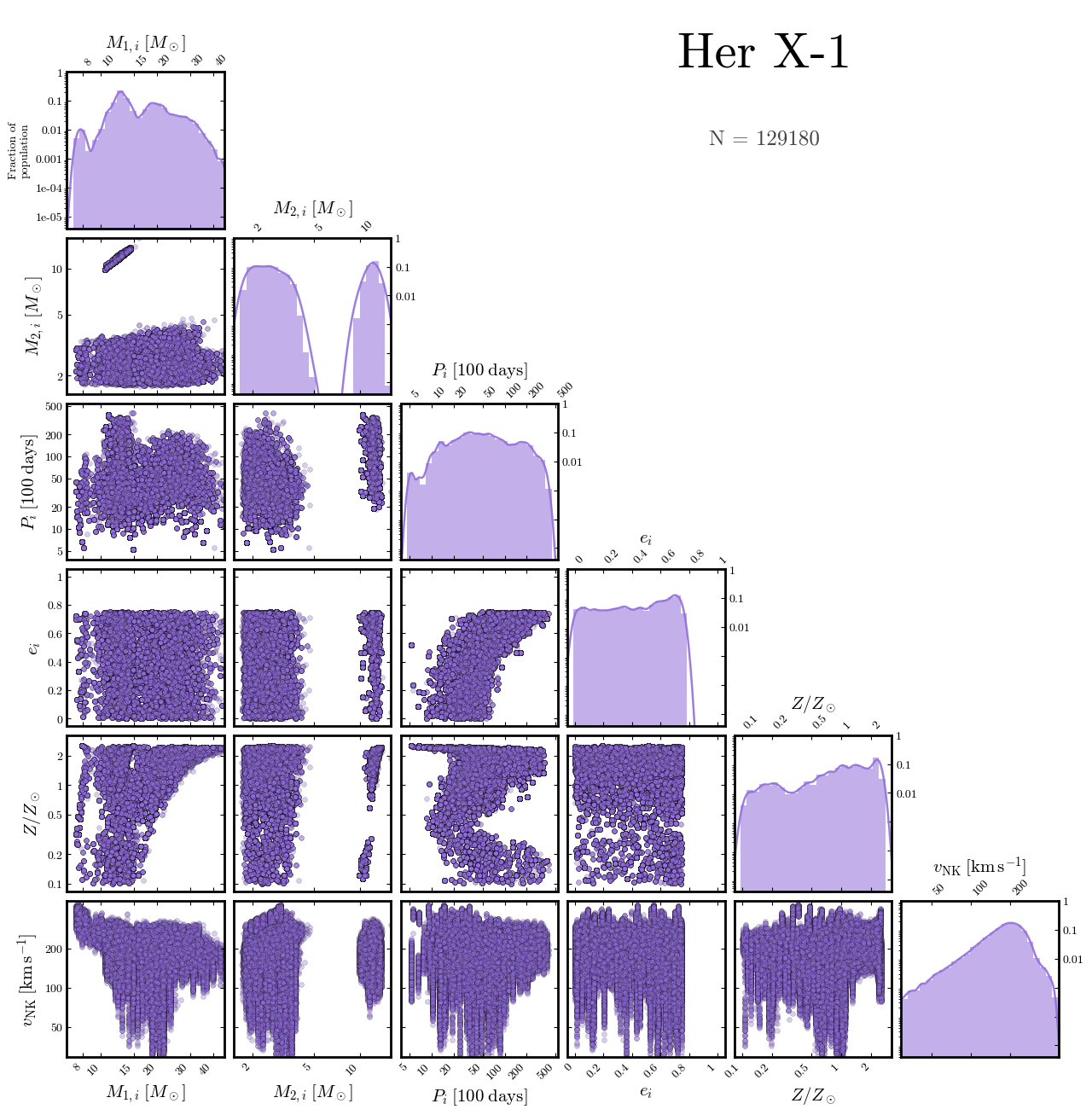}
    \caption{Corner plot highlighting key initial characteristics and acceptable natal kicks for simulated system analogues to Her X-1.}
    \label{Fig:HerX1_CornerPlot}
\end{figure*}
\begin{figure*}
    \centering
    \includegraphics[width=0.75\textwidth]{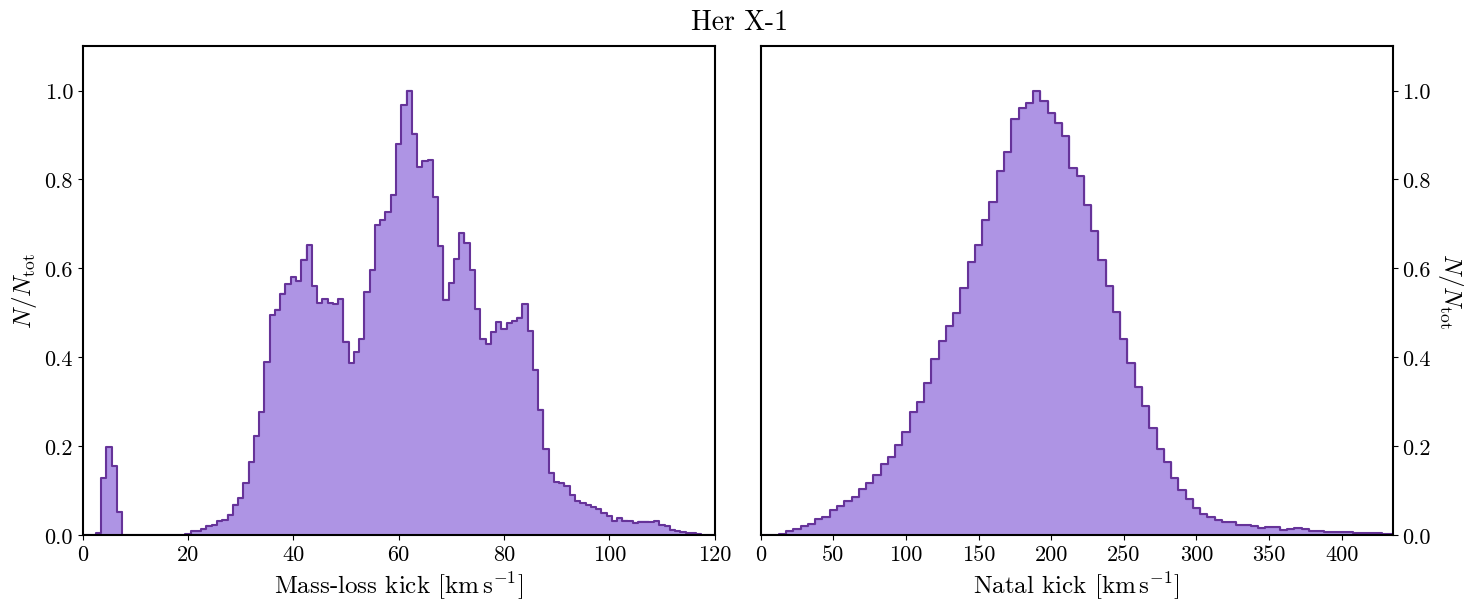}
    \caption{Distribution of mass-loss kicks ($\vmlk$) magnitude (left) and natal kick ($\vnk$) magnitude (right) for simulated system analogues to Her X-1.}
    \label{Fig:HerX1_KickHist}
\end{figure*}
\clearpage

\begin{figure*}
    \centering
    \includegraphics[width=0.75\textwidth]{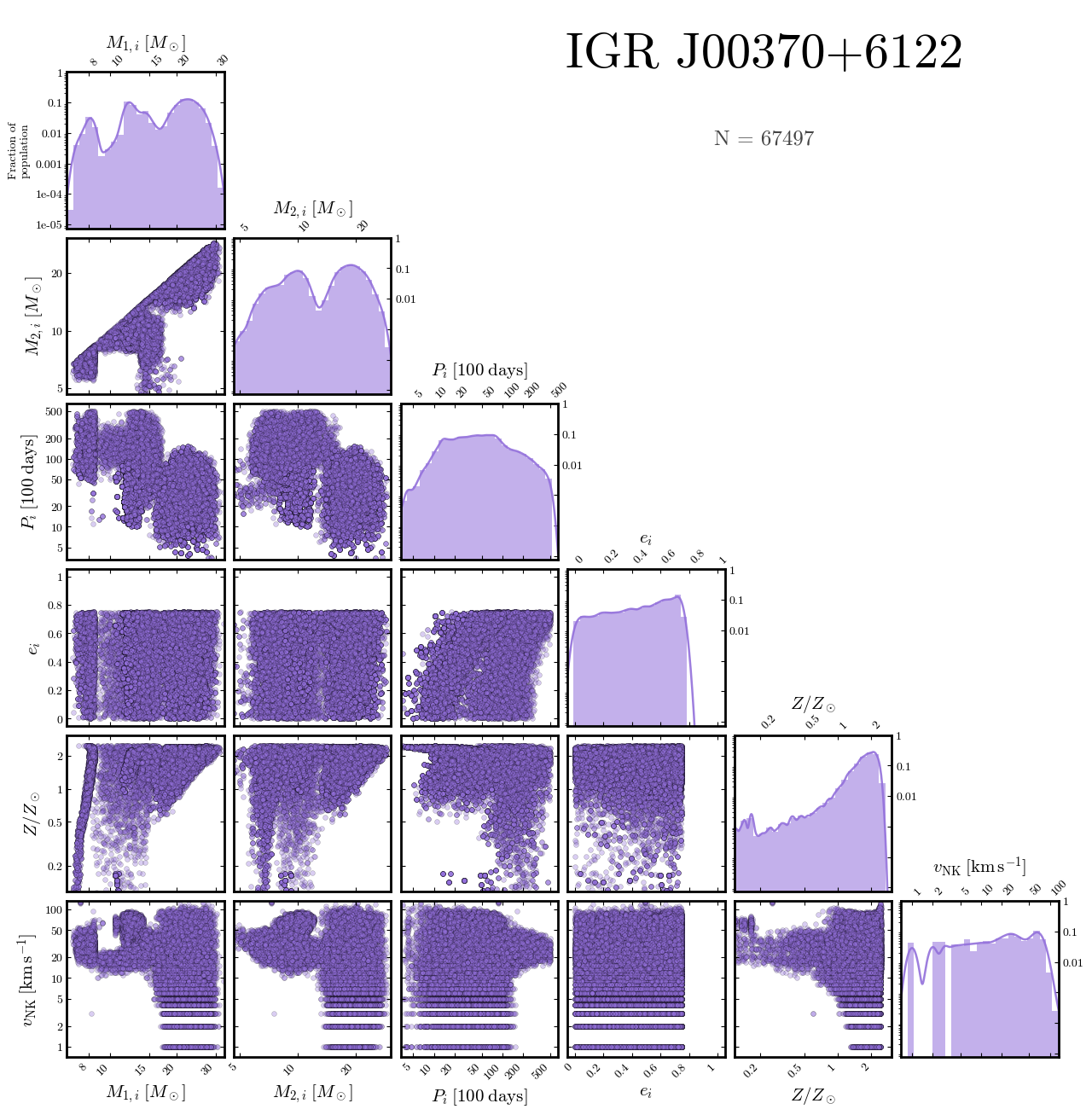}
    \caption{Corner plot highlighting key initial characteristics and acceptable natal kicks for simulated system analogues to IGR J00370+6122.}
    \label{Fig:J00370_CornerPlot}
\end{figure*}
\begin{figure*}
    \centering
    \includegraphics[width=0.75\textwidth]{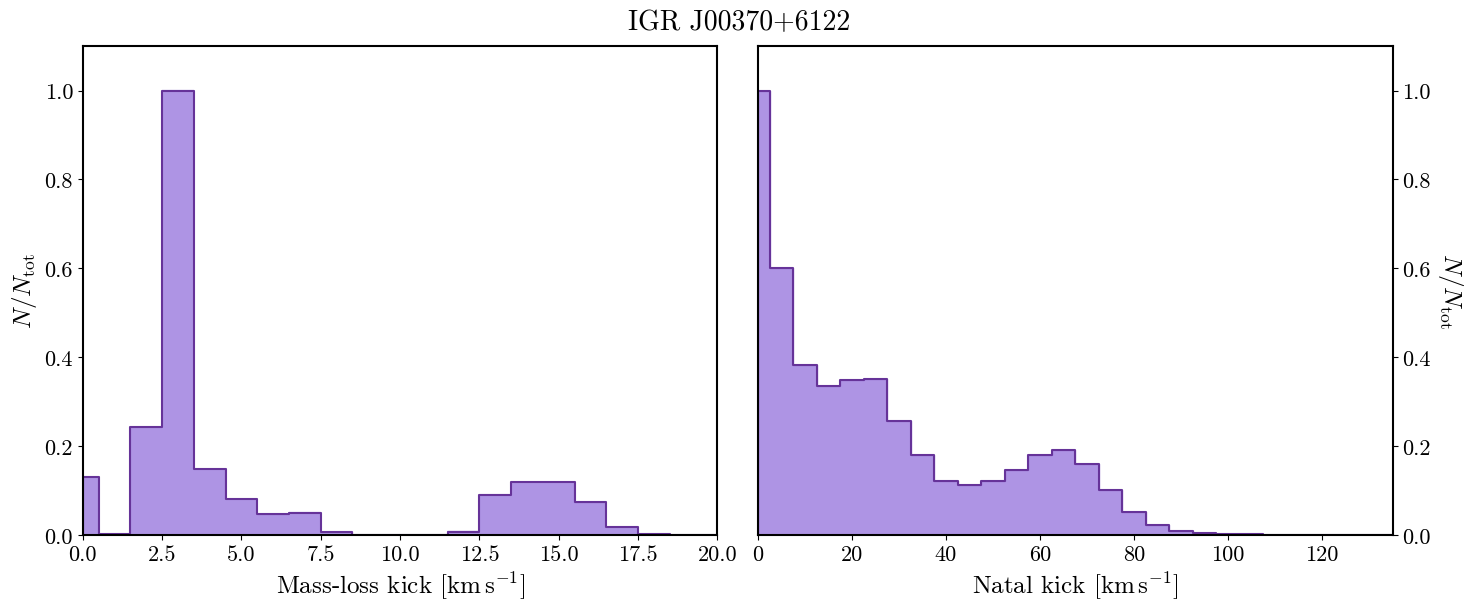}
    \caption{Distribution of mass-loss kicks ($\vmlk$) magnitude (left) and natal kick ($\vnk$) magnitude (right) for simulated system analogues to IGR J00370+6122.}
    \label{Fig:J00370_KickHist}
\end{figure*}
\clearpage

\begin{figure*}
    \centering
    \includegraphics[width=0.75\textwidth]{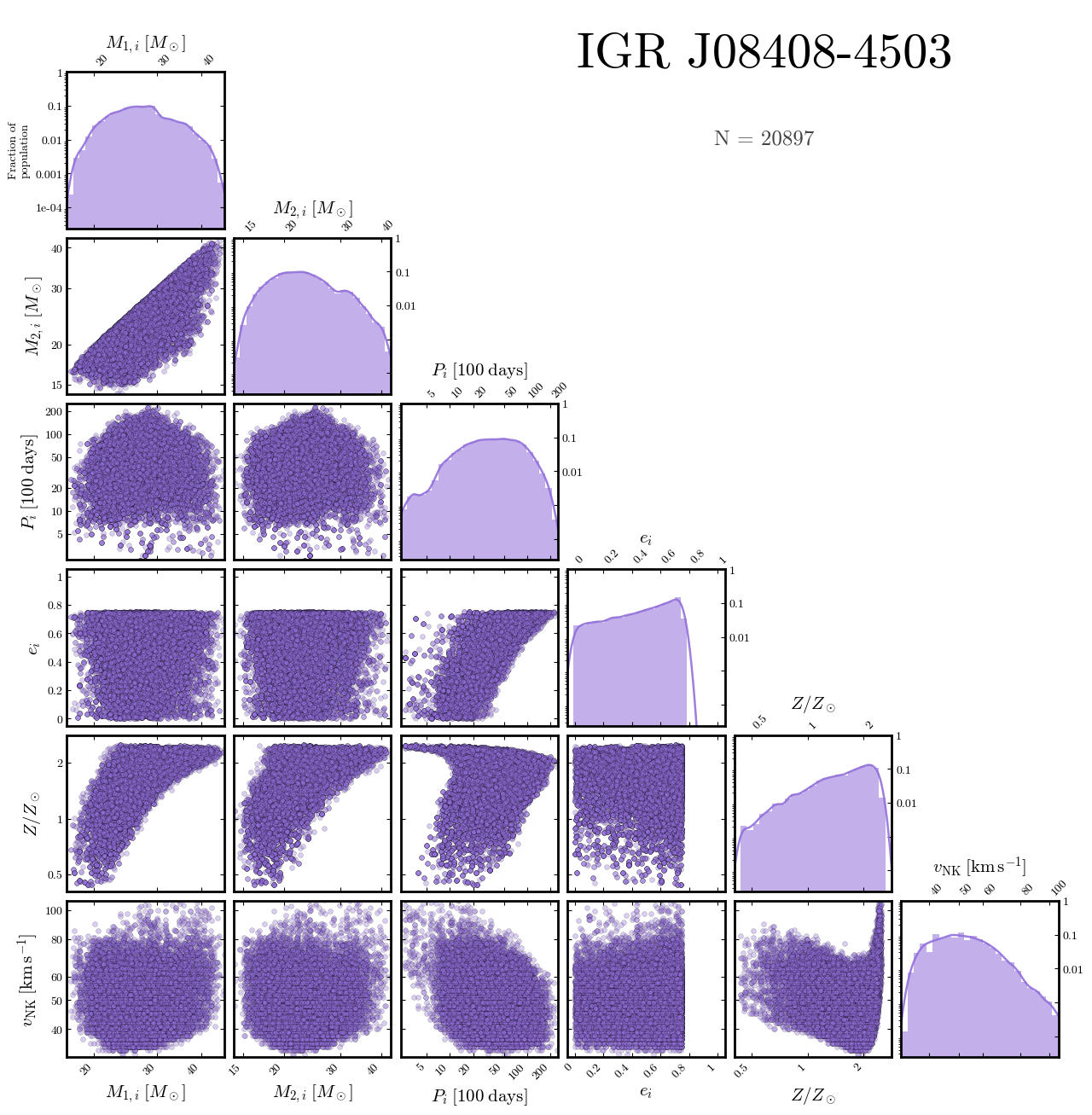}
    \caption{Corner plot highlighting key initial characteristics and acceptable natal kicks for simulated system analogues to IGR J08408-4503.}
    \label{Fig:J08408_CornerPlot}
\end{figure*}
\begin{figure*}
    \centering
    \includegraphics[width=0.75\textwidth]{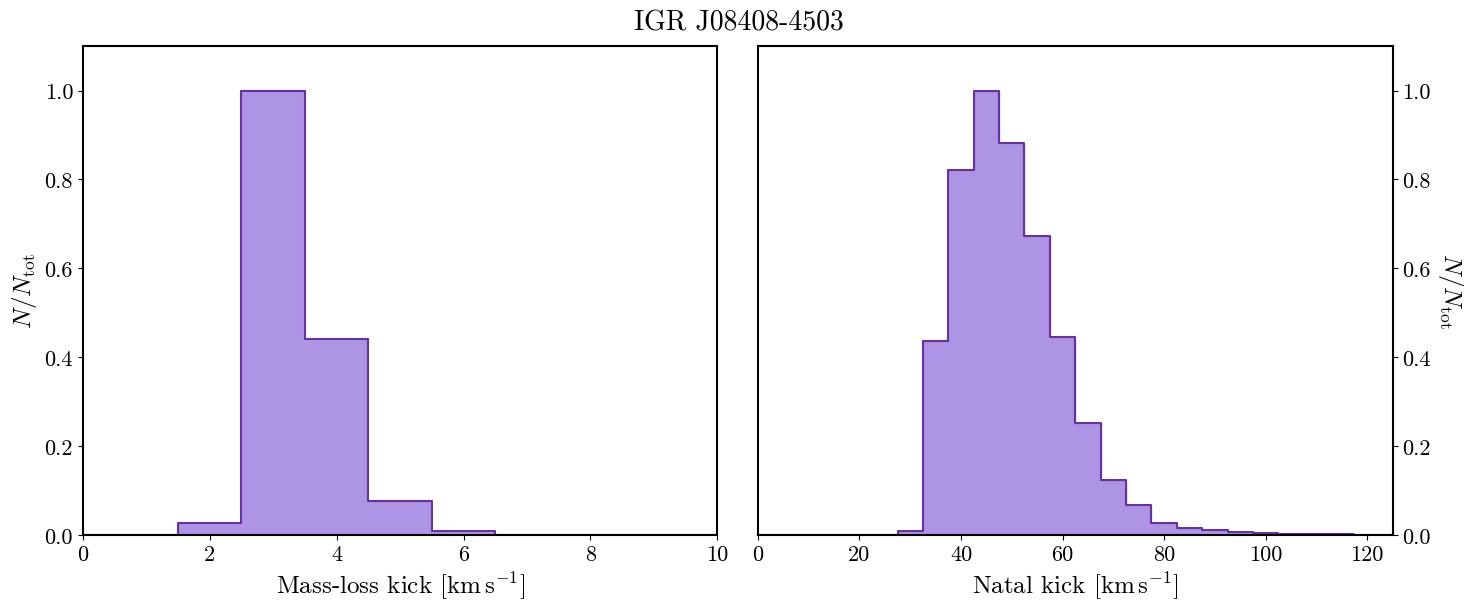}
    \caption{Distribution of mass-loss kicks ($\vmlk$) magnitude (left) and natal kick ($\vnk$) magnitude (right) for simulated system analogues to IGR J08408-4503.}
    \label{Fig:J08408_KickHist}
\end{figure*}
\clearpage

\begin{figure*}
    \centering
    \includegraphics[width=0.75\textwidth]{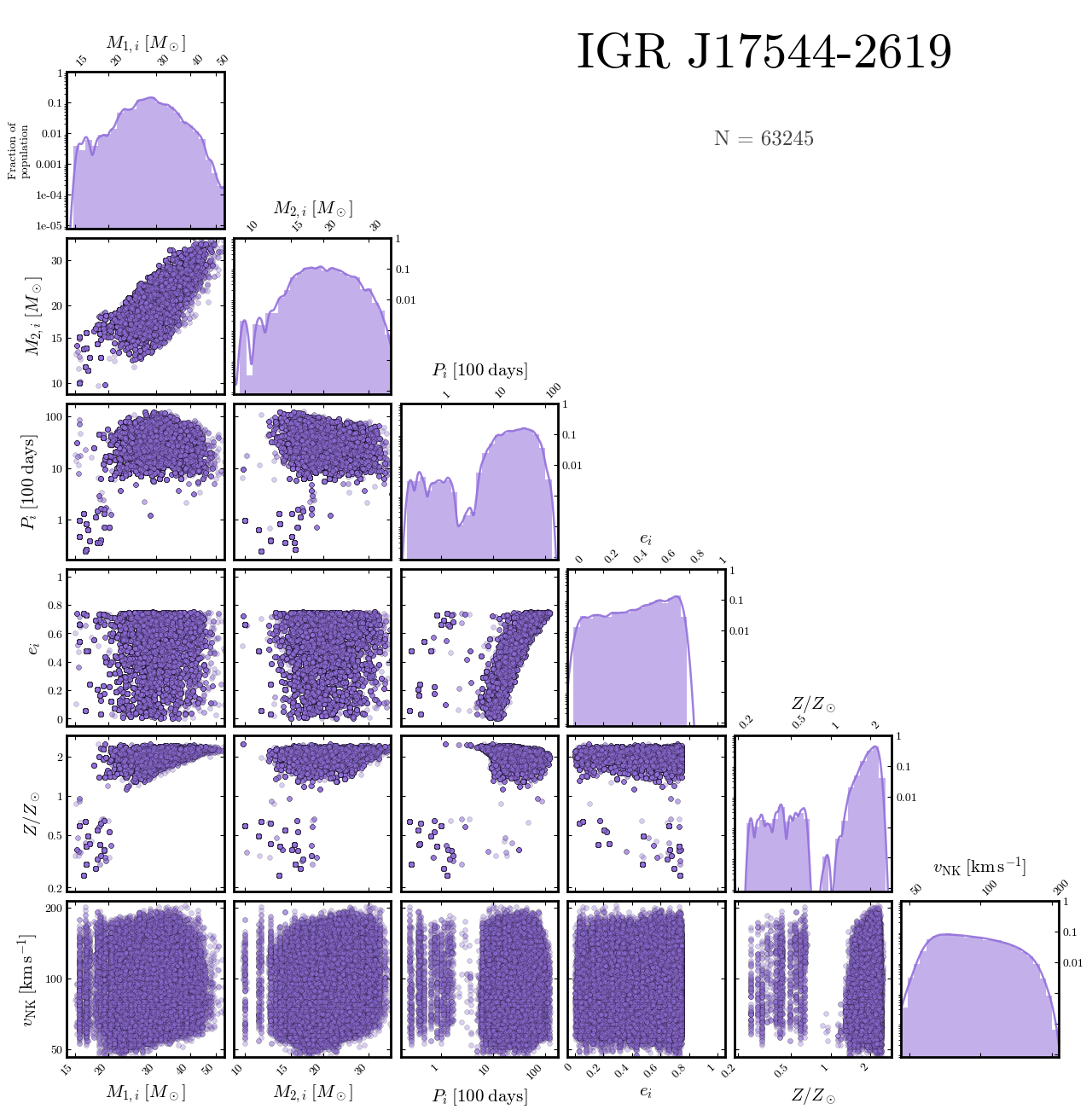}
    \caption{Corner plot highlighting key initial characteristics and acceptable natal kicks for simulated system analogues to IGR J17544-2619.}
    \label{Fig:J17544_CornerPlot}
\end{figure*}
\begin{figure*}
    \centering
    \includegraphics[width=0.75\textwidth]{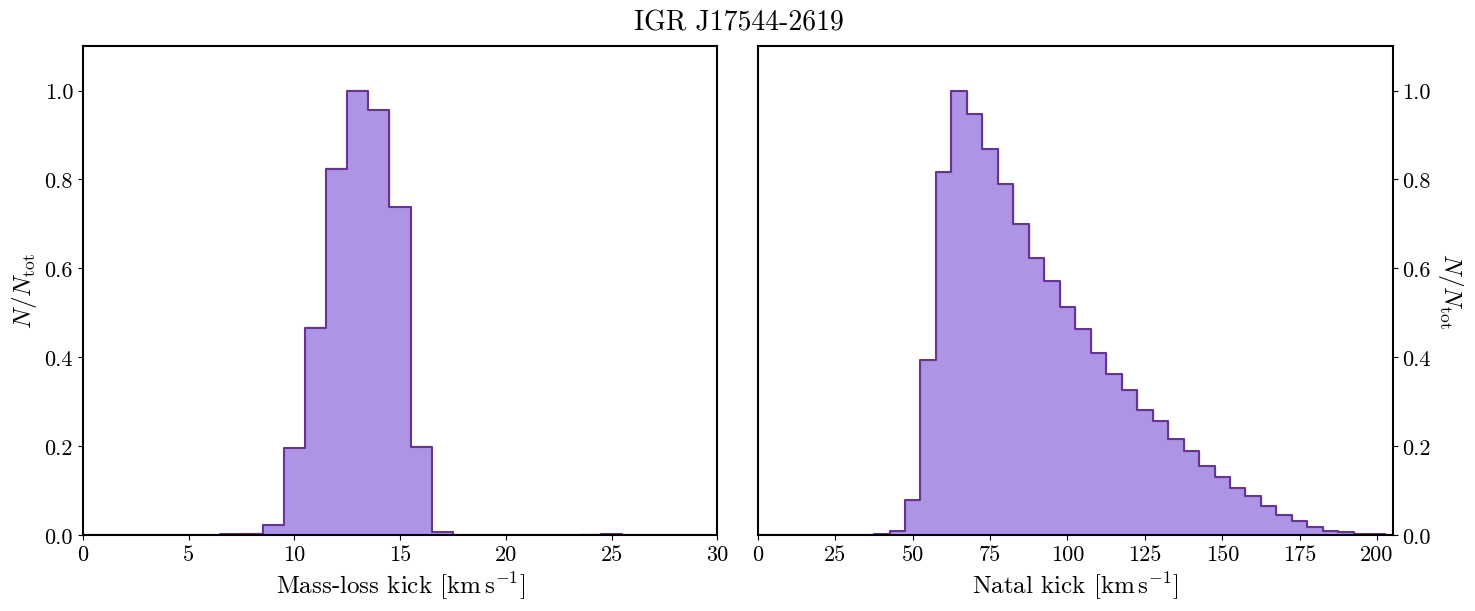}
    \caption{Distribution of mass-loss kicks ($\vmlk$) magnitude (left) and natal kick ($\vnk$) magnitude (right) for simulated system analogues to IGR J17544-2619.}
    \label{Fig:J17544_KickHist}
\end{figure*}
\clearpage

\begin{figure*}
    \centering
    \includegraphics[width=0.75\textwidth]{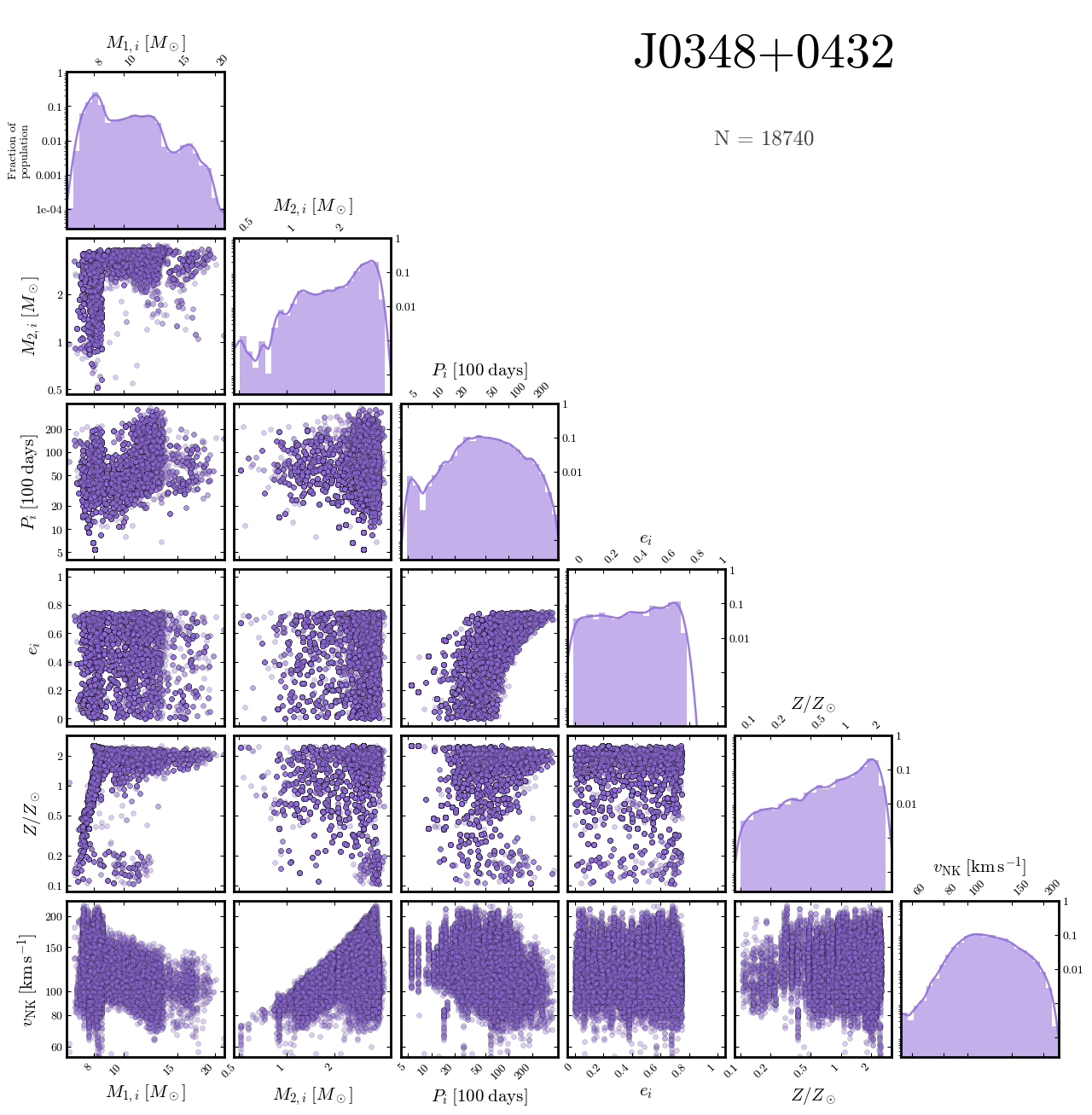}
    \caption{Corner plot highlighting key initial characteristics and acceptable natal kicks for simulated system analogues to J0348+0432.}
    \label{Fig:J0348_CornerPlot}
\end{figure*}
\begin{figure*}
    \centering
    \includegraphics[width=0.75\textwidth]{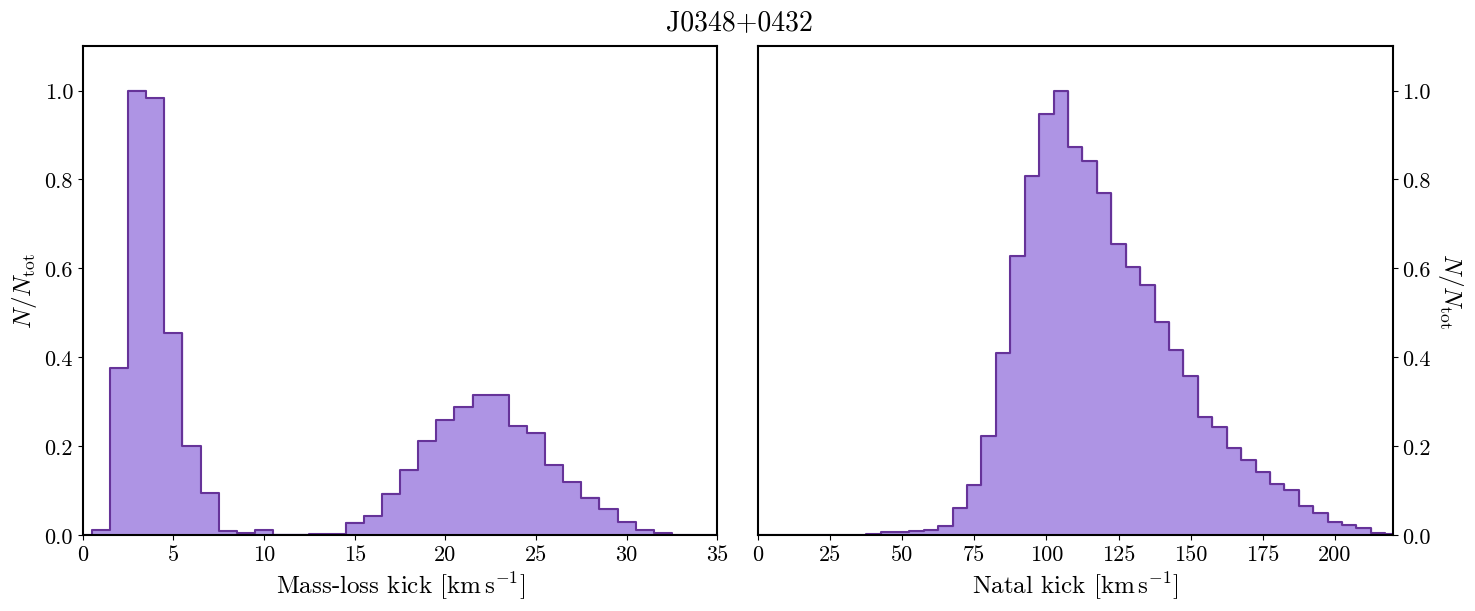}
    \caption{Distribution of mass-loss kicks ($\vmlk$) magnitude (left) and natal kick ($\vnk$) magnitude (right) for simulated system analogues to J0348+0432.}
    \label{Fig:J0348_KickHist}
\end{figure*}
\clearpage

\begin{figure*}
    \centering
    \includegraphics[width=0.75\textwidth]{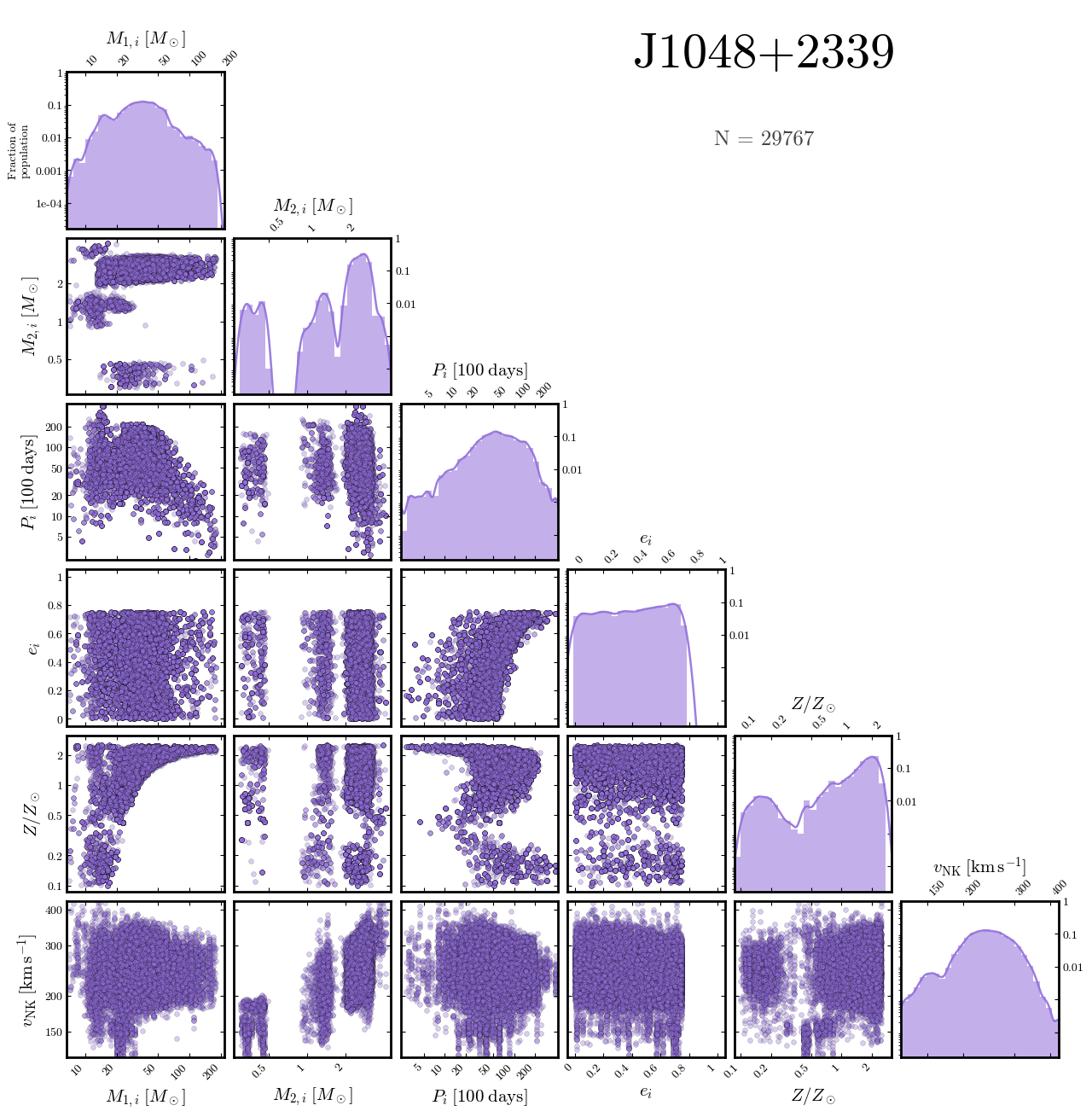}
    \caption{Corner plot highlighting key initial characteristics and acceptable natal kicks for simulated system analogues to J1048+2339.}
    \label{Fig:J1048_CornerPlot}
\end{figure*}
\begin{figure*}
    \centering
    \includegraphics[width=0.75\textwidth]{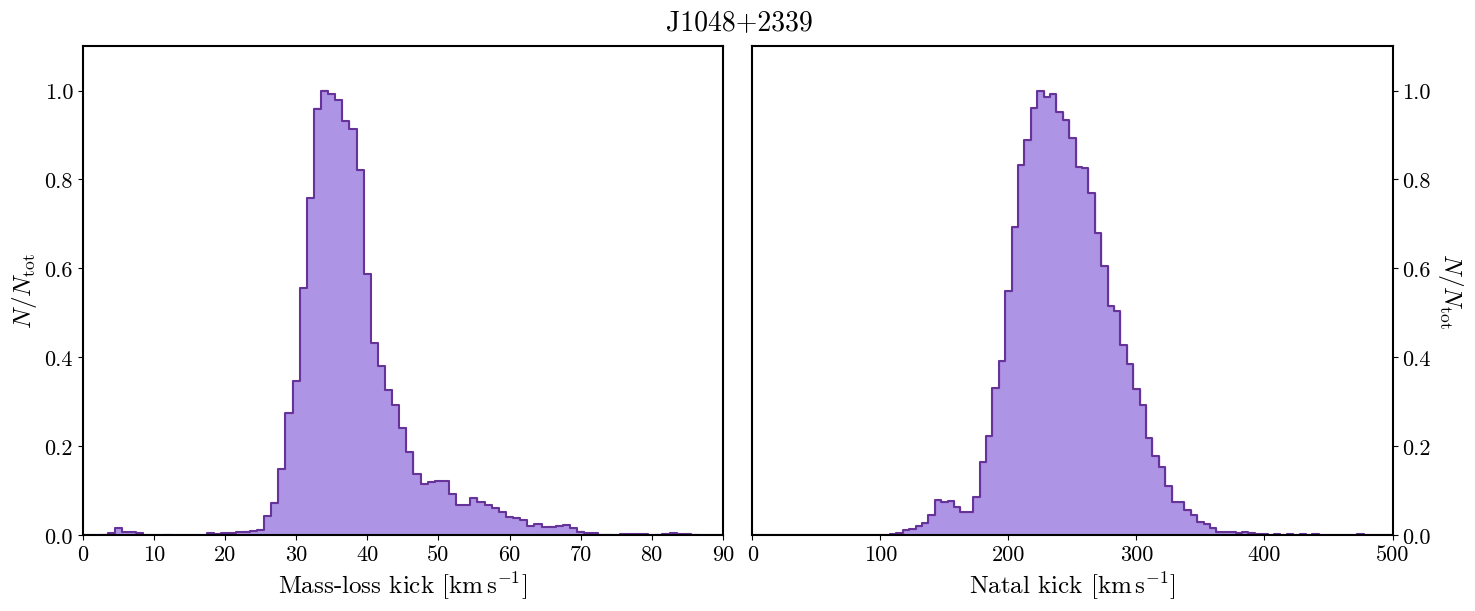}
    \caption{Distribution of mass-loss kicks ($\vmlk$) magnitude (left) and natal kick ($\vnk$) magnitude (right) for simulated system analogues to J1048+2339.}
    \label{Fig:J1048_KickHist}
\end{figure*}
\clearpage

\begin{figure*}
    \centering
    \includegraphics[width=0.75\textwidth]{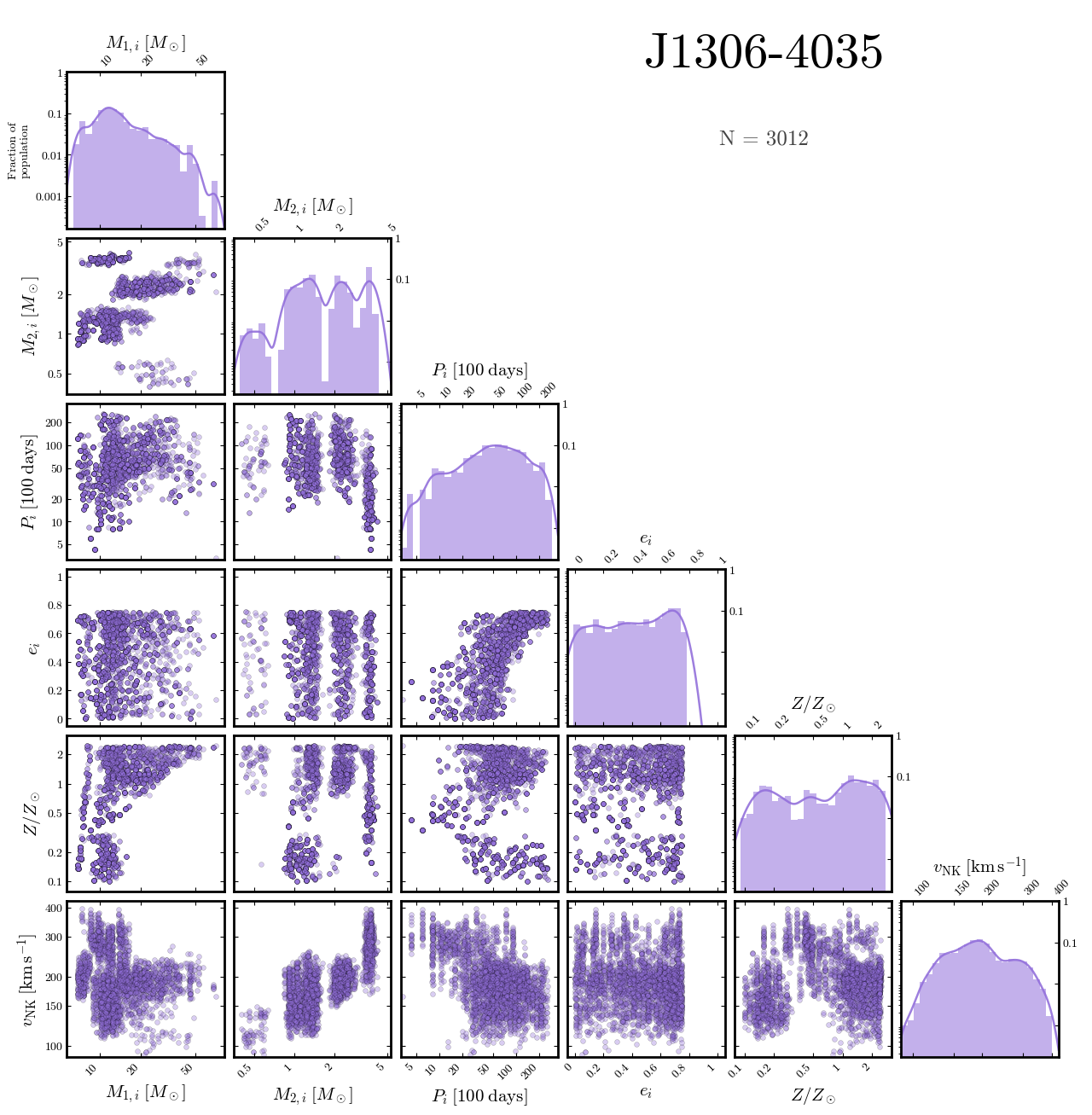}
    \caption{Corner plot highlighting key initial characteristics and acceptable natal kicks for simulated system analogues to J1306-4035.}
    \label{Fig:J1306_CornerPlot}
\end{figure*}
\begin{figure*}
    \centering
    \includegraphics[width=0.75\textwidth]{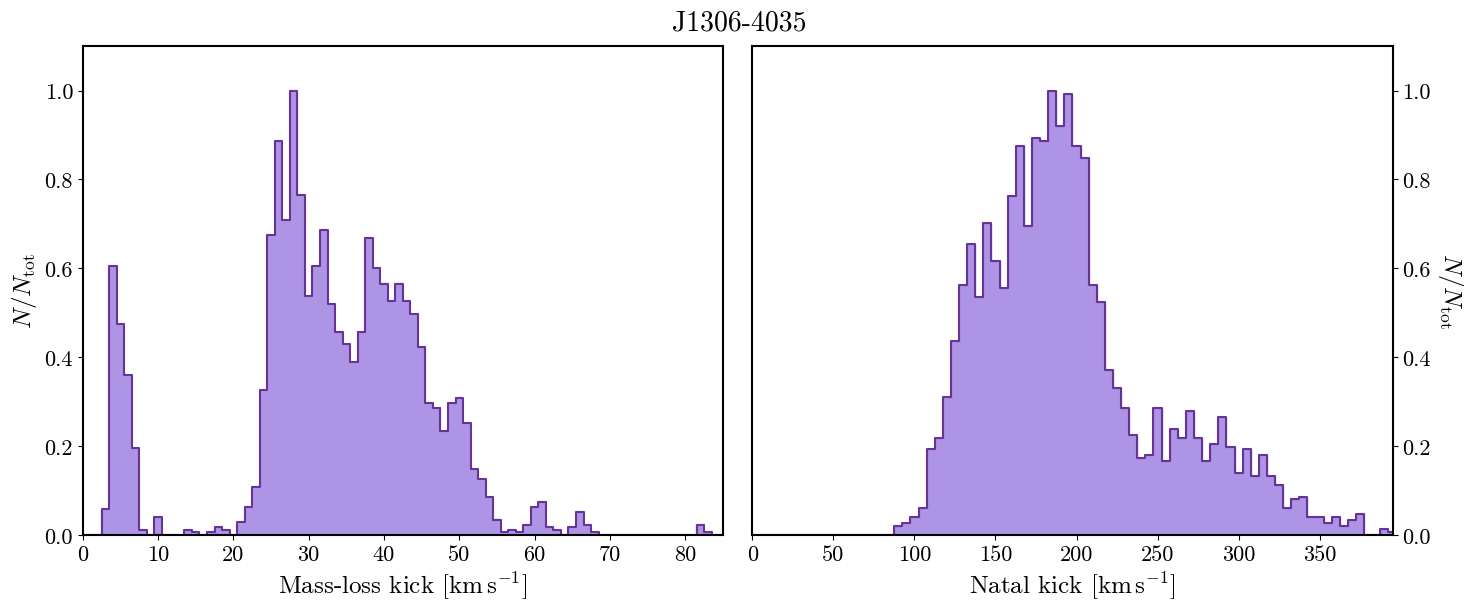}
    \caption{Distribution of mass-loss kicks ($\vmlk$) magnitude (left) and natal kick ($\vnk$) magnitude (right) for simulated system analogues to J1306-4035.}
    \label{Fig:J1306_KickHist}
\end{figure*}
\clearpage

\begin{figure*}
    \centering
    \includegraphics[width=0.75\textwidth]{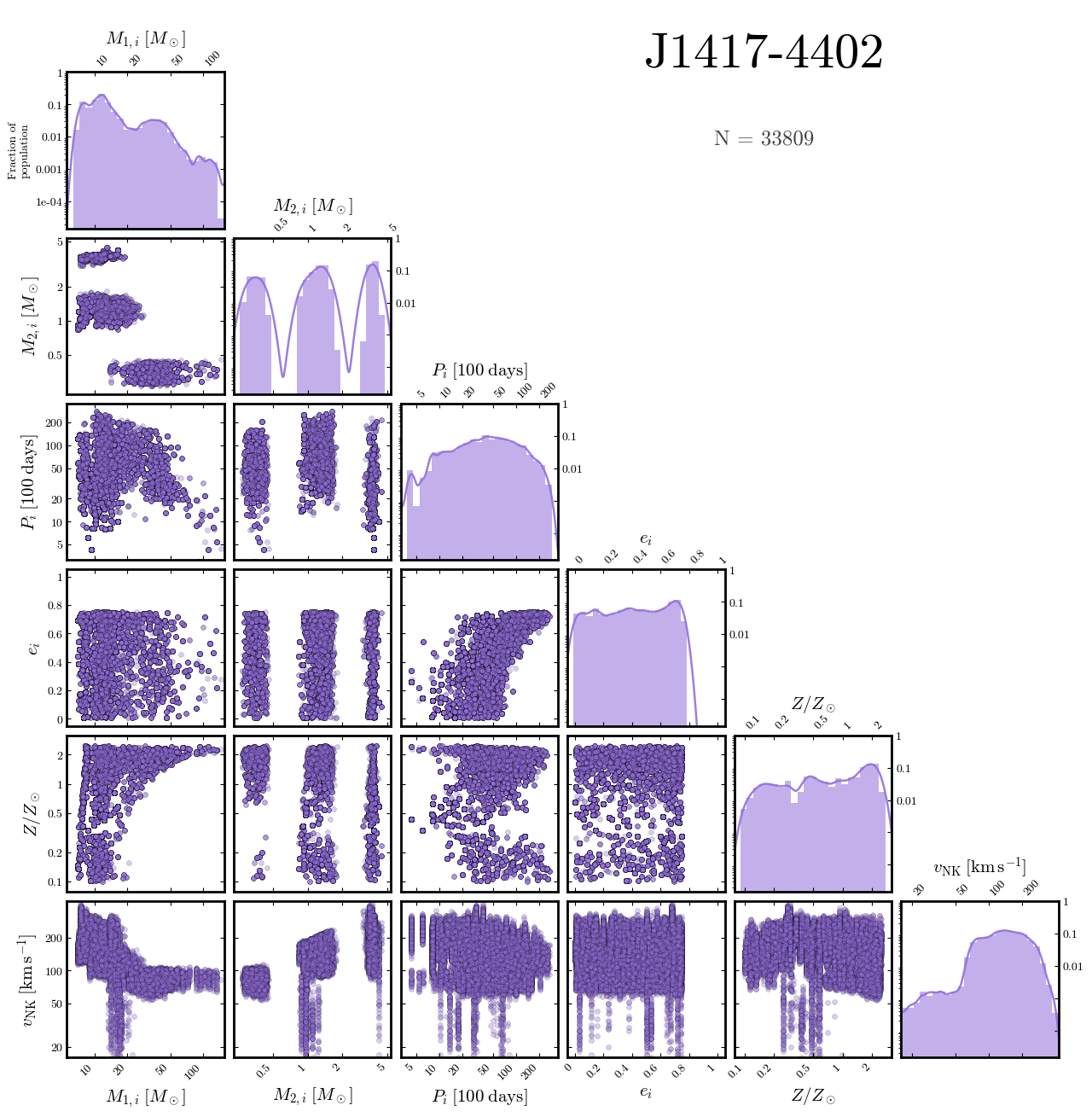}
    \caption{Corner plot highlighting key initial characteristics and acceptable natal kicks for simulated system analogues to J1417-4402.}
    \label{Fig:J1417_CornerPlot}
\end{figure*}
\begin{figure*}
    \centering
    \includegraphics[width=0.75\textwidth]{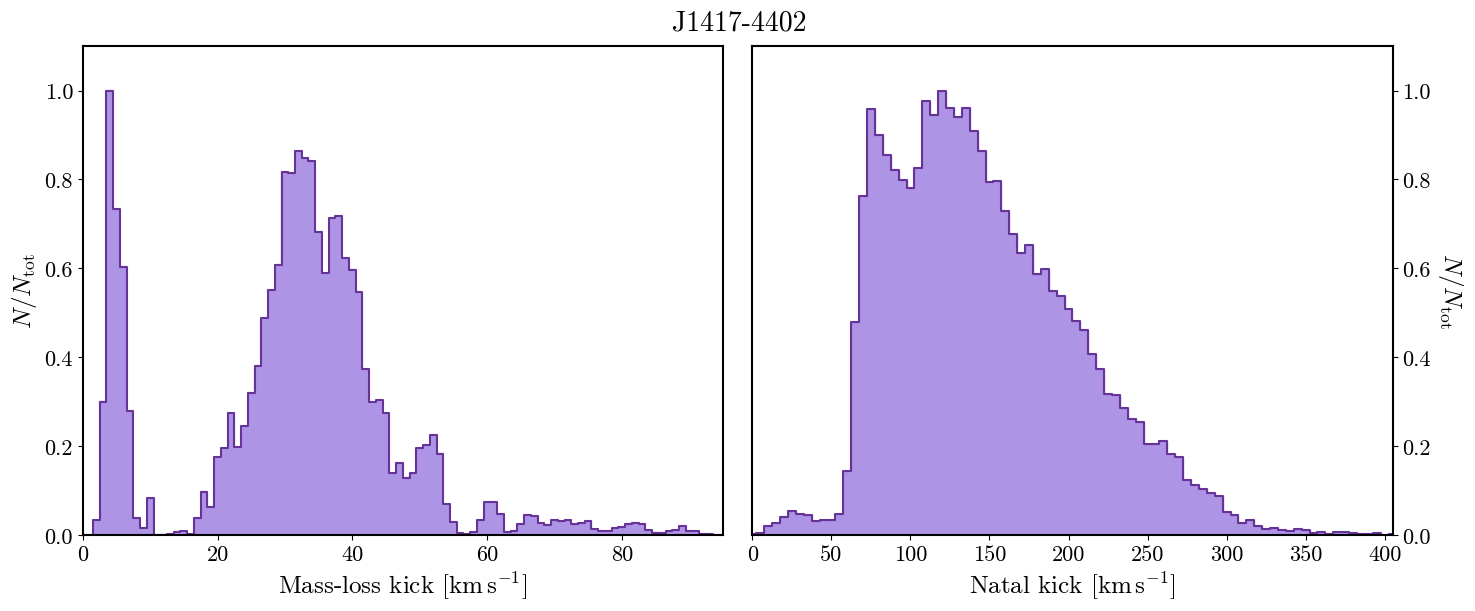}
    \caption{Distribution of mass-loss kicks ($\vmlk$) magnitude (left) and natal kick ($\vnk$) magnitude (right) for simulated system analogues to J1417-4402.}
    \label{Fig:J1417_KickHist}
\end{figure*}
\clearpage

\begin{figure*}
    \centering
    \includegraphics[width=0.75\textwidth]{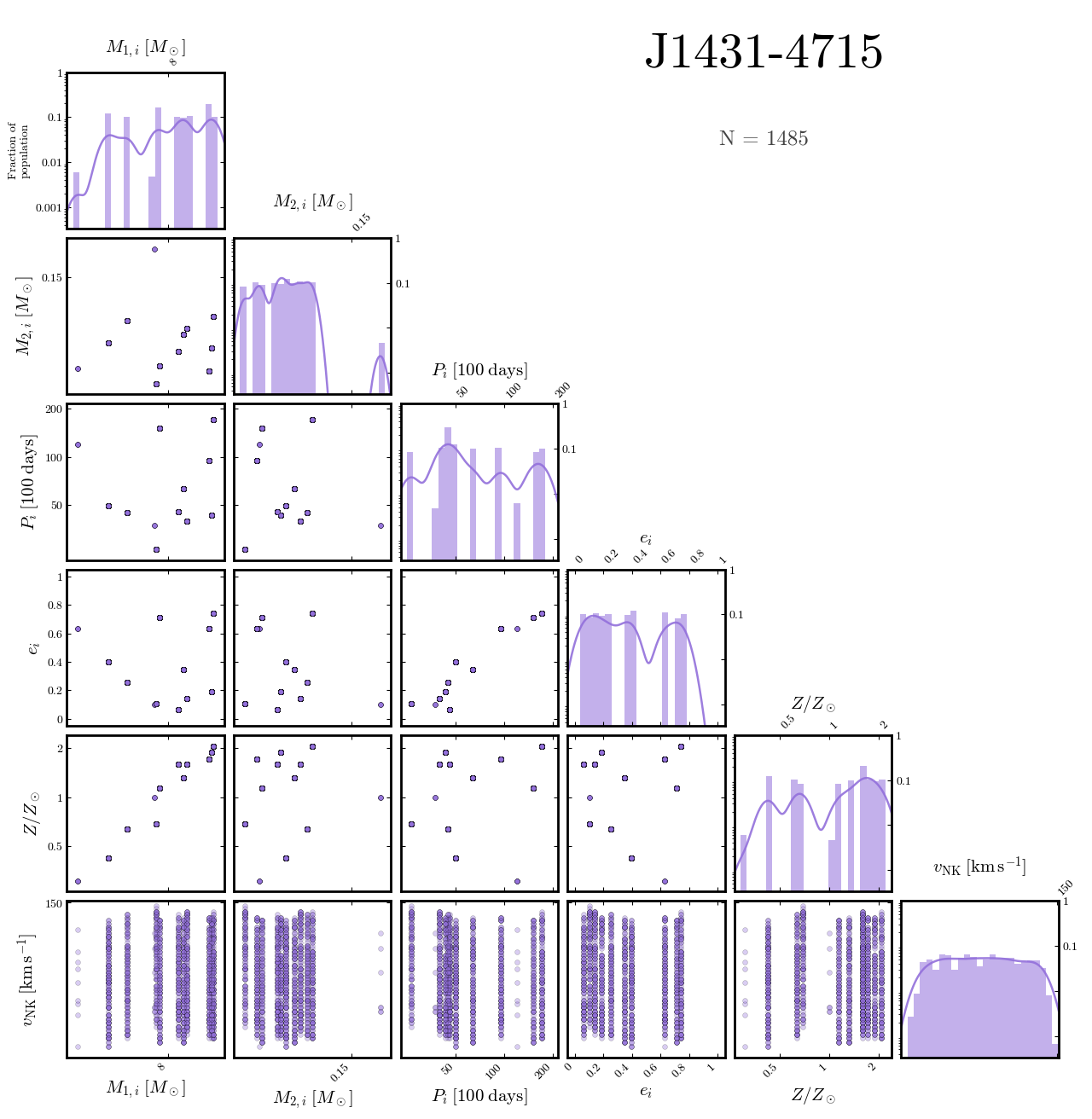}
    \caption{Corner plot highlighting key initial characteristics and acceptable natal kicks for simulated system analogues to J1431-4715.}
    \label{Fig:J1431_CornerPlot}
\end{figure*}
\begin{figure*}
    \centering
    \includegraphics[width=0.75\textwidth]{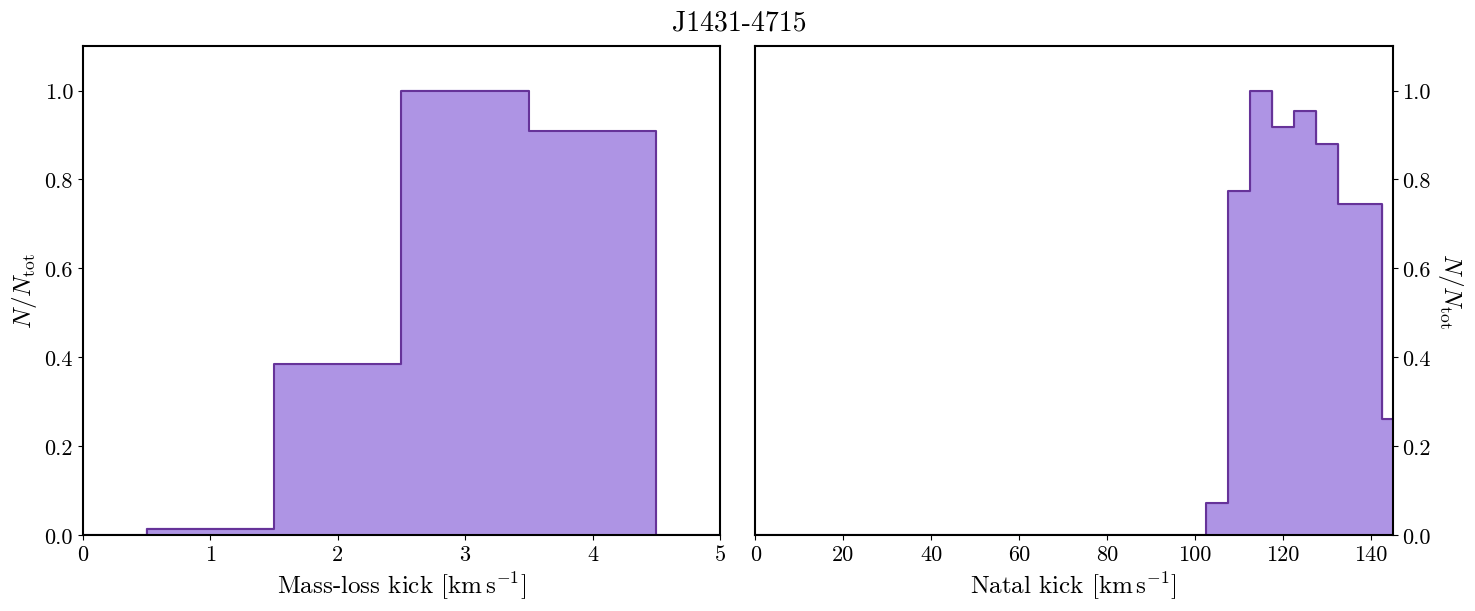}
    \caption{Distribution of mass-loss kicks ($\vmlk$) magnitude (left) and natal kick ($\vnk$) magnitude (right) for simulated system analogues to J1431-4715.}
    \label{Fig:J1431_KickHist}
\end{figure*}
\clearpage

\begin{figure*}
    \centering
    \includegraphics[width=0.75\textwidth]{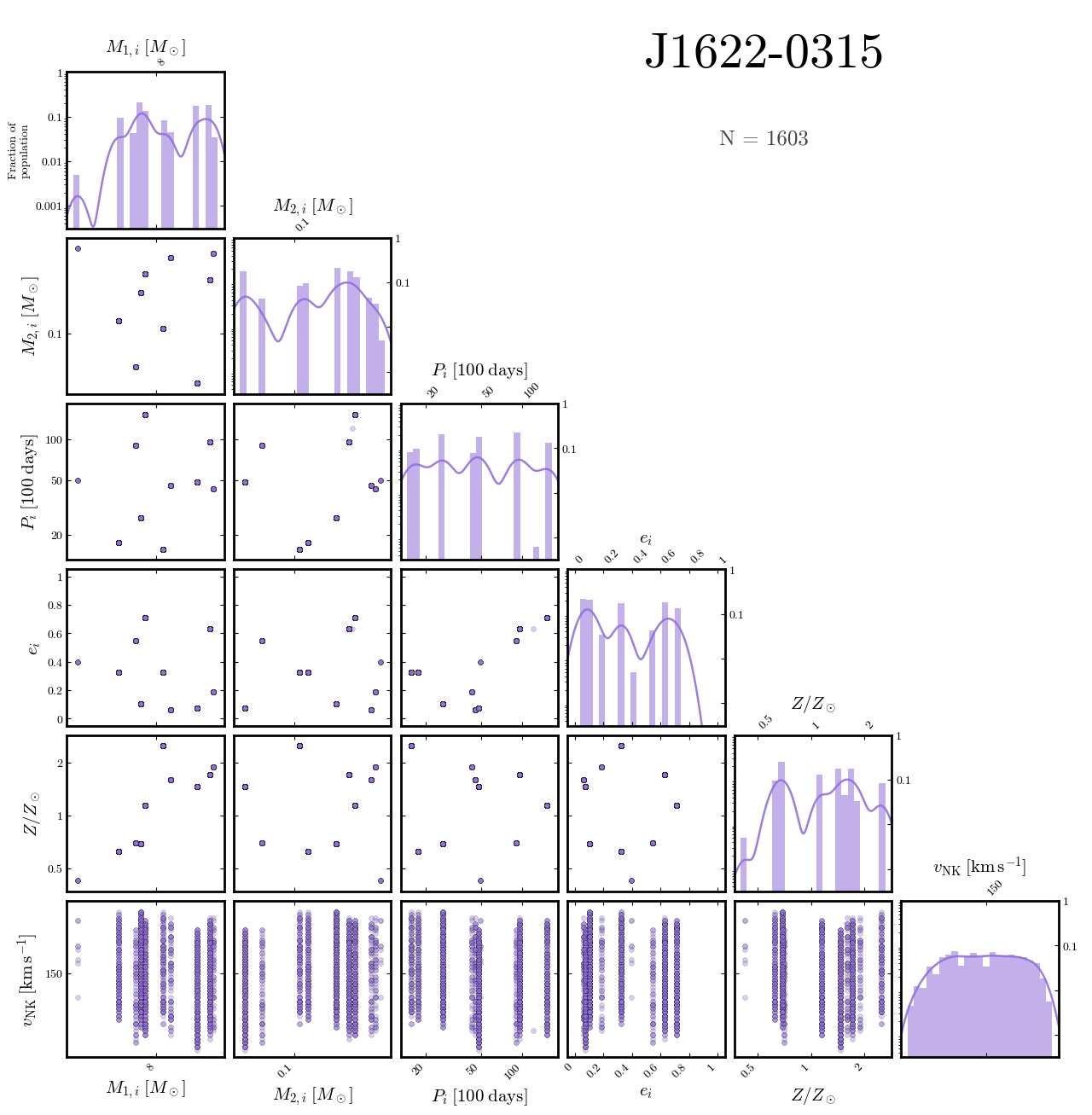}
    \caption{Corner plot highlighting key initial characteristics and acceptable natal kicks for simulated system analogues to J1622-0315.}
    \label{Fig:J1622_CornerPlot}
\end{figure*}
\begin{figure*}
    \centering
    \includegraphics[width=0.75\textwidth]{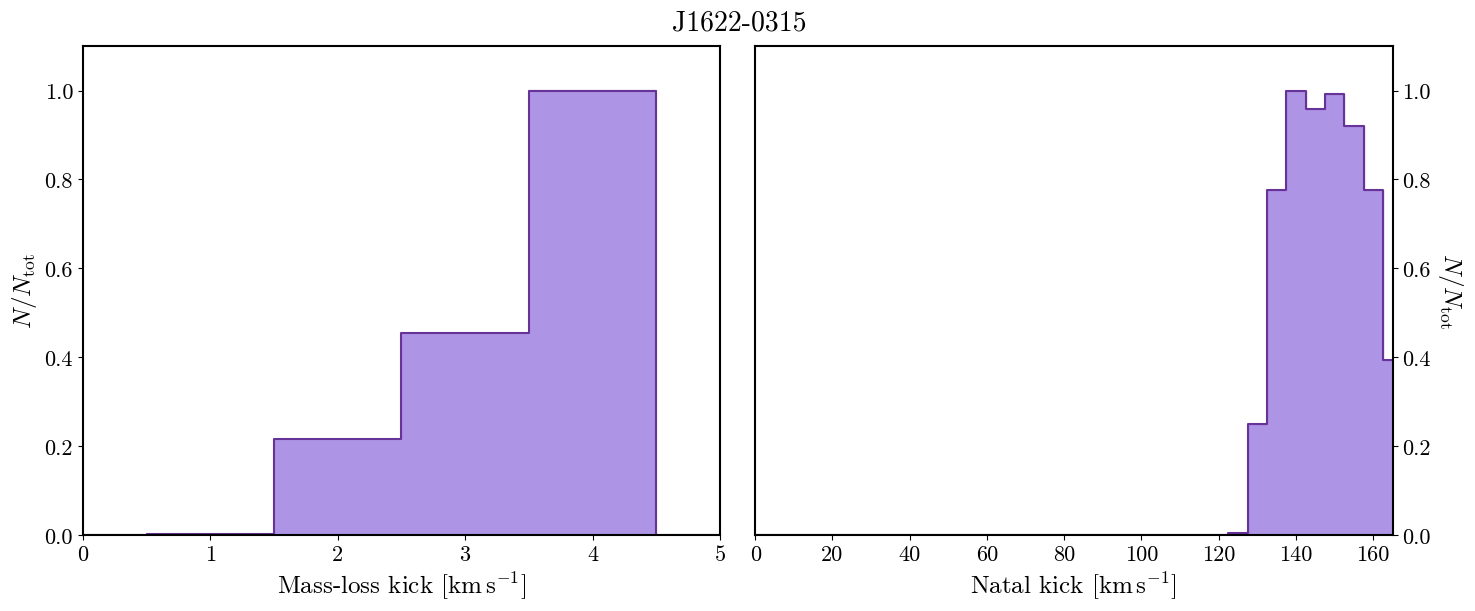}
    \caption{Distribution of mass-loss kicks ($\vmlk$) magnitude (left) and natal kick ($\vnk$) magnitude (right) for simulated system analogues to J1622-0315.}
    \label{Fig:J1622_KickHist}
\end{figure*}
\clearpage

\begin{figure*}
    \centering
    \includegraphics[width=0.75\textwidth]{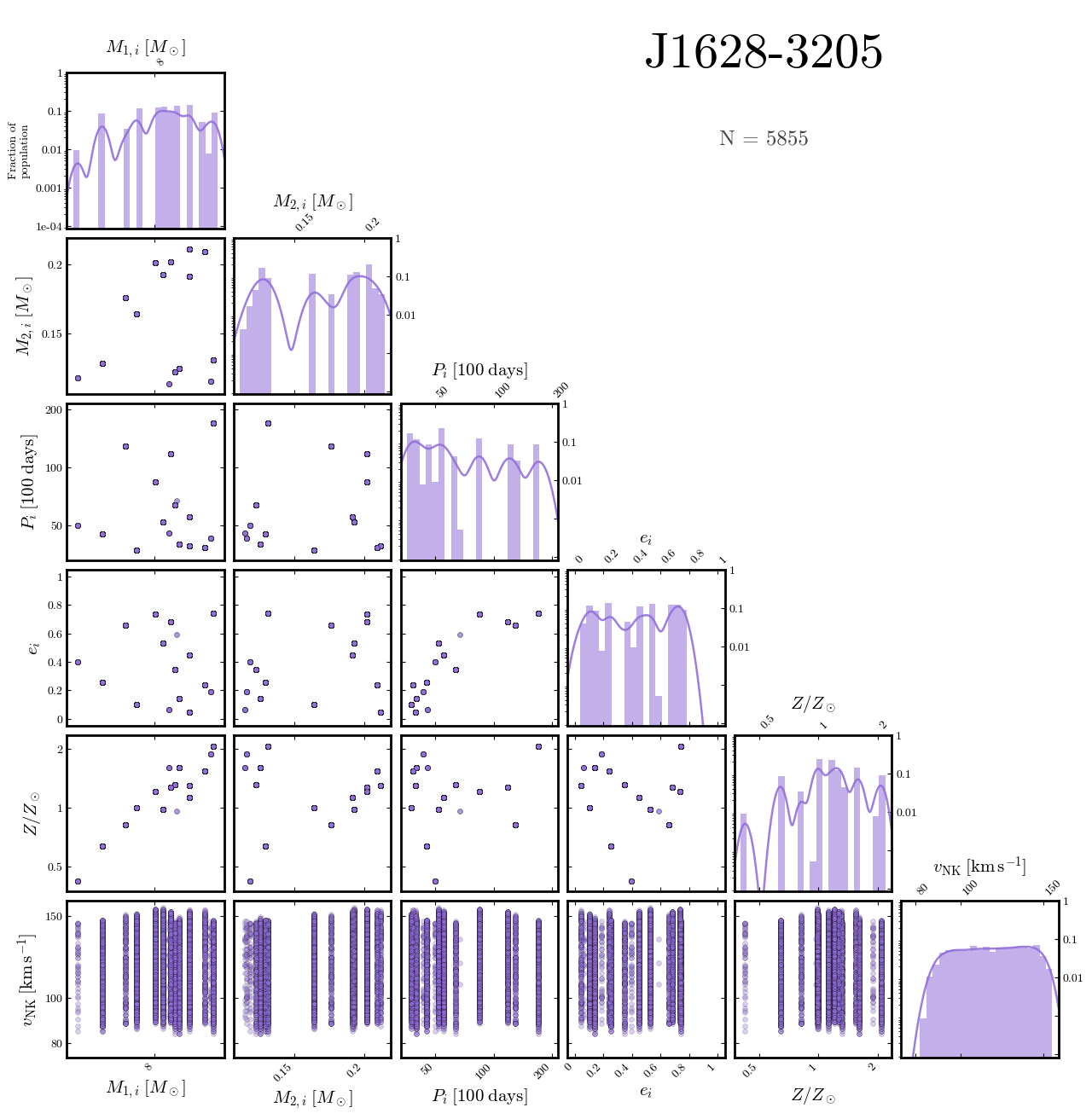}
    \caption{Corner plot highlighting key initial characteristics and acceptable natal kicks for simulated system analogues to J1628-3205.}
    \label{Fig:J1628_CornerPlot}
\end{figure*}
\begin{figure*}
    \centering
    \includegraphics[width=0.75\textwidth]{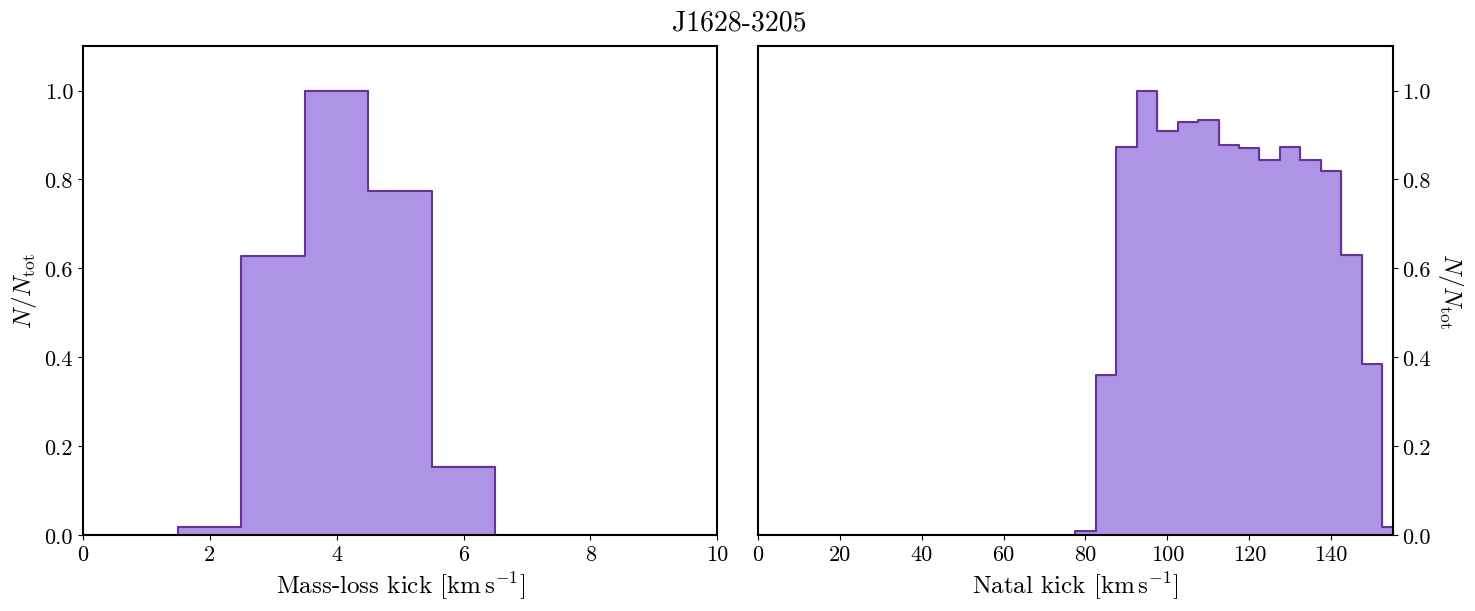}
    \caption{Distribution of mass-loss kicks ($\vmlk$) magnitude (left) and natal kick ($\vnk$) magnitude (right) for simulated system analogues to J1628-3205.}
    \label{Fig:J1628_KickHist}
\end{figure*}
\clearpage

\begin{figure*}
    \centering
    \includegraphics[width=0.75\textwidth]{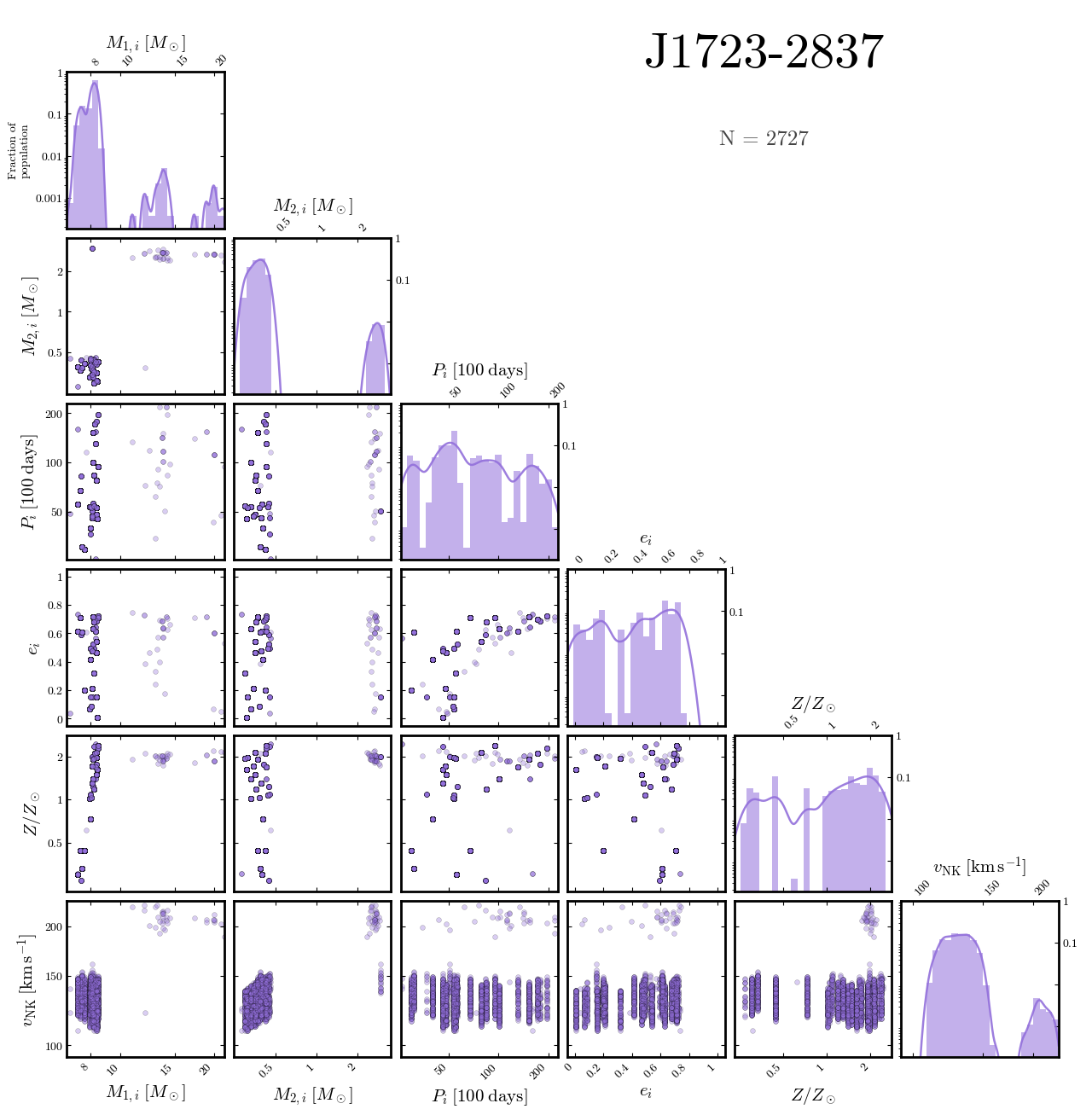}
    \caption{Corner plot highlighting key initial characteristics and acceptable natal kicks for simulated system analogues to J1723-2837.}
    \label{Fig:J1723_CornerPlot}
\end{figure*}
\begin{figure*}
    \centering
    \includegraphics[width=0.75\textwidth]{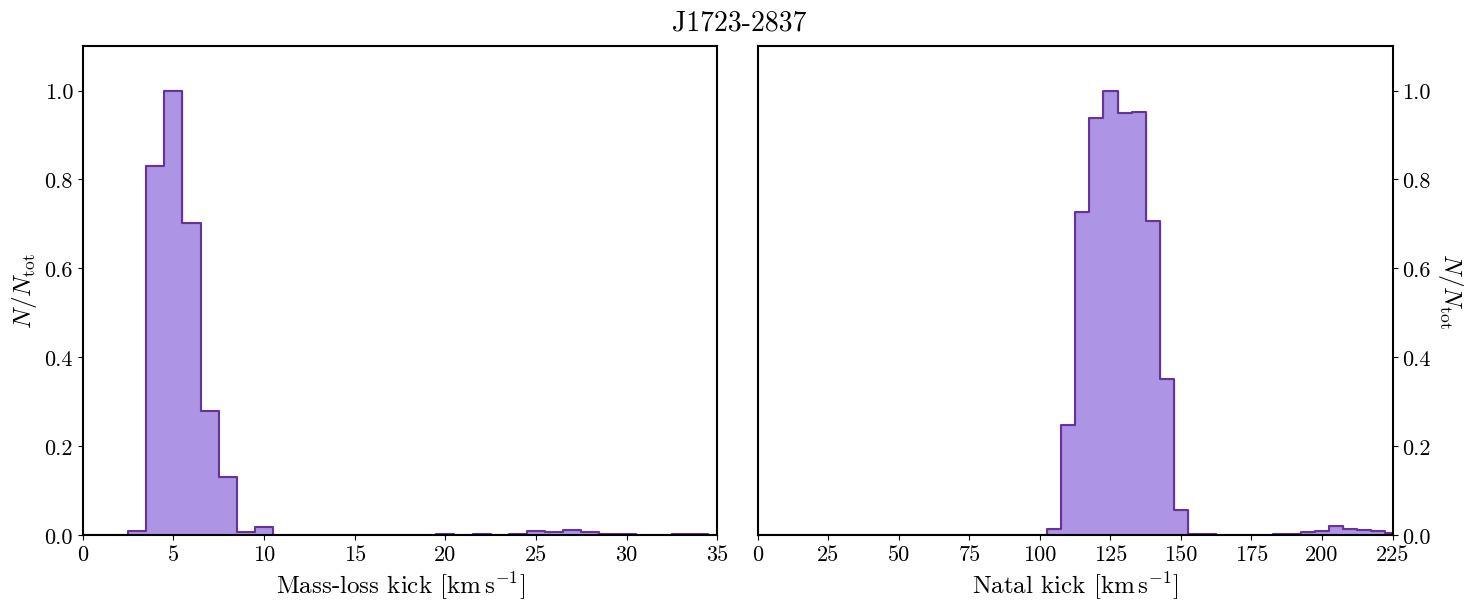}
    \caption{Distribution of mass-loss kicks ($\vmlk$) magnitude (left) and natal kick ($\vnk$) magnitude (right) for simulated system analogues to J1723-2837.}
    \label{Fig:J1723_KickHist}
\end{figure*}
\clearpage

\begin{figure*}
    \centering
    \includegraphics[width=0.75\textwidth]{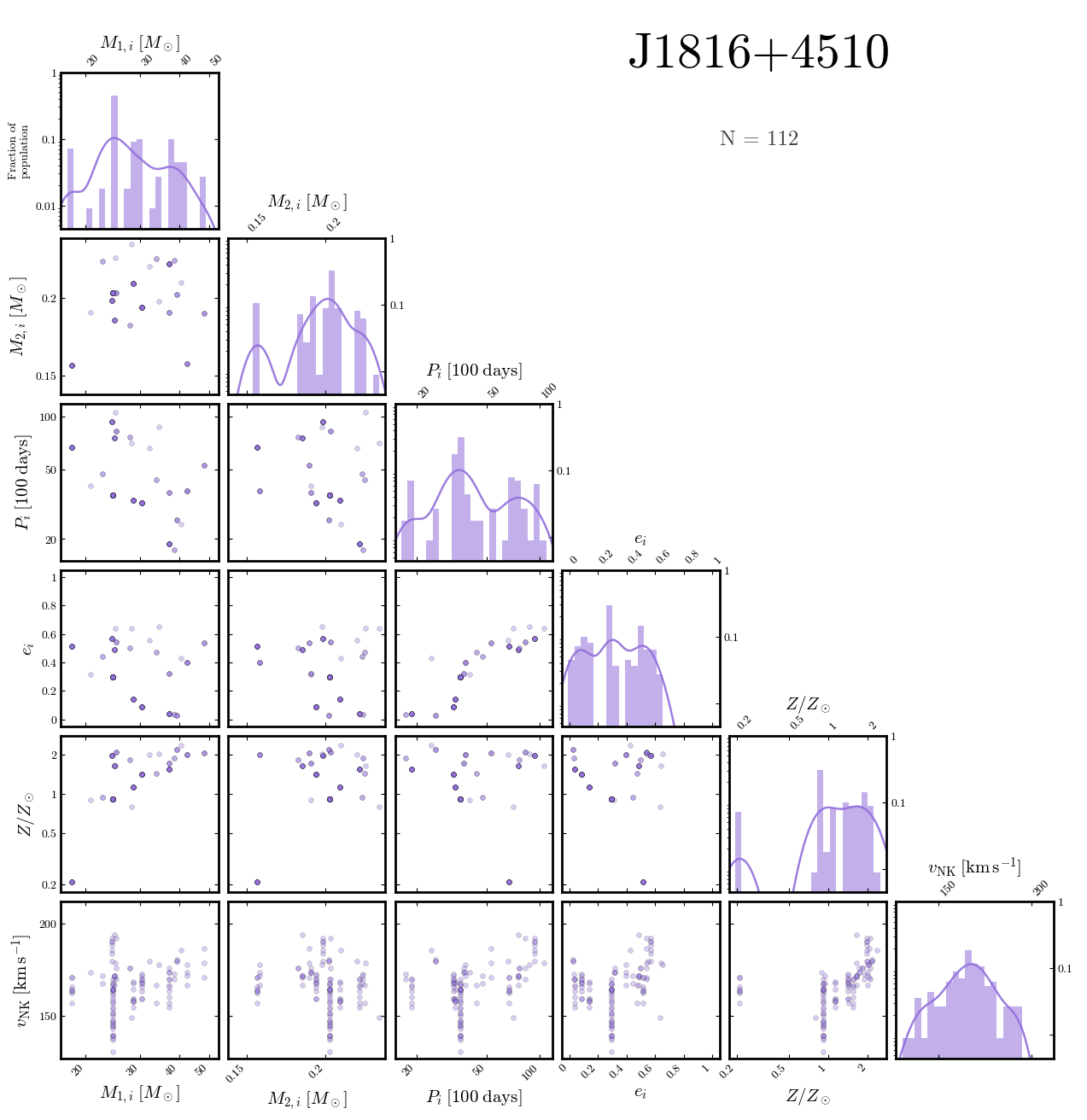}
    \caption{Corner plot highlighting key initial characteristics and acceptable natal kicks for simulated system analogues to J1816+4510.}
    \label{Fig:J1816_CornerPlot}
\end{figure*}
\begin{figure*}
    \centering
    \includegraphics[width=0.75\textwidth]{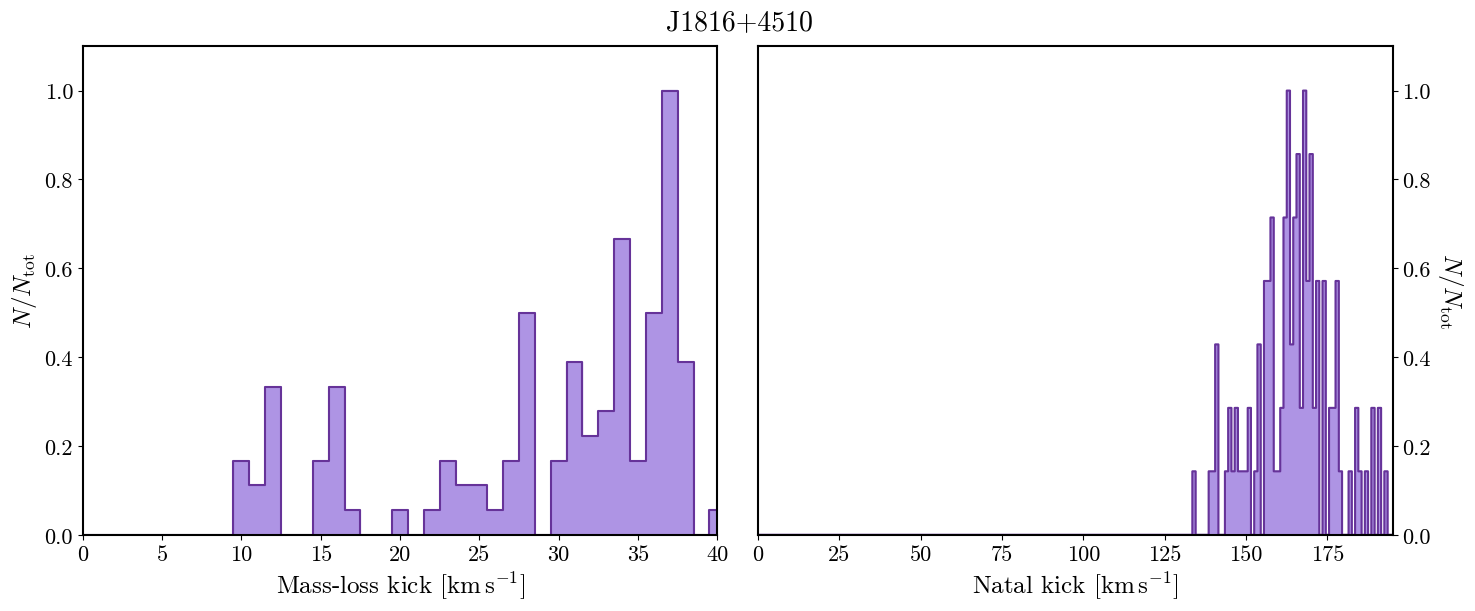}
    \caption{Distribution of mass-loss kicks ($\vmlk$) magnitude (left) and natal kick ($\vnk$) magnitude (right) for simulated system analogues to J1816+4510.}
    \label{Fig:J1816_KickHist}
\end{figure*}
\clearpage

\begin{figure*}
    \centering
    \includegraphics[width=0.75\textwidth]{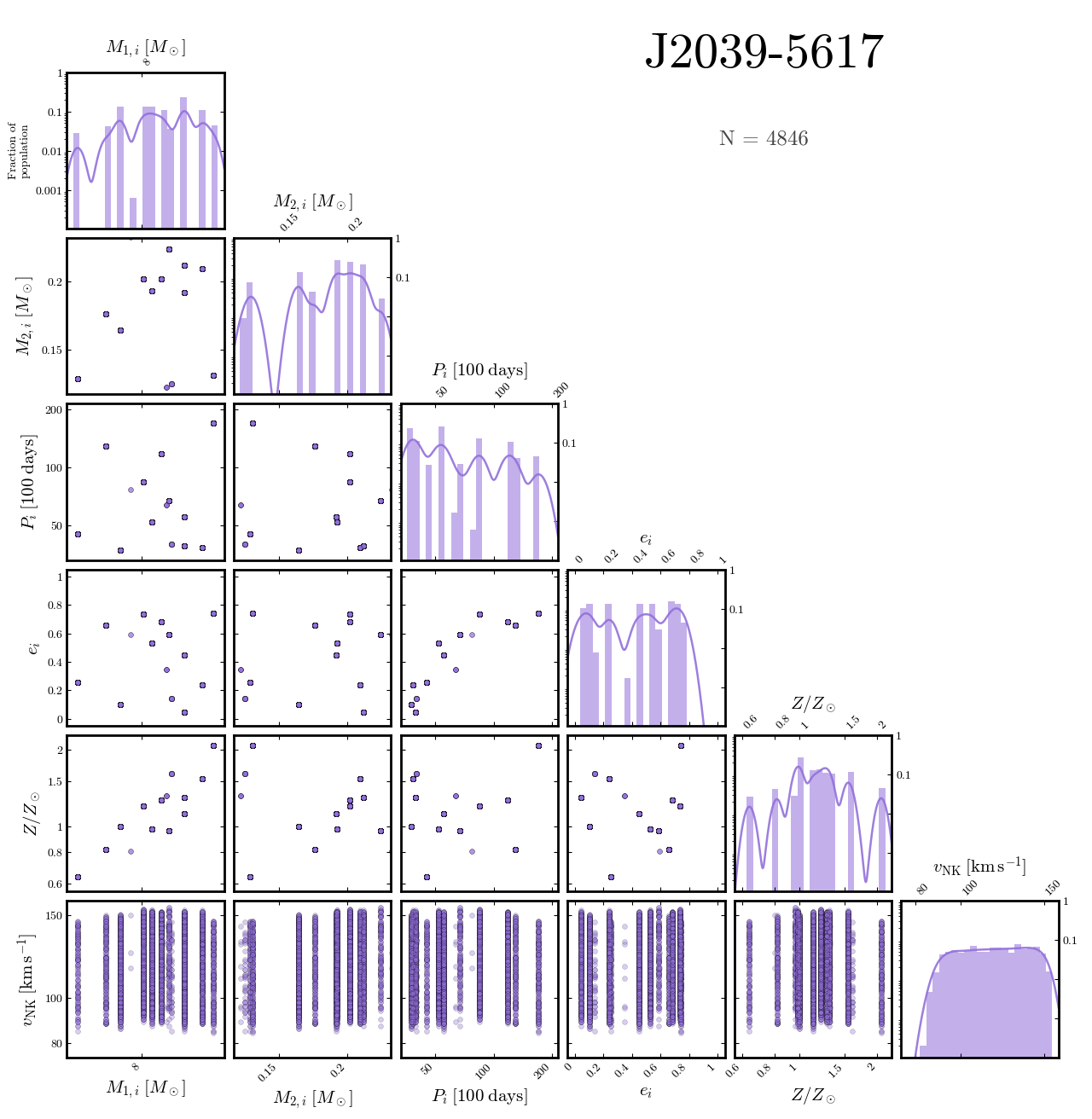}
    \caption{Corner plot highlighting key initial characteristics and acceptable natal kicks for simulated system analogues to J2039-5617.}
    \label{Fig:J2039_CornerPlot}
\end{figure*}
\begin{figure*}
    \centering
    \includegraphics[width=0.75\textwidth]{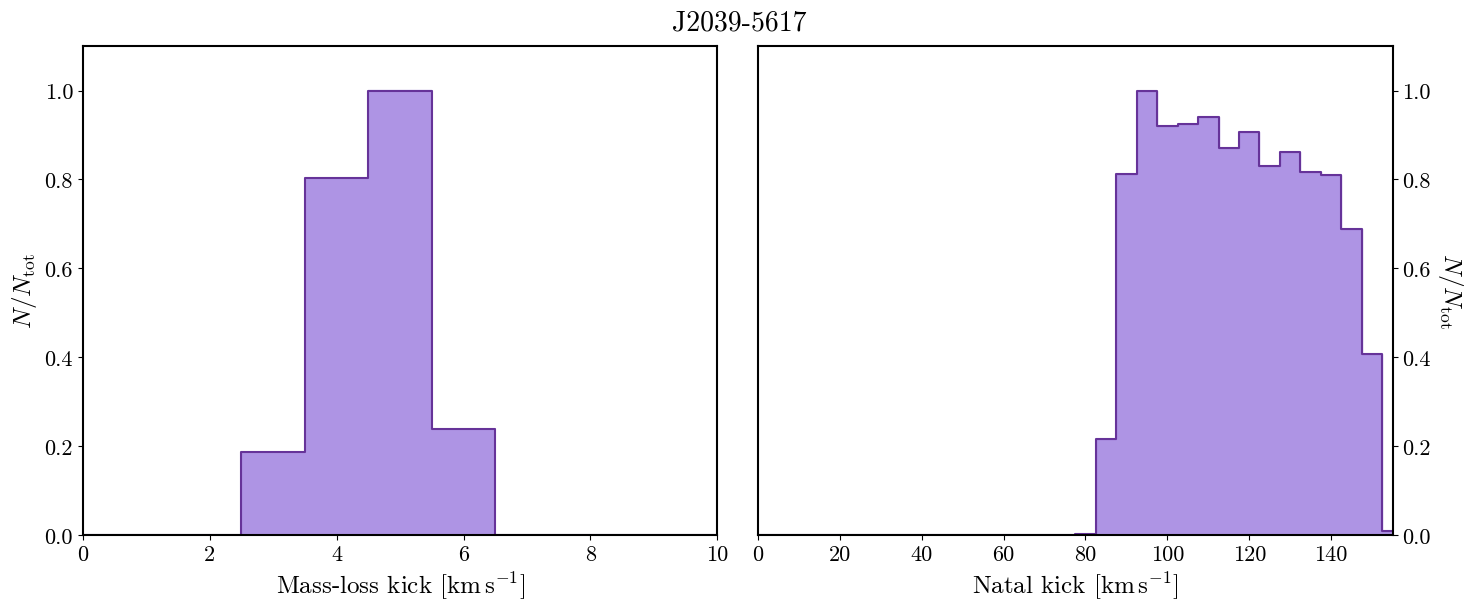}
    \caption{Distribution of mass-loss kicks ($\vmlk$) magnitude (left) and natal kick ($\vnk$) magnitude (right) for simulated system analogues to J2039-5617.}
    \label{Fig:J2039_KickHist}
\end{figure*}
\clearpage

\begin{figure*}
    \centering
    \includegraphics[width=0.75\textwidth]{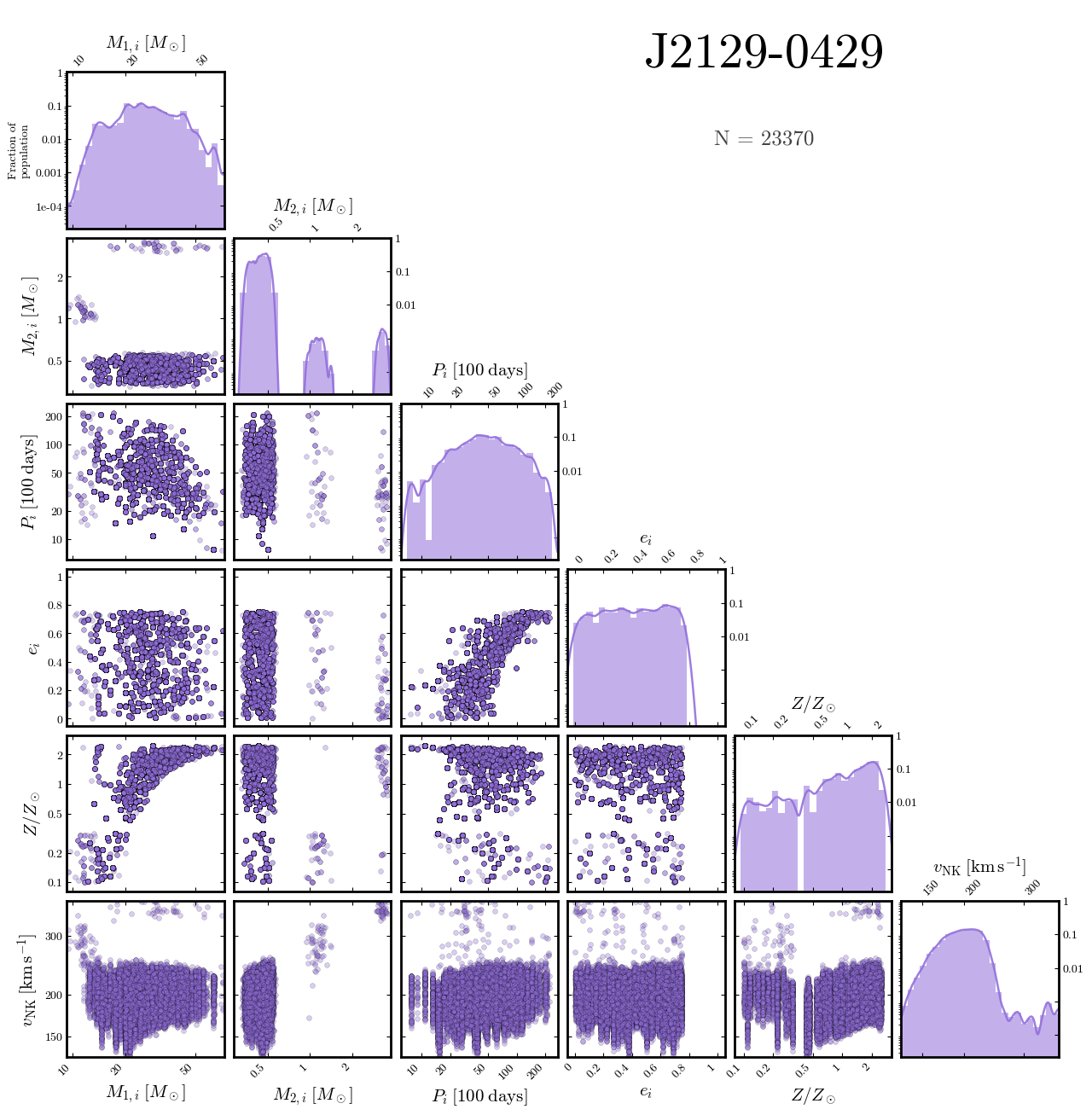}
    \caption{Corner plot highlighting key initial characteristics and acceptable natal kicks for simulated system analogues to J2129-0429.}
    \label{Fig:J2129_CornerPlot}
\end{figure*}
\begin{figure*}
    \centering
    \includegraphics[width=0.75\textwidth]{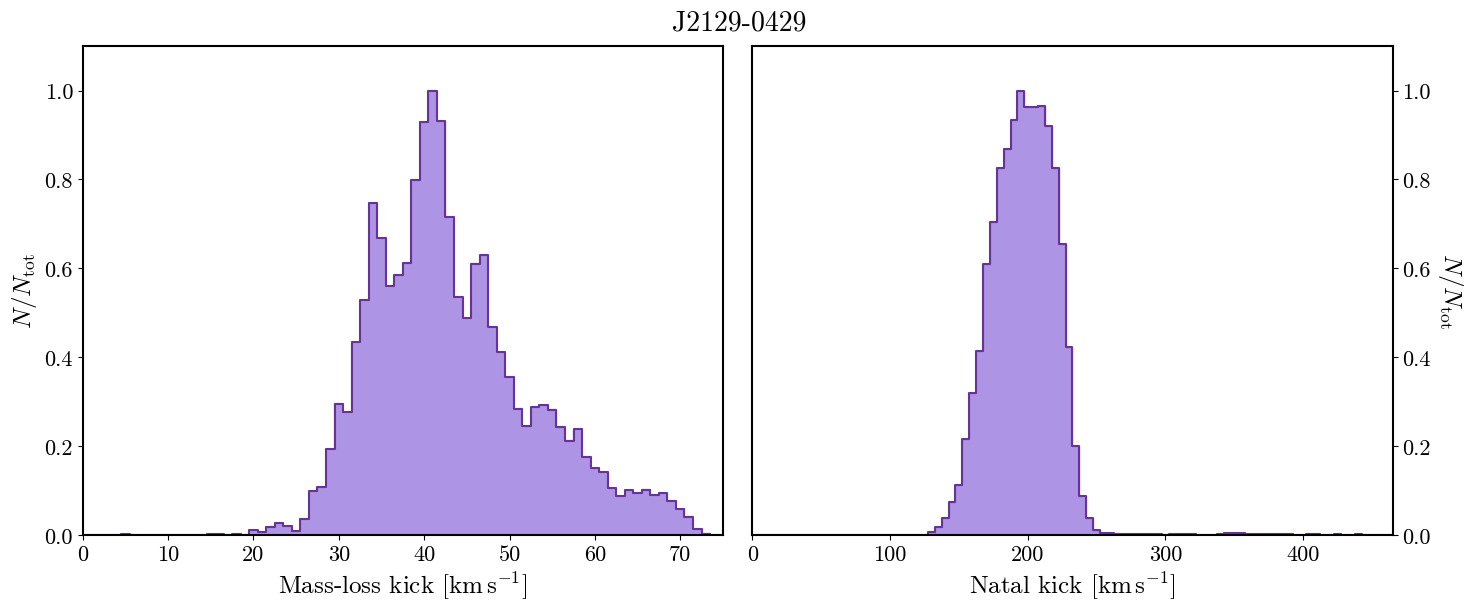}
    \caption{Distribution of mass-loss kicks ($\vmlk$) magnitude (left) and natal kick ($\vnk$) magnitude (right) for simulated system analogues to J2129-0429.}
    \label{Fig:J2129_KickHist}
\end{figure*}
\clearpage

\begin{figure*}
    \centering
    \includegraphics[width=0.75\textwidth]{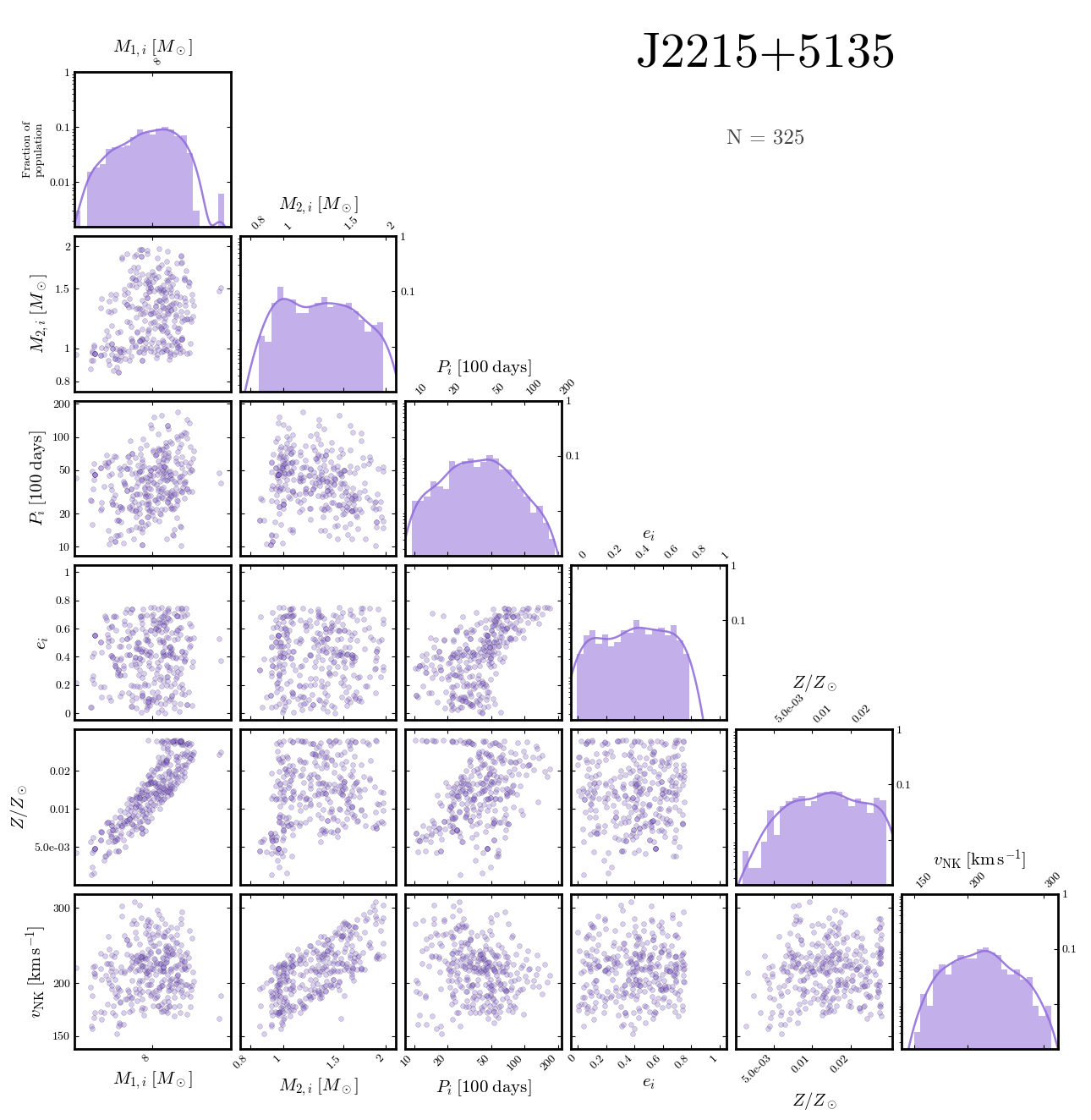}
    \caption{Corner plot highlighting key initial characteristics and acceptable natal kicks for simulated system analogues to J2215+5135.}
    \label{Fig:J2215_CornerPlot}
\end{figure*}
\begin{figure*}
    \centering
    \includegraphics[width=0.75\textwidth]{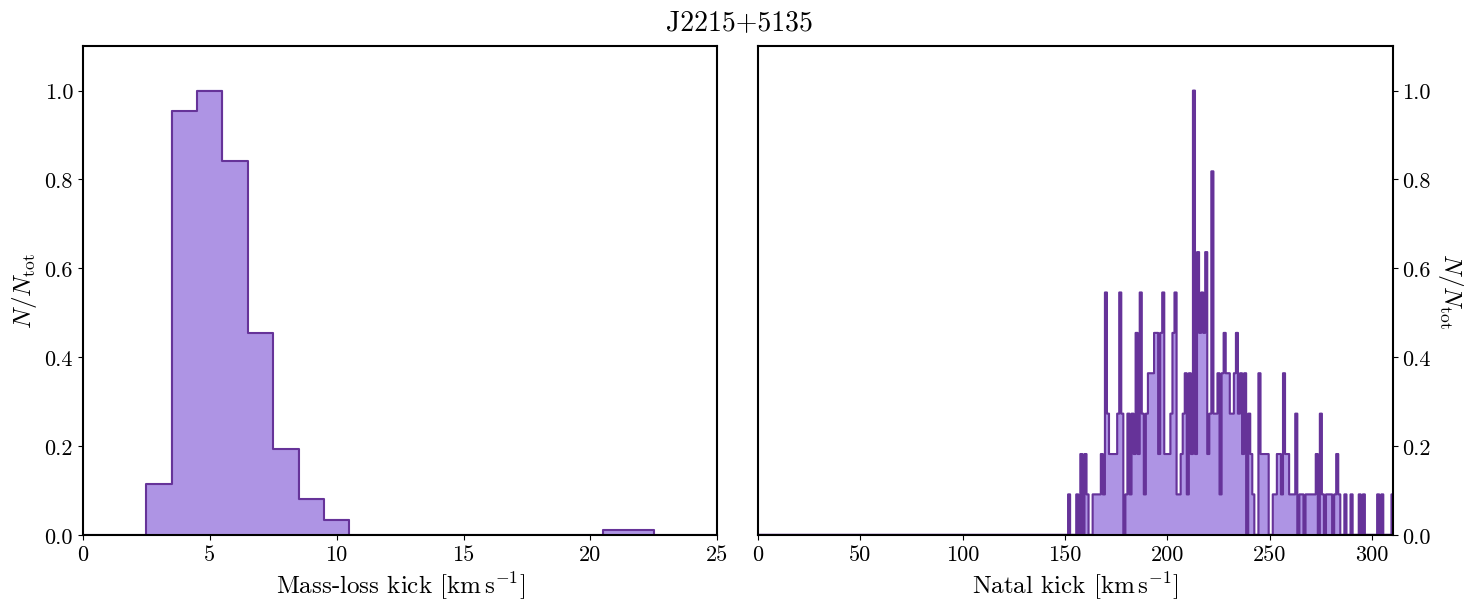}
    \caption{Distribution of mass-loss kicks ($\vmlk$) magnitude (left) and natal kick ($\vnk$) magnitude (right) for simulated system analogues to J2215+5135.}
    \label{Fig:J2215_KickHist}
\end{figure*}
\clearpage

\begin{figure*}
    \centering
    \includegraphics[width=0.75\textwidth]{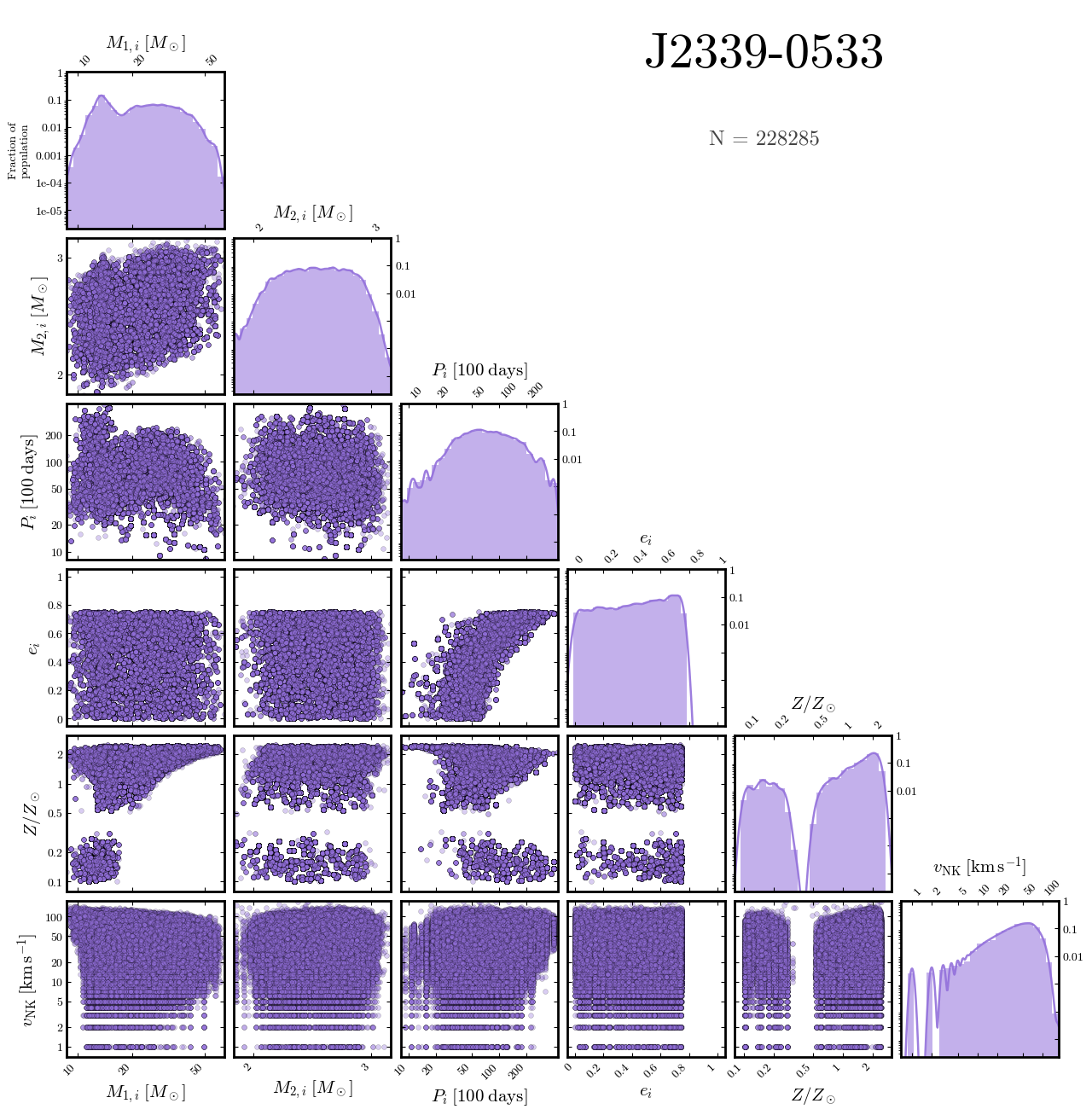}
    \caption{Corner plot highlighting key initial characteristics and acceptable natal kicks for simulated system analogues to J2339-0533.}
    \label{Fig:J2339_CornerPlot}
\end{figure*}
\begin{figure*}
    \centering
    \includegraphics[width=0.75\textwidth]{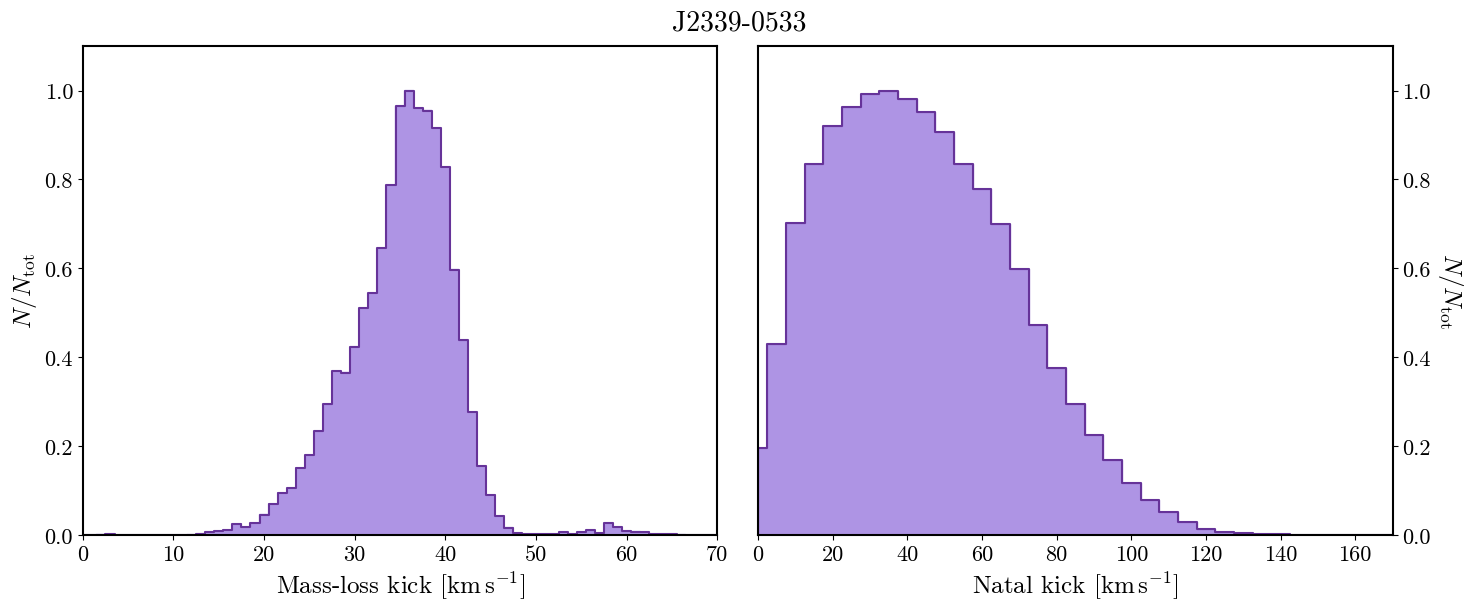}
    \caption{Distribution of mass-loss kicks ($\vmlk$) magnitude (left) and natal kick ($\vnk$) magnitude (right) for simulated system analogues to J2339-0533.}
    \label{Fig:J2339_KickHist}
\end{figure*}
\clearpage

\begin{figure*}
    \centering
    \includegraphics[width=0.75\textwidth]{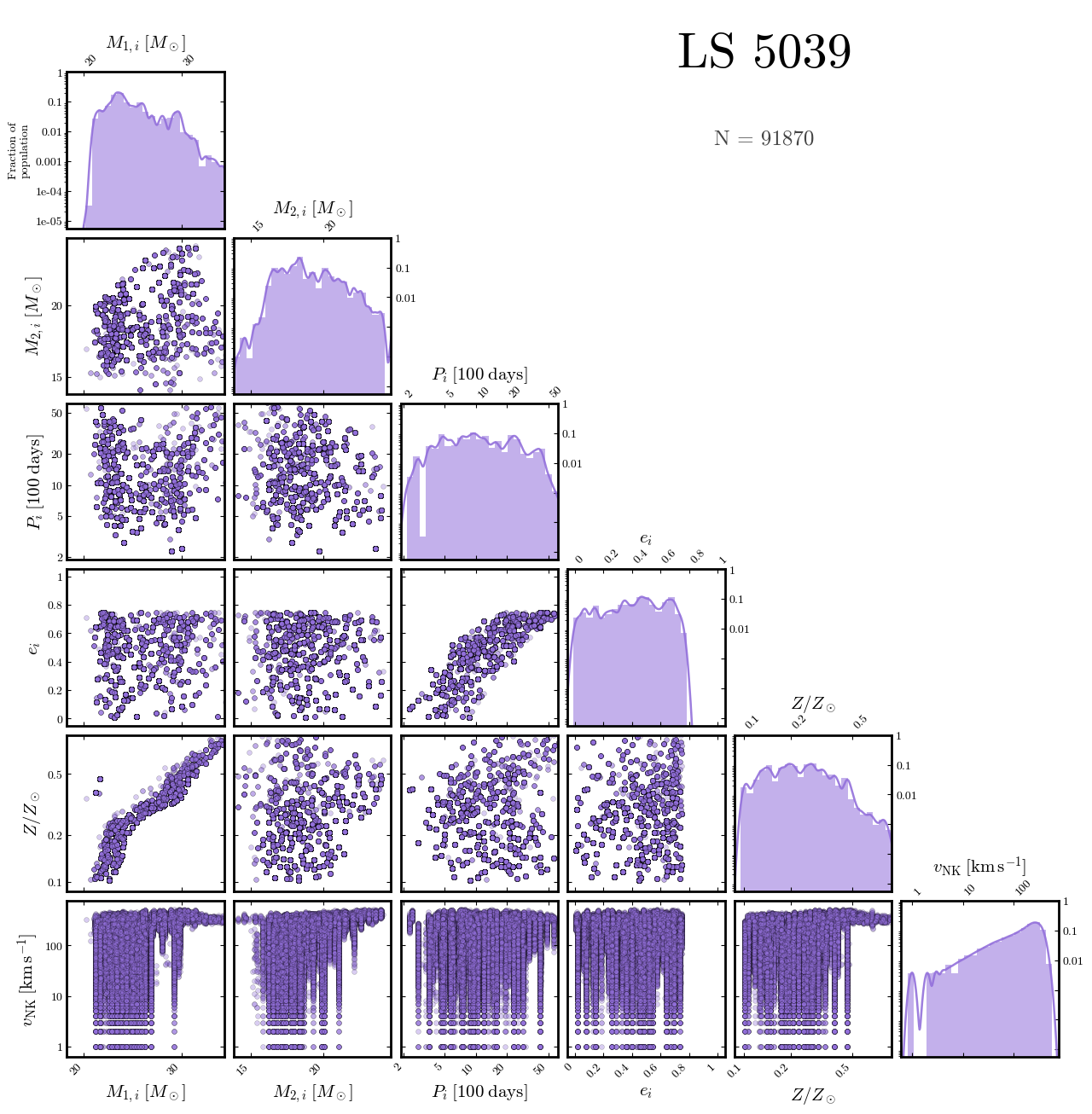}
    \caption{Corner plot highlighting key initial characteristics and acceptable natal kicks for simulated system analogues to LS 5039.}
    \label{Fig:LS5039_CornerPlot}
\end{figure*}
\begin{figure*}
    \centering
    \includegraphics[width=0.75\textwidth]{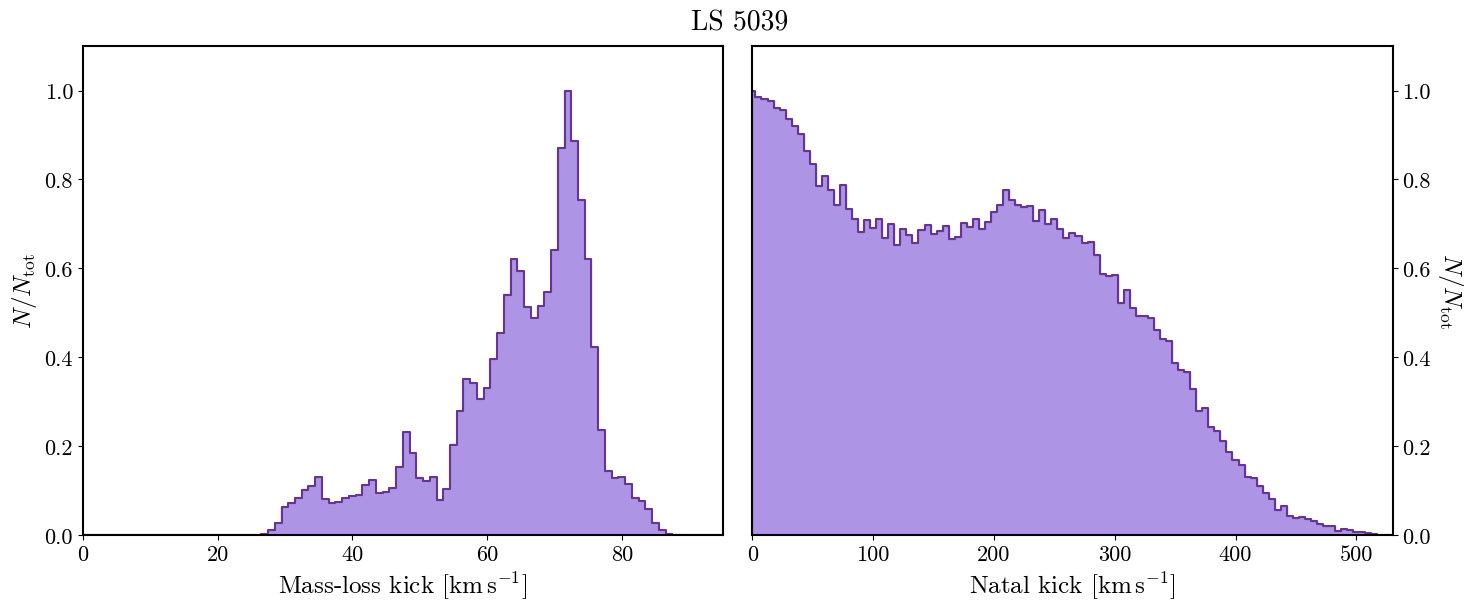}
    \caption{Distribution of mass-loss kicks ($\vmlk$) magnitude (left) and natal kick ($\vnk$) magnitude (right) for simulated system analogues to LS 5039.}
    \label{Fig:LS5039_KickHist}
\end{figure*}
\clearpage

\begin{figure*}
    \centering
    \includegraphics[width=0.75\textwidth]{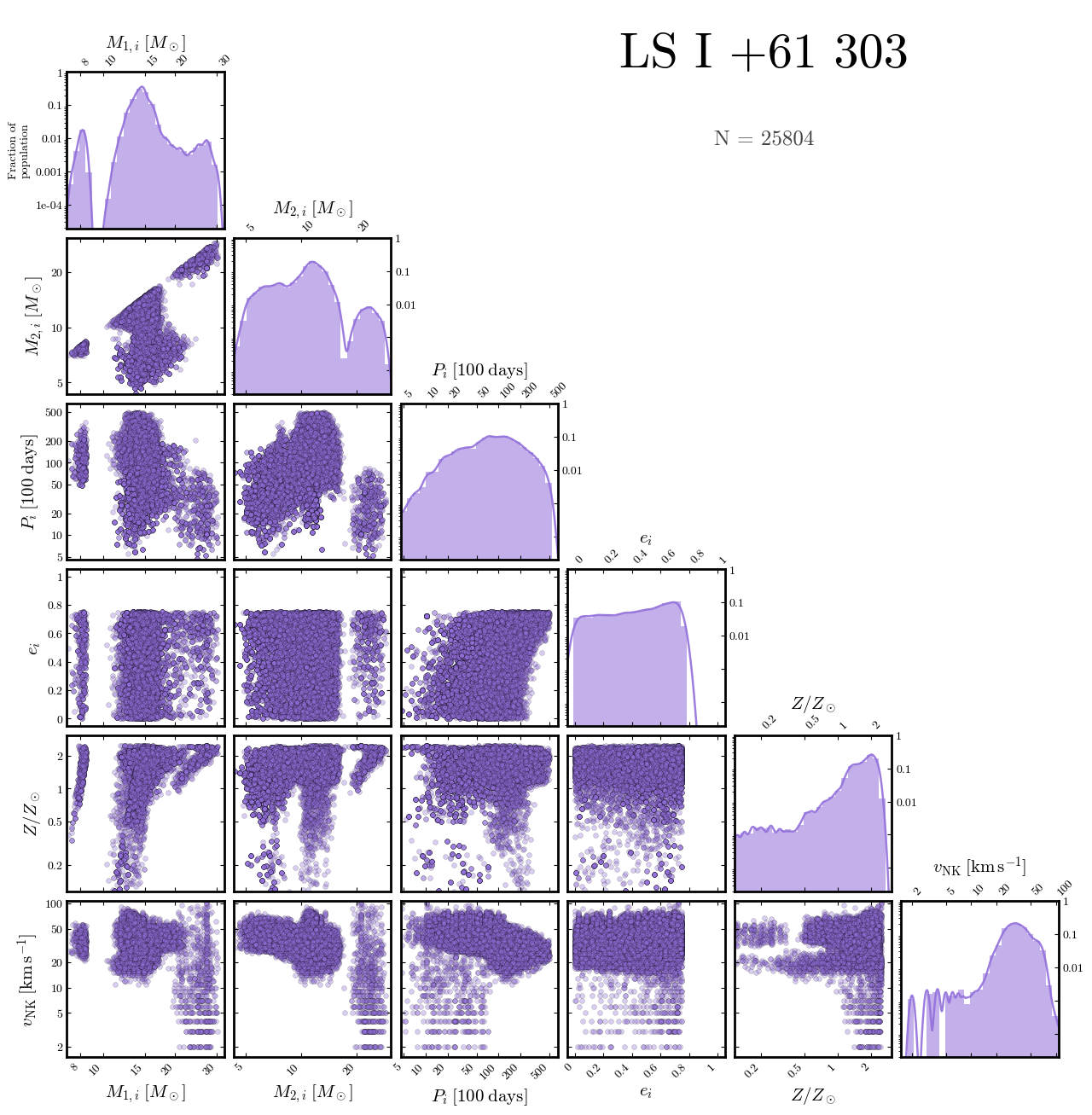}
    \caption{Corner plot highlighting key initial characteristics and acceptable natal kicks for simulated system analogues to LS I +61 303.}
    \label{Fig:LSI61_CornerPlot}
\end{figure*}
\begin{figure*}
    \centering
    \includegraphics[width=0.75\textwidth]{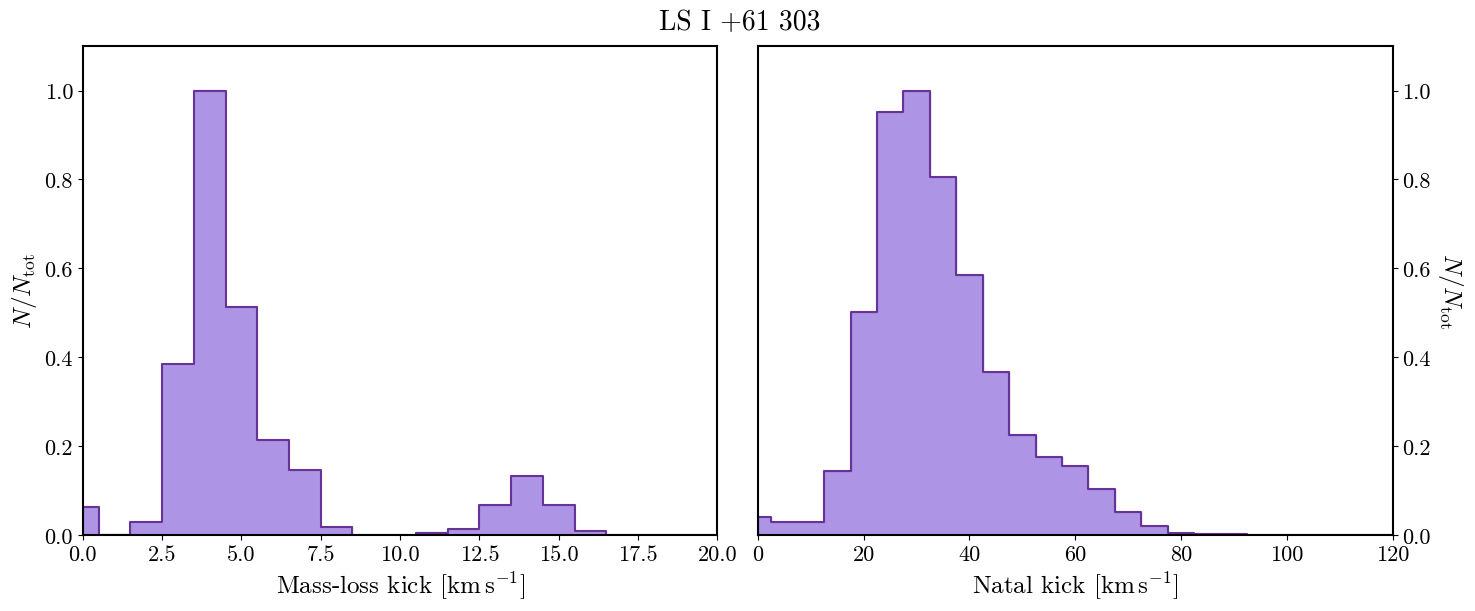}
    \caption{Distribution of mass-loss kicks ($\vmlk$) magnitude (left) and natal kick ($\vnk$) magnitude (right) for simulated system analogues to LS I +61 303.}
    \label{Fig:LSI61_KickHist}
\end{figure*}
\clearpage

\begin{figure*}
    \centering
    \includegraphics[width=0.75\textwidth]{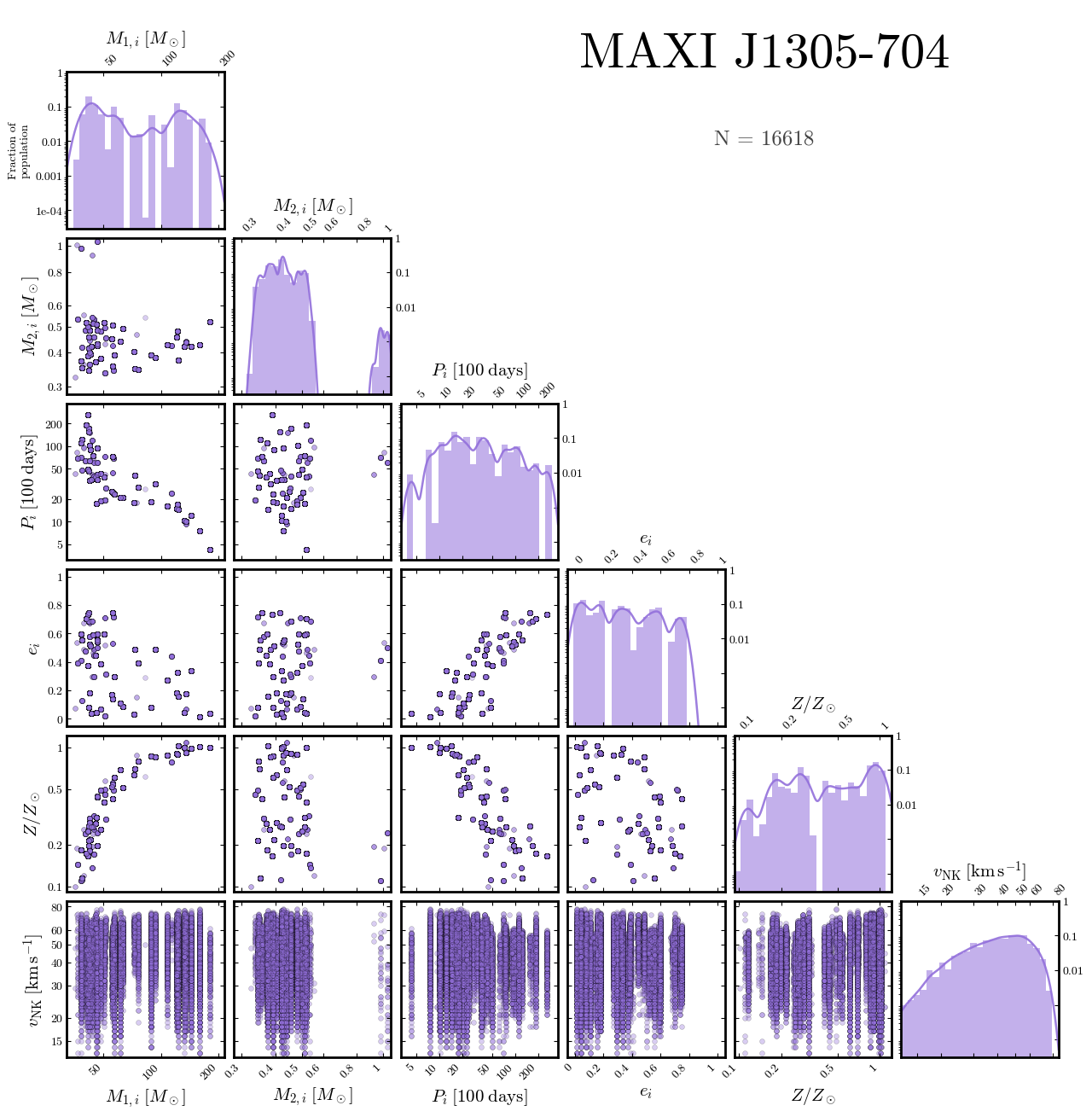}
    \caption{Corner plot highlighting key initial characteristics and acceptable natal kicks for simulated system analogues to MAXI J1305-704.}
    \label{Fig:J1305_CornerPlot}
\end{figure*}
\begin{figure*}
    \centering
    \includegraphics[width=0.75\textwidth]{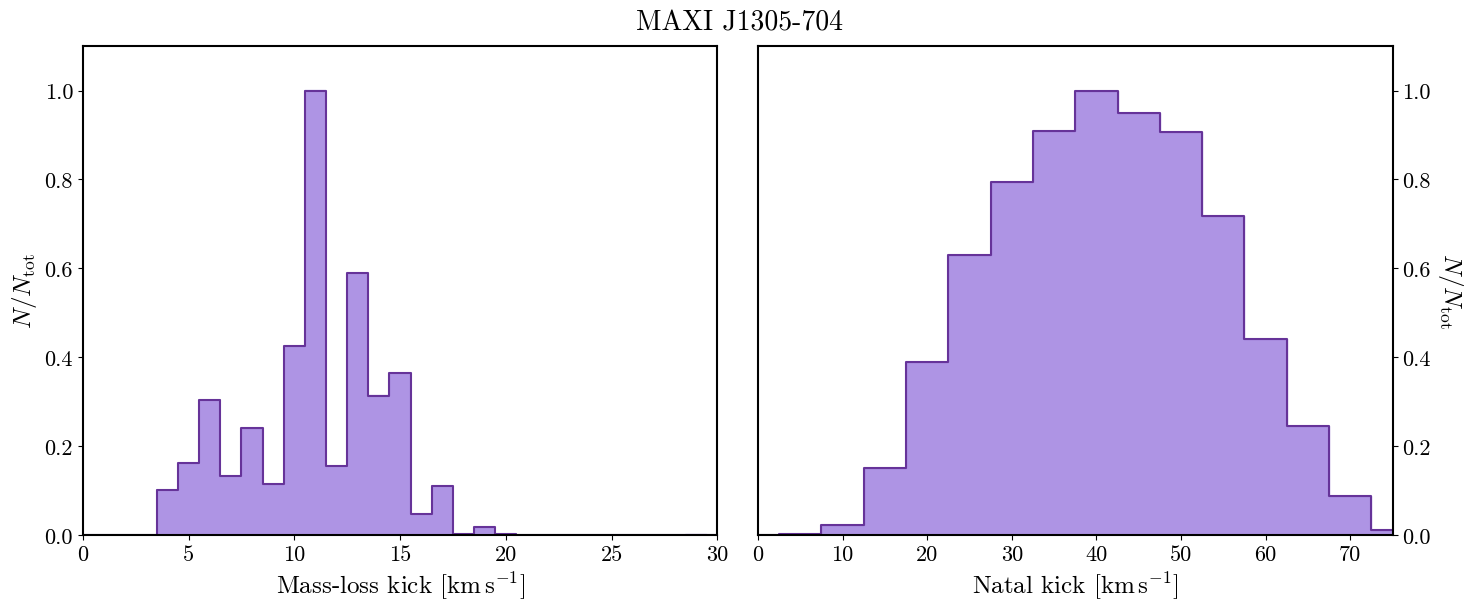}
    \caption{Distribution of mass-loss kicks ($\vmlk$) magnitude (left) and natal kick ($\vnk$) magnitude (right) for simulated system analogues to MAXI J1305-704.}
    \label{Fig:J1305_KickHist}
\end{figure*}
\clearpage

\begin{figure*}
    \centering
    \includegraphics[width=0.75\textwidth]{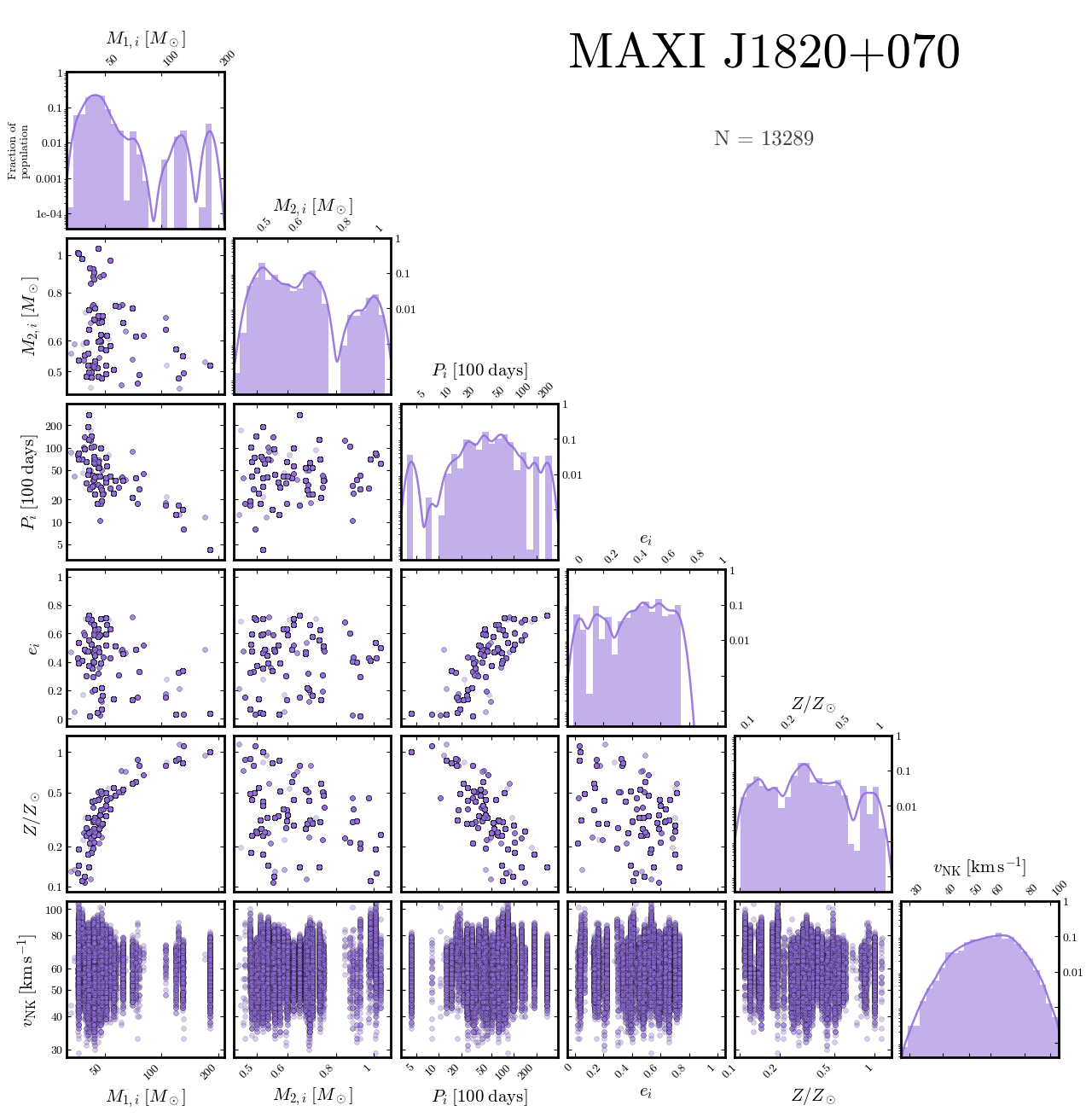}
    \caption{Corner plot highlighting key initial characteristics and acceptable natal kicks for simulated system analogues to MAXI J1820+070.}
    \label{Fig:J1820_CornerPlot}
\end{figure*}
\begin{figure*}
    \centering
    \includegraphics[width=0.75\textwidth]{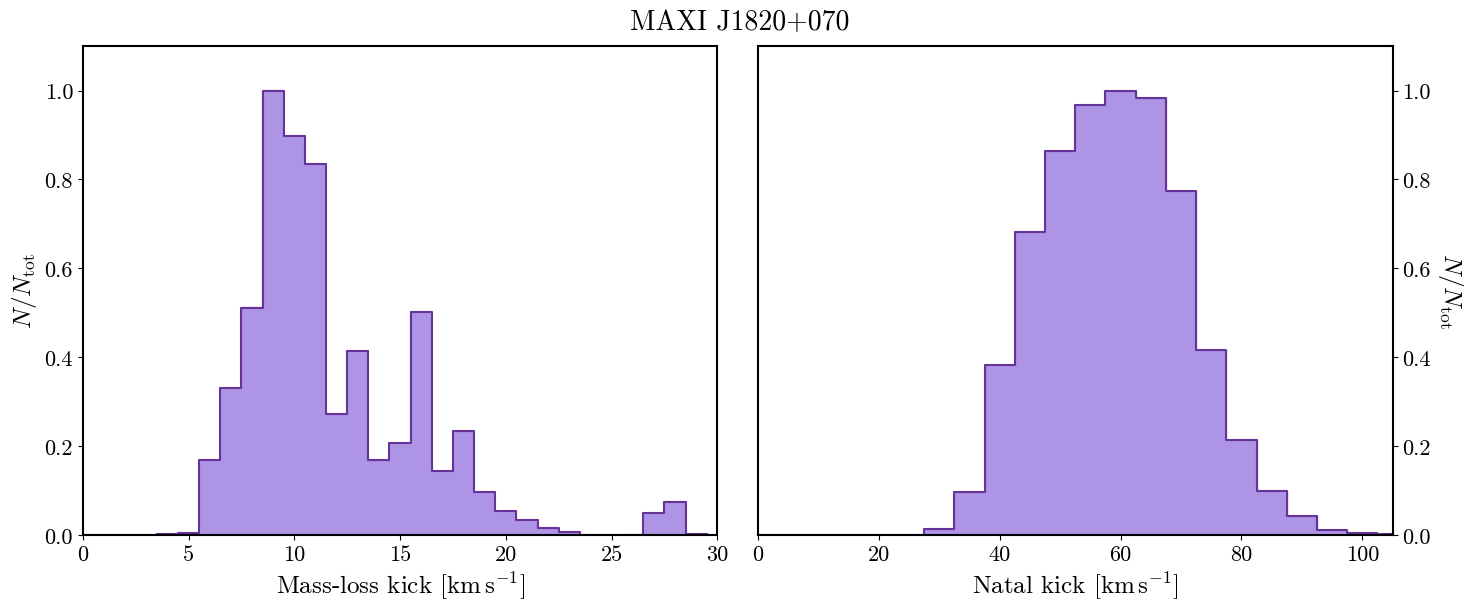}
    \caption{Distribution of mass-loss kicks ($\vmlk$) magnitude (left) and natal kick ($\vnk$) magnitude (right) for simulated system analogues to MAXI J1820+070.}
    \label{Fig:J1820_KickHist}
\end{figure*}
\clearpage

\begin{figure*}
    \centering
    \includegraphics[width=0.75\textwidth]{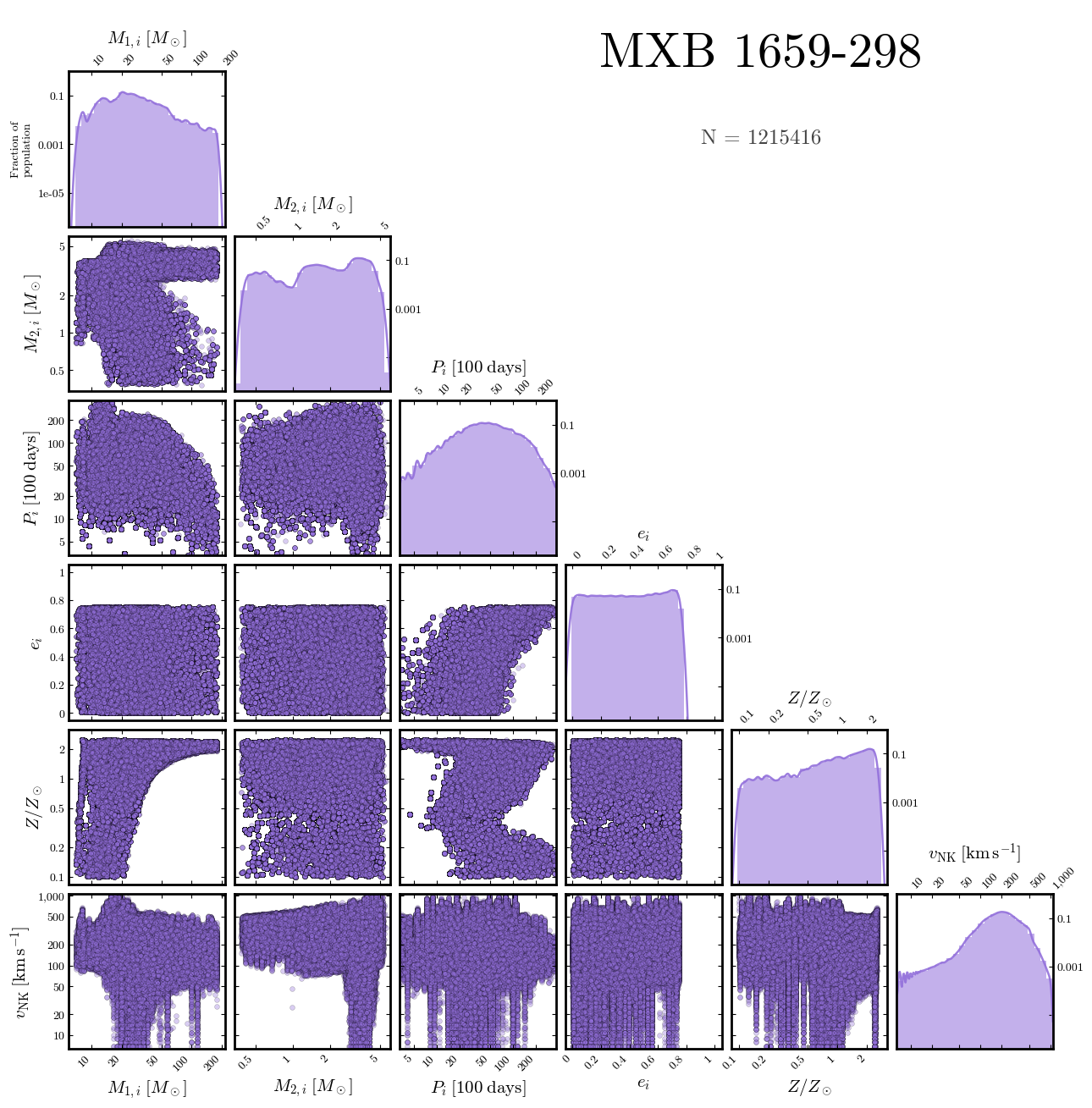}
    \caption{Corner plot highlighting key initial characteristics and acceptable natal kicks for simulated system analogues to MXB 1659-298.}
    \label{Fig:MXB1659_CornerPlot}
\end{figure*}
\begin{figure*}
    \centering
    \includegraphics[width=0.75\textwidth]{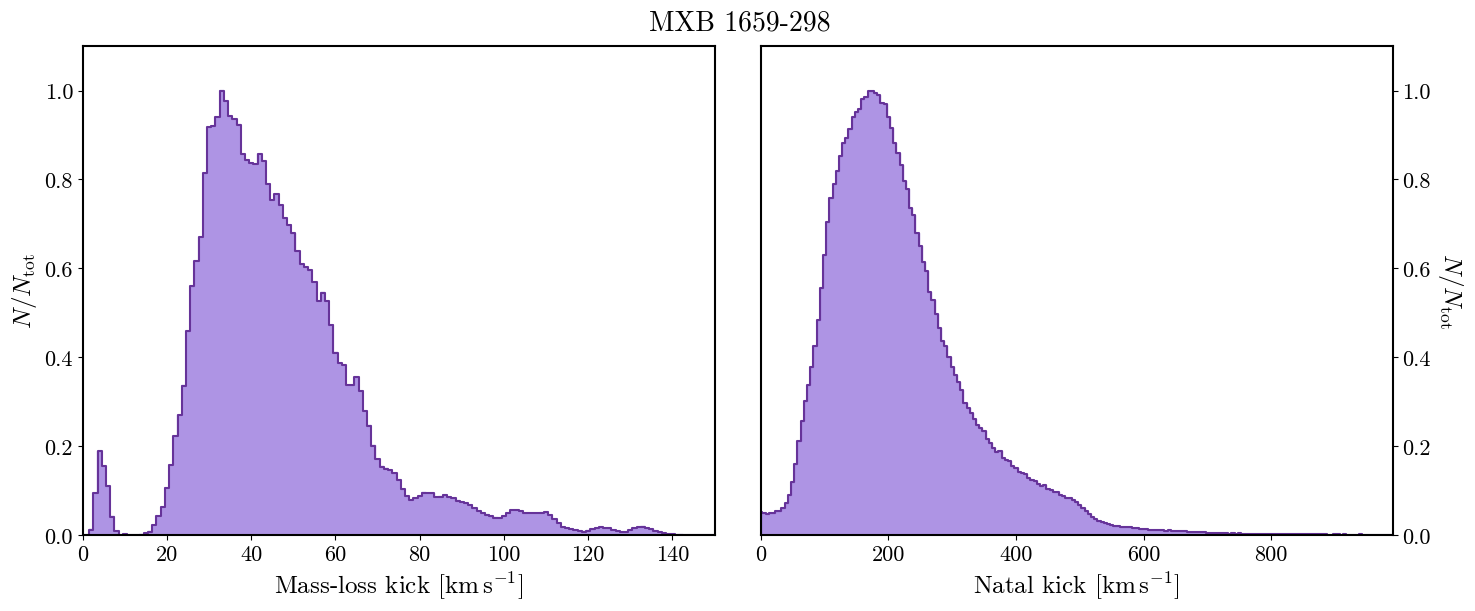}
    \caption{Distribution of mass-loss kicks ($\vmlk$) magnitude (left) and natal kick ($\vnk$) magnitude (right) for simulated system analogues to MXB 1659-298.}
    \label{Fig:MXB1659_KickHist}
\end{figure*}
\clearpage

\begin{figure*}
    \centering
    \includegraphics[width=0.75\textwidth]{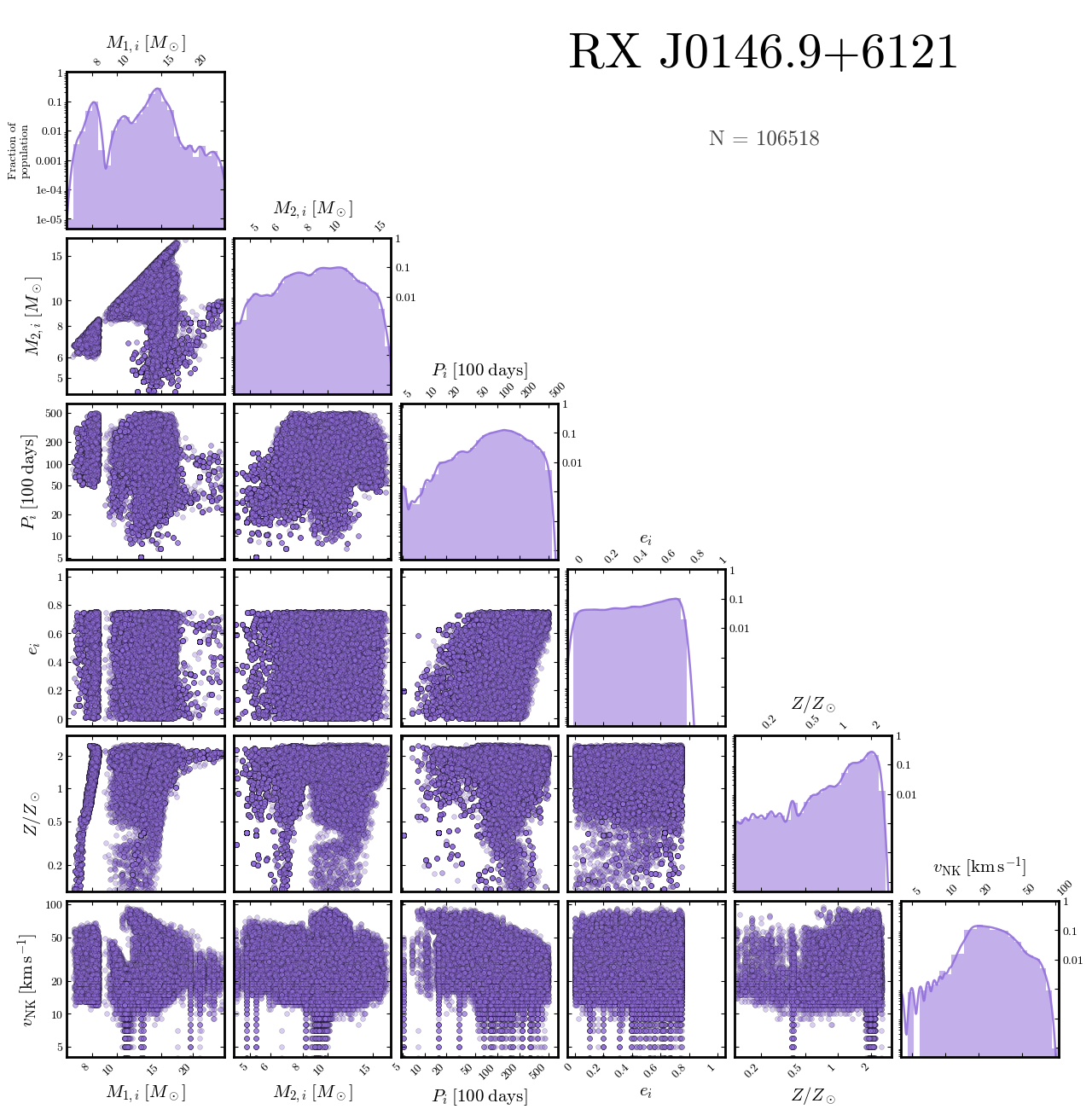}
    \caption{Corner plot highlighting key initial characteristics and acceptable natal kicks for simulated system analogues to RX J0146.9+6121.}
    \label{Fig:J0146_CornerPlot}
\end{figure*}
\begin{figure*}
    \centering
    \includegraphics[width=0.75\textwidth]{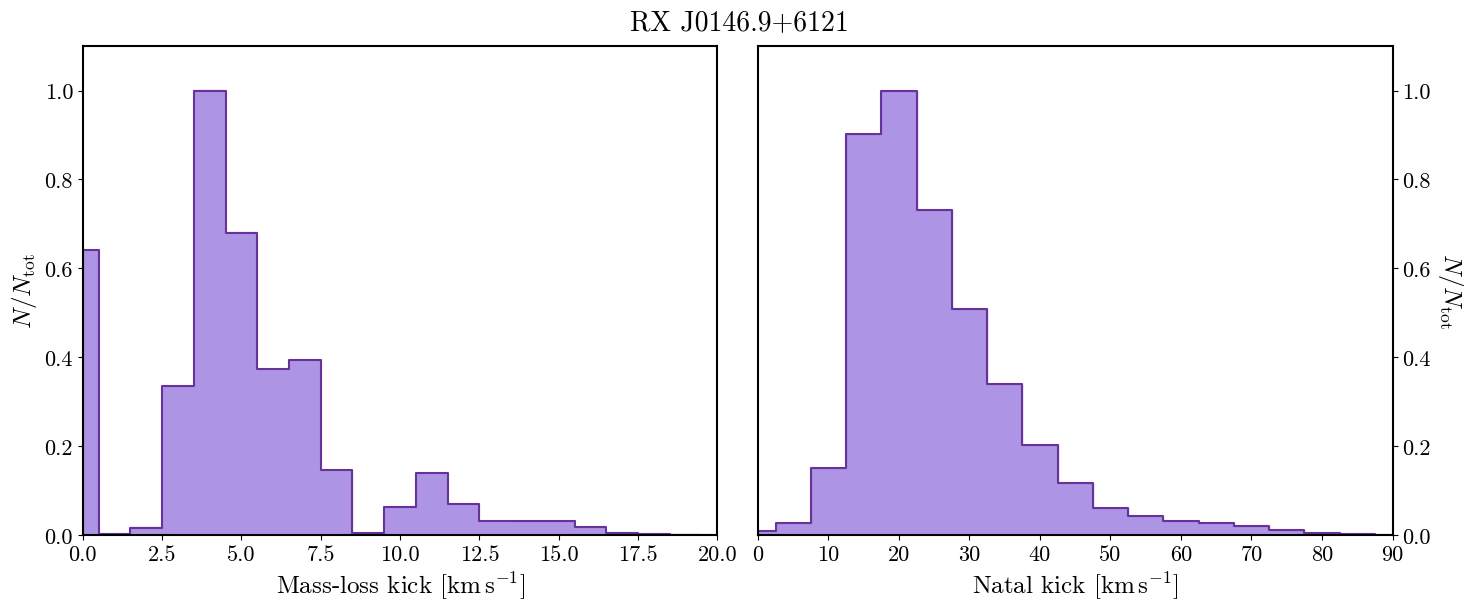}
    \caption{Distribution of mass-loss kicks ($\vmlk$) magnitude (left) and natal kick ($\vnk$) magnitude (right) for simulated system analogues to RX J0146.9+6121.}
    \label{Fig:J0146_KickHist}
\end{figure*}
\clearpage

\begin{figure*}
    \centering
    \includegraphics[width=0.75\textwidth]{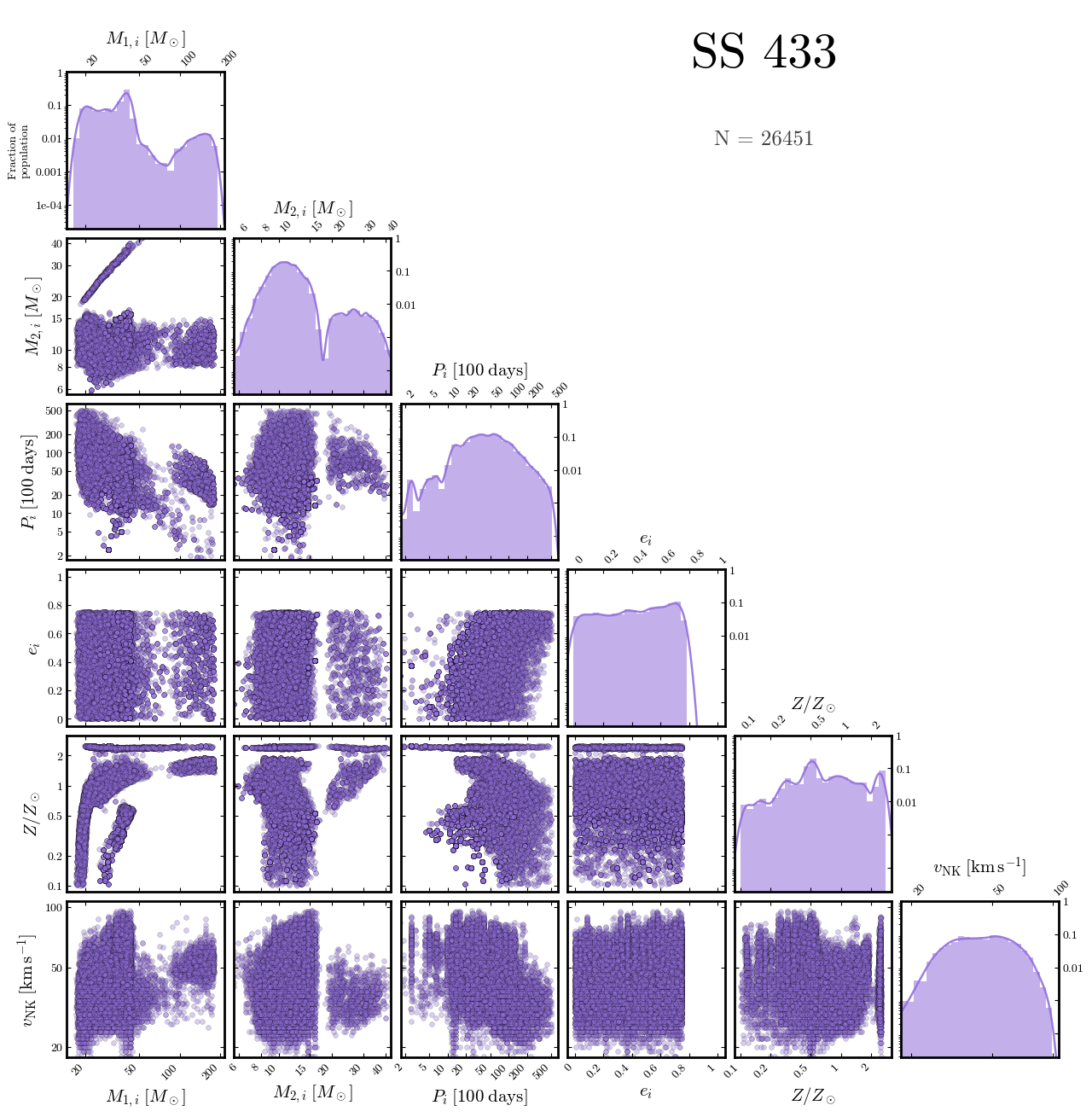}
    \caption{Corner plot highlighting key initial characteristics and acceptable natal kicks for simulated system analogues to SS 433.}
    \label{Fig:SS433_CornerPlot}
\end{figure*}
\begin{figure*}
    \centering
    \includegraphics[width=0.75\textwidth]{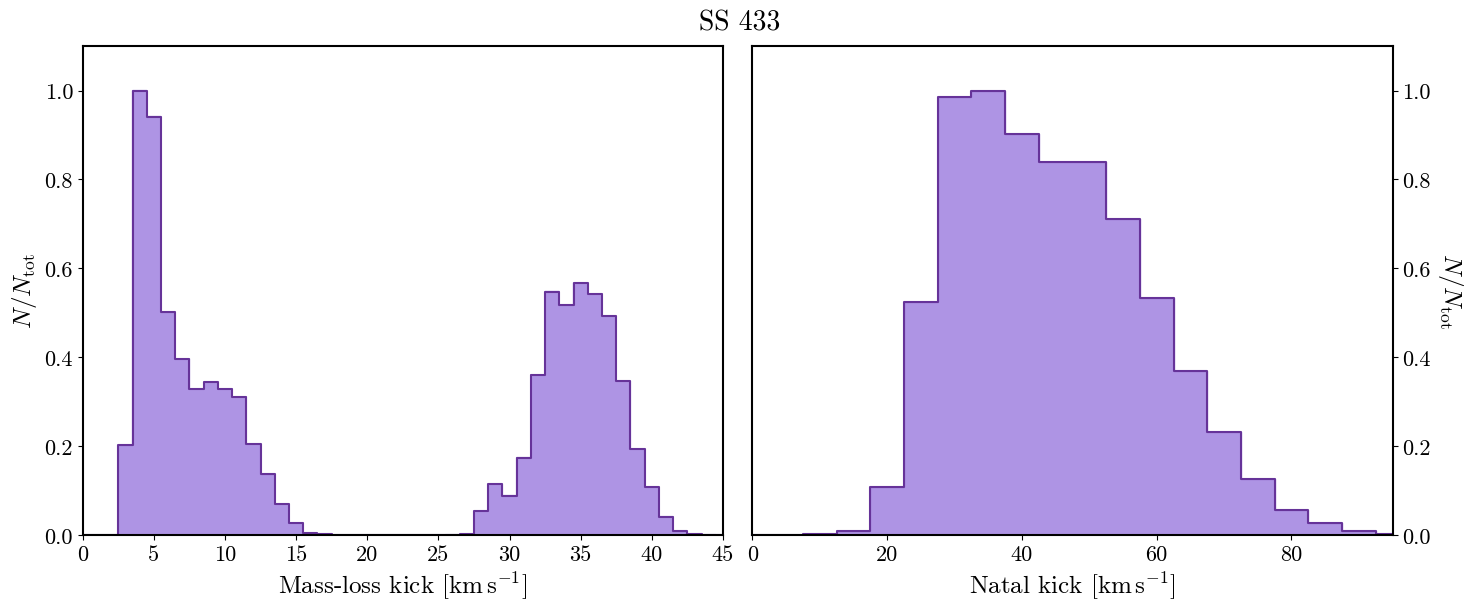}
    \caption{Distribution of mass-loss kicks ($\vmlk$) magnitude (left) and natal kick ($\vnk$) magnitude (right) for simulated system analogues to SS 433.}
    \label{Fig:SS433_KickHist}
\end{figure*}
\clearpage

\begin{figure*}
    \centering
    \includegraphics[width=0.75\textwidth]{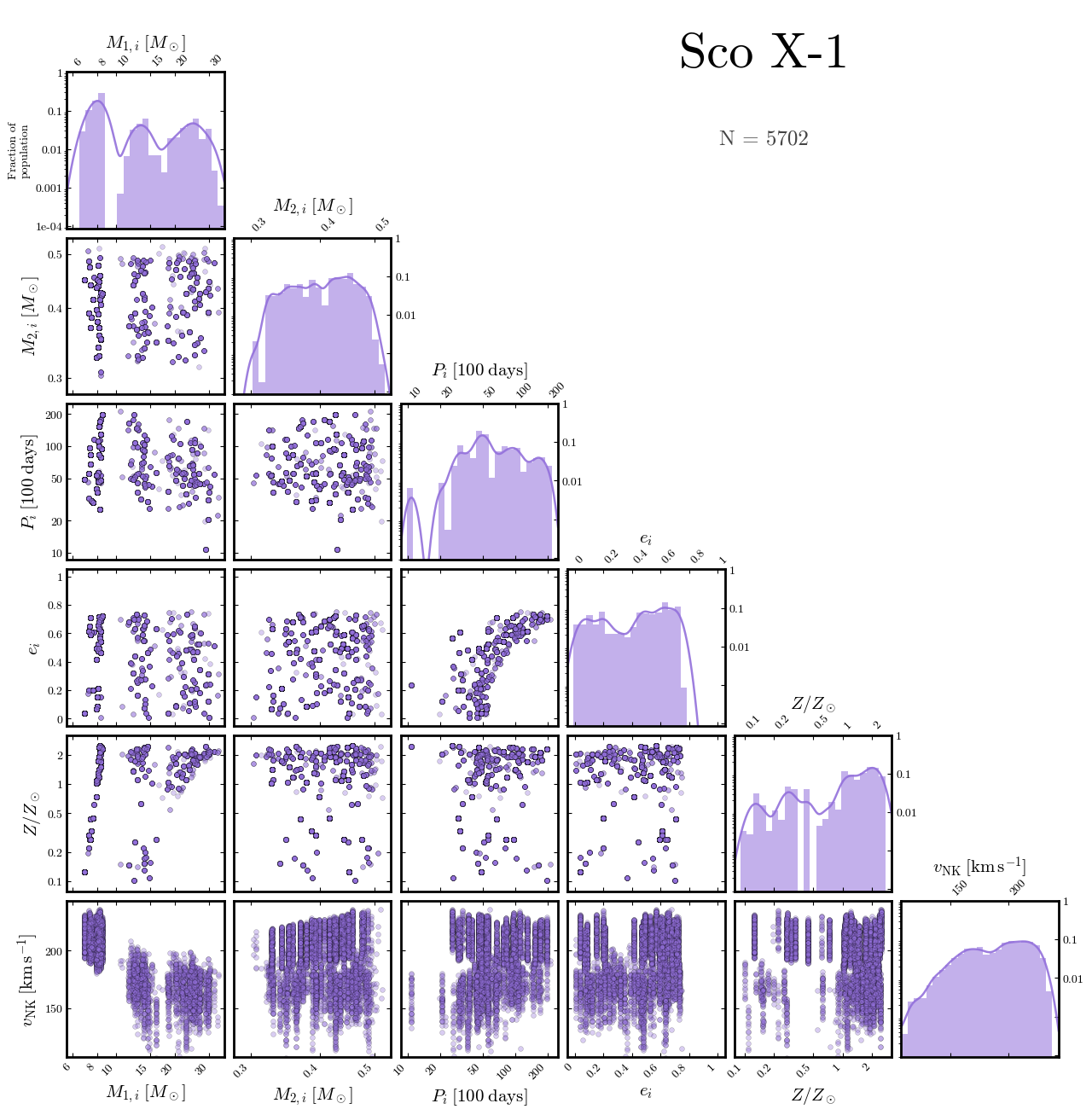}
    \caption{Corner plot highlighting key initial characteristics and acceptable natal kicks for simulated system analogues to Sco X-1.}
    \label{Fig:ScoX1_CornerPlot}
\end{figure*}
\begin{figure*}
    \centering
    \includegraphics[width=0.75\textwidth]{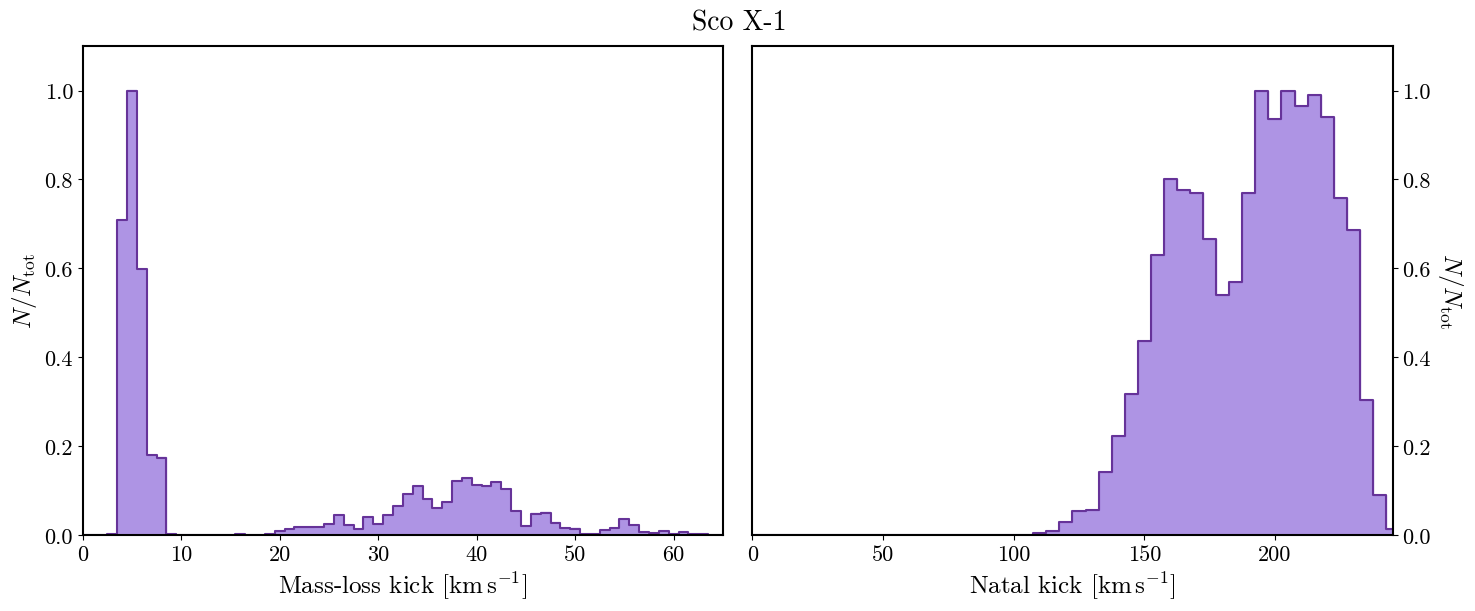}
    \caption{Distribution of mass-loss kicks ($\vmlk$) magnitude (left) and natal kick ($\vnk$) magnitude (right) for simulated system analogues to Sco X-1.}
    \label{Fig:ScoX1_KickHist}
\end{figure*}
\clearpage

\begin{figure*}
    \centering
    \includegraphics[width=0.75\textwidth]{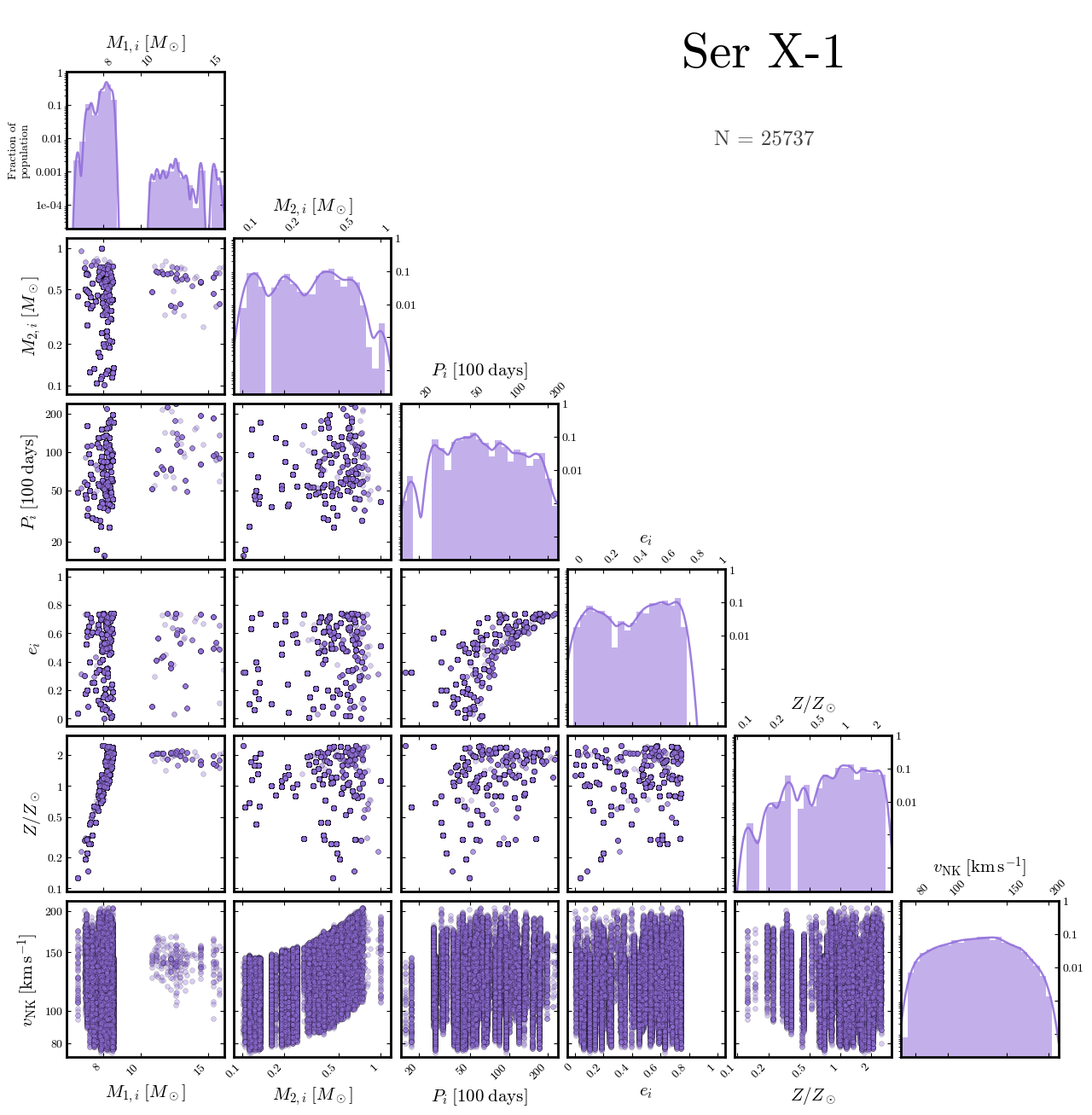}
    \caption{Corner plot highlighting key initial characteristics and acceptable natal kicks for simulated system analogues to Ser X-1.}
    \label{Fig:SerX1_CornerPlot}
\end{figure*}
\begin{figure*}
    \centering
    \includegraphics[width=0.75\textwidth]{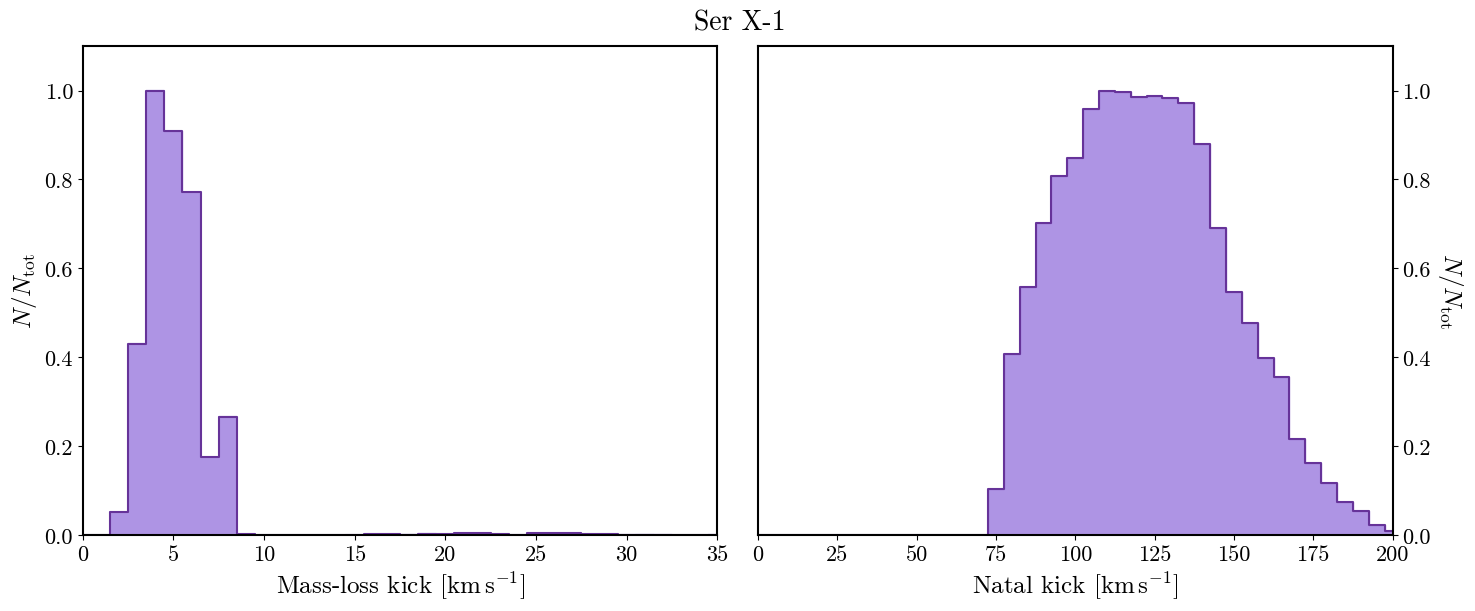}
    \caption{Distribution of mass-loss kicks ($\vmlk$) magnitude (left) and natal kick ($\vnk$) magnitude (right) for simulated system analogues to Ser X-1.}
    \label{Fig:SerX1_KickHist}
\end{figure*}
\clearpage

\begin{figure*}
    \centering
    \includegraphics[width=0.75\textwidth]{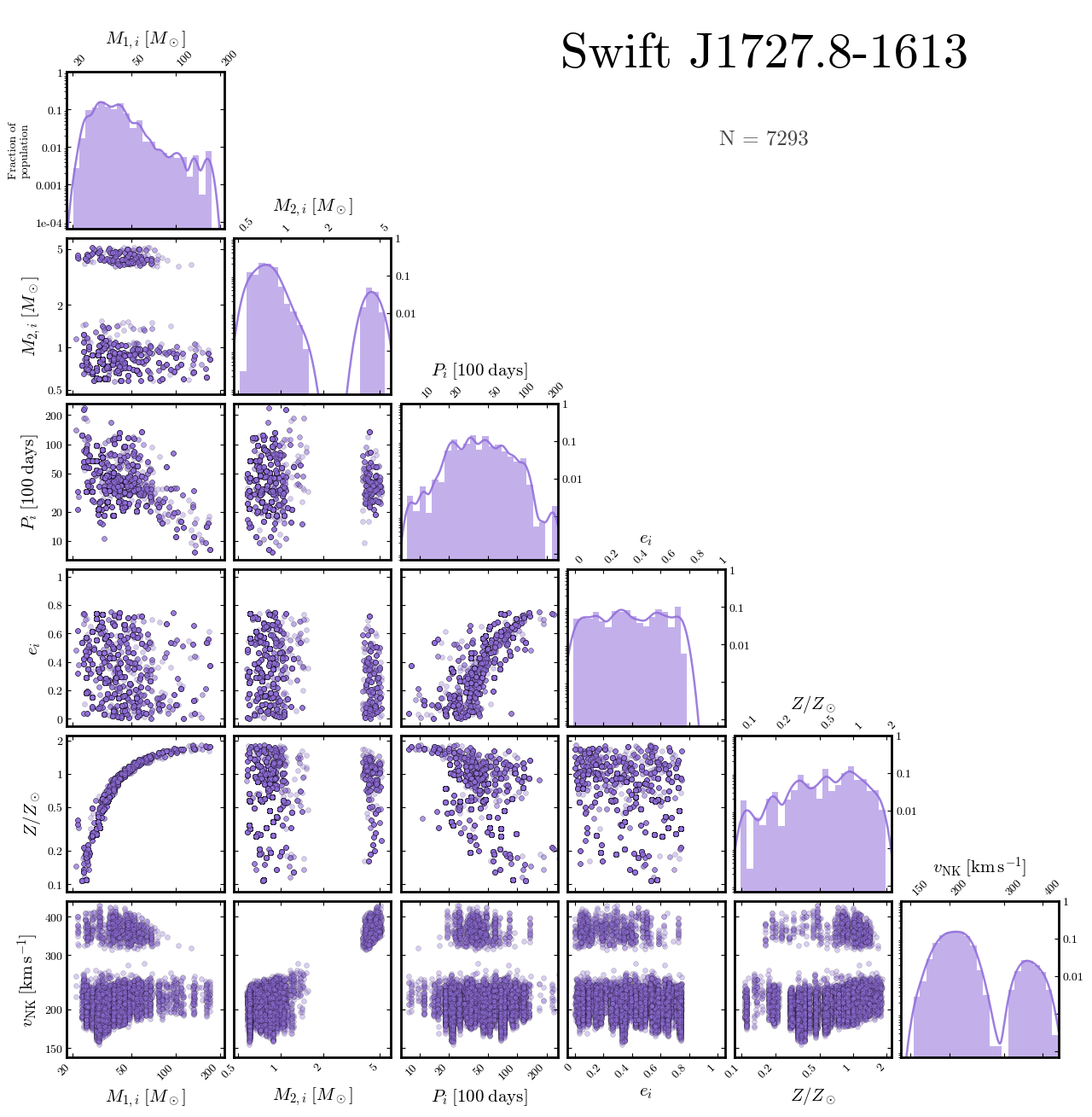}
    \caption{Corner plot highlighting key initial characteristics and acceptable natal kicks for simulated system analogues to Swift J1727.8-1613.}
    \label{Fig:J1727_CornerPlot}
\end{figure*}
\begin{figure*}
    \centering
    \includegraphics[width=0.75\textwidth]{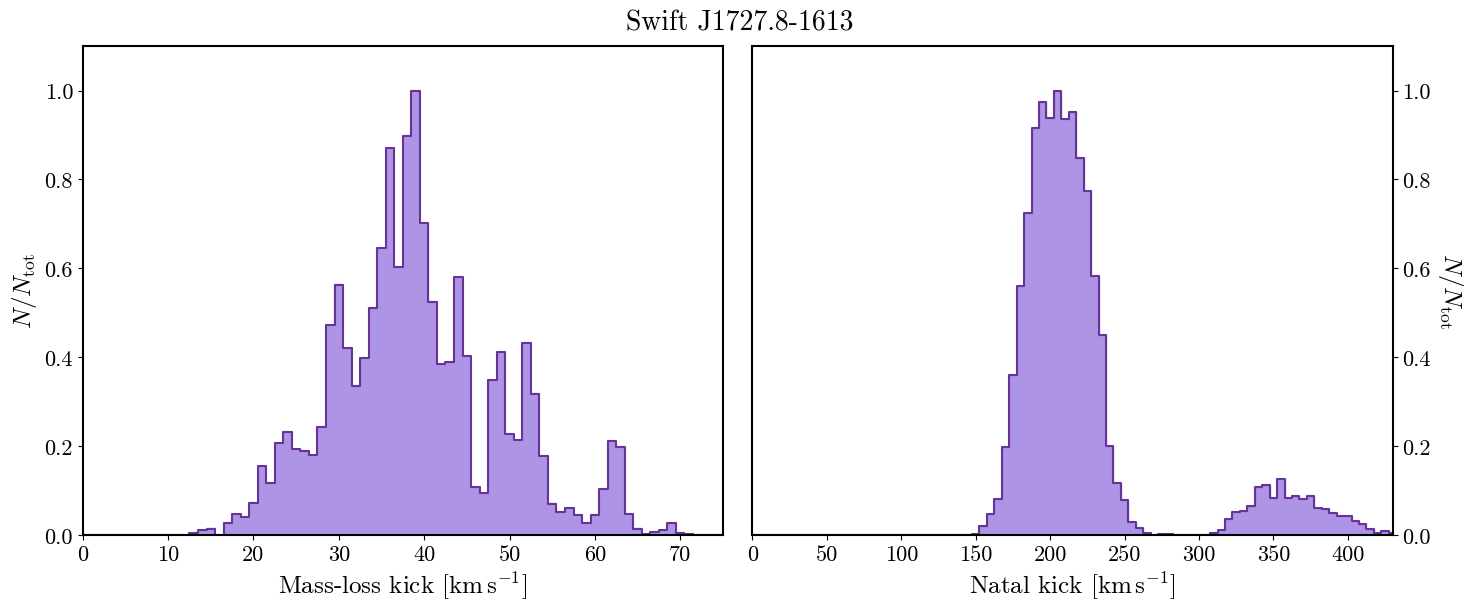}
    \caption{Distribution of mass-loss kicks ($\vmlk$) magnitude (left) and natal kick ($\vnk$) magnitude (right) for simulated system analogues to Swift J1727.8-1613.}
    \label{Fig:J1727_KickHist}
\end{figure*}
\clearpage

\begin{figure*}
    \centering
    \includegraphics[width=0.75\textwidth]{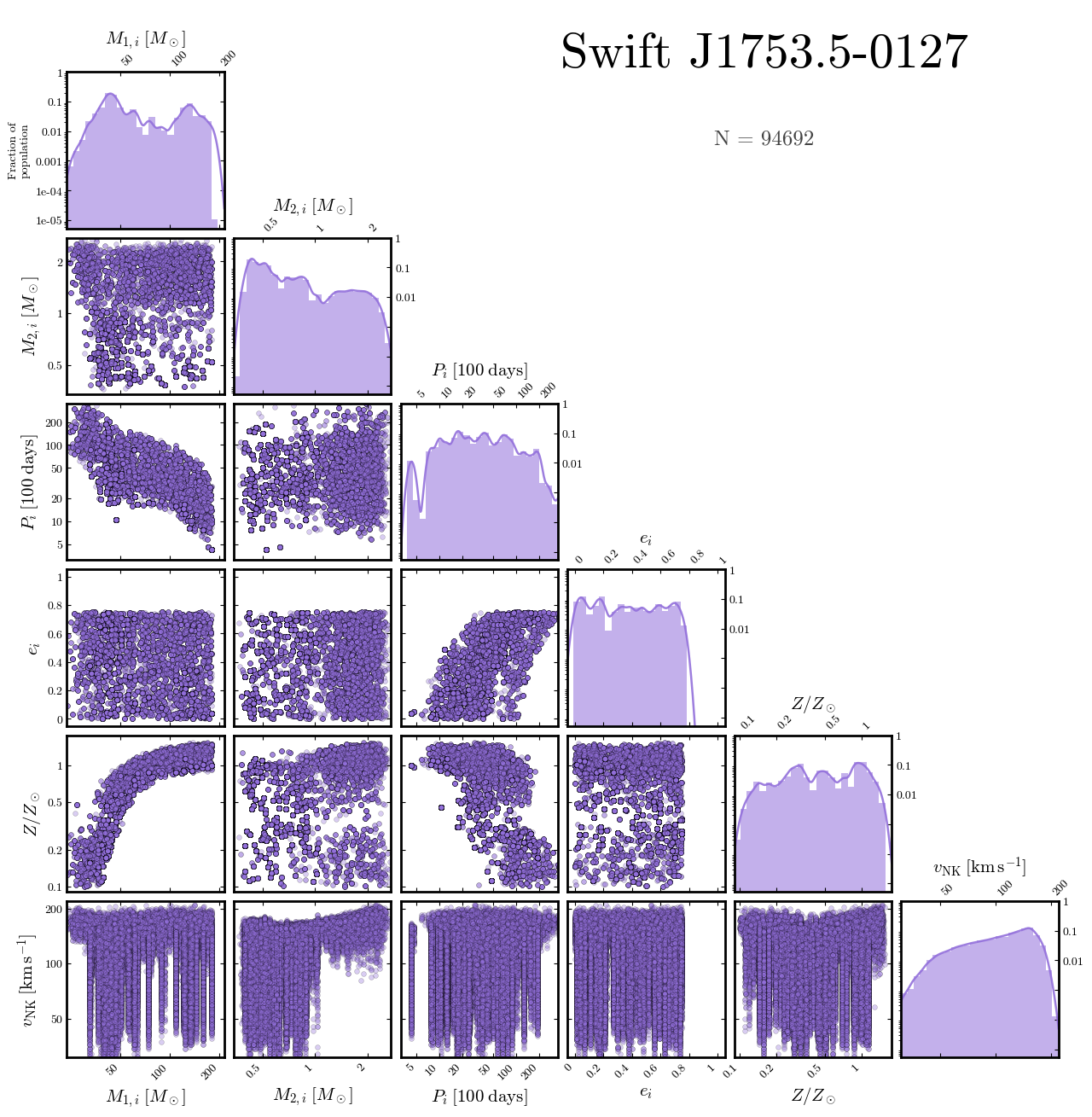}
    \caption{Corner plot highlighting key initial characteristics and acceptable natal kicks for simulated system analogues to Swift J1753.5-0127.}
    \label{Fig:J1753_CornerPlot}
\end{figure*}
\begin{figure*}
    \centering
    \includegraphics[width=0.75\textwidth]{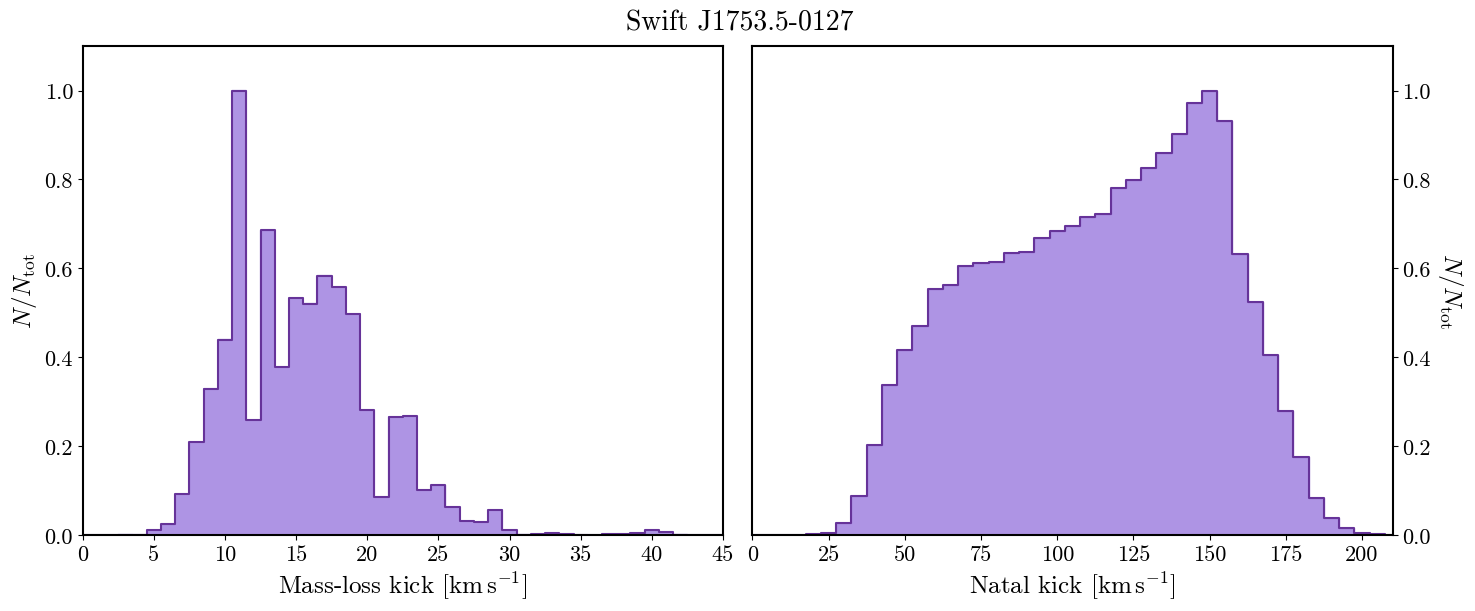}
    \caption{Distribution of mass-loss kicks ($\vmlk$) magnitude (left) and natal kick ($\vnk$) magnitude (right) for simulated system analogues to Swift J1753.5-0127.}
    \label{Fig:J1753_KickHist}
\end{figure*}
\clearpage

\begin{figure*}
    \centering
    \includegraphics[width=0.75\textwidth]{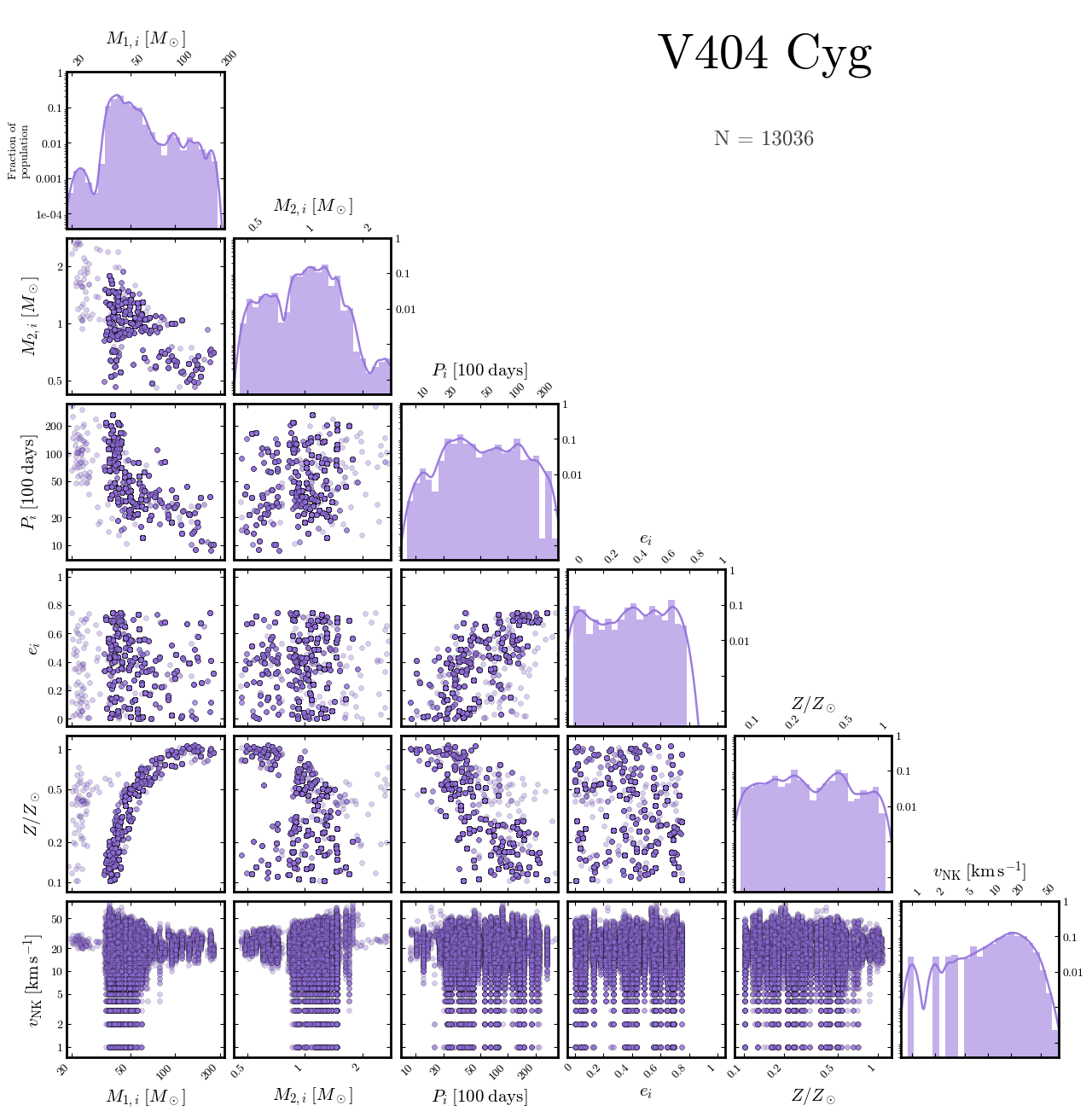}
    \caption{Corner plot highlighting key initial characteristics and acceptable natal kicks for simulated system analogues to V404 Cyg.}
    \label{Fig:V404Cyg_CornerPlot}
\end{figure*}
\begin{figure*}
    \centering
    \includegraphics[width=0.75\textwidth]{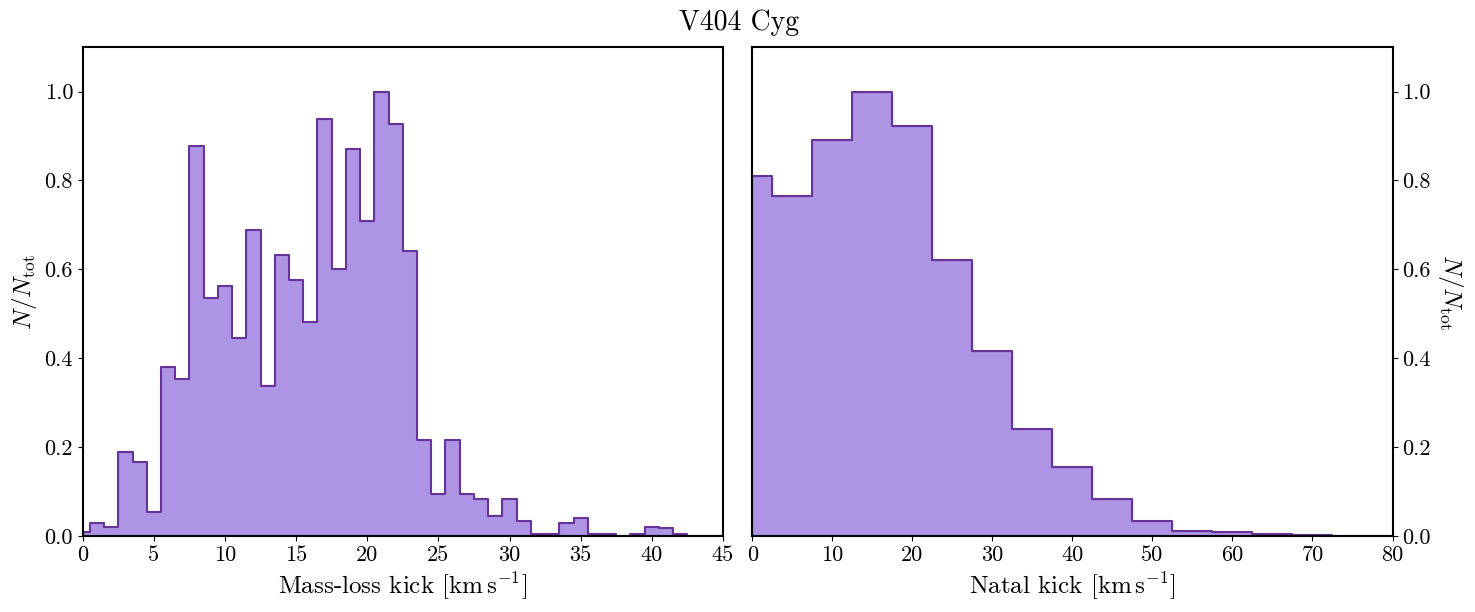}
    \caption{Distribution of mass-loss kicks ($\vmlk$) magnitude (left) and natal kick ($\vnk$) magnitude (right) for simulated system analogues to V404 Cyg.}
    \label{Fig:V404Cyg_KickHist}
\end{figure*}
\clearpage

\begin{figure*}
    \centering
    \includegraphics[width=0.75\textwidth]{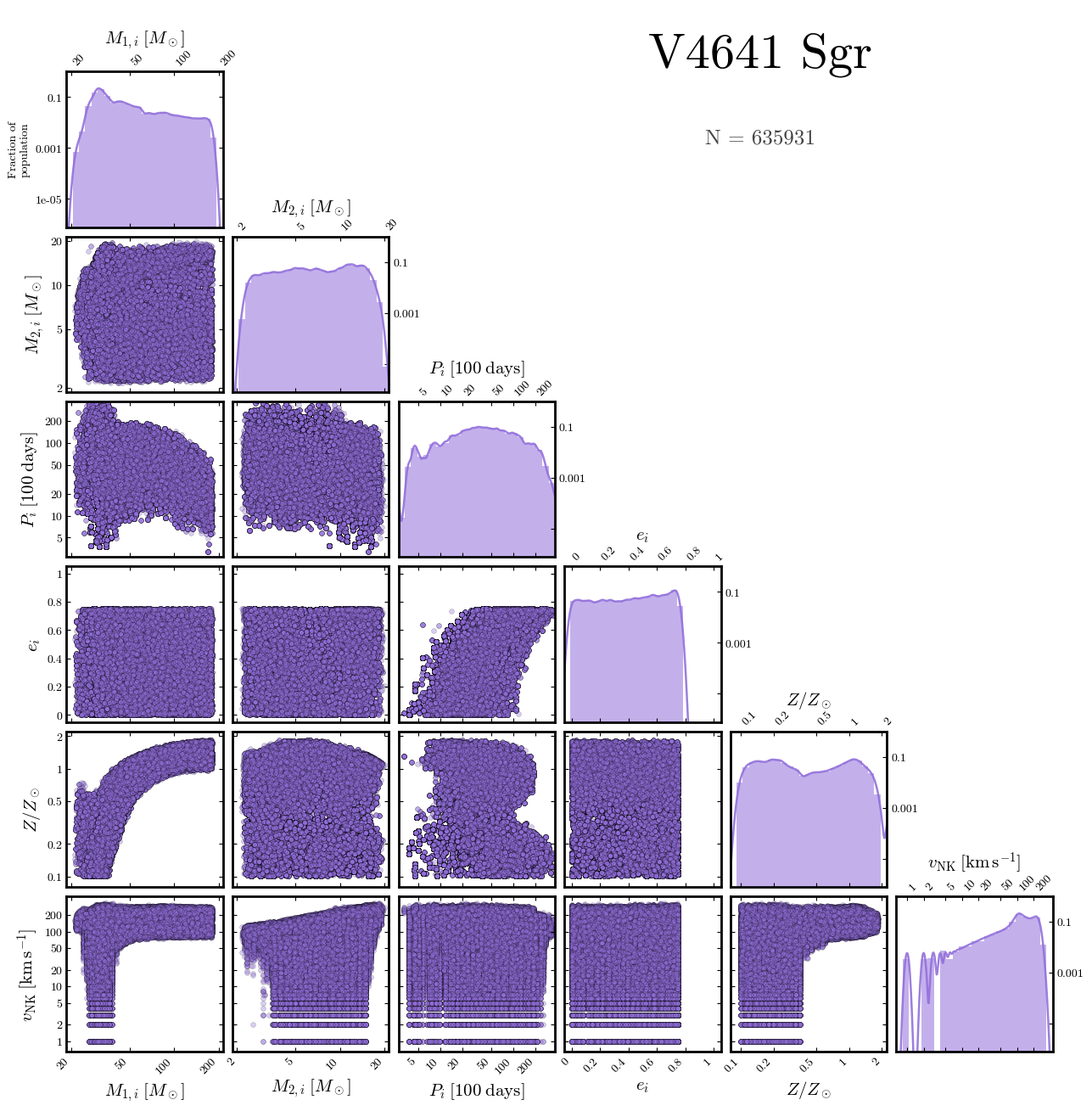}
    \caption{Corner plot highlighting key initial characteristics and acceptable natal kicks for simulated system analogues to V4641 Sgr.}
    \label{Fig:V4641_CornerPlot}
\end{figure*}
\begin{figure*}
    \centering
    \includegraphics[width=0.75\textwidth]{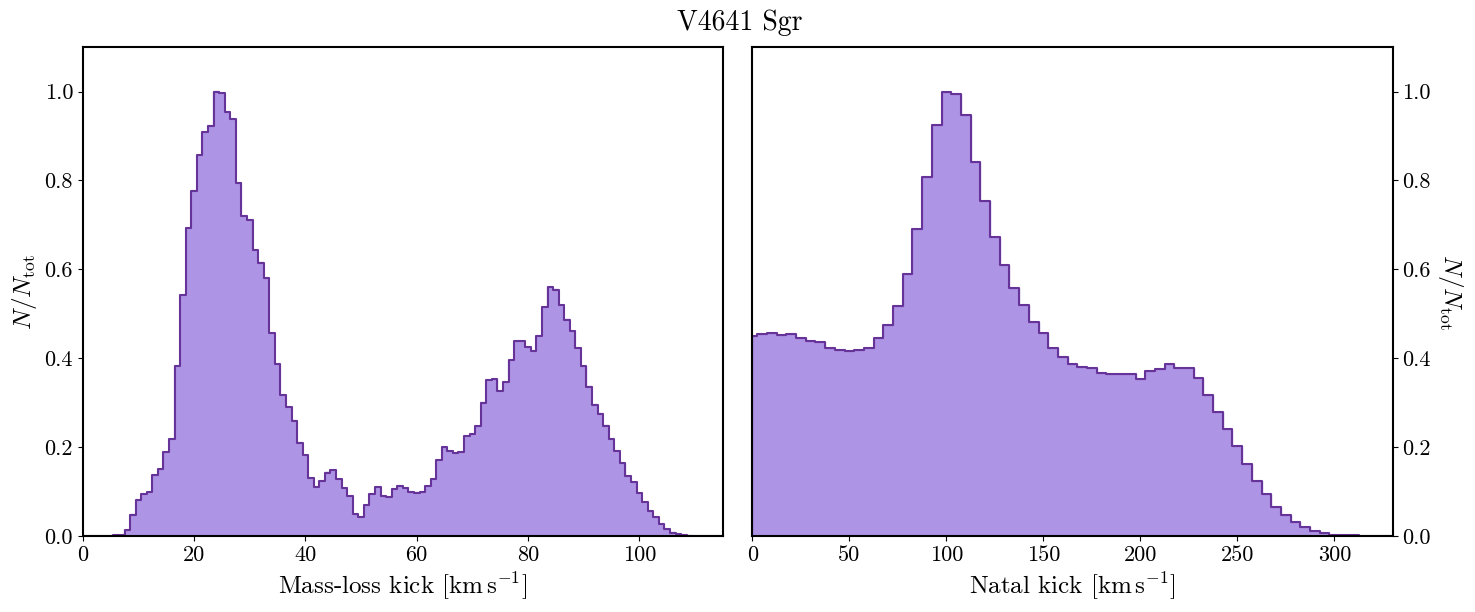}
    \caption{Distribution of mass-loss kicks ($\vmlk$) magnitude (left) and natal kick ($\vnk$) magnitude (right) for simulated system analogues to V4641 Sgr.}
    \label{Fig:V4641_KickHist}
\end{figure*}
\clearpage

\begin{figure*}
    \centering
    \includegraphics[width=0.75\textwidth]{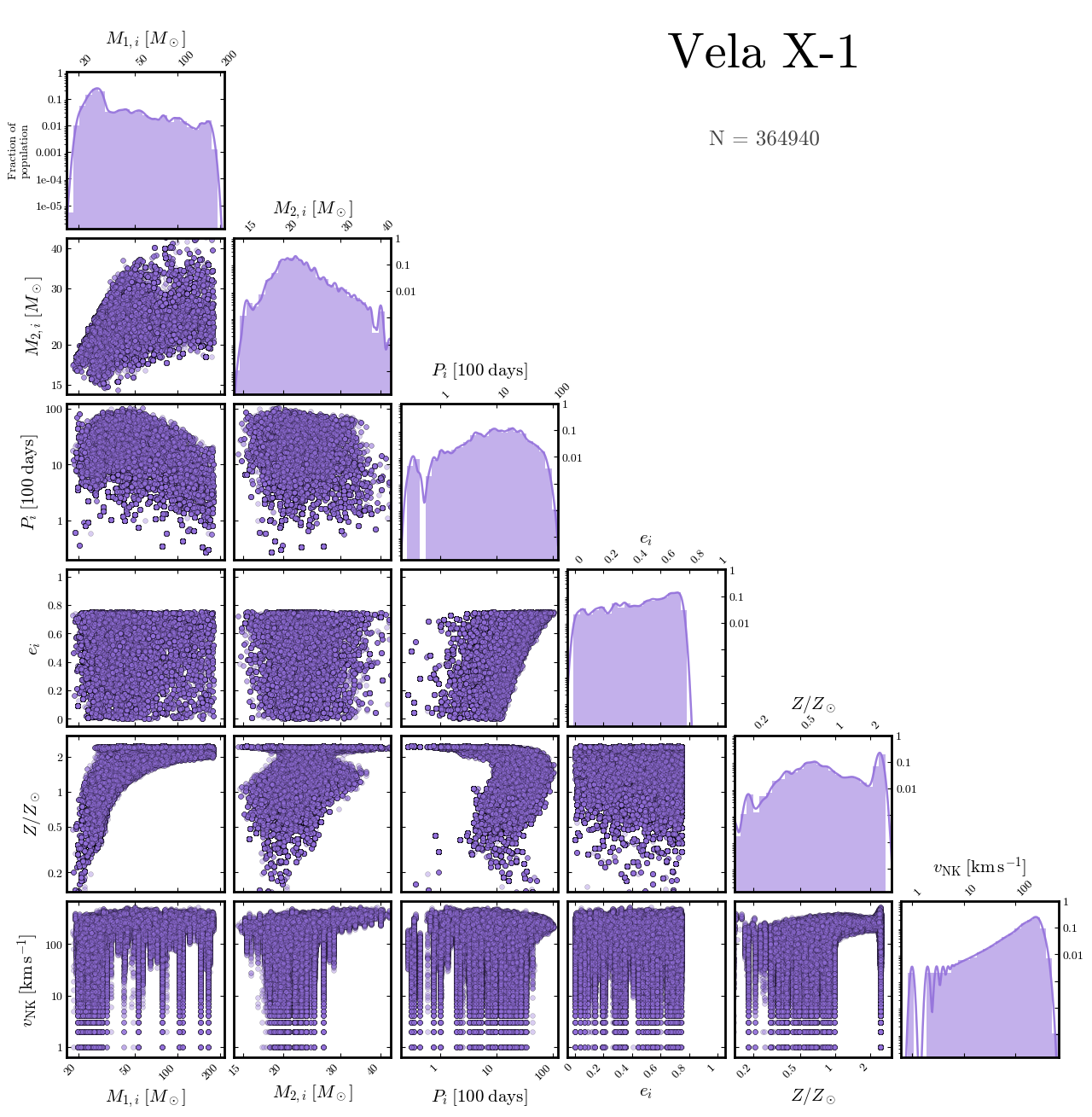}
    \caption{Corner plot highlighting key initial characteristics and acceptable natal kicks for simulated system analogues to Vela X-1.}
    \label{Fig:VelaX1_CornerPlot}
\end{figure*}
\begin{figure*}
    \centering
    \includegraphics[width=0.75\textwidth]{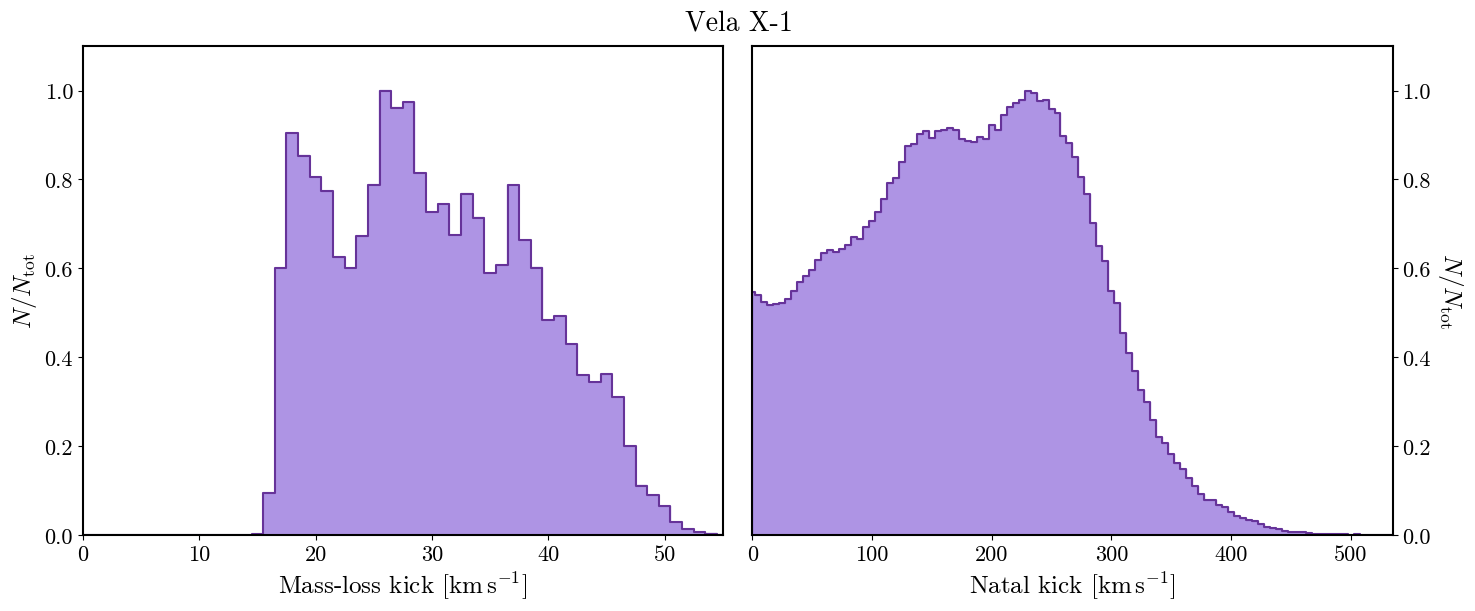}
    \caption{Distribution of mass-loss kicks ($\vmlk$) magnitude (left) and natal kick ($\vnk$) magnitude (right) for simulated system analogues to Vela X-1.}
    \label{Fig:VelaX1_KickHist}
\end{figure*}
\clearpage

\begin{figure*}
    \centering
    \includegraphics[width=0.75\textwidth]{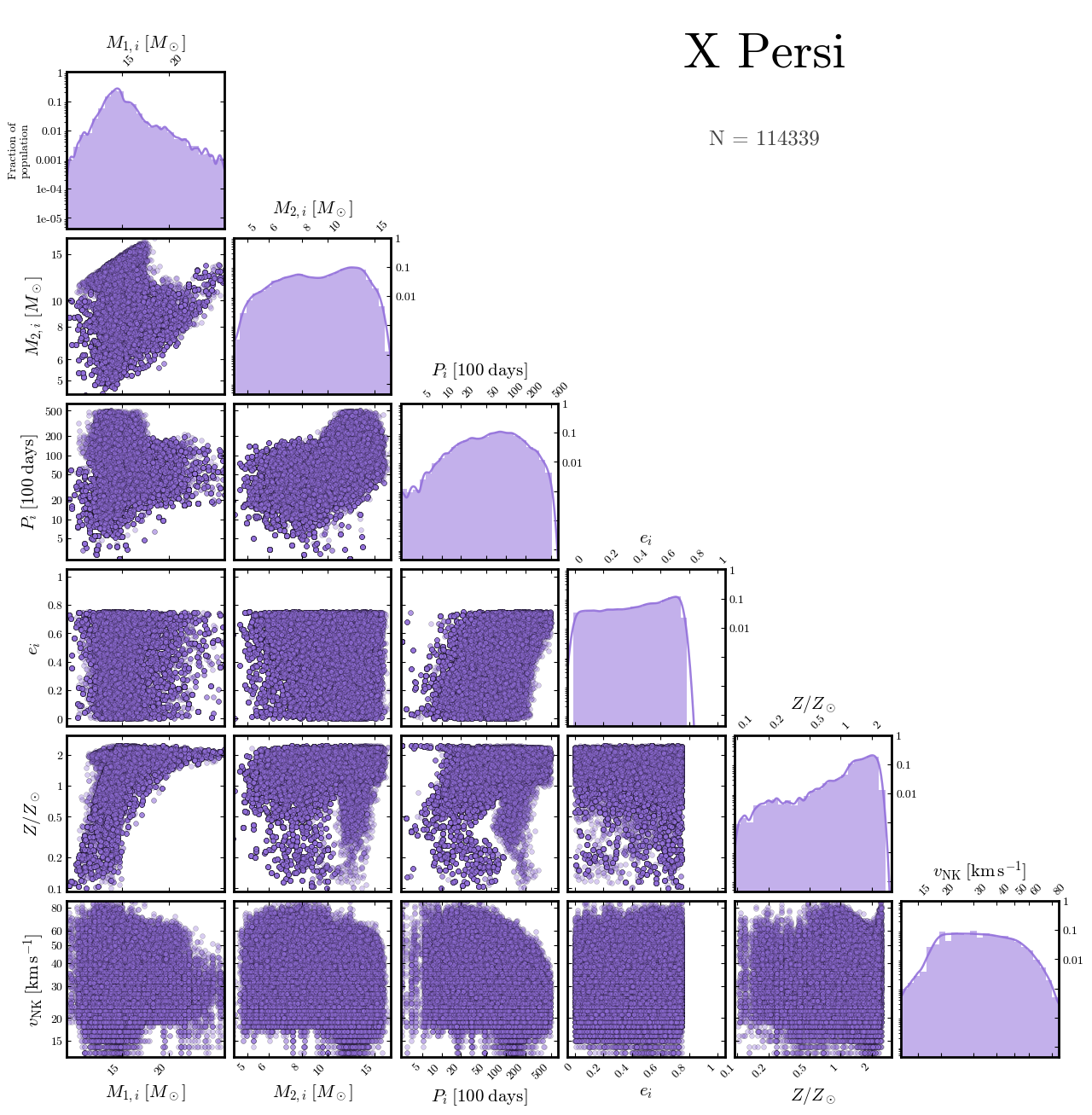}
    \caption{Corner plot highlighting key initial characteristics and acceptable natal kicks for simulated system analogues to X Persi.}
    \label{Fig:Xpersi_CornerPlot}
\end{figure*}
\begin{figure*}
    \centering
    \includegraphics[width=0.75\textwidth]{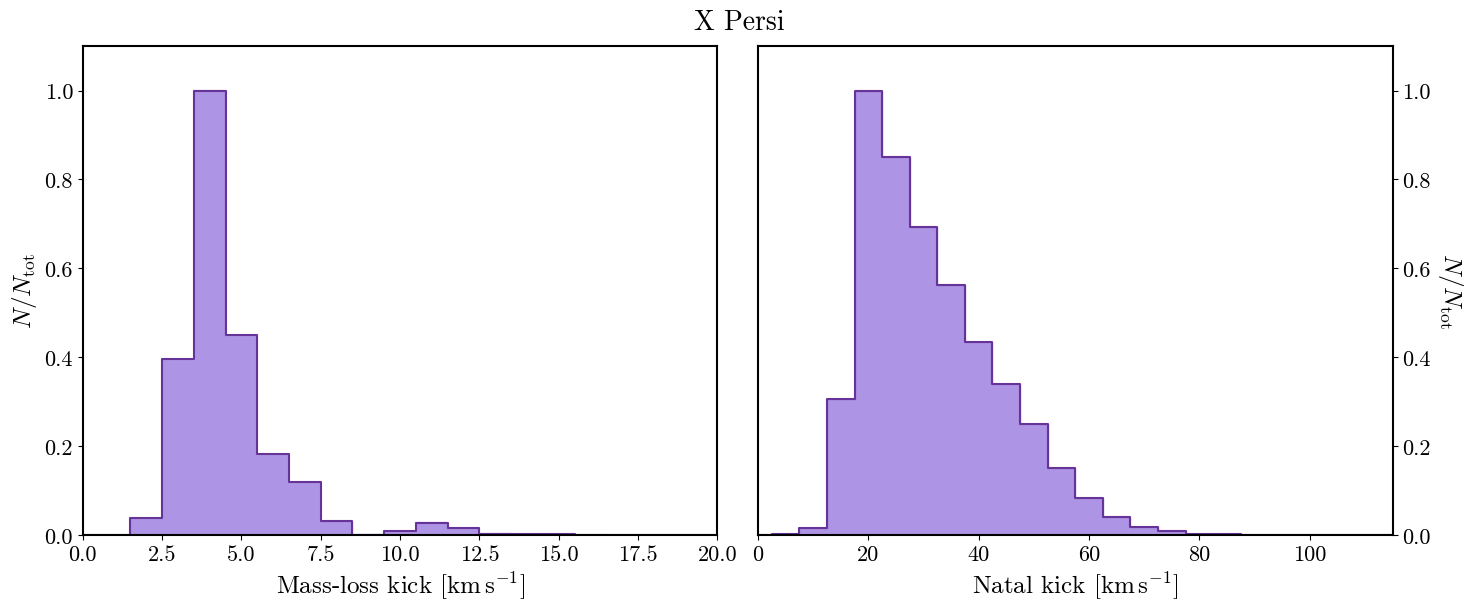}
    \caption{Distribution of mass-loss kicks ($\vmlk$) magnitude (left) and natal kick ($\vnk$) magnitude (right) for simulated system analogues to X Persi.}
    \label{Fig:Xpersi_KickHist}
\end{figure*}
\clearpage

\begin{figure*}
    \centering
    \includegraphics[width=0.75\textwidth]{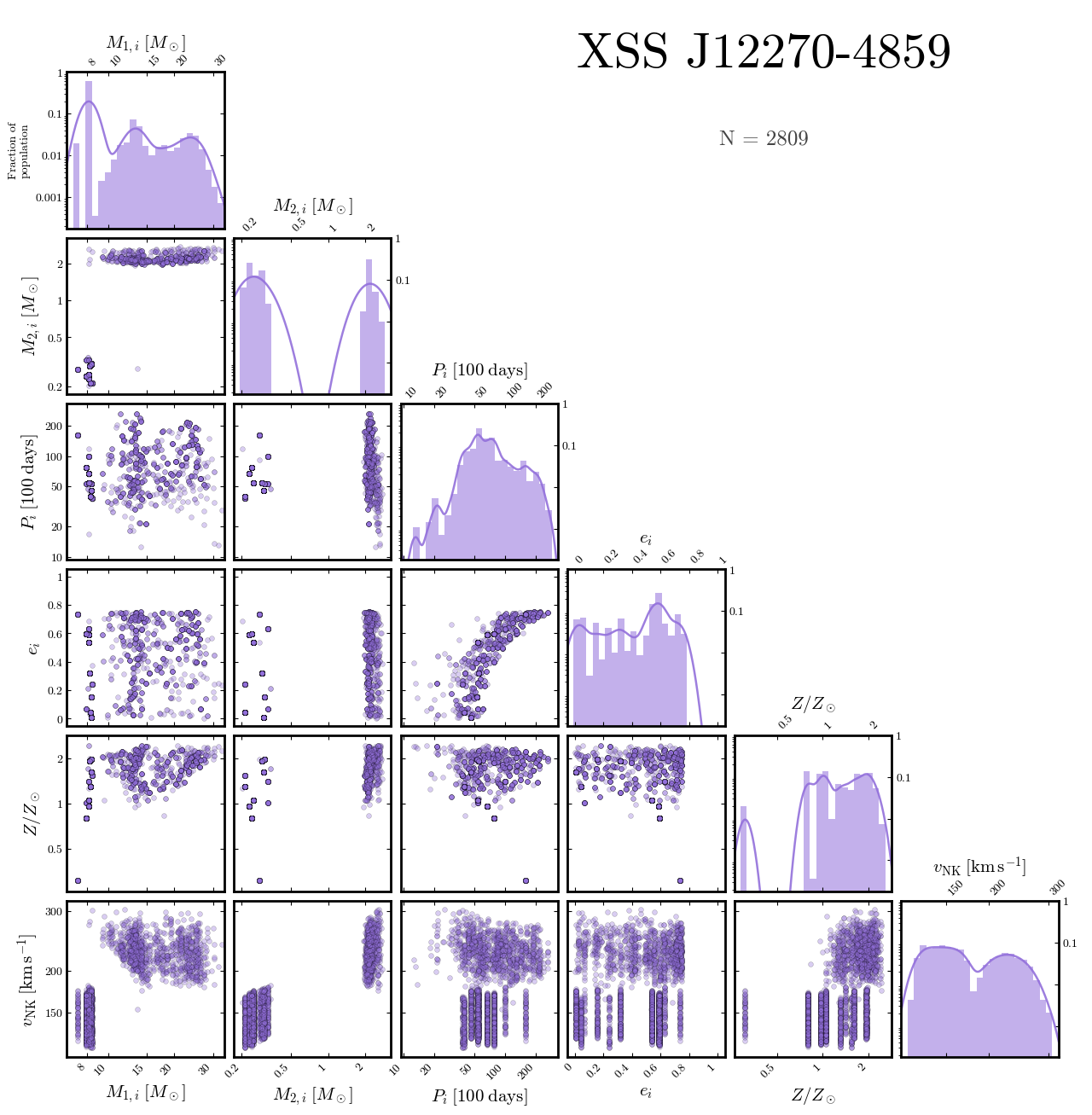}
    \caption{Corner plot highlighting key initial characteristics and acceptable natal kicks for simulated system analogues to XSS J12270-4859.}
    \label{Fig:J12270_CornerPlot}
\end{figure*}
\begin{figure*}
    \centering
    \includegraphics[width=0.75\textwidth]{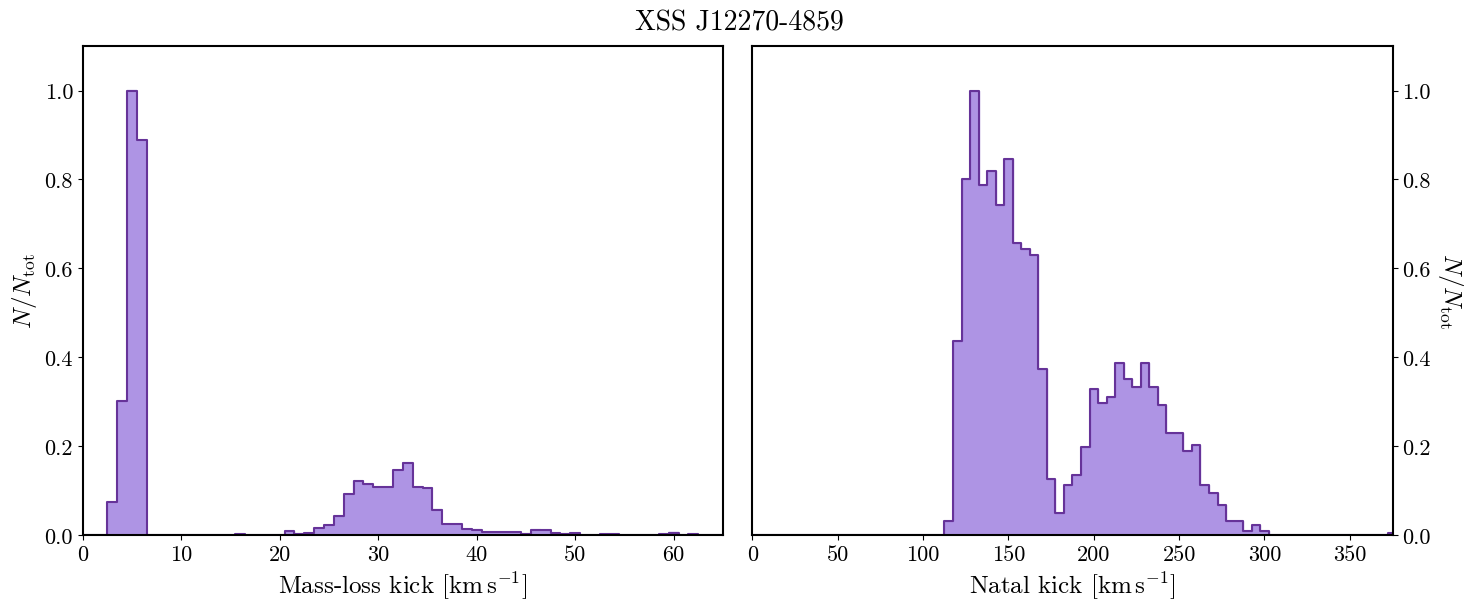}
    \caption{Distribution of mass-loss kicks ($\vmlk$) magnitude (left) and natal kick ($\vnk$) magnitude (right) for simulated system analogues to XSS J12270-4859.}
    \label{Fig:J12270_KickHist}
\end{figure*}
\clearpage

\begin{figure*}
    \centering
    \includegraphics[width=0.75\textwidth]{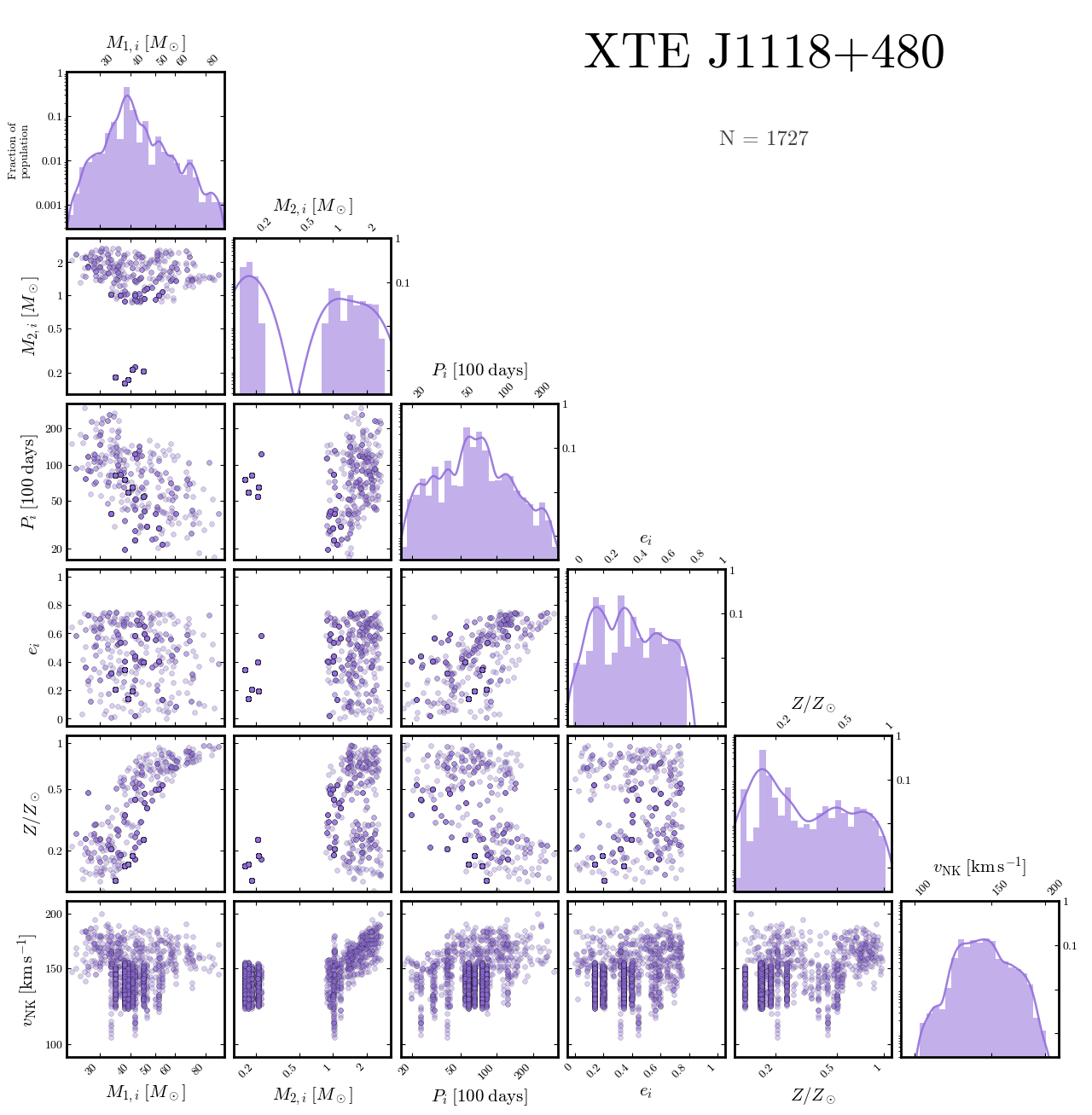}
    \caption{Corner plot highlighting key initial characteristics and acceptable natal kicks for simulated system analogues to XTE J1118+480.}
    \label{Fig:J1118_CornerPlot}
\end{figure*}
\begin{figure*}
    \centering
    \includegraphics[width=0.75\textwidth]{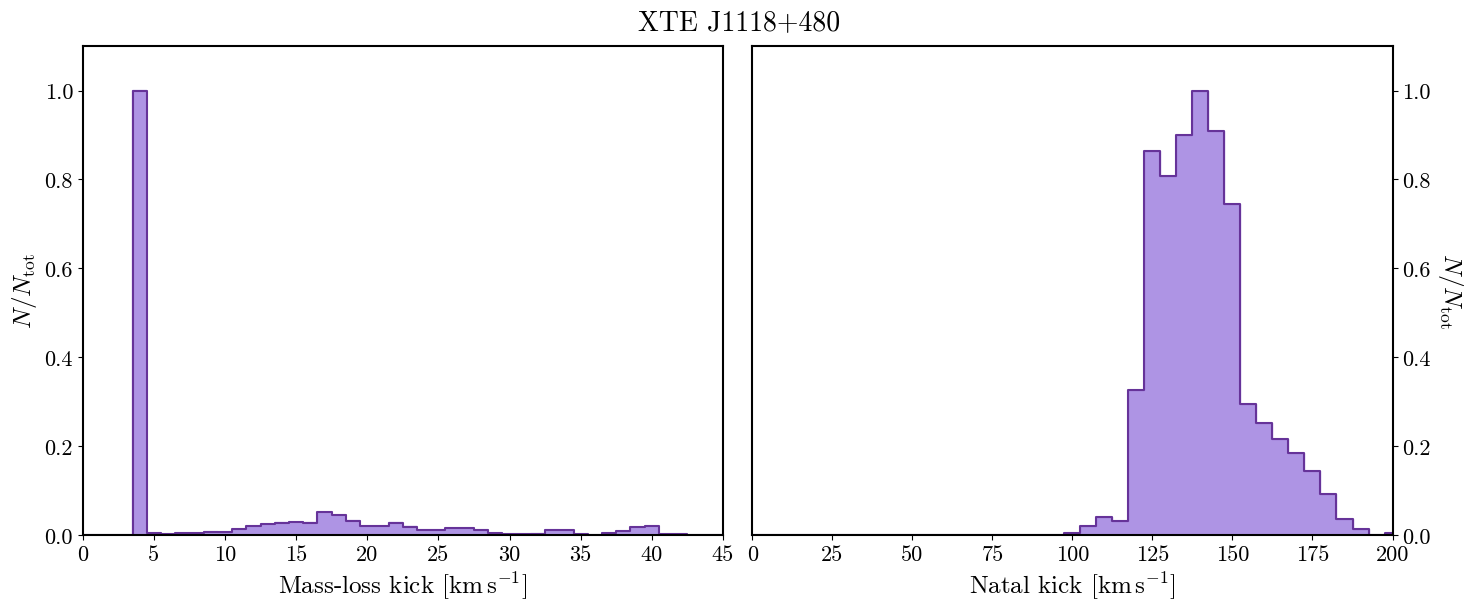}
    \caption{Distribution of mass-loss kicks ($\vmlk$) magnitude (left) and natal kick ($\vnk$) magnitude (right) for simulated system analogues to XTE J1118+480.}
    \label{Fig:J1118_KickHist}
\end{figure*}
\clearpage

\begin{figure*}
    \centering
    \includegraphics[width=0.75\textwidth]{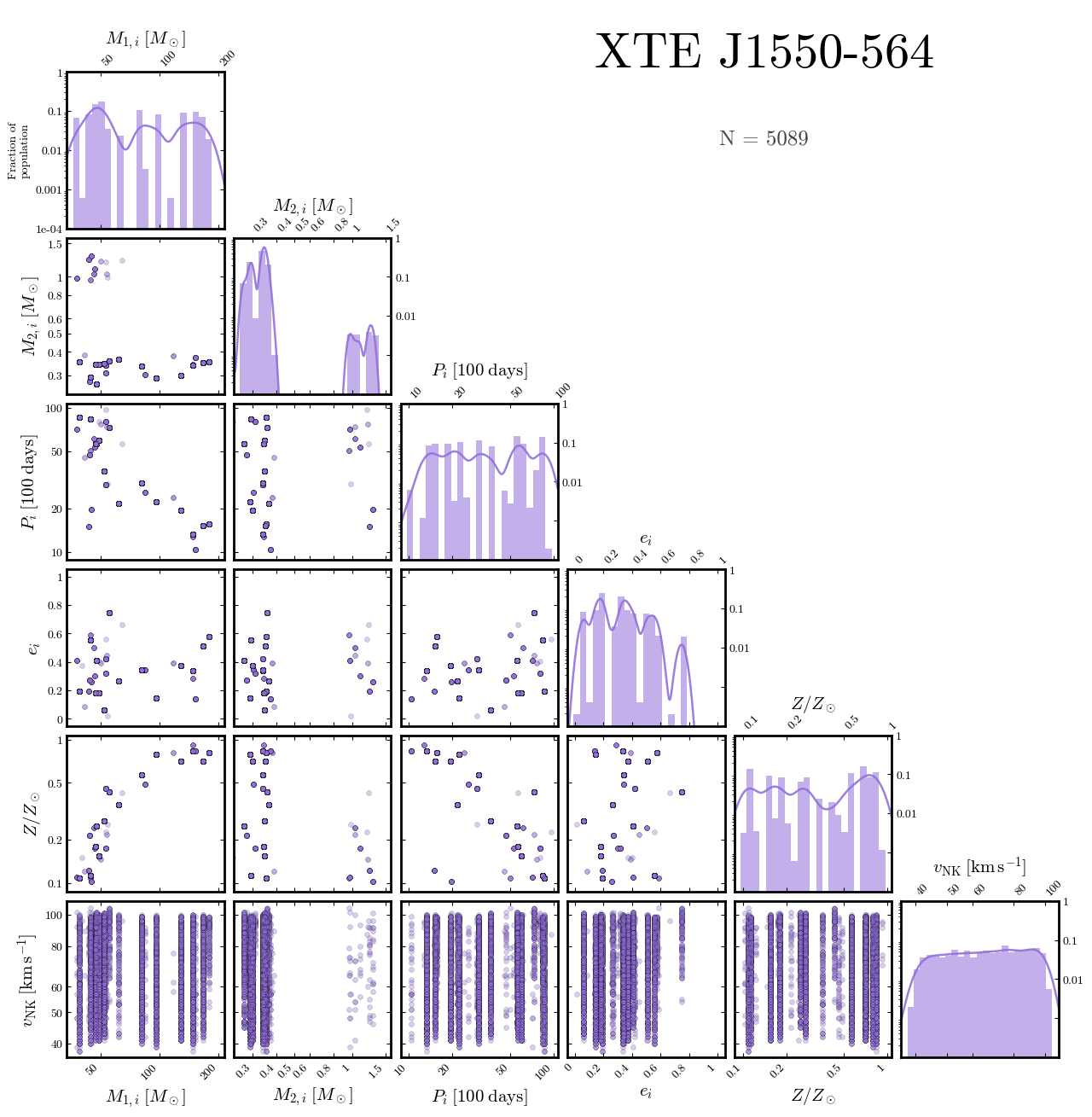}
    \caption{Corner plot highlighting key initial characteristics and acceptable natal kicks for simulated system analogues to XTE J1550-564.}
    \label{Fig:J1550_CornerPlot}
\end{figure*}
\begin{figure*}
    \centering
    \includegraphics[width=0.75\textwidth]{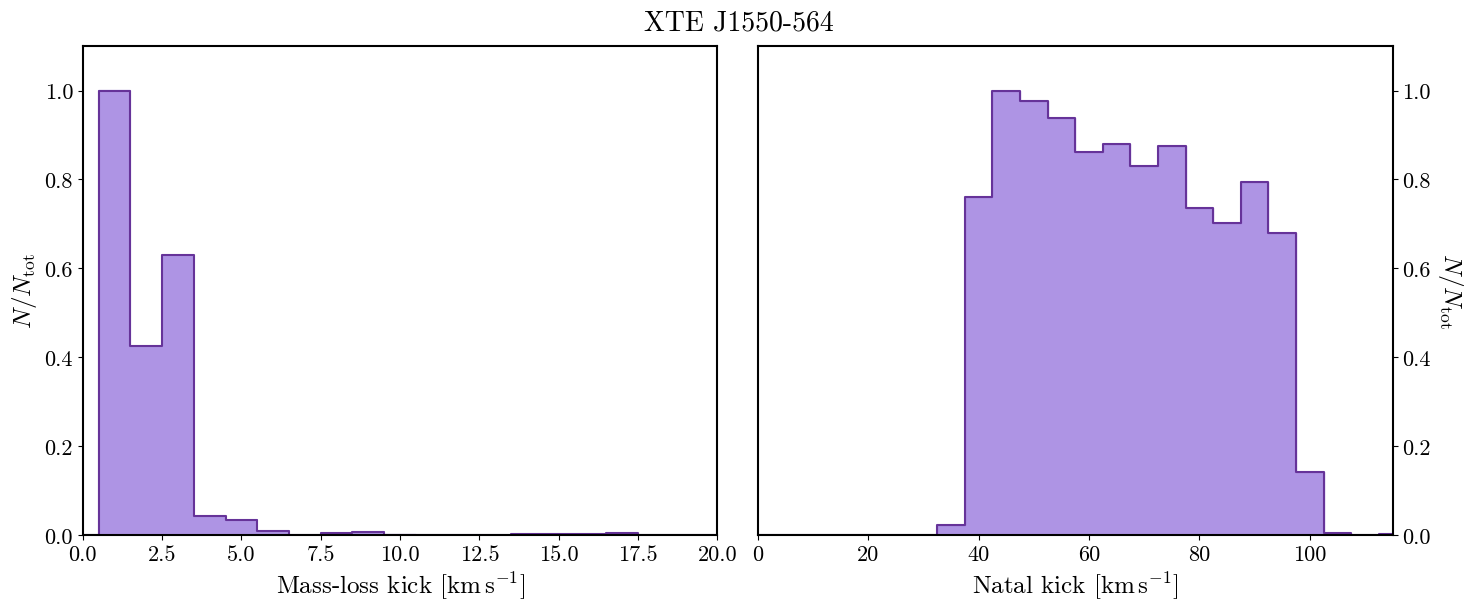}
    \caption{Distribution of mass-loss kicks ($\vmlk$) magnitude (left) and natal kick ($\vnk$) magnitude (right) for simulated system analogues to XTE J1550-564.}
    \label{Fig:J1550_KickHist}
\end{figure*}
\clearpage

\begin{figure*}
    \centering
    \includegraphics[width=0.75\textwidth]{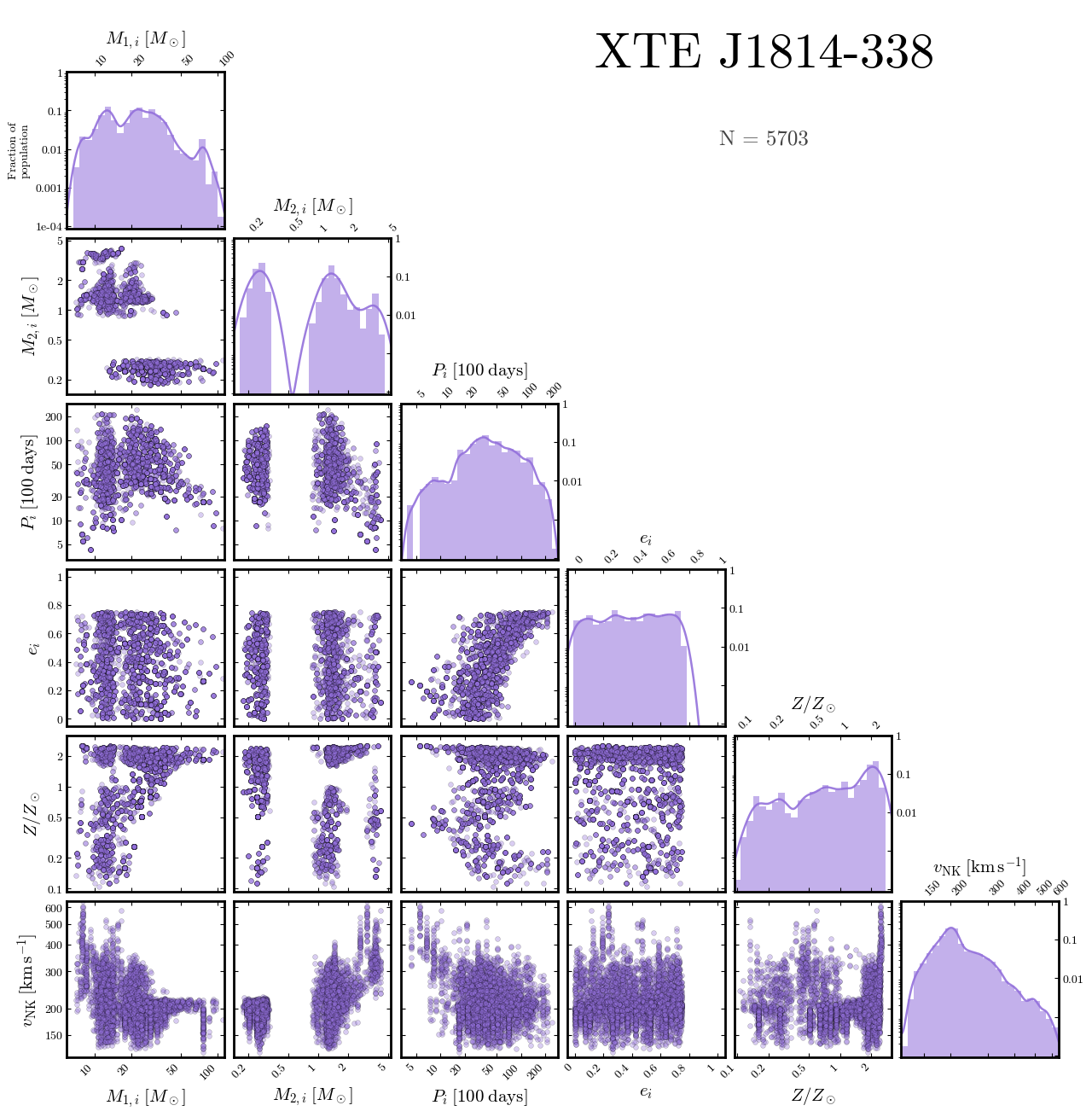}
    \caption{Corner plot highlighting key initial characteristics and acceptable natal kicks for simulated system analogues to XTE J1814-338.}
    \label{Fig:J1814_CornerPlot}
\end{figure*}
\begin{figure*}
    \centering
    \includegraphics[width=0.75\textwidth]{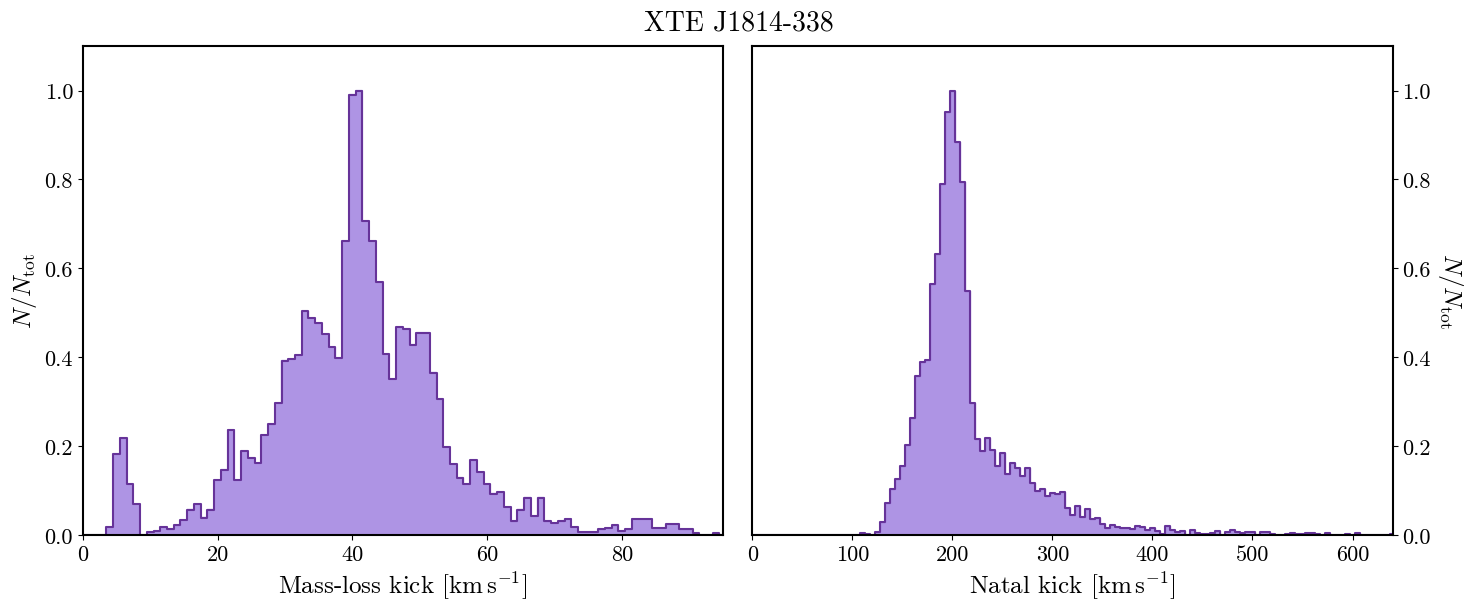}
    \caption{Distribution of mass-loss kicks ($\vmlk$) magnitude (left) and natal kick ($\vnk$) magnitude (right) for simulated system analogues to XTE J1814-338.}
    \label{Fig:J1814_KickHist}
\end{figure*}
\clearpage

\bsp	% typesetting comment

\label{lastpage}

\end{document}